\documentclass{aa} %referee instead of openany for a single column
\usepackage{natbib}
\bibpunct{(}{)}{;}{a}{}{,}

\usepackage[utf8]{inputenc}
\usepackage[T1]{fontenc}
\usepackage[title]{appendix}
\usepackage{subcaption}
\usepackage{txfonts}
\usepackage{graphicx}
\usepackage{color}
\usepackage{multirow}

\usepackage{rotating}
\usepackage{pifont}

\usepackage{float}

\begin{document} 

    \title{Multi-frequency characterisation of remnant radio galaxies \\ in the Lockman Hole field}

    \author{N. Jurlin\inst{1,}\inst{2,}\thanks{jurlin@astro.rug.nl},
          M. Brienza \inst{3,}\inst{4},
          R. Morganti \inst{1,}\inst{2},
          Y. Wadadekar\inst{5},
          C. H. Ishwara-Chandra\inst{5},
          N. Maddox\inst{6}, and
          V. Mahatma\inst{7}}

    \institute{Kapteyn Astronomical Institute, University of Groningen, PO Box 800, 9700 AV, Groningen, The Netherlands
    \and ASTRON, Netherlands Institute for Radio Astronomy, Oude Hoogeveensedijk 4, 7991 PD, Dwingeloo, The Netherlands
    \and Dipartimento di Fisica e Astronomia, Università di Bologna, Via P. Gobetti 93/2, I-40129, Bologna, Italy
    \and INAF - Istituto di Radio Astronomia, Via P. Gobetti 101, I-40129 Bologna, Italy
    \and National Centre for Radio Astrophysics, TIFR, Post Bag 3, Ganeshkhind, Pune 411007, India
    \and Faculty of Physics, Ludwig-Maximilians-Universit\"at, Scheinerstr. 1, 81679 Munich, Germany
    \and Th\"uringer Landessternwarte, Sternwarte 5, 07778 Tautenburg, Germany}

    \date{\today}

    \abstract
  % context heading (optional)
  {Remnant radio galaxies represent an important phase in the life-cycle of radio active galactic nuclei. It is suggested that in this phase, the jets have switched off and the extended emission is fading rapidly. This phase is not well-studied due to the lack of statistical samples observed at both low and high frequencies.}
  % aims heading (mandatory)
  {In this work, we study a sample of 23 candidate remnant radio galaxies previously selected using the Low Frequency Array at 150 MHz in the Lockman Hole field. We examine their morphologies and study their spectral properties to confirm their remnant nature and revise the morphological and spectral criteria used to define the initial sample.}
  % methods heading (mandatory)
  {We present new observations with the Karl G. Jansky Very Large Array at 6000 MHz at both high and low resolution. These observations allowed us to observe the presence or absence of cores and study the spectral curvature and steepness of the spectra of the total emission expected at these high frequencies for the remnant candidates.}
  % results heading (mandatory)
  {We confirm 13 out of 23 candidates as remnant radio sources. This corresponds to 7\% of the full sample of active, restarted, and remnant candidates from the Lockman Hole field. Surprisingly, only a minority of remnants reside in a cluster (23\%). The remnant radio galaxies show a range of properties and morphologies. The majority do not show detection of the core at 6000 MHz and their extended emission often shows ultra-steep spectra (USS). However, there are also remnants with USS total emission and a detection of the core at 6000 MHz, possibly indicating a variety of evolutionary stages in the remnant phase. We confirm the importance of the combination of morphological and spectral criteria and this needs to be taken into consideration when selecting a sample of remnant radio sources.}
  % conclusions heading (optional)
  {}

    \keywords{Surveys - radio continuum : galaxies - galaxies : active}
\authorrunning{Jurlin et al.}
\titlerunning{Characterising remnant radio galaxies}
\maketitle
	
%________________________________________________________________
\section{Introduction}
\label{sec:introduction}
Jetted active galactic nuclei (AGN) are known to go through recurrent phases of activity (see \citealt{2015aska.confE.173Kapinska_review} and \citealt{2017NatAs...1..596Morganti} for extensive reviews). 
They start from a young phase, likely represented by gigahertz peak spectrum sources, compact steep spectrum sources, and high frequency peakers (\citealt{1998PASP..110..493Odea, 2000A&A...363..887DallacasaHFP, 2012ApJ...760...77An_Baan, 2016AN....337....9Orienti, 2020arXiv200902750OdeaSaikia_review}), and move to an evolved active phase. The active phase typically lasts tens of millions of years (e.g. \citealt{2000ApJ...544..671Wan_double_radio_sources, 2015ApJ...806...59TurnerShabala}) and it can be followed by a remnant phase when the jet activity stops or substantially decreases (e.g. \citealt{1994A&A...285...27KomissarovGubanov,2007A&A...470..875Parma2007,2011A&A...526A.148Murgia2011,2015A&A...583A..89Shulevski2015,2017A&A...606A..98B} (hereafter MB17); \citealt{2018MNRAS.475.4557Mahatma_remnants, 2021A&A...648A...9Morganti_resolvedsi}). Current models suggest that, for about 70\% of cases, the duration of the remnant phase is less than 50\% of the previous active phase and that the source total ages are between 5 $\times$ $10^7$ and $10^8$ years (see Fig. 6 in MB17).

As a result of the central engine turning off, studies performed so far on AGN remnants suggest that in these sources, all the typical signatures of activity --- core, jets, and hotspots --- have disappeared or are strongly dimmed. At the same time, the extended lobe radio emission is expected to become diffuse and fainter as a consequence of plasma ageing and expansion. Because of the absence of strong observational signatures, identifying remnant radio galaxies is a complex task, which depends on many parameters. One of the main drivers of the duration of the remnant radio emission visibility is radiative cooling, which is proportional to the plasma magnetic field. Synchrotron cooling and inverse-Compton scattering of cosmic microwave background photons cause a steepening of the spectrum at gigahertz (GHz) frequencies making these sources difficult to detect in many of the available surveys, for example, Very Large Array (VLA) Faint Images of the Radio Sky at Twenty-cm survey (FIRST; \citealt{1995ApJ...450..559BeckerFIRST}), National Radio Astronomy Observatory (NRAO) VLA Sky Survey (NVSS; \citealt{1998AJ....115.1693Condon}), and Karl G. Jansky VLA Sky Survey (VLASS; \citealt{2020PASP..132c5001LacyVLASS}).
This steepening is characterised by a spectral index $\alpha$ larger than $\sim$ 1.2 ($\alpha$ is defined as $S_{\nu} \propto \nu^{-\alpha}$; see \citealt{1970ranp.book.....Pacholczyk} and \citealt{1994A&A...285...27KomissarovGubanov} for more details), as a result of electrons ageing and the cessation of the new particles fuelling the radio lobes. Furthermore, adiabatic expansion of the lobes contributes additional energy losses, decreasing the source luminosity across the whole spectrum. Therefore, steep spectrum sources such as remnants are easier to detect at low radio frequencies ($\nu$ $<$ 1400 MHz) where the remnant radio plasma stays visible for a longer time. However, the combination with high-frequency ($\nu$ $>$ 1400 MHz) data is crucial to confirm their remnant nature by detecting the curvature of the spectra that appears first at shorter wavelengths. Other factors that may influence the spectral energy distribution and the duration of the remnant phase are the plasma composition \citep{2018MNRAS.476.1614Croston}, the mixing between different particle populations \citep{2018MNRAS.473.4179TurnerRAISEII}, and the surrounding environment \citep{2015ApJ...806...59TurnerShabala, 2018MNRAS.480.5286Yates}. However, the latter factors are not measured easily and are, except for the environment, not considered in this paper.\\ 

The observational study of remnant radio galaxies has for long been focused on either single objects, mostly selected by chance (e.g. \citealt{1987MNRAS.227..695Cordey, 1993AJ....105..769Harris, 2004A&A...427...79Jamrozy, 2015MNRAS.447.2468Hurley-Walker, 2015MNRAS.453.2438Tamhane, 2016A&A...585A..29Brienza, 2017A&A...600A..65Shulevski17, 2019PASA...36...16Duchesne, 2020MNRAS.496.3381RandriamanakotoZara}), or very few statistical samples, often targeting remnant candidates selected based on a single specific property (e.g. \citealt{2007A&A...470..875Parma2007, 2011A&A...526A.148Murgia2011,2012ApJS..199...27Saripalli}).
So far, the majority of the studied remnants were selected based on their ultra-steep spectrum (USS; where $\alpha$ $\geq$ 1.2) or spectral curvature (SPC, defined as $\alpha_{high}$-$\alpha_{low}$; \citealt{2007A&A...470..875Parma2007, 2011A&A...526A.148Murgia2011}).

The availability of the LOw Frequency ARray (LOFAR; \citealt{2013A&A...556A...2VanHaarlem}) telescope allows us to expand the study of remnant radio galaxies thanks to its MHz observing frequency and its high sensitivity to low-surface brightness (SB) structures.
So far, two systematic studies of remnants have been carried out with LOFAR: one by MB17 and the other one by \citet{2018MNRAS.475.4557Mahatma_remnants}, who further investigated the sample of remnant candidates selected by \citet{2016MNRAS.462.1910Hardcastle}. 
Both studies highlight the importance of using morphological information for the selection of candidate remnant radio galaxies. \citet{2021A&A...648A...9Morganti_resolvedsi} complemented the work done by MB17 in the Lockman Hole field (LH; \citealt{1986ghg..conf...75Jahoda}), studying the resolved spectral properties of the sources in this field.

Recent modelling of the remnant population, assuming radiative and dynamical evolution models, has further advanced our understanding of the remnant phase. Modelling studies based on the results obtained from LOFAR observations by MB17, \citet{2017MNRAS.471..891Godfrey}, \citet{2018MNRAS.475.2768Hardcastle} and \citet{2020MNRAS.496.1706Shabala} all show that remnants fade rapidly even at MHz frequencies due to the termination of injection of new relativistic electrons combined with radiative and adiabatic losses of the previously existing populations (see e.g. Fig 2 in \citealt{2020MNRAS.496.1706Shabala}). Additionally, the results from MB17 and \citet{2017MNRAS.471..891Godfrey} suggest that most of the observed remnant radio galaxies in the LH region are at the beginning of their remnant phase, while USS radio sources likely represent older remnants (see also \citealt{2018MNRAS.475.4557Mahatma_remnants} and \citealt{2021A&A...648A...9Morganti_resolvedsi}).

The studies so far have given a view of some of the typical properties of remnant radio galaxies, such as their amorphous shape and low SB of the extended emission. However, a variety of properties are observed. For example, \citet{1988A&A...199...73Giovannini} and \citet{2018MNRAS.475.4557Mahatma_remnants} claim that all genuine remnants are expected to lack a visible radio core at all frequencies, while \citet{2010evn..confE..89ParmaVLBI}, \citet{2011A&A...526A.148Murgia2011} and \citet{2012ApJS..199...27Saripalli} consider a weak radio core as a sign that the source is `dying', and therefore, it is not yet completely turned off but considered a remnant. 
\citet{2016A&A...585A..29Brienza} studied one remnant candidate, dubbed blob1, selected based on its unusually large angular extension with amorphous shape and low-SB emission. It has a very weak central compact core, but a lack of other compact features. 
From the variety of observed properties, it becomes clear that statistical samples are needed to solve the disagreement found in the literature on which properties genuinely define remnant radio sources and furthermore, disentangle various evolution stages of the selected remnants.\\

Based on the knowledge of remnant radio galaxies, the aim of this work is to characterise both the presence of cores and the spectral properties of the extended emission of the remnant candidates selected from LOFAR data at 150 MHz by MB17. We also aim to re-evaluate the criteria used to select them. 
To achieve this, we extend the analysis of MB17 using the Karl G. Jansky VLA. We obtained both high- and low-spatial resolution observations at 6000 MHz to identify the presence of weak cores, and measure the spectral index or curvature of the radio spectrum, respectively.
In addition, we also present optical identifications of the host galaxies and study their properties and their environments. The latter analysis is particularly important as it allows us to explore the possibility that the expansion and fading of the remnant emission slow down in a dense environment, making it possible to observe the remnant phase for a longer time (e.g. \citealt{2011A&A...526A.148Murgia2011, 2007A&A...476...99Giacintucci, 2015A&A...583A..89Shulevski2015}).
Finally, based on the refined sample of remnants, we re-estimate their fraction in the total sample of radio galaxies in the field and discuss its implication for the life cycle of radio galaxies. This is possible due to the sample of radio sources in the LH field presented by \citet{2020A&A...638A..34Jurlin}, where radio galaxies with angular sizes $>$ 60$^{\prime\prime}$ are divided into candidate remnants, restarted, and normal active sources. While the criteria to select restarted candidates are presented in that paper, the remnant sample is a sub-sample of the remnants studied in this work and originally selected by MB17.

The paper is structured as follows. In Sect.~\ref{sec:sample} we describe the sample studied in this work and the criteria used to select it. Section~\ref{sec:LOFAR_and_other_radio_data} presents data from complementary radio surveys used in the analysis. Ancillary optical and infrared data and the process of the optical identification are given in Sect.~\ref{sec:optical_identification}. In the same section, we also analyse environments of the remnant candidates. New observations at 6000 MHz and the data reduction process are described in Sect.~\ref{sec:Observations_at_6000_and_data_reduction}. The results obtained from these new observations at 6000 MHz and a summary of radio properties of the sample in the broad frequency range 150 MHz - 6000 MHz are given in Sect.~\ref{sec:results_radio}. In Sect.~\ref{sec:revised sample} we present the criteria used to discriminate between remnant and active radio sources in the sample, and the results of it.
We discuss the occurrence and the impact of remnants on the overall life cycle of radio galaxies based on the new, revised sample in Sect.~\ref{sec:discussion}. Finally, in Sect.~\ref{sec:conclusions} we present the conclusions and future prospects.
The cosmology adopted throughout the paper assumes a flat universe with the following parameters: $\rm H_{0} = 70$ $\rm km$ $\rm s^{-1}$ $\rm Mpc^{-1}$, $\rm \Omega_{\Lambda} =0.7$, $\rm \Omega_{M} =0.3$.

\section{The sample} \label{sec:sample}
MB17 selected a sample of 23 remnant candidates with angular sizes $>$ 60${^{\prime\prime}}$ from the LOFAR image of the LH field at 150 MHz with 18$^{\prime\prime}$ resolution (hereafter `LOFAR18') published by \citet{2016MNRAS.463.2997Mahony} and the catalogue based on the LOFAR image of the same field with a resolution of 45$^{\prime\prime}$. A summary of the selection criteria, both spectral and morphological, used by MB17 is given below. 

In particular, they considered as remnant candidates all sources with at least one of the following properties. Firstly, they used ultra-steep integrated spectrum criterion ($\alpha^{1400}_{150}$ $>$ 1.2; USS), computed using values obtained from the cross-match of LOFAR 45${^{\prime\prime}}$ (MB17) and NVSS catalogues. Secondly, they considered all sources that satisfy the spectral curvature criterion (0.5 $\leq$ $\alpha^{325}_{150}$ $<$ 1 and $\alpha^{1400}_{325}$ $\geq$ 1.5; SPC). Thirdly, they used the core prominence criterion (CP $\lesssim$ $\times$ $10^{-4}$), where CP describes the flux density of a radio galaxy core relative to its total emission. This was computed using LOFAR18 total flux densities with FIRST upper limits for the core emission: CP = S$_{\rm core~(FIRST)}$ / S$_{\rm total~(LOFAR18)}$. Finally, with the morphological criterion, they selected sources showing relaxed morphology with low SB ($\rm SB_{150~MHz}$ $<$ 50 mJy arcmin$\rm^{-2}$), absence of compact features in the LOFAR18 image and absence of compact features above 3$\rm \sigma$ in the FIRST image at 1400 MHz with $5^{\prime\prime}$ resolution.

Using the aforementioned selection, 23 sources were identified as candidate remnant radio galaxies in the LH area covering $\sim$ 30 deg$^{\rm 2}$ (see MB17 for details regarding the selection process and the data used). LOFAR18 images and 3$\sigma_{local}$ contours of the 23 remnant candidates can be seen in the first column of Fig.~\ref{fig:remnants1}.
%-----------------------------------------------------------------

\section{LOFAR and other available radio data}
\label{sec:LOFAR_and_other_radio_data}

Here we expand on the analysis from MB17 by making use of the newly available LH image provided as part of the LOFAR Two metre Sky Survey (LoTSS) deep fields data release and presented by \citet{2021A&A...648A...1Tasse_DEEP_FIELDS}.
This image is deeper because it was obtained with a longer integration time of 96 hours, and has higher spatial resolution of 6$^{\prime\prime}$. Hereafter, we refer to this image as `LOFAR6'. Furthermore, the LOFAR6 image (unlike the LOFAR18 image presented by \citealt{2016MNRAS.463.2997Mahony}) has been obtained with an improved pipeline, which makes use of direction-dependent calibration (see \citealt{2021A&A...648A...1Tasse_DEEP_FIELDS} for a full description of the procedure).
This increased the image dynamic range and reduced the number of artefacts around bright sources, allowing for a detection of low-SB structures. The image reaches a root mean square (rms) noise level of rms $\lesssim$ 23 $\rm \mu$Jy beam$^{-1}$.

Within the LOFAR6 image, we could appreciate many more details of the morphology of our remnant candidates while fully recovering their total flux density. The quality of the image was also crucial for the identification of the optical host galaxy (see Sect.~\ref{sec:optical_identification/approach_and_results}).
The LOFAR6 images of the 23 remnant candidates are shown in the second column of Fig.~\ref{fig:remnants1}. Inspection of the LOFAR6 image allowed us to immediately reject one remnant candidate (J1034+6003) from our sample, which was revealed to be an active wide-angle tail (WAT) radio galaxy. The LOFAR6 image also allowed for a better characterisation of the source J1101+5926. This source was selected as a remnant candidate based on its morphology, assuming that the two radio components, that can be seen in MB17, are lobes of one source. The LOFAR6 image suggests that the two radio components might be unrelated. We comment on this source further in Sect.~\ref{sec:optical_identification/approach_and_results}.

To study the integrated spectrum of the sources we also use the Giant Metrewave Radio Telescope (GMRT) data at 325 MHz (Wadadekar et al., in prep). These data were taken over 11 nights between February 15 and March 20, 2013, with 23 different pointing centres in snapshot mode. The field centre of the entire mosaic of 23 pointings is at R.A. 11h 03m, Dec. $\rm +59^\circ$ $\rm 44^\prime$, and cover approximately $6 \times 6$ deg$^2$ of the LH region. Eighteen out of 22 remnant candidates are located in the area observed with GMRT.
The rms noise, measured far from any bright sources, has a nearly uniform value of $\sim$ 50 $\mu$Jy beam$^{-1}$ over most of the mosaic image. The resolution of the image is 10$^{\prime\prime}$. For a detailed description of the observations and the data reduction procedure, see Wadadekar et al (in prep). 

We also use the images from the NVSS survey at 1400 MHz. The resolution of the images are 45$^{\prime\prime}$ and the value of rms noise is 0.45 mJy beam$^{-1}$. Furthermore, we use the images from the FIRST survey at 1400 MHz with an angular resolution of 5$^{\prime\prime}$ and a typical rms noise of 0.15 mJy\textbf{,} and a deep mosaic at 1400 MHz observed with the Westerbork Synthesis Radio Telescope (WSRT) covering an area of 6.6 $\rm deg^{2}$ \citep{2018MNRAS.481.4548Prandoni}, with a resolution of 12$^{\prime\prime}$ reaching an rms noise of 11 $\rm \mu$Jy beam$^{-1}$. The WSRT image only covers three of our 22 candidate remnant radio sources. Also included in the analysis is the VLASS, at 3000 MHz with a resolution of 2.5$^{\prime\prime}$ and reaching rms noise of 120 $\rm \mu$Jy. Our remnant candidates were observed with the first campaign of the first epoch of the VLASS (VLASS1.1). For objects in VLASS1.1 with flux densities below $\sim$ 1 Jy, the peak flux densities are systematically underestimated by $\sim$ 15\% as described in their documentation\footnote{https://science.nrao.edu/science/surveys/vlass/vlass-epoch-1-quick-look-users-guide}. Therefore, we correct our VLASS core flux density measurements by 15\%.
For the LOFAR (both LOFAR6 and LOFAR18), GMRT, WSRT and NVSS measurements, we assume 11\% \citep{2019A&A...622A...1Shimwell}, 8\% \citep{2012MNRAS.419.1136Basu}, 10\% \citep{2014MNRAS.445.1213Stroe} and 5\%, as flux density scale errors, respectively.

\section{Optical identification} \label{sec:optical_identification}
Because of the amorphous morphology and the lack of strong cores, the identification of the optical host galaxy of remnant radio sources is a difficult task. For example, this is illustrated by the work of \citet{2018MNRAS.475.4557Mahatma_remnants}. The latter authors report that 10\% of their 40 remnant candidates were misidentified in the H-ATLAS catalogue \citep{2016MNRAS.462.1910Hardcastle} prior to detecting the radio cores at GHz frequency. Moreover, after the jet switches off, the AGN remnant plasma can move from the original host position influenced by the surrounding environment making the identification even more challenging. One example where this occurrence is clearly observed is the source 3C338 \citep{2007ApJ...659..225Gentile}. Another example was recently reported by \citet{2019PASA...36...16Duchesne}. 
In this section, we describe the optical and infrared data used and the approach we adopted to identify the host galaxies of the remnant candidates in our sample.

\subsection{Optical and infrared data}
\label{sec:LH_data/optical_and_infrared}
We performed the optical identifications of the candidate remnant radio galaxies using images from the Sloan Digital Sky Survey Data Release 16 (SDSS DR16, \citealt{2017AJ....154...28Blanton_SDSS)}), the LoTSS Deep Fields catalogue by \citealt{2021A&A...648A...3Kondapally} listing the optical identifications of the radio sources in the LOFAR6 image of the LH, and the catalogue by \citealt{2021A&A...648A...4Duncan}, which expands the \citet{2021A&A...648A...3Kondapally} catalogue by including multi-wavelength data and galaxy redshifts. The reported redshifts in these catalogues are from spectroscopy when available and photometric when spectra are not available.

In the catalogue presented by \citet{2021A&A...648A...3Kondapally}, the counterparts were identified using a combination of the Likelihood Ratio method and visual classification.
Photometric redshifts were estimated as described in \citet{2021A&A...648A...4Duncan} using deeper photometry than available in the SDSS and the Wide-field Infrared Survey Explorer (WISE; \citealt{2010AJ....140.1868Wright,2011ApJ...731...53Mainzer}).

Since the aforementioned catalogues cover only the central $\sim$ 10.73 $\rm deg^2$ of the entire LOFAR6 image, we searched for the optical identifications of all the candidate remnant radio galaxies by visually inspecting the SDSS DR16 images. The LoTSS Deep field catalogues were then used to confirm the associations and redshift information. We describe the procedure in the following section.

To estimate the stellar masses of the remnant candidates (see Sect.~\ref{sec:optical_identification/approach_and_results}), we used the WISE catalogue. WISE mapped the sky at 3.4, 4.6, 12, and 22 $\rm \mu$m (W1, W2, W3, W4) with an angular resolution of 6.1$^{\prime\prime}$, 6.4$^{\prime\prime}$, 6.5$^{\prime\prime}$, and 12.0$^{\prime\prime}$, in the four bands respectively. In particular, we used the AllWISE Source Catalogue (\citealt{2014yCat.2328....0Cutri14allWISE}). The latter catalogue provides enhanced photometric sensitivity and accuracy, and improved astrometric precision compared to the WISE All-Sky Release Catalogue (\citealt{2012yCat.2311....0CutriallSky}).

\subsection{Approach and results} \label{sec:optical_identification/approach_and_results}

\begin{figure*}
    \centering
    \includegraphics{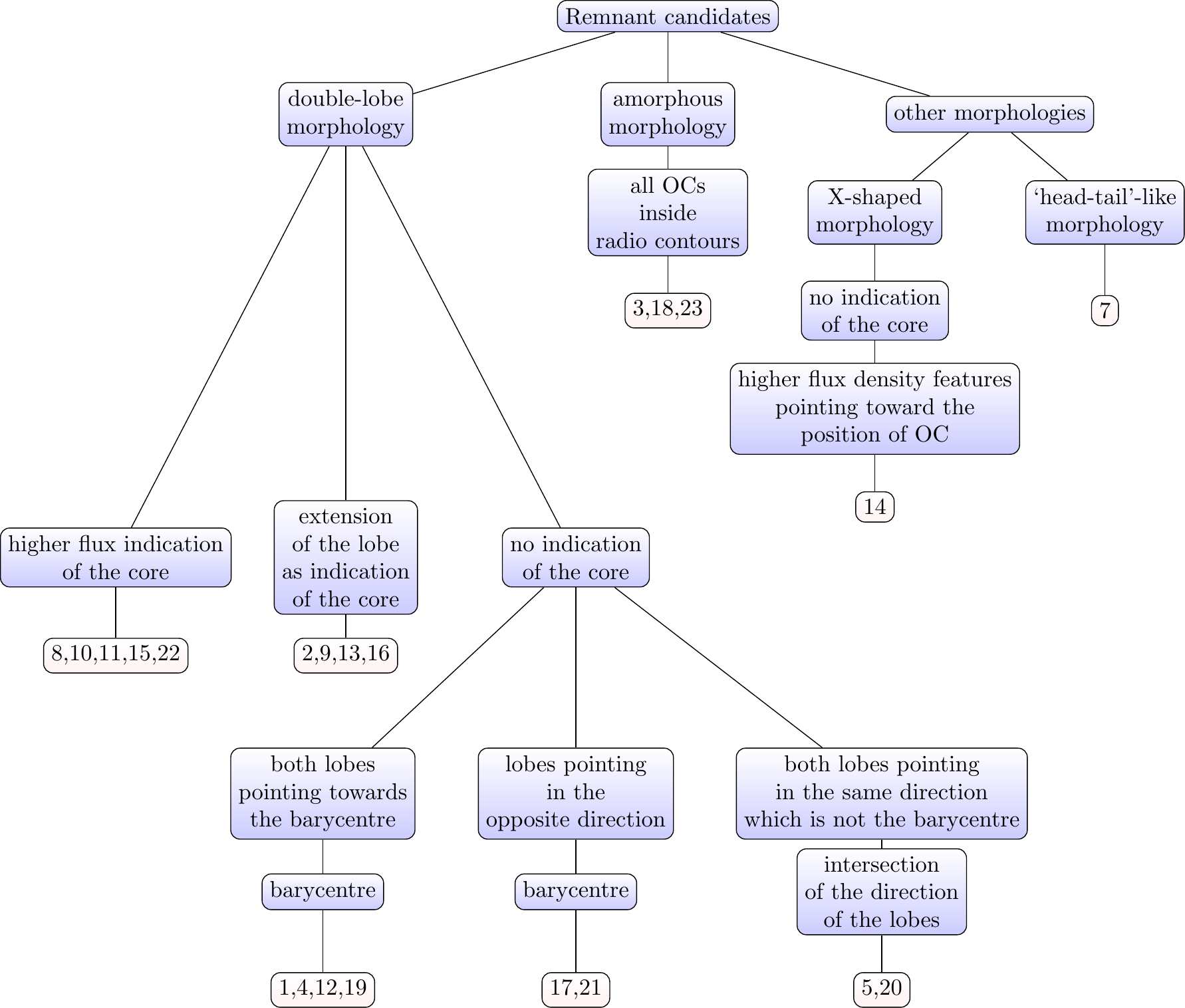}
    \caption{Flow chart showing the process used to identify the optical host galaxy of the sources in the sample of 22 candidate remnant radio galaxies. Numbers represent the source ID as listed in Table~\ref{tab:optical_table}. This flow chart presents the identification criteria used for a range of morphologies of remnant radio galaxies, so as to make the OC identification reproducible in future studies.}
    \label{fig:flow_chart_OC}
\end{figure*}

The first step in the identification was done using the morphology of the source in the LOFAR6 image. 
We divided our sources into three groups, (i) double or reminiscing of double-lobed morphology; (ii) amorphous morphology, and (iii) all the other types of morphologies. More details about the steps taken are described in the flow chart in Fig.~\ref{fig:flow_chart_OC}. In this section, we give a brief overview of the procedure and comment on the specific cases for which the optical identification process using SDSS DR16 images was challenging. It is important to point out that all the identifications done with the procedure described in this section were later confirmed with detections of radio cores in the 6000 MHz observations. We discuss this in Sect.~\ref{sec:results_radio}.

For the sources with a double-lobed morphology which show hints of radio emission in the region between the two lobes, we looked for a possible optical counterpart (OC) in this central region. For the sources with detached radio lobes, we searched for the OC either at the barycentre of the overall emission or at the intersection of the two lobe axes. The latter approach was used in case both lobes point towards the same direction, which is off-set from the barycentre.

In this process, we realised that the source J1101+5926, mentioned in Sect.~\ref{sec:LOFAR_and_other_radio_data}, has a candidate optical host galaxy in both radio `lobes'. 
This suggested that the two radio components are not two lobes of a single source but rather two unrelated sources. We, therefore, decided to reject the southern radio component from the sample and to retain the northern source as a candidate remnant due to its remnant-like features (low-SB, no core detection at 1400 MHz, and relaxed morphology). In Fig.~\ref{fig:remnants1} we show only the northern source that remained in the sample of remnant candidates. A similar case is the source J1029+5842, where we identified an OC in what was previously considered the southern `lobe'. We discuss the source J1029+5842 further in Sect.~\ref{results_radio/radio_cores}. 

Other double sources in which the OC identification was not trivial are noted here. Source J1046+5647 has a bright star at its centre, obscuring our view towards its optical host galaxy. Source J1101+5603 has an asymmetric structure in the inner region of the lobes and, therefore, it is challenging to predict a position of the radio core and host galaxy. Only one OC was detected between the two lobes and was considered as the host galaxy. Finally, source J1102+5857 has two possible optical identifications, and we chose the most central one as the most probable host.

A similar approach was also used for the sources with (ii) amorphous morphology, and (iii) other morphologies (including the X-shaped source J1052+5636 and source J1036+5540).
We considered all the optical galaxies within the 3$\sigma_{\rm local}$ radio contours to be possible host galaxies. When possible, we followed the higher flux density features as an indication of the position of the radio core. In the case of the source J1029+5857, there are two optical galaxies, very close to each other, as possible hosts. J1108+5831 has three possible OCs. For both aforementioned sources, we chose the optical galaxy located between the two higher flux density regions of the fossil lobes.

\bigskip

%\subsubsection{Results of the optical identification} \label{sect:results_optical}
In conclusion, we have identified an SDSS OC for 18 out of 22 remnant candidates. Out of these 18 candidates, 15 have a single high probability OC, two have two possible OCs, and one source has three possible OCs. We show the optical $r$-band SDSS DR12 maps in the third column of Fig.~\ref{fig:remnants1}. For 13 of the most secure OCs, we have SDSS photometric redshifts\footnote{http://skyserver.sdss.org/dr16/}, and the remaining five objects have SDSS spectra available. A more detailed study of the optical spectra of the remnant candidates from MB17 is now in progress using dedicated observations (Jurlin et al., in prep). Redshifts of the candidate remnant radio galaxies are listed in Table~\ref{tab:optical_table}.

Out of 22 candidates in the sample, four are also part of the LoTSS Deep fields catalogue by \citealt{2021A&A...648A...3Kondapally}, which consists of radio detections and their assigned OC: J1042+5920, J1045+5631, J1052+5636 and J1057+5847. For all four sources, the coordinates of the OC are consistent with our SDSS match, though the value of the redshift is different for three of the four sources. In particular, one of these four identifications is located in a region where the photometry is reported to be less reliable (J1052+5636). We decided to use the redshift from the LoTSS Deep fields catalogue only for the source J1045+5631 due to its unreliable SDSS photometry.

The LoTSS deep field catalogue of optical host galaxies by \citet{2021A&A...648A...4Duncan} allowed us to get an OC with related redshift estimate also for the three out of four sources in our sample not detected in neither SDSS nor LoTSS Deep fields catalogue with radio detections and their assigned OC \citep{2021A&A...648A...3Kondapally}. These redshift estimates are indicated with `$lp$' in Table~\ref{tab:optical_table} due to the limited photometric data of these galaxies. Therefore, radio and optical properties derived using these redshift values should be considered with care.

\begin{figure}
\center
    \includegraphics[width=8cm]{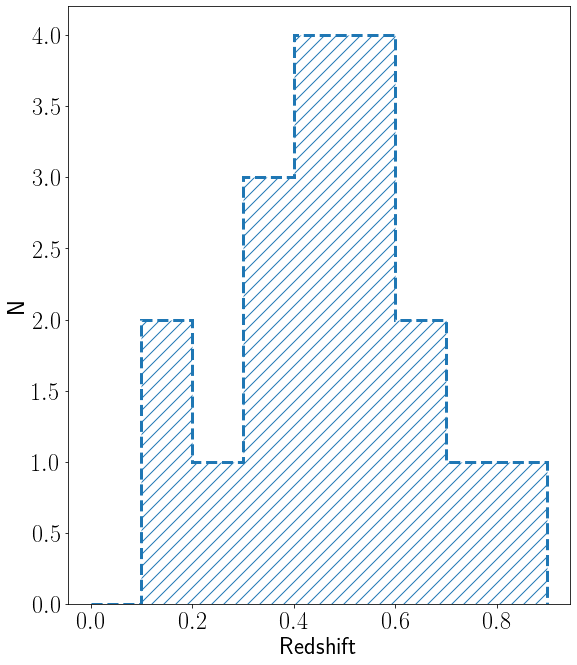}
\caption{Redshift distribution of the 18 remnant candidates with reliable redshifts (see Table~\ref{tab:optical_table} and Sect.~\ref{sec:optical_identification/approach_and_results}).}
\label{fig:redshift}
\end{figure}

In conclusion, only source J1029+5842 has no detected OC in the position expected from the described procedure, in neither SDSS nor the LoTSS Deep fields catalogues (see Sect.~\ref{results_radio/radio_cores} for a further discussion of this candidate). Reliable redshifts are in the range [0.1092, 0.83053] and we show the distribution in Fig.~\ref{fig:redshift}.

The stellar masses of the SDSS counterparts were presented in \citet{2020A&A...638A..34Jurlin}. Here we only re-calculated the stellar mass of the source J1045+5631, due to the more reliable redshift value found in the LH Deep field catalogue. For the approach used to estimate stellar masses, we refer to \citet{2020A&A...638A..34Jurlin}.
The derived stellar masses of remnant candidates span in the range [0.07, 40] $\times$ $10^{11}$ (see Table~\ref{tab:optical_table}) and we discuss them in Sect.~\ref{sec:discussion}.

\subsection{Environment} \label{sec:optical_identification/environment}
The environment is an important parameter to consider when studying remnant radio galaxies, as it can influence the way the plasma evolves and distributes in the surrounding medium. In particular, a dense intra-cluster medium is thought to be able to confine the radio plasma making it visible for a longer time.
In order to understand whether the remnant radio galaxies studied in this paper are associated to a cluster of galaxies, we adopted the approach presented in \citet{2019A&A...622A..10Croston_environment} and made use of two SDSS cluster catalogues with well-calibrated richness estimators – the DR8 RedMaPPer catalogue \citep{2014ApJ...785..104Rykoff} and the DR8 photo-z cluster catalogue of \citet{2012ApJS..199...34WenHanLiu}. The \citet{2014ApJ...785..104Rykoff} RedMaPPer catalogue contains $\sim$ 86 clusters over the area covered by our remnant candidates in the LH region, selected using a red-sequence finding method optimised to minimise scatter on the mass-richness relation. The \citet{2012ApJS..199...34WenHanLiu} cluster catalogue extends to a somewhat lower richness, with $\sim$ 338 clusters over the area covered by our remnant candidates, and is based on an iterative method incorporating photometric redshift selection and a friends-of-friends method. Below, we report the names and redshifts as listed in the \citet{2012ApJS..199...34WenHanLiu} cluster catalogue, and with `$+RedMaPPer$', we indicate that the cluster is also present in the \citet{2014ApJ...785..104Rykoff} catalogue and we list the corresponding redshift.

We cross-matched the catalogues mentioned above with our sample of candidate remnant radio galaxies using a search radius corresponding to 1500 kpc at the redshift of each source. We did not put a constraint in redshift in the initial cross-match. Therefore, we report here all the matches based only on the coordinates of the sources and clusters, and we list and comment on the corresponding redshifts.

The three sources, J1054+5505, J1102+5857 and J1108+5831, belong to the clusters J105401.2+550611 (z = 0.4), J110252.6+585731$^{+RedMaPPer}$ (z = 0.34; z$^{RedMaPPer}$ = 0.36 $\pm$ 0.02) and J110804.3+583141 (z = 0.15), as they are located at the same redshift and have a projected distance from the cluster centre of 2, 124 and 1 kpc, respectively. 
In addition, there are two other sources, J1045+5631 and J1052+5636, which are matched with the clusters J104519.1+564242 and J105249.7+563154$^{+RedMaPPer}$ with a separation of 442 and 1138 kpc, respectively. The redshifts of these two clusters are z = 0.2 and z = 0.38 (z$^{RedMaPPer}$ = 0.39 $\pm$ 0.02). The SDSS photometric redshifts of the two corresponding remnant candidates are significantly different, being equal to 0.034 $\pm$ 0.100 and 0.25 $\pm$ 0.03. The redshift listed in the LoTSS Deep field catalogue, for the source J1045+5631 is 0.11. Due to the large separation from the associated galaxy cluster centre, we do not consider these two sources to be a part of the cluster.

With the approach described above, we conclude that only three out of 22 candidate remnant radio galaxies reside in a cluster (see Table~\ref{tab:optical_table}). There is only one Abell cluster (Abell 1132; \citealt{2018MNRAS.473.3536Wilber}) in the LH, and none of the remnant candidates is a part of it. We discuss the results on the environment of remnant radio galaxies in Sect.~\ref{sec:discussion}.\\

\begin{table}
\centering
\caption{Candidate remnant radio sources.}
\resizebox{0.5\textwidth}{!}{\begin{tabular}{c c c c c}\hline\hline
ID  & Source name       & Redshift                                                  & $\rm M_{*}$                                       & Cluster   \\
    &                   &                                                           & [$\rm \times 10^{11}$ $\rm M_{\sun}$]             & flag      \\ \hline \hline
1   & J102818+560811    & 0.52 $\pm$ 0.05$\rm ^{p}$                                 & 2 $\pm$ 1                                         & 0         \\ \vspace{0.5mm}
2   & J102842+575122    & 0.4 $\pm$ 0.1$\rm ^{p}$                                   & 0.1 $\pm$ 0.1                                     & 0         \\ \vspace{0.5mm}
3   & J102905+585721    & 0.53 $\pm$ 0.03$\rm ^{p}$                                 & 8 $\pm$ 3                                         & 0         \\ \vspace{0.5mm}
4   & J102917+584208    & -                                                         & -                                                 & 0         \\ \vspace{0.5mm}
5   & J103132+591549    & 1.8686$\rm_{~0}^{4.69}$ $\rm ^{lp}$                       & -                                                 & 0         \\ \vspace{0.5mm}
7   & J103602+554007    & 0.68 $\pm$ 0.07$\rm ^{p}$                                 & 6 $\pm$ 4                                         & 0         \\ \vspace{0.5mm}
8   & J103641+593702    & 0.47 $\pm$ 0.03$\rm ^{p}$                                 & -                                                 & 0         \\ \vspace{0.5mm}
9   & J103805+601150    & 0.73 $\pm$ 0.05$\rm ^{p}$                                 & 27 $\pm$ 15                                       & 0         \\ \vspace{0.5mm}
10  & J104208+592030    & 0.5099 $\pm$ 0.0001$\rm ^{s}$                             & 12 $\pm$ 3                                        & 0         \\ \vspace{0.5mm}
11  & J104516+563148    & 0.1092$\rm_{~0}^{~0.18}$ $\rm ^{LD}$                      & 0.57$\rm_{~0}^{~1.69}$                            & 0         \\ \vspace{0.5mm}
12  & J104646+564744    & 0.3032$\rm_{~0.25}^{~0.36}$ $\rm ^{lp}$                   & -                                                 & 0         \\ \vspace{0.5mm}
13  & J104732+555007    & 0.57 $\pm$ 0.04$\rm ^{p}$                                 & 3 $\pm$ 2                                         & 0         \\ \vspace{0.5mm}
14  & J105230+563602    & 0.25 $\pm$ 0.03$\rm ^{p}$                                 & 0.9 $\pm$ 0.3                                     & 0         \\ \vspace{0.5mm}
15  & J105402+550554    & 0.38249 $\pm$ 0.00006$\rm ^{s}$                           & -                                                 & 1         \\ \vspace{0.5mm}
16  & J105554+563532    & 2.1752$\rm_{~0}^{~5.07}$ $\rm ^{lp}$                      & -                                                 & 0         \\ \vspace{0.5mm}
17  & J105703+584721    & 0.83053 $\pm$ 0.00005$\rm ^{s}$                           & 4 $\pm$ 2                                         & 0         \\ \vspace{0.5mm}
18  & J105729+591128    & 0.47 $\pm$ 0.02$\rm ^{p}$                                 & 6 $\pm$ 3                                         & 0         \\ \vspace{0.5mm}
19  & J110108+560330    & 0.4 $\pm$ 0.1$\rm ^{p}$                                   & -                                                 & 0         \\ \vspace{0.5mm}
20  & J110136+592602    & 0.49 $\pm$ 0.05$\rm ^{p}$                                 & 9 $\pm$ 11                                        & 0         \\ \vspace{0.5mm}
21  & J110255+585740    & 0.3391 $\pm$ 0.0001$\rm ^{s}$                             & 2.1 $\pm$ 0.8                                     & 1         \\ \vspace{0.5mm}
22  & J110420+585409    & 0.66 $\pm$ 0.05$\rm ^{p}$                                 & 40 $\pm$ 50                                       & 0         \\ \vspace{0.5mm}
23  & J110806+583144    & 0.15431 $\pm$ 0.00003$\rm ^{s}$                           & 0.07 $\pm$ 0.06                                   & 1         \\ \hline \hline
    \end{tabular}}
    \tablefoot{
    \small
    ID listed in Col. 1 is used in the flow chart showing the process of the optical identification. In Col. 2 are the source names in J2000 coordinates; Col. 3 notes the redshift of the optical host galaxy, where `p' indicates SDSS photometric and `s' SDSS spectroscopic redshift, `LD' and `$\rm{lp}$' come from the LoTSS Deep Fields catalogues, where `lp' indicates that the redshift is obtained from the limited photometry - indicating less reliable redshifts; Col. 4 shows stellar masses; Col. 5 lists environment flags where `1' means that the source is in a cluster.\\
    In the LoTSS Deep Fields catalogues, errors on the redshift are not symmetrical and therefore we quote minimum and maximum redshift values from the respective catalogues. The maximum and minimum value of the $\rm M_{*}$ for the source J1045+5631 is calculated using the maximum and minimum redshift for this source.\\ 
    Sources without the optical and/or infrared identification do not have a measurement of the stellar mass (see \citealt{2020A&A...638A..34Jurlin}).}
    \label{tab:optical_table}
\end{table}

%__________________________________________________________________

\section{Observations at 6000 MHz and data reduction}
\label{sec:Observations_at_6000_and_data_reduction}
To expand the investigation of the sample to higher frequencies and confirm the remnant nature of the candidates, we conducted follow-up observations at 6000 MHz with the VLA.
We observed all 22 candidate remnant radio galaxies in our sample using the telescope in A-configuration to investigate the presence of a radio core.
We could observe only a subset of the ten sources in D configuration, due to limits in the allocated time. The D configuration was chosen to detect extended emission that allowed us to measure the total flux densities of remnant candidates. We chose to observe the ten sources that showed relaxed morphologies with low SB and which do not display an USS below 1400 MHz.
In the following sections, we describe the observations and data reduction procedures that we used to produce the final images at 6000 MHz.

\subsection{VLA 6000 MHz data: A-configuration} \label{sec:LH_data/VLA_A_observations}
The 22 targets were observed with the broad-band C-band receiver in 3-bit mode using the telescope in A-configuration on May 14, 2018.
The field of view of the observations was big enough to ensure the investigation of any other compact component within the radio source extent ($\sim$ 8$^{\prime}$).
The sampling time was set to 2 seconds, and four polarisation products were recorded (RR, LL, RL and LR). The total bandwidth, equal to 4000 MHz in the range 3976-8024 MHz, was divided by default into 32 sub-bands of 128 MHz with 64 frequency channels each. All targets were observed for 9 minutes interleaved by 1.5-minute observations of a phase calibrator (J1035+5628). We used the quasar 3C286 as a primary calibrator. It was observed at the beginning and the end of the observing block for a total amount of 32 minutes. The observational setup is summarised in Table~\ref{tab:vla_data}. 

The data were calibrated by the observatory using the CASA VLA pipeline (version 41154 Pipeline-CASA51-P2-B) based on CASA 5.1.2-4 r40000. A sample of the inspection plots provided by the observatory was inspected to verify the quality of the calibration. As we did not identify any major problem in the calibration performed by the pipeline, we proceeded with the imaging using CASA (version 5.3.0) and the CLEANing algorithm \citep{1974A&AS...15..417Hogbom}. We produced full-band images centred at 6000 MHz and 1000 MHz-band images centred at 4500, 5500, 6500 and 7500 MHz to investigate the in-band spectral variations.
We produced the final images using Briggs weighting with robust=0 \citep{1995AAS...18711202Briggs}. The final images have a resolution of about 0.3$^{\prime\prime}$ $\times$ 0.2$^{\prime\prime}$. The noise in the full band images is in the range 9-10 $\rm \mu$Jy beam$^{-1}$ in agreement with the expected theoretical thermal noise (8 $\rm \mu$Jy beam$^{-1}$) computed using the VLA Exposure Calculator\footnote{https://obs.vla.nrao.edu/ect/} and an order of magnitude improvement in sensitivity compared to FIRST. The noise in the 1000 MHz bandwidth images is between 10 and 20 $\rm \mu$Jy beam$^{-1}$.

We show the final full-band images in the fourth column of Fig.~\ref{fig:remnants1}. Image parameters are listed in Table~\ref{tab:VLA_data_reduction_table}.    

\subsection{VLA 6000 MHz data: D-configuration}
\label{sec:LH_data/VLA_D_observations}
On September 21 and 27, 2018, we observed ten remnant candidates using the telescope in the D-configuration.
Observations were made with the broad-band C-band system in 3-bit mode (4000 MHz bandwidth from 3976-8024 MHz). Exposure times were set to 20 minutes resulting in a theoretical noise of 6 $\rm \mu$Jy beam$^{-1}$. As for the A-configuration, the observations were targeted at the candidate host galaxy for each source. We used the quasar 3C286 as the primary flux calibrator and observed it at the beginning and the end of the scheduling block for a total time of 24.5 minutes. The source J1035+5628 was used as a phase calibrator, observed for 1.5 minutes every 10 minutes. These details are summarised in Table~\ref{tab:vla_data}. We used the data flagged by the observatory pipeline as a starting point and performed further manual flagging and calibration, following the standard approach. Initial images were then produced by CLEANing dirty images centred on each target. Phase and amplitude self-calibration was performed for all the observed sources. We produced the final full-band images using natural weighting and have a final resolution of about 15$^{{\prime}{\prime}}$ $\times$ 12$^{{\prime}{\prime}}$ and an rms noise in the range 8-18 $\rm \mu$Jy beam$^{-1}$. 

The largest angular size of the sources in our sample is 3.5$^{\prime}$, meaning that all of the sources in our sample are smaller than 4$^{\prime}$, which is the largest observable scale with the VLA in D-array in C-band. Therefore, we expect to recover the full flux density of any given source at this frequency.

As for the A-array, in addition to the full-band images, we also produced 1000 MHz-band images to investigate the in-band spectral variations. Their final resolution is 19$^{{\prime}{\prime}}$ $\times$ 16$^{{\prime}{\prime}}$ and the noise is in the range 30-40 $\rm \mu$Jy beam$^{-1}$.

For the flux density scale error at 6000 MHz, we assume a conservative value of 5\%\ (\citealt{2017ApJS..230....7Perley}). The final full-band images are shown in the fifth column of Fig.~\ref{fig:remnants1}. Image parameters are listed in Table~\ref{tab:VLA_data_reduction_table}.

\begin{table}
\caption{Set-up for the VLA observations centred at 6000 MHz.}
\label{tab:vla_data}
\centering                                      
\begin{tabular}{l l}          
\hline\hline
    VLA project code & 18A-274 \\
    Primary calibrator & 3C~286 \\ 
    Phase calibrator & J1035+5628 \\ 
    Bandwidth [MHz] & 4000-8000 \\
    Field of view [$^{\prime}$] & 8  \\\hline \hline
    VLA A-array & \\ \hline
    Date of the observations & 14-May-2018\\ 
    Observing time per source & 9 min \\ 
    Number of sources observed & 22\\ \hline \hline
    VLA D-array &  \\ \hline 
    \multirow{2}{*}{Date of the observations} & 21-Sep-2018 \\& 27-Sep-2018 \\
    Observing time per source & 20 min \\
    Number of sources observed & 10 \\ \hline \hline
\end{tabular}
\end{table}

%\FloatBarrier
\begin{table*}[h!]
\caption{VLA A and D array properties of the full-band images centred at 6000 MHz.}
\label{tab:VLA_data_reduction_table}
\centering
\resizebox{0.9\textwidth}{!}{\begin{tabular} {l c c c c c c}
\hline\hline
 Source name         & \multicolumn{2}{c}{beam A array}    & rms noise A array & \multicolumn{2}{c}{beam D array}    & rms noise D array  \\
                    & [arcsec $\times$ arcsec] & [deg] & [$\rm \mu$Jy beam$^{-1}$] & [arcsec $\times$ arcsec] & [deg] & [$\rm \mu$Jy beam$^{-1}$]  \\ \hline
J102818+560811    & $0.32 \times 0.22$ & $-14.6$    & 10 &  $14.52 \times 12.70$   & $-33.4$      & 11\\
J102842+575122    & $0.28 \times 0.23$ & $-5.2 $    & 10 &  $9.80 \times 8.08 $    & $-16.1$      & 9.2\\
J102905+585721    & $0.33 \times 0.23$ & $-10.9$    & 10 &  $13.58 \times 11.62$   & $-48.7$      & 8.1\\ 
J102917+584208    & $0,32 \times 0.24$ & $-56.8$    & 10 &  -   & -      & -\\
J103132+591549    & $0.33 \times 0.24$ & $-58.9$    & 10 &  -   & -      & -\\
%6   &J103414+600333    & -                  & -          & -    & -    & -      & -\\
J103602+554007    & $0.32 \times 0.23$ & $-63.0$    & 10 &  -   & -      & -\\
J103641+593702    & $0.29 \times 0.23$ & $-16.7$    & 10 &  $14.51 \times 12.63$   & $-33.2$      &  10 \\
J103805+601150    & $0.30 \times 0.23$ & $-37.7$    & 10 &  -   & -      & -\\
J104208+592030    & $0.29 \times 0.23$ & $-22.2$    & 10 &  -   & -      & -\\
J104516+563148    & $0.29 \times 0.23$ & $-26.1$    & 10 &  $15.94 \times 11.97$   & $-25.1$      & 10\\ 
J104646+564744    & $0.33 \times 0.24$ & $-62.7$    & 10 &  -   & -      & -\\ 
J104732+555007    & $0.28 \times 0.23$ & $-9.2$     & 10 &  $15.19 \times 11.27$   & $-55.7$      & 9.4\\
J105230+563602    & $0.29 \times 0.24$ & $-38.6$    & 9 &  $16.32 \times 11.44$   & $-59.1$      & 12\\
J105402+550554    & $0.28 \times 0.23$ & $+8.9$     & 10 &  $14.69 \times 12.67$   & $-21.3$     & 18\\
J105554+563532    & $0.33 \times 0.23$ & $-66.0$    & 10 &  -   & -      & -\\
J105703+584721    & $0.30 \times 0.24$ & $-39.6$    & 10 &  -   & -      & -\\
J105729+591128    & $0.34 \times 0.23$ & $-66.3$    & 10 &  $15.88 \times 11.86$   & $-60.2$      & 7.6\\ 
J110108+560330    & $0.29 \times 0.23$ & $-24.4$    & 10 &  -   & -      & -\\
J110136+592602    & $0.34 \times 0.24$ & $-65.8$    & 10 &  -   & -      & -\\
J110255+585740    & $0.30 \times 0.24$ & $-40.7$    & 10 &  -   & -      & -\\
J110420+585409    & $0.31 \times 0.24$ & $-43.6$    & 10 &  -   & -      & -\\
J110806+583144    & $0.29 \times 0.23$ & $-25.0$    & 9 &  $14.59 \times 11.98$   & $-46.7$      & 9.2\\   \hline\hline  
\end{tabular}}
\end{table*}

%--------------------------------------------------------------------

\section{Results from the radio data} \label{sec:results_radio}
In this section, we use the radio data presented in Sect.~\ref{sec:Observations_at_6000_and_data_reduction} to perform a detailed investigation of the properties of the remnant candidates and confirm their remnant nature.
For a detailed description of each source, see Sect.~\ref{sec:description_of_individual_sources}.

\subsection{Detection of radio cores and their spectral index}
\label{results_radio/radio_cores}
With the new data at 6000 MHz in A configuration, we detected a core structure above 3$\rm \sigma_{local}$ in 14 out of 22 sources.
All 14 newly discovered radio cores at 6000 MHz confirmed the proposed OC described in Sect.~\ref{sec:optical_identification/approach_and_results}.

Interestingly, all three of the sources showing amorphous morphology, J1029+5857, J1057+5911 and J1108+5831, are among the eight candidate remnant radio sources with no identification of the core.
The additional four sources with no identification of the core have a double-lobe morphology and only one of them does not have an SDSS optical or infrared counterpart (J1046+5647).
Finally, the remaining source without detection of the core is the remnant candidate J1029+5842. As mentioned in Sect.~\ref{sec:optical_identification/approach_and_results}, the source J1029+5842 has a potential optical host galaxy at the centre of the southern component, indicating that the two `lobes' might rather be two independent sources (as can be seen in Fig.~\ref{fig:remnants1}). At 6000 MHz we detected cores at the centre of both lobes, confirming this scenario. Because of this, we decided to reject the source J1029+5842 as a remnant radio galaxy and do not consider it in the following analysis.

We measured the flux density of the radio cores as the maximum-pixel value at the position of the 6000 MHz detection. All detected radio cores have peak flux densities below 1 mJy. In particular, all of the core radio luminosities are below 2 $\times$ 10$^{24}$ W Hz$^{-1}$ at 6000 MHz, spanning in the range of radio luminosities [0.1, 13.9] $\times$ $10^{23}$ W Hz$^{-1}$.
The 6000 MHz core flux densities are reported in column 4 of Table~\ref{tab:core_flux_and_SI} and the respective radio luminosities are reported in column 7 of Table~\ref{tab:derived_radio_values}.

Using the same method, we measured the core peak flux density in the FIRST image at 1400 MHz and in the VLASS image at 3000 MHz.  In case the core is not detected, we computed a 3$\sigma_{\rm local}$ upper limit. Peak flux density measurements are reported in Table.~\ref{tab:core_flux_and_SI}. These measurements allowed us to construct the radio spectra of the cores shown in Fig.~\ref{fig:spectral_index_core}. We also present in-band measurements from 4000 to 8000 MHz with light blue squares in Fig.~\ref{fig:spectral_index_core}. These in-band measurements are consistent with the 6000 MHz flux densities. As can be seen in that figure, the majority of the measurements represent upper limits (indicated with triangle symbols). In seven cases, no core is detected at any frequency used in this work. Only three remnant candidates have detection of the core at 1400 and 6000 MHz, allowing to derive its spectral index, $\alpha_{1400}^{6000}$. In addition, three more remnant candidates have detection of the core at 3000 and 6000 MHz, allowing to derive its spectral index, $\alpha_{3000}^{6000}$.

The derived core spectral indices are flat ($\alpha_{1400}^{6000}$ and $\alpha_{3000}^{6000}$ being in the range [-0.48, 0.28] and [-0.6, 0.4], respectively). This is consistent with the typical spectral indices of cores in radio galaxies as result of self-absorption (see e.g. \citealt{1979ApJ...232...34Blandford,1984A&A...139...55Feretti, 2008MNRAS.390..595Mullin}).
Some scatter in the flux density variations measured at different frequencies might be due to intrinsic variability given the fact that the observations were carried out at different epochs.

\begin{small}
\begin{table*}[h]
\centering
\caption{Core flux density measurements and spectral indices of the core of 21 remnant candidates.}
\resizebox{0.8\textwidth}{!}{\begin{tabular}{l c c c c c c }\hline\hline
Source name  & $\rm S_{core;~1400~MHz}$ & $\rm S_{core;~3000~MHz}$ & $\rm S_{core;~6000~MHz}$  & $\alpha^{3000}_{1400}$ & $\alpha^{6000}_{3000}$    & $\alpha^{6000}_{1400}$ \\
                 & $\rm _{(FIRST)}$ [mJy]                  & $\rm _{(VLASS)}$ [mJy]                      & $\rm _{(VLA)}$ [mJy] \\ \hline
J102818+560811    & <0.43                & 0.58 $\pm$ 0.03   & 0.87 $\pm$ 0.04   & <-0.38            & -0.6 $\pm$ 0.1    & <-0.48   \\
J102842+575122     & <0.36               & <0.123            & <0.03             & -                 & -                 & -                 \\
J102905+585721     & <0.39               & <0.119            & <0.03             & -                 & -                 & -                 \\
J103132+591549     & <0.42               & 0.50 $\pm$ 0.03   & 0.48 $\pm$ 0.02   & <-0.22            & 0.1 $\pm$ 0.1     & <-0.09  \\
J103602+554007     & <0.36               & <0.122            & 0.124 $\pm$ 0.006 & -                 & <0.0              & <0.73   \\
J103641+593702     & <0.40               & <0.121            & 0.23 $\pm$ 0.01   & -                 & <-0.9             & <0.38   \\
J103805+601150     & <0.40               & <0.120            & 0.169 $\pm$ 0.008 & -                 & <-0.5             & <0.58   \\
J104208+592030    & 0.48 $\pm$ 0.02     & 0.82 $\pm$ 0.04   & 0.54 $\pm$ 0.03   & -0.69 $\pm$ 0.09  &  0.6 $\pm$ 0.1    & -0.08 $\pm$ 0.05   \\
J104516+563148    & 0.55 $\pm$ 0.03     & 0.45 $\pm$ 0.02   & 0.37 $\pm$ 0.02   &  0.26 $\pm$ 0.09  &  0.3 $\pm$ 0.1    & 0.28 $\pm$ 0.05   \\
J104646+564744    & <0.47               & <0.127            & <0.03             & -                 & -                 & -                 \\
J104732+555007    & 0.53 $\pm$ 0.03     & 0.55 $\pm$ 0.03   & 0.59 $\pm$ 0.03   & -0.06 $\pm$ 0.09  &  -0.1 $\pm$ 0.1   & -0.08 $\pm$ 0.05   \\
J105230+563602    & <0.28               & <0.137            & 0.121 $\pm$ 0.006 & -                 & <0.2              & <0.57  \\
J105402+550554    & <0.25               & <0.104            & 0.142 $\pm$ 0.007 & -                 & <-0.5             & <0.38  \\
J105554+563532    & <0.31               & <0.156            & 0.125 $\pm$ 0.006 & -                 & <0.3              & <0.62  \\
J105703+584721    & <0.50               & 0.52 $\pm$ 0.03   & 0.38 $\pm$ 0.02   & <-0.05            & 0.4 $\pm$ 0.1     & <0.18  \\
J105729+591128    & <0.32               & <0.117            & <0.03             & -                 & -                 & -                 \\
J110108+560330    & <0.46               & <0.128            & <0.03             & -                 & -                 & -                 \\
J110136+592602    & <0.41               & <0.143            & 0.28 $\pm$ 0.01   & -                 & <-1.0             & <0.27  \\
J110255+585740    & <0.37               & <0.114            & <0.03             & -                 & -                 & -                 \\
J110420+585409    & <0.33               & <0.110            & 0.123 $\pm$ 0.006 & -                 & <-0.2             & <0.68  \\
J110806+583144    & <0.28               & <0.097            & <0.03             & -                 & -                 & -                 \\            
 \hline \hline
    \end{tabular}}
    \tablefoot{
    \small
    In Col. 1 are the source names in J2000 coordinates; Col. 2, 3 and 4 are core flux densities measured directly from images at 1400, 3000 and 6000 MHz, respectively;  Cols. 5, 6, and 7 are the spectral indices of the core emission calculated between 1400 MHz and 3000 MHz, 3000 and 6000 MHz and 1400 and 6000 MHz, respectively. Mark `-' indicates unreliable measurements..}
    \label{tab:core_flux_and_SI}
\end{table*}
\end{small}

\subsection{Extended radio emission and spectral index}
\label{results_radio/extended_emission}
We detected extended emission in eight out of the ten observed sources at 6000 MHz in the D-configuration.
To reconstruct the integrated radio spectrum, we measured the total flux density directly from the images at all available frequencies (150 MHz LOFAR18; 325 MHz; 1400 MHz WSRT; 6000 MHz) using the 3 $\times$ $\sigma_{\rm LOFAR18}$ contours as a reference. The total flux density of the sources in the 1400 MHz NVSS image was measured following its 3 $\times$ $\sigma_{\rm local}$ contours, due to its lower resolution of 45${^{\prime\prime}}$. In case of non-detection, we computed the 3$\sigma_{\rm local}$ upper limit from a set of ten flux density measurements surrounding the expected location of the source using a box with the same size as the source.
The obtained flux densities are listed in Table~\ref{tab:total_flux_SI_SB}. Figure~\ref{fig:SI_total_plots} also shows in-band measurements from 4000 to 8000 MHz with light blue squares. Based on these values, we constructed the spectra shown in Fig.~\ref{fig:SI_total_plots} and computed the spectral indices $\alpha^{6000}_{150}$, $\alpha^{1400}_{150}$ and $\alpha^{6000}_{1400}$ presented in Table~\ref{tab:total_flux_SI_SB}. Two of the three sources covered by the WSRT mosaic have flux densities in agreement with the NVSS values. The third source (J1057+5911) is not detected at 1400 MHz, but the WSRT upper limit measurement allowed us to put a tighter constraint on its total flux density at 1400 MHz, and therefore, we used this measurement for the calculation of its spectral index and the total radio luminosity presented in the following sections.

For two of the ten sources observed at 6000 MHz, the spectral index is found to be ultra-steep up to 6000 MHz ($\alpha^{6000}_{150}$ $>$ 1.2; sources J1028+5751 and J1029+5857).
We recalculated the spectral index $\alpha^{1400}_{150}$, compared to the one presented in MB17 (see Sect.~\ref{sec:sample}), using the measurements directly from the LOFAR18 and NVSS images\footnote{Except for the source J1057+5911, for which we used total flux density upper limit obtained directly from the WSRT image, as mentioned earlier in this section.}. We found that seven sources have USS already at low frequencies (5 of these were already identified as US by MB17). We also note that sources J1045+5631 and J1047+5550 show a significant flattening of their spectrum at a high frequency (see Sect.~\ref{sec:discussion}).

Finally, we computed the radio luminosity and linear sizes of the sources and list them in Table~\ref{tab:derived_radio_values}. The total radio luminosities at 6000 MHz, and at 1400 and 150 MHz were k-corrected using the spectral indices presented in Table~\ref{tab:total_flux_SI_SB} ($\rm \alpha^{6000}_{150}$ and $\rm \alpha^{1400}_{150}$, respectively). 

We measured the angular sizes of the radio sources in the full sample along the maximum extent of the source using the 3$\sigma_{\rm local}$ contours of the 18$^{\prime\prime}$ image as a reference. 
The projected linear sizes of all remnant candidates are given in Table~\ref{tab:derived_radio_values} together with their uncertainties propagated from redshift uncertainties.
The median value of the sizes is 601 kpc and nine remnant candidates have linear sizes greater than 0.7 Mpc and therefore, can be classified as giant radio galaxies (\citealt{1974Natur.250..625WillisGRG, 1999MNRAS.309..100Ishwara-Chandra_Saikia_GRG, 2020A&A...635A...5DabhadeGRG}).

\subsection{Core prominence}
\label{results_radio/core_prominence}
The CP of a radio galaxy is a useful tool to investigate its nuclear activity. Typical values found in the literature are in the range 0.1-0.001 for sources of low and intermediate radio luminosities in the B2 sample, while powerful sources in the 3CRR sample show a mean radio CP of 3 $\times$ $\rm 10^{-4}$ \citep{1988A&A...199...73Giovannini, 2008MNRAS.390..595Mullin}. Radio core flux densities are observed to correlate with the total flux density of a source. This trend is also illustrated by the anti-correlation between the ratio of radio luminosity of the core and extended emission, and total radio luminosity found by \citet{1990A&A...227..351DeRuiter} for a sample of low luminosity active radio galaxies. FRI radio galaxies have higher CP values than FRII radio galaxies, while the most extreme CP values have been found in compact FR0 radio galaxies (CP > 0.1, see \citealt{2015A&A...576A..38Baldi}).

We computed the CP at 1400 MHz to allow for a better comparison with the literature findings. For this, we used the core flux density at 6000 MHz (assuming a spectral index equal to 0, justified by the results in Sect.~\ref{sec:LH_data/VLA_A_observations}) and the total flux densities at 1400 MHz, measured directly from the NVSS image. Therefore, CP is defined as CP = S$_{\rm core~(6000~MHz)}$ / S$_{\rm total~(1400~MHz;~NVSS)}$.

\begin{figure} [!h]
\center
    \includegraphics[width=0.4\textwidth]{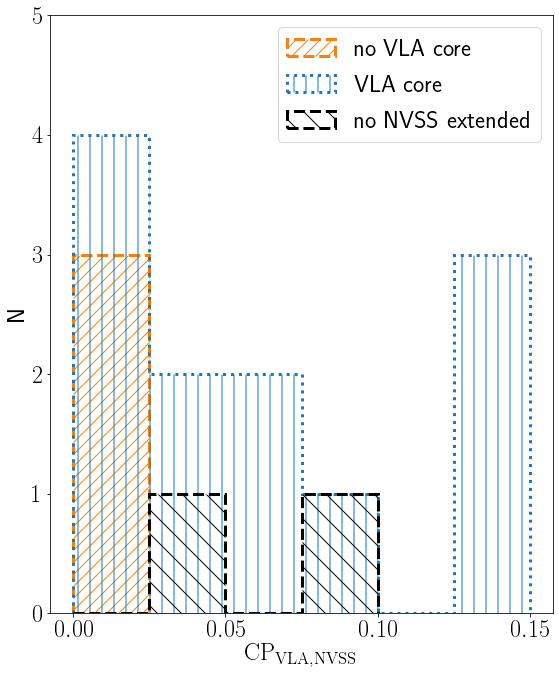}
    \caption{Histogram showing 17 CP values or limits (see Table~\ref{tab:derived_radio_values}) of the candidate remnant radio sources divided into sources with the detection of the core and the extended emission (blue); those representing upper limits of the CP (orange); and those representing lower limits of the CP (black). Core flux densities are measured at 6000 MHz while total flux densities at 1400 MHz from the NVSS images. More details can be found in Sect~\ref{results_radio/core_prominence}.}
\label{fig:cp_histogram}
\end{figure}

We found CP values in the range [0.0019, 0.15] as shown in the histogram in Fig.~\ref{fig:cp_histogram}. To further investigate these CP values and how they relate to the total radio luminosities of our sources, we put them in the context of the relation derived by \citet{1990A&A...227..351DeRuiter} as described below.

%%%%%%%%% DE RUITER

\begin{figure} [!h]
    \includegraphics[width=0.48\textwidth]{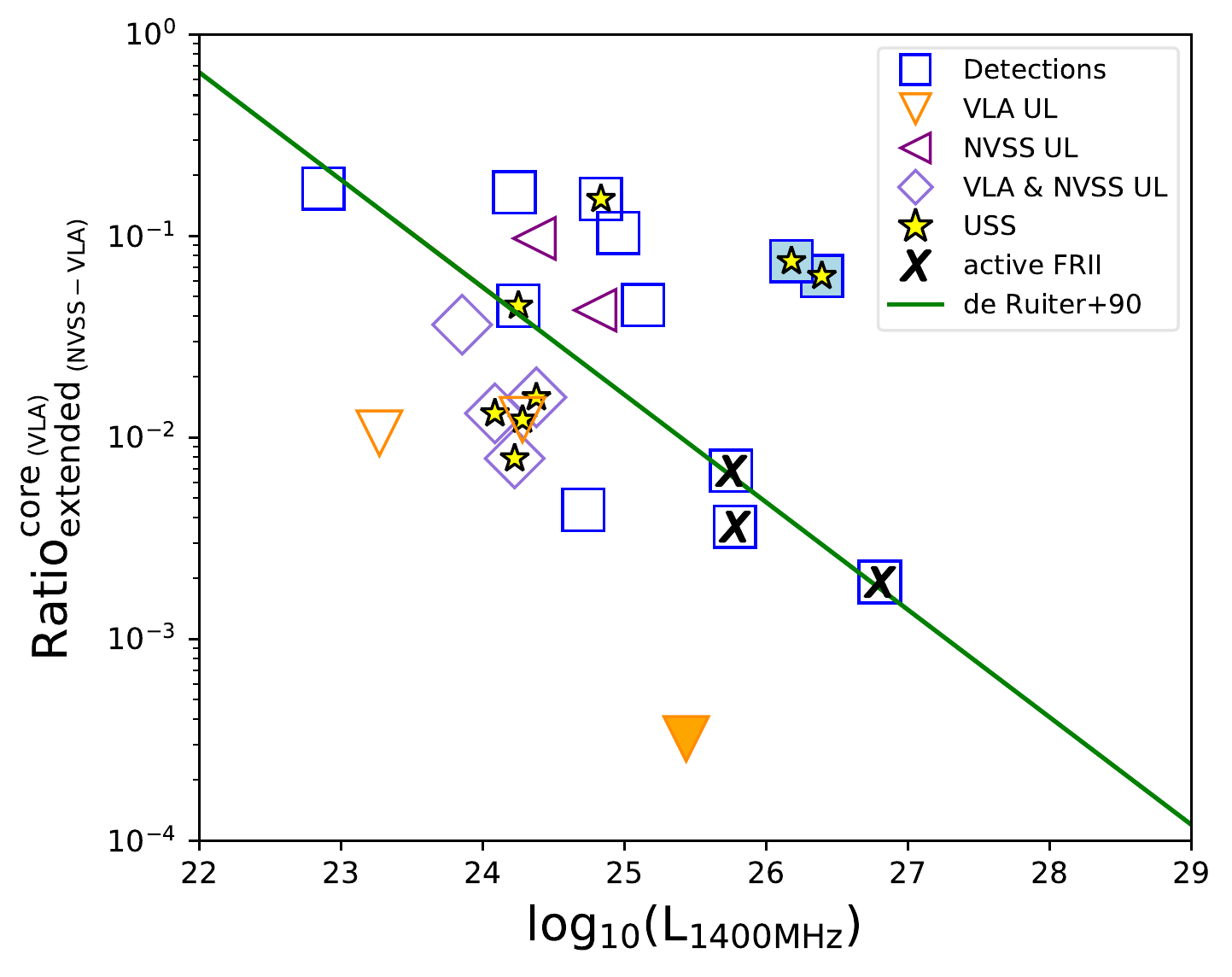}
    \caption{Ratio of radio luminosity of the core and of extended emission as a function of total radio luminosity at 1400 MHz of 21 remnant candidates. The green line represents the relation from \citet{1990A&A...227..351DeRuiter}, corrected for $\rm H_{0}$. Markers filled with colour are the ones with less reliable redshift, affecting the total radio luminosity (see Table~\ref{tab:optical_table}). Points filled with a yellow star symbol represent sources with USS. Sources with `\ding{55}' written inside them are those with observed hotspots at 150 MHz (see Sect.~\ref{results_radio/core_prominence}). `UL' in the legend stands for the `upper limit value'.}
\label{fig:deRuiter_plot}
\end{figure}

In Fig.~\ref{fig:deRuiter_plot} we show the derived relation presented by \citet{1990A&A...227..351DeRuiter} and include our measurements of the ratio of radio luminosity of the core and extended emission\footnote{Extended radio emission is derived by subtracting the flux density of the core from the total flux density of the source; Ratio = S$_{\rm core~(6000~MHz)}$ / (S$_{\rm total~(NVSS)}$ - S$_{\rm core~(6000~MHz)})$.} as a function of total radio luminosity. The green line represents the relation found by \citet{1990A&A...227..351DeRuiter} for 3C and B2 sources, scaled for the different $\rm H_{0}$ parameter used in this work. 
Sources below the line show a very reduced rate of nuclear activity with respect to typical active radio galaxies of the same power. We note that the line from \citet{1990A&A...227..351DeRuiter} is derived using median values, and therefore, some scatter around it is expected. We discuss this figure further in Sects.~\ref{sec:revised sample} and~\ref{sec:discussion}.

\begin{small}
\begin{table*} [!h]
\caption{Total flux density measurements and spectral indices of the total emission of 21 remnant candidates.}
\label{tab:total_flux_SI_SB}
\centering
\resizebox{\textwidth}{!}{\begin{tabular} { l c c c c c c c c c c p{5cm}}
\hline\hline
Source name&  $\rm S_{total;~150MHz}$ & $\rm S_{total;~325MHz}$ & $\rm S_{total;~1400MHz}$ & $\rm S_{total;~6000~MHz}$   & $\alpha^{1400}_{150}$& $\alpha^{6000}_{1400}$ & $\alpha^{6000}_{150}$\\
                          &$\rm _{(LOFAR18)}$ [mJy]            & $\rm _{(GMRT)}$[mJy]             & $\rm _{(NVSS)}$ [mJy]                   &  [mJy]                    & & & \\ \hline
J102818+560811   & 54 $\pm$ 6       & 32 $\pm$ 3      & 9.3 $\pm$ 0.5         & 3.2 $\pm$ 0.2     & 0.79 $\pm$ 0.05   & 0.73 $\pm$ 0.05    &0.77 $\pm$ 0.03        \\
J102842+575122   & 33 $\pm$ 4       & -                 & <2.31                 & <0.192            & >1.19             & -                 &>1.39*                \\
J102905+585721   & 31 $\pm$ 4       & -                 & <1.92                 & <0.014            & >1.25*            & -                 &>2.10*                \\
J103132+591549   & 121 $\pm$ 13     & 47 $\pm$ 4        & 8.0 $\pm$ 0.4         & -                 & 1.22 $\pm$ 0.05*  &-                  &-                     \\   
J103602+554007   & 30 $\pm$ 3       & 11.8 $\pm$ 0.9    & <3.01                 & -                 & >1.04             &-                  &-                     \\
J103641+593702   & 36 $\pm$ 4       & 13 $\pm$ 1        & <2.61                 & 0.98 $\pm$ 0.05   & >1.17             & <0.68             &0.98 $\pm$ 0.03       \\
J103805+601150   & 190 $\pm$ 21     & 73 $\pm$ 6        & 25 $\pm$ 1            & -                 & 0.91 $\pm$ 0.05   &-                  &-                     \\
J104208+592030  & 172 $\pm$ 19      & 48 $\pm$ 4        & 12.5 $\pm$ 0.6        & -                 & 1.17 $\pm$ 0.05   &-                  &-                     \\
J104516+563148  & 18 $\pm$ 2        & 7.2 $\pm$ 0.6     & 2.5 $\pm$ 0.1       & 1.41 $\pm$ 0.07   & 0.88 $\pm$ 0.05   & 0.39 $\pm$ 0.05   & 0.69 $\pm$ 0.03        \\
J104646+564744  & 770 $\pm$ 80      & 325 $\pm$ 26      & 93 $\pm$ 5            & -                 & 0.94 $\pm$ 0.05   &-                  &-                      \\
J104732+555007  & 89 $\pm$ 10       & 27 $\pm$ 2        & 4.5 $\pm$ 0.2         & 1.63 $\pm$ 0.08   & 1.34 $\pm$ 0.05*  & 0.69 $\pm$ 0.05   & 1.08 $\pm$ 0.03       \\
J105230+563602  & 195 $\pm$ 21      & 74 $\pm$ 6        & 28 $\pm$ 1            & 6.3 $\pm$ 0.3     & 0.87 $\pm$ 0.05   & 1.02 $\pm$ 0.05   & 0.93 $\pm$ 0.03       \\
J105402+550554  & 49 $\pm$ 5        & -                 & 3.3 $\pm$ 0.2         & 0.75 $\pm$ 0.04   & 1.21 $\pm$ 0.05*   & 1.01 $\pm$ 0.05   & 1.13 $\pm$ 0.03      \\
J105554+563532  & 88 $\pm$ 10       & 22 $\pm$ 2        & 1.79 $\pm$ 0.09       & -                 & 1.75 $\pm$ 0.05*  & -                 &-                      \\
J105703+584721  & 1581 $\pm$ 170    & 650 $\pm$ 50      & 200 $\pm$ 10          & -                 & 0.93 $\pm$ 0.05   &-                  &-                      \\
J105729+591128  & 9 $\pm$ 1       & 3.3 $\pm$ 0.3     & <0.86$\rm ^{WSRT}$    & 0.189 $\pm$ 0.009 & >1.04             & <1.04             & 1.04 $\pm$ 0.03         \\
J110108+560330  & 114 $\pm$ 13      & 34 $\pm$ 3        & 2.5 $\pm$ 0.1         & -                 & 1.71 $\pm$ 0.05*  &-                  &-                      \\
J110136+592602  & 13 $\pm$ 2        & 5.3 $\pm$ 0.4     & 2.0 $\pm$ 0.1       & -                 & 0.85 $\pm$ 0.05   &-                  &-                      \\
J110255+585740  & 99 $\pm$ 11       & 31 $\pm$ 3        & <3.84                & -                 & >1.46*            &-                  &-                      \\
J110420+585409  & 236 $\pm$ 26      & 103 $\pm$ 8       & 34 $\pm$ 2        & -                 & 0.86 $\pm$ 0.05   &-                  &-                      \\ 
J110806+583144  & 27 $\pm$ 3        & 13 $\pm$ 1        & 2.9 $\pm$ 0.1       & 0.66 $\pm$ 0.03   & 1.00 $\pm$ 0.05   & 1.02 $\pm$ 0.05   & 1.01 $\pm$ 0.03      \\
\hline \hline
\end{tabular}}
\tablefoot{
    \small
    In Col. 1 are the source names in J2000 coordinates; Col. 2, 3, 4 and 5 are total flux densities measured directly from images at 150, 325, 1400 and 6000 MHz, respectively;  Cols. 6, 7, and 8 are the spectral indices of the total emission calculated between 150 and 1400 MHz, 1400 and 6000 MHz, and 150 and 6000 MHz, respectively. Mark `-' indicates unreliable measurements, and mark `*' indicates USS values.\\
Flux densities are measured inside 3 $\times$ $\sigma_{\rm 150~MHz;18^{\prime\prime}}$ contours from all of the images except for the NVSS image at 1400 MHz, where due to the low resolution of the image, we used 3 $\times$ $\sigma_{\rm {local}}$.\\
Total flux density at 1400 MHz is measured directly from the NVSS image for all the sources except for the source J1057+5911. The measurement of the flux density for this source from the WSRT image provides a tighter constraint and is noted with `WSRT'.}
\end{table*}
\end{small}

%\begin{small}
\begin{table*}[h]
\caption[]{Properties of the candidate remnant radio sources, as derived in this work.}
\label{tab:derived_radio_values}
\centering
\resizebox{\textwidth}{!}{\begin{tabular}{l c c c c c c c}\hline\hline
Source name& CP$\rm ^{VLA} _{NVSS}$   & SPC & \textbf{$\rm L_{~total;~150~MHz}$}  & \textbf{$\rm L_{~total;~1400~MHz}$} & \textbf{$\rm L_{~total;~6000~MHz}$}& \textbf{$\rm L_{~core;~6000~MHz}$}     & size     \\
        &                           &   \textbf{($\alpha^{6000}_{1400}$ - $\alpha^{1400}_{150}$)}     &   [$\times$ 10$^{23}$ W Hz$\rm^{-1}$] &   [$\times$ 10$^{23}$ W Hz$\rm^{-1}$]    & [$\times$ 10$^{23}$ W Hz$\rm^{-1}$] & [$\times$ 10$^{23}$ W Hz$\rm^{-1}$]  & [kpc] \vspace{0.5mm} \\ \hline \hline
J102818+560811   & 0.094 $\pm$ 0.007    & -0.05 $\pm$ 0.08  &524 $\pm$ 128      & 91 $\pm$ 20       & 31 $\pm$ 7        & 6 $\pm$ 1         & 565 $\pm$ 27              \\ \vspace{0.5mm}
J102842+575122   & -                    & -                 &173 $\pm$ 159      & <12               &<1.1               &<0.2               & 601 $\pm$ 142             \\ \vspace{0.5mm}
J102905+585721   & -                    & -                 &391 $\pm$ 69       & <24               &<0.24              &<0.22              & 517 $\pm$ 15              \\ \vspace{0.5mm}
J103132+591549   & 0.060 $\pm$ 0.004    & -                 &37169 $\pm$ 4089   & 2460 $\pm$ 120    & -                 &41 $\pm$ 2         & 817$\rm_{~97}^{~628}$     \\ \vspace{0.5mm}
J103602+554007   & >0.041               & -                 &636 $\pm$ 175      & <63               & -                 & 1.5 $\pm$ 0.3     & 871 $\pm$ 37              \\ \vspace{0.5mm}
J103641+593702   & >0.089               & <-0.49            &318 $\pm$ 58       & <24               &8 $\pm$ 1          & 1.3 $\pm$ 0.2     & 783 $\pm$ 26              \\ \vspace{0.5mm}
J103805+601150   & 0.0068 $\pm$ 0.0005  & -                 &4305 $\pm$ 874     & 560 $\pm$ 100     &   -               & 2.3 $\pm$ 0.4     & 984 $\pm$ 27              \\ \vspace{0.5mm}
J104208+592030   & 0.043 $\pm$ 0.003    & -                 &1859 $\pm$ 204     & 135 $\pm$ 7       &   -               & 3.6 $\pm$ 0.2     & 1192.7 $\pm$ 0.2          \\ \vspace{0.5mm}
J104516+563148   & 0.15 $\pm$ 0.01      & -0.49 $\pm$ 0.08  &5.4 $\pm$ 0.6      & 0.75 $\pm$ 0.04   & 0.42 $\pm$ 0.02   & 0.1 $\pm$ 0.2     & 319$\rm_{~160}^{~486}$    \\ \vspace{0.5mm}
J104646+564744   &<0.0003               & -                 & 2243 $\pm$ 247    & 272 $\pm$ 14      &   -               & <0.070            & 517$\rm_{~451}^{~ 580}$   \\ \vspace{0.5mm}
J104732+555007   & 0.132 $\pm$ 0.009    & -0.65 $\pm$ 0.08  & 1355 $\pm$ 274    & 68 $\pm$ 12       &   22 $\pm$ 4      & 4.9 $\pm$ 0.7     & 1385 $\pm$ 45             \\ \vspace{0.5mm}
J105230+563602   & 0.0044 $\pm$ 0.0003  & 0.14 $\pm$ 0.08   &360 $\pm$ 100      & 51 $\pm$ 13       &   12 $\pm$ 3      & 0.18 $\pm$ 0.04   & 786 $\pm$ 66              \\ \vspace{0.5mm}
J105402+550554   & 0.043 $\pm$ 0.003    & -0.2 $\pm$ 0.08   &261 $\pm$ 29       & 17.8 $\pm$ 0.9    &   4.0 $\pm$ 0.2   & 0.52 $\pm$ 0.03   & 530.92 $\pm$ 0.05         \\ \vspace{0.5mm}
J105554+563532   & 0.070 $\pm$ 0.005    & -                 &103820 $\pm$ 11420 & 1500 $\pm$ 80     &   -               & 13.9 $\pm$ 0.7    & 695$\rm_{~84}^{~524}$     \\ \vspace{0.5mm}
J105703+584721   & 0.0019 $\pm$ 0.0001  & -                 &49900 $\pm$ 5489   & 6320 $\pm$ 320    &   -               & 6.9 $\pm$ 0.4     & 1200.47 $\pm$ 0.02        \\ \vspace{0.5mm}
J105729+591128   & -                    & <0                &73 $\pm$ 11        & <7.2              &   1.6 $\pm$ 0.2   & <0.17             & 428 $\pm$ 10              \\ \vspace{0.5mm}
J110108+560330   &<0.012                & -                 &878 $\pm$ 801      & 19 $\pm$ 17       &   -               & <0.13             & 586 $\pm$ 124             \\ \vspace{0.5mm}
J110136+592602   & 0.14 $\pm$ 0.01      & -                 &128 $\pm$ 35       & 17 $\pm$ 4        &   -               &1.7 $\pm$ 0.4      & 584 $\pm$ 33              \\ \vspace{0.5mm}
J110255+585740   &-                     & -                 &434 $\pm$ 48       & <16.8             &   -               & <0.086            & 524.0 $\pm$ 0.1           \\ \vspace{0.5mm}
J110420+585409   & 0.0036 $\pm$ 0.0003  & -                 &4128 $\pm$ 923     & 600 $\pm$ 120     &   -               & 1.4 $\pm$ 0.2     & 1035 $\pm$ 36             \\ \vspace{0.5mm}
J110806+583144   &<0.010                & 0.02 $\pm$ 0.08   &17 $\pm$ 2         & 1.86 $\pm$ 0.09   &0.43  $\pm$ 0.02   & <0.017            & 214.40 $\pm$ 0.03         \\  \hline \hline
\end{tabular}}
\tablefoot{
\small
In Col. 1 are the source names in J2000 coordinates; Col. 2 notes the CP values (see Sect.~\ref{results_radio/core_prominence});  Col. 3 lists SPC values, where mark `*' indicates sources with significant SPC.; in Col. 4, 5 and 6 are the radio luminosities of the total emission at 150 and 6000 MHz, and radio luminosity of the core at 6000 MHz, respectively. Col. 7 lists projected linear sizes of radio emission in kiloparsecs, measured inside the 3$\sigma_{\rm local}$ contours of the LOFAR18 image.}
\end{table*}
%\end{small}

\section{Revised sample of remnant radio galaxies}
\label{sec:revised sample}

In this section, we combine all the new data described above to perform a revision of the classification of these sources with the final goal of compiling a more robust sample.
The following criteria are used to refine the classification:\\

\noindent
(i) total integrated radio spectrum steeper than 1.2 between 150 MHz and 6000 MHz or part of this range of frequency (USS criterion);\\
(ii) SPC > 0.5 (where SPC = $\alpha^{6000}_{1400}$ - $\alpha^{1400}_{150}$; SPC criterion); \\  
(iii) no 3$\sigma$ detection of the core in the 6000 MHz images (NC criterion), and/or CP lower than expected for active galaxies (see Fig.~\ref{fig:deRuiter_plot}; CP criterion).\\ 

In particular, the first two criteria recall the spectral criteria used in MB17 but are here expanded to a much larger and sensible frequency range. Spectral steepening or spectral curvature, due to radiative losses in the remnant phase, is expected to become particularly evident at frequencies higher than 1400 MHz. Concerning the third criterion, we decided to confirm as remnants not only those sources lacking a core detection (as done by MB17 and \citealt{2018MNRAS.475.4557Mahatma_remnants}) but also sources that have a very low CP for their total power, as presented in Fig.~\ref{fig:deRuiter_plot} with the relation by \citet{1990A&A...227..351DeRuiter}.
Indeed, the simultaneous presence of USS emission or emission with a strong SPC, combined with a low-luminosity radio core suggests that the nucleus is still active even if at a low level,
but the large scale radio jets might have switched off, resulting in the ageing of the large scale emission. This can be explained either
by the possibility that some sources never completely switch off, or that we are witnessing the start of a new phase of activity (see Sect.~\ref{sec:discussion} for more discussion). The SB criterion, although it worked well to identify a number of remnant candidates in the initial selection by MB17, was found to be the least constrained and therefore, it was not used in this selection.

Following these criteria, we confirmed the remnant nature of 13 out of 21 candidates. For convenience, the parameters used are summarised in Table~\ref{tab:summary_remnant_candidates}. In particular, eight remnants have been confirmed based on criterion (i), while the others are derived from a combination of the absence of core and low CP. Interestingly, no candidates were selected based on criterion (ii).
Remnants are indicated with `1' in the last column of Table~\ref{tab:summary_remnant_candidates}. 
Hereafter, we only refer to these 13 objects as remnants.\\

It is worth commenting on the fact that our new sensitive data at 6000 MHz clearly show the limits of selecting remnants using a flat value of CP for every total radio luminosity of the source, as done by MB17. While the CP criterion was considered to account for young remnants still showing hotspots in the LOFAR6 image (see Fig.~\ref{fig:remnants1}), it can easily introduce a number of false positives. Powerful FRII radio galaxies are indeed known to show very low levels of CP so that their cores can only be identified using very sensitive observations. Based on this we rejected sources J1038+6011, J1057+5847 and J1104+5854, marked with `\ding{55}' in Fig.~\ref{fig:deRuiter_plot}, from our final remnant sample\footnote{Remnant candidate J1104+5854 is located below the relation derived by \citet{1990A&A...227..351DeRuiter} and would otherwise be confirmed as a remnant based on the CP criterion presented in this section.}. 
As a final remark, the lack of high-frequency observation of the large scale emission for some of the objects in the sample prevents us from completely ruling out for these objects the possibility they are remnants, although all the other parameters seem to consistently indicate they are active radio sources.

\begin{small}
\begin{table*}
\centering
\caption{Summary of the parameters used to confirm the remnant nature of 21 candidate remnant radio sources.}
\resizebox{\textwidth}{!}{\begin{tabular}{c c c c c c c c c}\hline\hline
Source name & \multicolumn{2}{c}{\textbf{6000 MHz} detection}          & \multicolumn{3}{c}{USS extended SI}   & curvature     &\textbf{selection criteria} & Final\\
            & core & total    & $^{1400} _{150}$ & $^{6000} _{1400}$ & $^{6000} _{150}$   & extended SI      &  & sample \\ \hline \hline
J102818+560811      &   $\checkmark$&$\checkmark$   &x&x&x                                          &x                  & -             & 0 \\
J102842+575122      &   x           & x             &x& - &$\checkmark$                             &x                  & NC, USS       & 1 \\
J102905+585721      &   x           & x             &$\checkmark$& - &$\checkmark$                  &-                  & NC, USS       & 1 \\
J103132+591549      &   $\checkmark$& -             &$\checkmark$&-&-                               &-                  & USS           & 1 \\
J103602+554007      &   $\checkmark$& -             &x&-&-                                          &-                  & -             & 0 \\
J103641+593702      &   $\checkmark$&$\checkmark$   &x&x&x                                          &x                  & -             & 0 \\
J103805+601150      &   $\checkmark$& -             &x&-&-                                          &-                  & -             & 0 \\
J104208+592030      &   $\checkmark$& -             &x&-&-                                          &-                  & -             & 0 \\
J104516+563148      &   $\checkmark$&$\checkmark$   &x&x&x                                          &x                  & CP            & 1 \\
J104646+564744      &   x           & -             &x&-&-                                          &-                  & NC, CP        & 1 \\
J104732+555007      &   $\checkmark$&$\checkmark$   &$\checkmark$&x&x                               &x                  & USS           & 1 \\
J105230+563602      &   $\checkmark$&$\checkmark$   &x&x&x                                          &x                  & CP            & 1 \\
J105402+550554      &   $\checkmark$&$\checkmark$   &$\checkmark$&x&x                               &x                  & USS           & 1 \\
J105554+563532      &   $\checkmark$& -             &$\checkmark$&-&-                               &-                  & USS           & 1 \\
J105703+584721      &   $\checkmark$& -             &x&-&-                                          &-                  & -             & 0 \\
J105729+591128      &   x           &$\checkmark$   &x&x&x                                          &x                  & NC            & 1 \\
J110108+560330      &   x           & -             &$\checkmark$&-&-                               &-                  & NC, USS, CP   & 1 \\
J110136+592602      &   $\checkmark$& -             &x&-&-                                          &-                  & -             & 0 \\
J110255+585740      &   x           & -             &$\checkmark$&-&-                               &-                  & NC, USS       & 1 \\
J110420+585409      &   $\checkmark$& -             &x&-&-                                          &-                  & -             & 0 \\
J110806+583144      &   x           &$\checkmark$   &x&x&x                                          &x                  & NC, CP        & 1 \\ \hline \hline
    \end{tabular}}
    \tablefoot{
    \small
     In Col. 1 are the source names in J2000 coordinates; Col. 2 notes the detection of the core at 6000 MHz; Col. 3 notes the detection of the extended emission at 6000 MHz; Cols. 4, 5 and 6 are spectral indices of the total emission $\alpha^{1400}_{150}$, $\alpha^{6000}_{1400}$ and $\alpha^{6000}_{150}$; Col. 7 says whether the SPC > 0.5 is present in the spectrum of remnant candidates between 150 and 6000 MHz; Col. 8 lists criteria used to confirm the source as a remnant in this work. The abbreviations used to represent the criteria in Col. 8 are explained in Sect.~\ref{sec:revised sample}. Listed in Col. 9 is a flag where `1' indicates that the source is confirmed a remnant, `0' candidate is not a remnant. Mark `-' stands for sources that were not observed in D-array at 6000 MHz and therefore denote the values that are missing.}
    \label{tab:summary_remnant_candidates}
\end{table*}
\end{small}

\section{Discussion}
\label{sec:discussion}

Remnant radio galaxies have long been considered to have amorphous morphologies, low-SB extended emission and absence of compact features. The radio spectrum of their total emission is expected to be curved or ultra-steep due to the ageing of the plasma from the previous activity and no replenishment of newly accelerated particles. However, previous studies (\citealt{2011A&A...526A.148Murgia2011,2012ApJS..199...27Saripalli}; MB17) have suggested a more complex picture, which is confirmed by the current study. Here we discuss the selection criteria and the properties of the final remnants in detail in the context of current knowledge of the population of active and remnant radio galaxies.

\subsection{Selection criteria}
By combining the data at 150 MHz with higher frequency data up to 6000 MHz, we have been able to confirm 13 remnant radio galaxies (hereafter, `remnants') in the sample of 21 remnant candidates. This is a result of a combination of morphological and spectral criteria presented in Sect.~\ref{sec:revised sample} that are suggested to cover remnant radio galaxies with different properties, which may point to different phases of their evolution.

As we can see from Table~\ref{tab:summary_remnant_candidates}, eight of the remnants in the final sample have an USS below 1400 MHz or in the frequency range from 150 to 6000 MHz. The remaining five remnants without USS have no core detection at 6000 MHz, or low measurements or very tight limits on their CP. Three out of the latter five sources do not have a detection of the core at 6000 MHz, and two also have a very diffuse and amorphous morphology (J1057+5911 and J1108+5831 with the SB values of 20 mJy arcmin$^{-2}$ and 70 mJy arcmin$^{-2}$, respectively, measured from the LOFAR18 images). This demonstrates that the `classic' selection based on USS cannot be used to identify all remnant radio galaxies, however, it remains a very good proxy for this class of sources. As already discussed in MB17 remnants with low CP and non-USS may represent the youngest tail of the remnant population. We also note that in two sources (J1045+5631 and J1047+5550) we see a flattening of the integrated spectrum towards higher frequencies. As can be seen from Fig.~\ref{fig:remnants1}, in these sources central emission dominates the total flux density. Source J1047+5550 has USS at low frequencies, confirming its remnant nature, while the flat central emission and detection of the core, both at 6000 MHz, might point to a restart of the nuclear activity.

Out of seven remnants without a core detection, four have USS, thus representing the classical prototype of a remnant radio galaxy. 
One source worth noting is J1055+5635, whose optical host remained unidentified, despite a detection of the core at 6000 MHz. While the combination of the USS and the absence of an optical match may suggest that the source represents an active high-redshift radio galaxy, a recent work on its resolved spectral properties suggests that its spectral index is extremely steep even for a high redshift radio source supporting the remnant interpretation (for a detailed discussion see \citealt{2021A&A...648A...9Morganti_resolvedsi}).\\  

We confirm that some of the remnants (including some classified based on the USS emission) do show the presence of some level of nuclear radio activity (J1031+5915, J1045+5631, J1047+5550, J1052+5636, J1054+5505 and J1055+5635). The presence of cores (mean core radio luminosity of remnants being $\rm log_{10}$ $(L^{\rm core}_{6000~MHz}$ / W Hz$\rm ^{-1})$ = 23) means that not all remnants have a very low CP. This expands previous findings (e.g. \citealt{2010evn..confE..89ParmaVLBI,2011A&A...526A.148Murgia2011} and \citealt{2016A&A...585A..29Brienza}) and suggests that the nuclear activity in these sources might never completely switch off but only dims. Alternatively, we might be witnessing the beginning of a new phase of jet activity while the USS remnant plasma is still visible (see e.g. \citealt{2012ApJS..199...27Saripalli, 2020A&A...638A..34Jurlin, 2021A&A...648A...9Morganti_resolvedsi}). Discerning between these two scenarios is not a trivial task. How weak the core radio luminosity should be in order to classify a source as a remnant is not well assessed, and it is a strong function of frequency. The mean CP of the remnants with core detections at 6000 MHz in our sample is 0.08 when considering the measurement of the total flux density at 1400 MHz, and 0.006 when considering the measurement of the total flux density at 150 MHz. The latter value is comparable to the mean CP value of sources with core detections at 6000 MHz equal to $\sim$ 0.005 considering the measurement of the total flux density at 150 MHz, reported by \citet{2018MNRAS.475.4557Mahatma_remnants}.

A useful tool that can be used in this respect is the relation between the CP and the radio luminosity. Sources that have USS and very low levels of CP for the total power (see Fig.~\ref{fig:deRuiter_plot}), might be interpreted as dying sources (J1101+5603, marked with a star below the green line). On the contrary, ultra-steep sources with high levels of CP for their total power might be hinting at a restarting jet activity in the nuclear regions (e.g. sources J1031+5915, J1047+5550, J1054+5505 and J1055+5635, marked with stars above the green line).
Overall, the morphological selection based on the core prominence has been revealed to be complex, due to the scatter around the relation presented by \citet{1990A&A...227..351DeRuiter}, and due to the lack of high spatial resolution and deep images at 1400 MHz.

%%RADIO PROPERTIES
\subsection{Properties of the remnant radio sources and their host galaxies}
The 13 remnants show a variety of morphologies, from amorphous to double-lobed, which likely depends on the progenitor, as well as on the plasma dynamical evolution after the jets switch off.
The observed linear sizes are in the range [214.40, 1385] kpc, including \textbf{3} giant radio galaxies (linear size > 0.7 Mpc; sources J1031+5915, J1047+5550 and J1052+5636). We note that the sample is biased towards radio galaxies with larger sizes due to the cut in the angular size of 60$^{\prime\prime}$ applied in the selection of the initial sample by MB17. This also influences the total radio luminosities of the sources, which is for the majority of the sources (18 out of 21) above L$_{\rm 150~MHz}$ $>$ $10^{25}$ W Hz$\rm ^{-1}$ as can be seen in Table~\ref{tab:derived_radio_values}.
Given the fact that the source luminosity is decreasing in the remnant phase due to radiative and expansion losses (see e.g. Fig. 2 in \citealt{2020MNRAS.496.1706Shabala}), we infer that during their active phase all remnants in our sample must have had a luminosity well above the classical FRI-FRII dividing line \citep{1994ASPC...54..319OwenLedlow_FRIandII}. While more recent work by \citealt{2019MNRAS.488.2701Mingo} showed that such dichotomy is much less strict than previously thought, we can still conclude that the majority of the 13 remnants come from a population of powerful active radio sources with double morphologies. We note that the three amorphous remnant radio galaxies (J1029+5857, J1057+5911 and J1108+5831), which all interestingly lack a core detection at 6000 MHz, are among the least powerful ones, with radio luminosities lower than the median value of the remnant sample (median value of radio luminosities of remnants is log$_{\rm 10}$(L$_{\rm 150~MHz}/\rm W~Hz^{-1}$) = {26}). Based on their radio luminosities and morphology, we suggest that these sources could have been powerful FRI radio sources, during their active time. This conclusion is in agreement with the progenitor analysis of remnant radio sources presented by \citet{2021A&A...648A...9Morganti_resolvedsi}.

\smallskip

As far as the properties of the host galaxies are concerned, \citet{2020A&A...638A..34Jurlin} already found that the remnant hosts do not show any significant difference with respect to a parent sample of active radio galaxies. Using our updated sample of remnants we revised the mean value of stellar masses, which is now 1.6 $\times$ $10^{11}$ $\rm M_{\sun}$. This is still comparable to the value found for active radio sources in the sample presented by \citet{2020A&A...638A..34Jurlin} of 2.8 $\times$ $10^{11}$ $\rm M_{\sun}$ and confirms that both remnant and active radio galaxies come from the same parent population. 

\subsection{Environment of the remnant radio galaxies}
Based on the redshift of the optical host galaxy of the remnants in our sample, we could study the environments in which they reside. Only three out of 13 remnants in our final sample (J1054+5505, J1102+5857 and J1108+5831) are located in one of the clusters listed in the \citet{2012ApJS..199...34WenHanLiu} and/or \citet{2014ApJ...785..104Rykoff} catalogues. Sources J1054+5505 and J1108+5831 have a distance from the cluster centre of 2 and 1 kpc, respectively, and are therefore associated to the central galaxy of the cluster. The third remnant lies instead at 124 kpc from the cluster centre. We do not find any common properties characterising all our remnant sources located in clusters. For example, source J1054+5505 has its core detected at 6000 MHz, while J1102+5857 and J1108+5831 do not. The spectral indices of these three cluster sources are not remarkable when compared with the rest of the sample.

The fact that the majority of the remnant radio sources in our sample does not reside in a cluster environment is unexpected. Indeed \citet{2011A&A...526A.148Murgia2011} suggested that the higher fraction of remnants observed in their sample was due to the ability of the plasma to remain confined for longer times in such dense environments. We stress, however, that a proper study of the impact of the environment, such as on the fraction and properties of the remnants, would require dedicated X-ray data, which are not available at this moment.

\bigskip
\subsection{Occurrence and implications for the life-cycle}
With the revised sample of remnant radio galaxies defined in this work, we can not only investigate further the properties of individual remnants but also confirm their number, which is important for the overall understanding of the radio galaxy life cycle. 
In a recent study, briefly described in Sect.~\ref{sec:introduction}, \citet{2020A&A...638A..34Jurlin} selected 158 radio galaxies in the Lockman Hole extragalactic field and divided them into three samples: active sources, and remnant and restarted candidates. As mentioned in Sect.~\ref{sec:introduction}, the remnant sample presented there is a sub-sample of the sample studied in this work. Eleven out of 13 remnant radio galaxies presented in this study are inside the sample of \citet{2020A&A...638A..34Jurlin}. Therefore, the fraction of confirmed remnant radio sources in the Lockman Hole field is $\sim$ 7\%.
In addition to these 11 remnants, from the study of the resolved spectral index in a region of the LH, \citet{2021A&A...648A...9Morganti_resolvedsi} found that there are two more remnant radio galaxies in the sample of sources presented by \citet{2020A&A...638A..34Jurlin}. This results in a fraction of $\sim$ 8\% remnant radio sources in the LH field. The fraction of remnant radio galaxies found in this work is in agreement with the study of remnants in the Herschel-ATLAS field by \citet{2018MNRAS.475.4557Mahatma_remnants}.\\
The general agreement from the results of these studies provides tight constraints for the modelling of radio source evolution. 
Indeed, using dynamical models of radio galaxies, \citet{2020MNRAS.496.1706Shabala} found that such a small remnant fraction (together with the fraction of restarted presented in \citealt{2020A&A...638A..34Jurlin}) can only be explained by a dominant population of short-lived jets.
The finding of \citet{2020MNRAS.496.1706Shabala} is in agreement with the results from \citet{1994ASPC...61..165B, 2017MNRAS.471..891Godfrey} and \citet{2018MNRAS.475.2768Hardcastle} showing that remnant radio lobes fade quickly.

%__________________________________________________________________
\section{Conclusions}
\label{sec:conclusions}
Thanks to LOFAR, in recent years we have been able to select statistical samples of candidate remnant radio galaxies with various properties, tracing their different remnant stage.
In this study, we used information over a broad range of frequencies (from 150 to 6000 MHz) to confirm the remnant nature of the candidates selected by MB17.
Here we list the main conclusions from this work:
\begin{itemize}
    \item We observe USS, an indication of remnant radio emission, even in sources with a faint detection of the core. Therefore, non-detection of the core should be carefully considered as a selection criterion for remnants as it may erroneously reject good candidates (see Sects.~\ref{results_radio/radio_cores} and \ref{results_radio/extended_emission}, and Table~\ref{tab:summary_remnant_candidates}).
    \item We find that the low core prominence (CP) is an important parameter to select remnant radio sources. However, when using this criterion one should remember that an anti-correlation exists between the CP and total source radio luminosity for active radio galaxies. This implies that high power radio galaxies (FRII sources) could be erroneously classified as remnants unless very sensitive observations are used to search for the core emission.
    \item We confirm 13 out of the 23 sources as remnant radio sources. This corresponds to the fraction of $\sim$ 7\% considering all of the sources with angular sizes $>$ $60^{\prime\prime}$ in the LH field. This suggests that the remnant phase is even shorter than previously thought (see Sects.~\ref{sec:revised sample} and \ref{sec:discussion}). 
    \item Optical identification of remnant radio sources remains a challenging task, requiring visual inspection. Using low-frequency images, we presented a method (see Sect.~\ref{sec:optical_identification}) which successfully identified optical host galaxies.
    \item All of the remnant radio galaxies in our sample show high radio luminosity (L$_{\rm 150~MHz}$ $>$ $10^{25}$ W Hz$\rm ^{-1}$) and the majority has a double-lobe morphology, indicative of powerful FRII progenitors (see Sects.~\ref{results_radio/extended_emission} and \ref{sec:discussion}). However, it is important to note that the sample is biased towards radio galaxies with large sizes and high radio luminosities due to the cut in the angular size of 60$^{\prime\prime}$ applied in the selection of the initial sample by MB17.
    \item Only a minority of remnants in our sample reside in a cluster (23\%). This is interesting and unexpected as remnant radio galaxies in low-density environments were long claimed to become invisible more quickly than those in cluster environment due to the quick expansion of the plasma (see Sects.~\ref{sec:optical_identification/environment} and \ref{sec:discussion}).
\end{itemize} 

In order to confirm the results from this paper, we need to develop automatic criteria which minimise the need for visual inspection. Then the criteria can be applied to much bigger areas of the sky, allowing us to sample a statistically significant number of remnant candidates and draw more robust conclusions.
Once the sample is selected, we stress the importance of obtaining a multi-wavelength picture of the remnants, which is crucial to study their properties and environments in more detail.
Larger systematic studies of resolved low-frequency radio spectra, such as the one recently presented by \citet{2021A&A...648A...9Morganti_resolvedsi} are needed to confirm and expand our knowledge of remnant and restarted radio galaxies.

\begin{acknowledgements}
The authors would like to thank the anonymous referee for constructive comments that helped to improve the quality of this paper. 
MB acknowledges support from the ERC-Stg DRANOEL, no 714245 and support from INAF under the PRIN SKA/CTA project `FORECaST'.
YW and CHIC acknowledges the support of the Department of Atomic Energy, Government of India, under the project 12-R\&D-TFR-5.02-0700.
NM acknowledges support from the Bundesministerium f{\"u}r Bildung und Forschung (BMBF) D-MeerKAT award 05A20WM4.
LOFAR, the Low Frequency Array designed and constructed by ASTRON (Netherlands Institute for Radio Astronomy), has facilities in several countries, that are owned by various parties (each with their own funding sources), and that are collectively operated by the International LOFAR Telescope (ILT) foundation under a joint scientific policy. The authors thank the staff of the GMRT that made the observations used in this paper possible. GMRT is run by the National Centre for Radio Astrophysics of the Tata Institute of Fundamental Research.
This publication makes use of data products from the Wide-field Infrared Survey Explorer, which is a joint project of the University of California, Los Angeles, and the Jet Propulsion Laboratory/California Institute of Technology, and NEOWISE, which is a project of the Jet Propulsion Laboratory/California Institute of Technology. WISE and NEOWISE are funded by the National Aeronautics and Space Administration.
This research has made use of the NASA/IPAC Extragalactic Database (NED), which is operated by the Jet Propulsion Laboratory, California Institute of Technology, under contract with the National Aeronautics and Space Administration. This research made use of APLpy, an open-source plotting package for Python hosted at http://aplpy.github.com.
\end{acknowledgements}

%-------------------------------------------------------------------

 \bibliographystyle{aa} % style aa.bst
\bibliography{Yourfile} % your references Yourfile.bib

\appendix
\section{Figures of candidate remnant radio galaxies}
Here we show the LOFAR18 and LOFAR6 images, along with the SDSS DR12 $r$-band images, where the OC is marked with a green square. We also show the images of the core and the total emission at 6000 MHz.

\begin{figure*} [h]
\caption{Radio and optical images of candidate remnant radio galaxies. The colourmaps from left to right are: LOFAR18, LOFAR6, SDSS DR12 $r$-band, VLA 6000 MHz A array, VLA 6000 MHz D-array. LOFAR18 contours are shown in the first column in purple, LOFAR6 contours are shown in the following three columns in black, and contours from the FIRST images at 1400 MHz are shown in all five columns in orange. VLA 6000 MHz D-array contours are shown in the fifth column in grey. The green square represents the position of the optical host galaxy.}
\minipage{\textwidth}
        \includegraphics[width=0.195\linewidth] {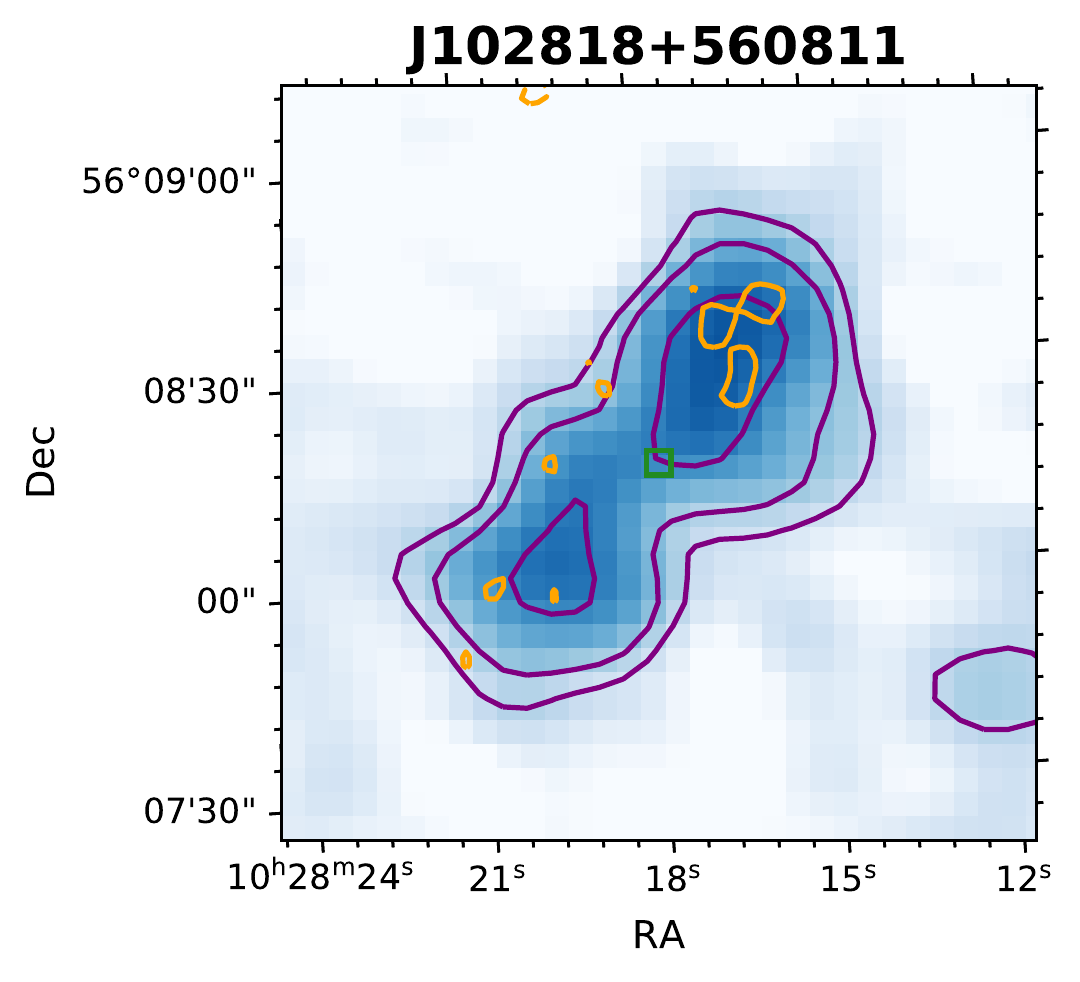}  
        \includegraphics[width=0.195\linewidth] {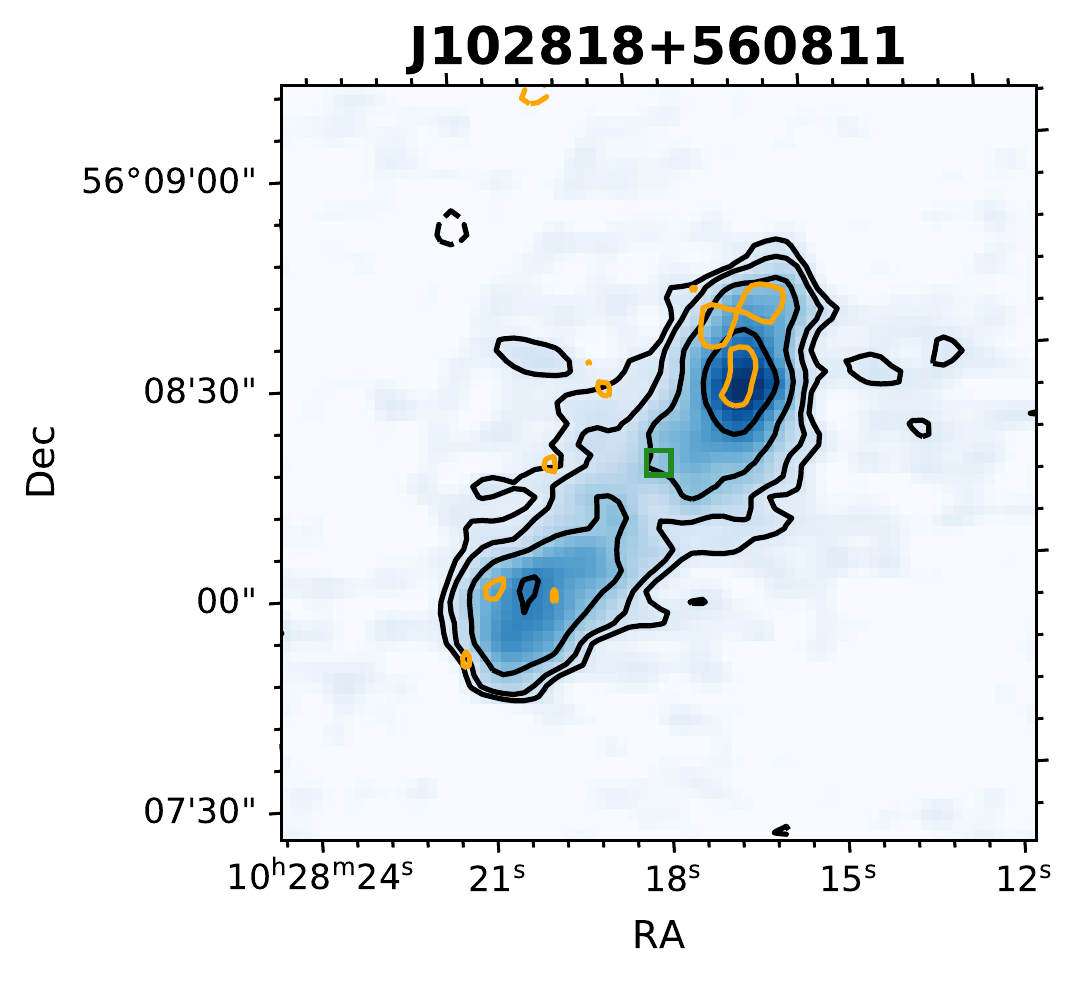} 
        \includegraphics[width=0.195\linewidth] {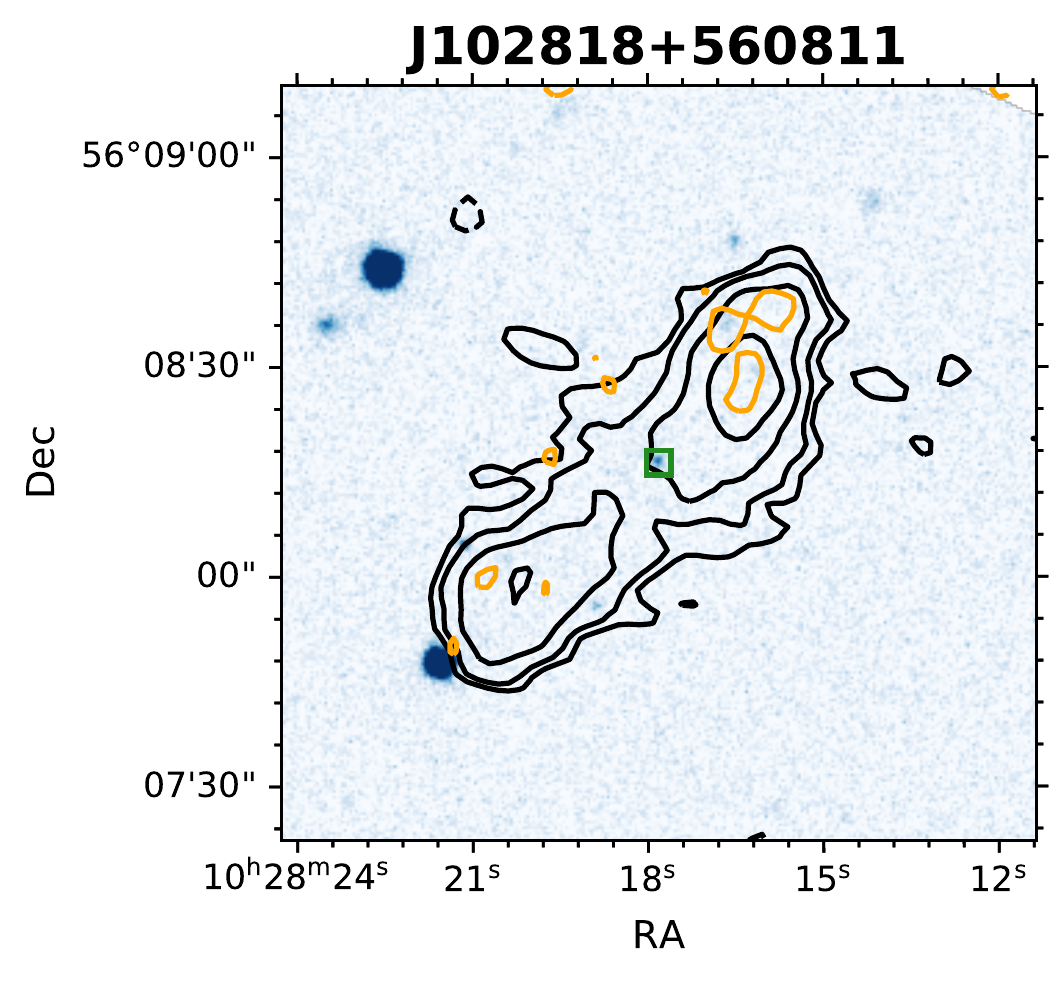}
        \includegraphics[width=0.195\linewidth] {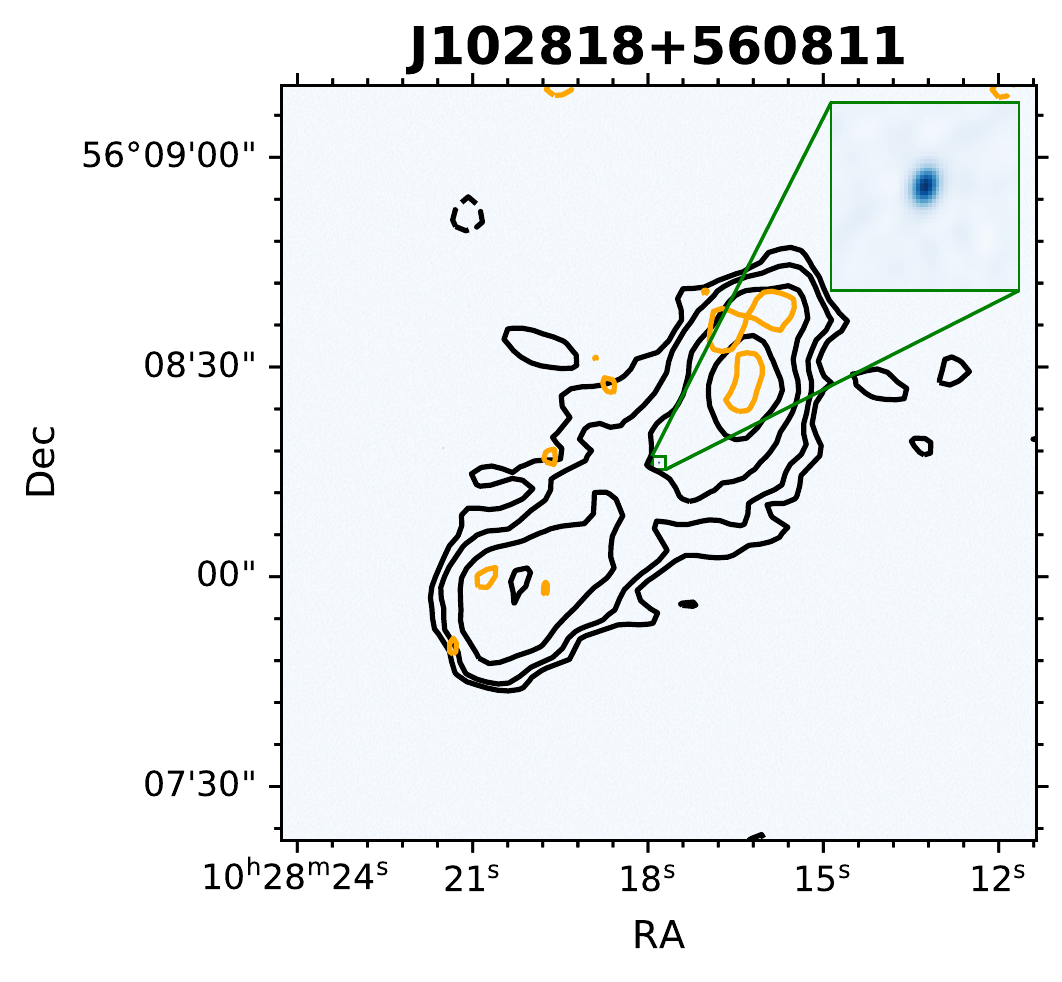}
 	    \includegraphics[width=0.195\linewidth] {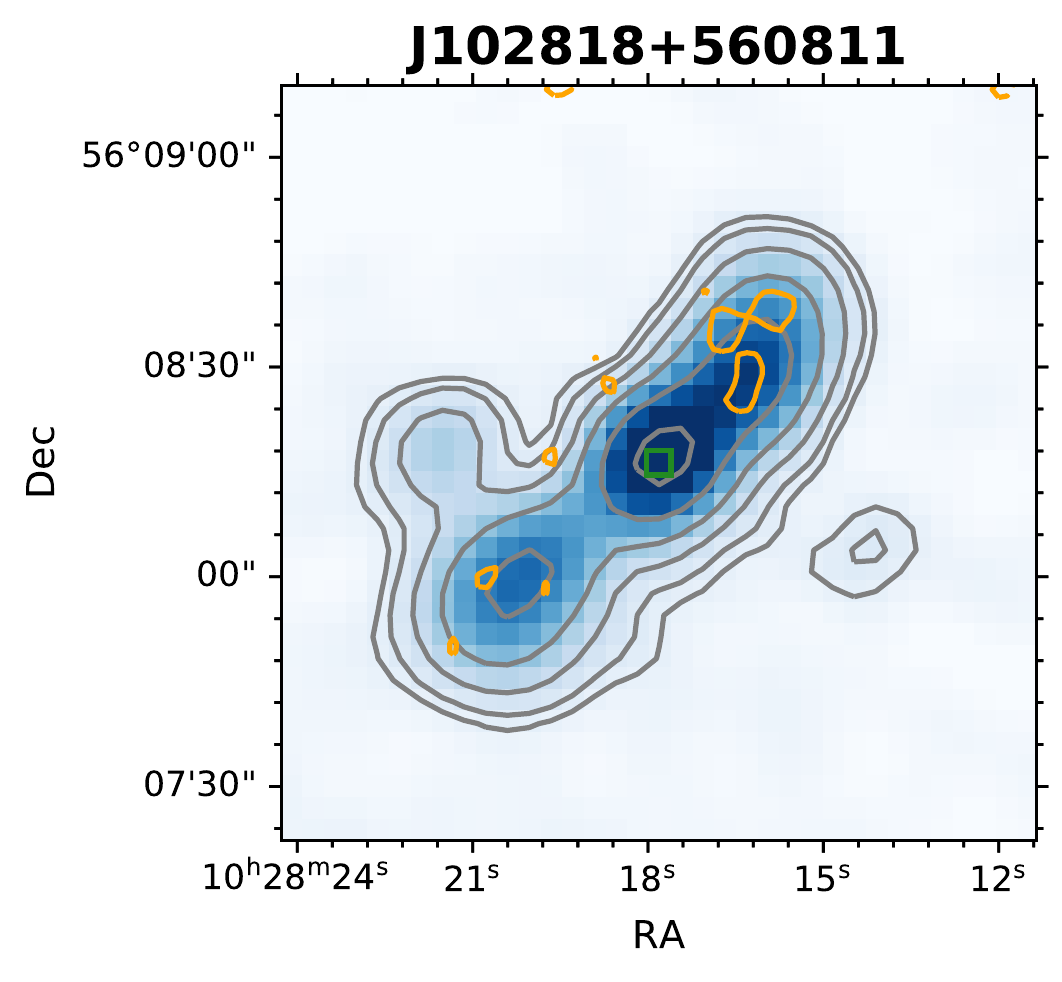}
\endminipage \hfill
\minipage{\textwidth}
        \includegraphics[width=0.195\linewidth] {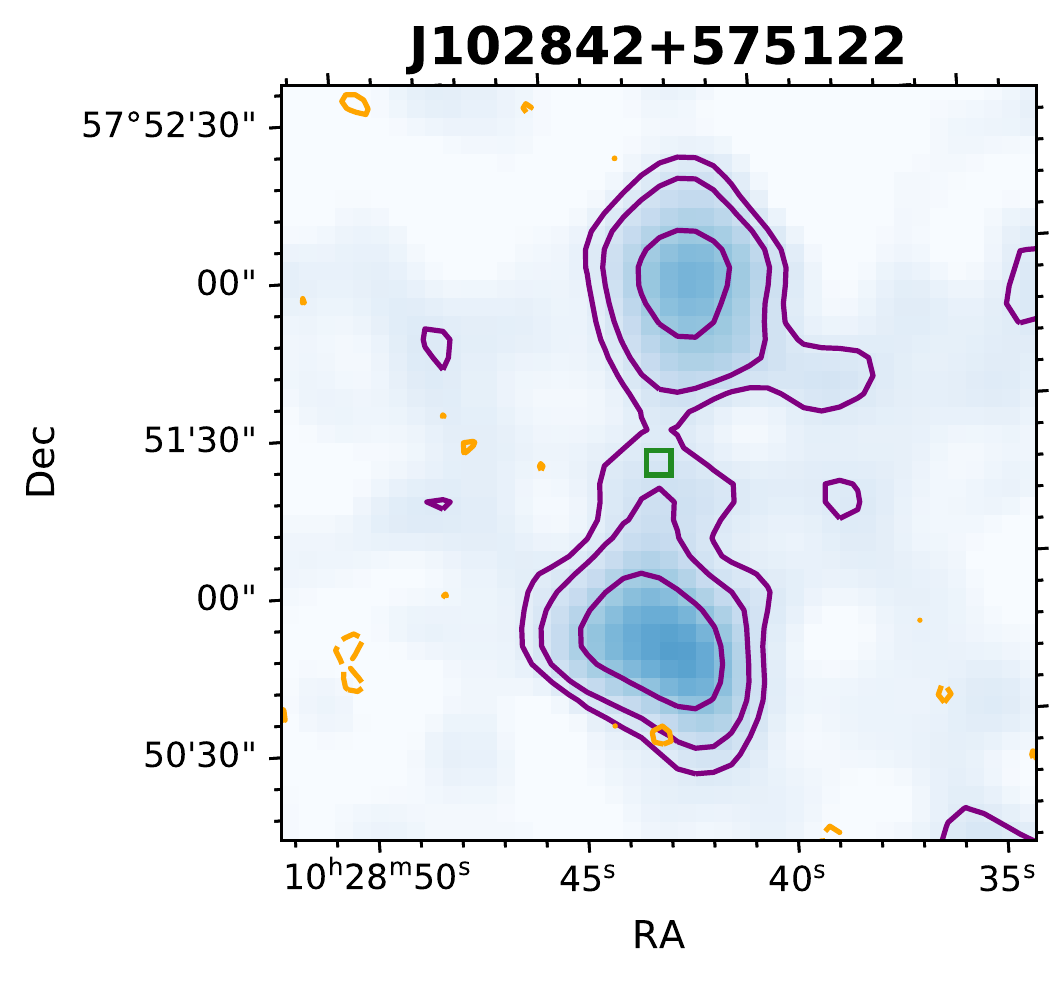}
        \includegraphics[width=0.195\linewidth] {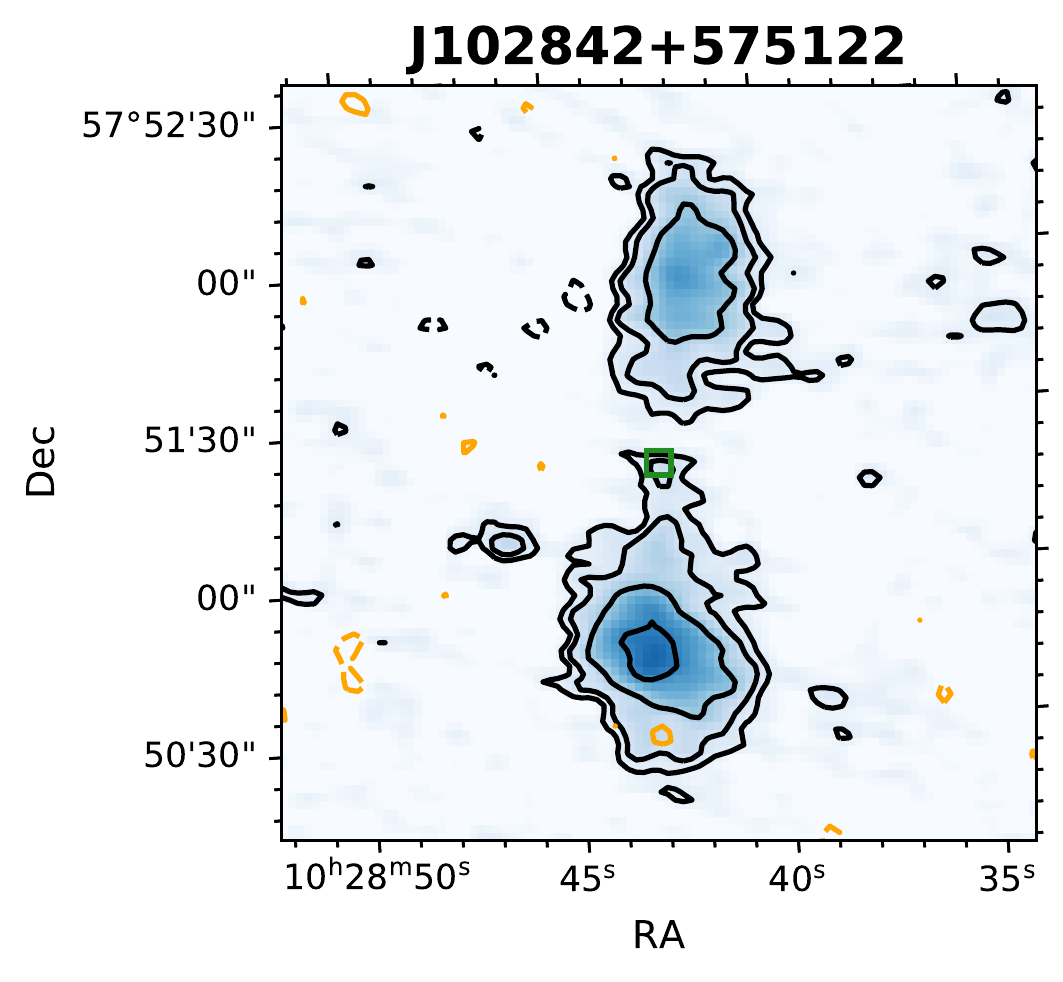} 
        \includegraphics[width=0.195\linewidth] {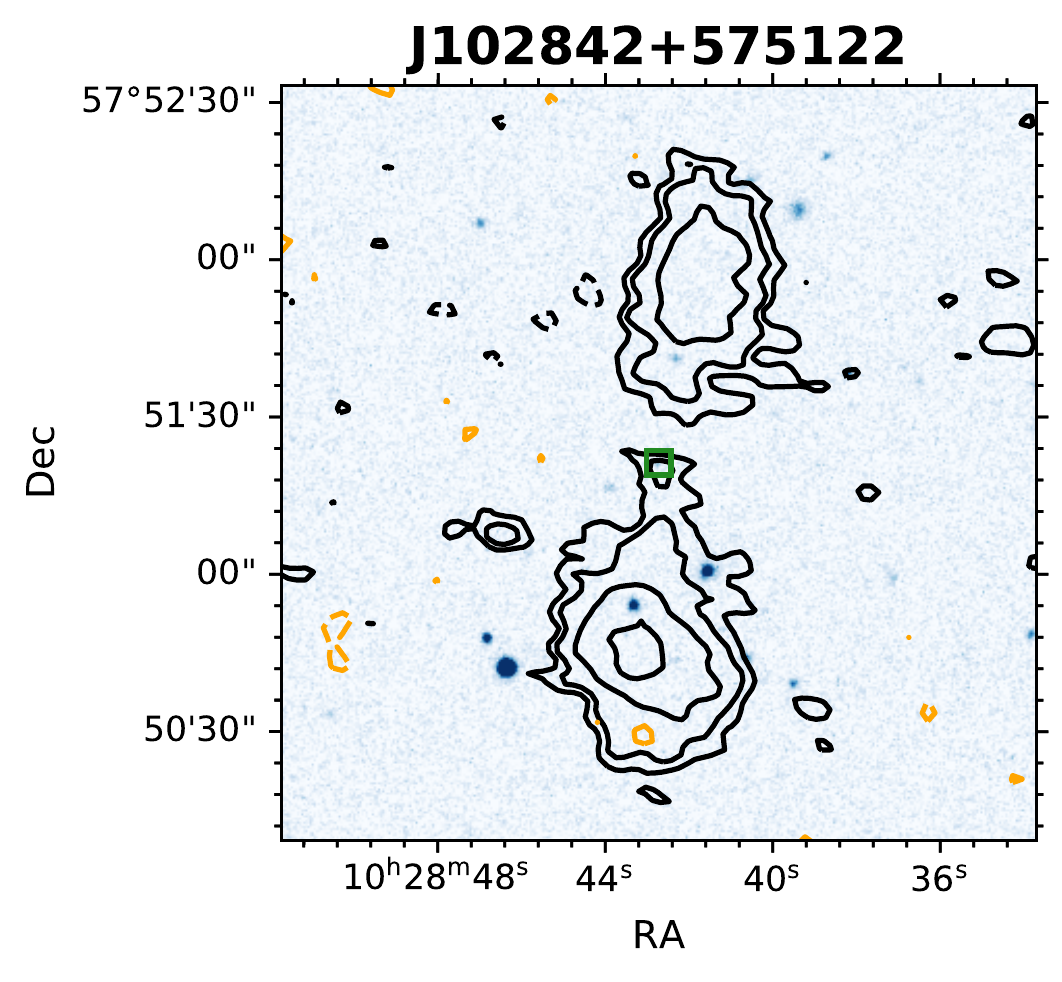}
        \includegraphics[width=0.195\linewidth] {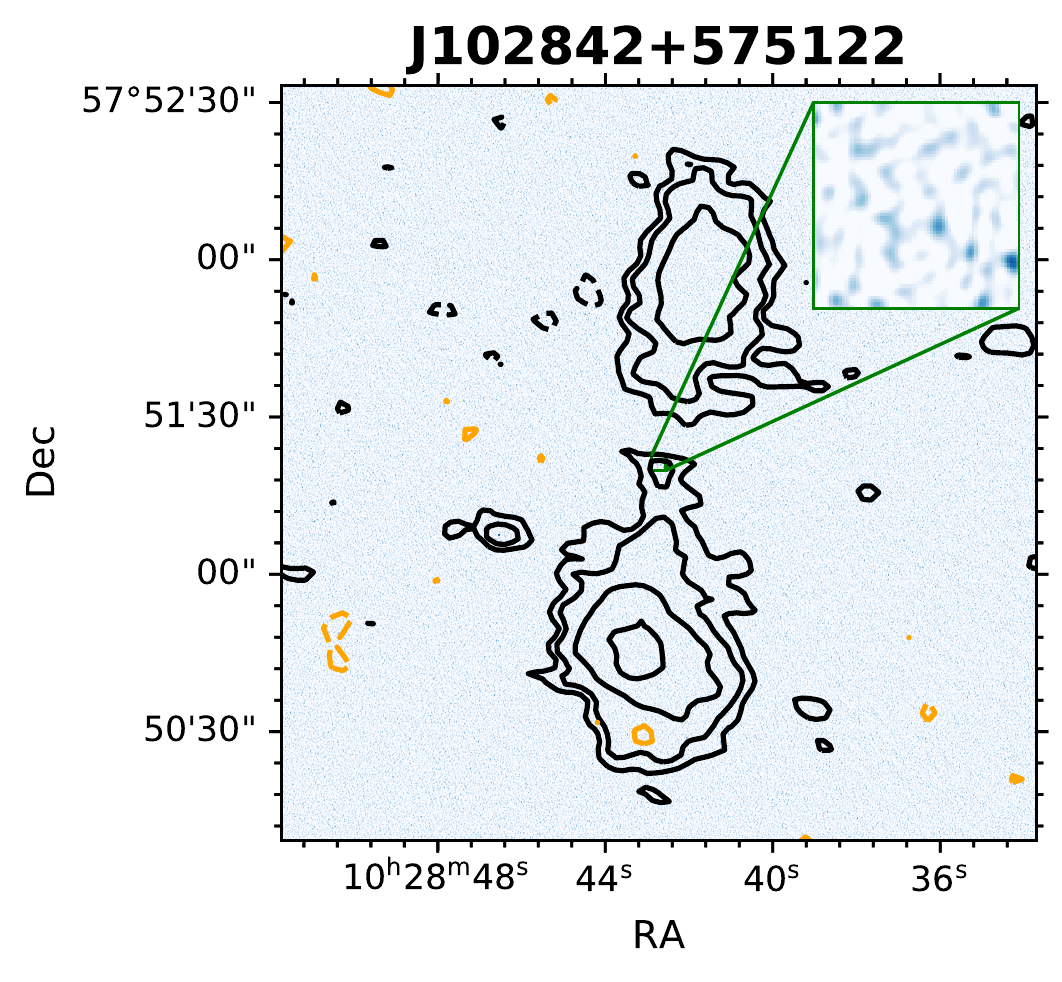}
\endminipage \hfill
\minipage{\textwidth}
        \includegraphics[width=0.195\linewidth] {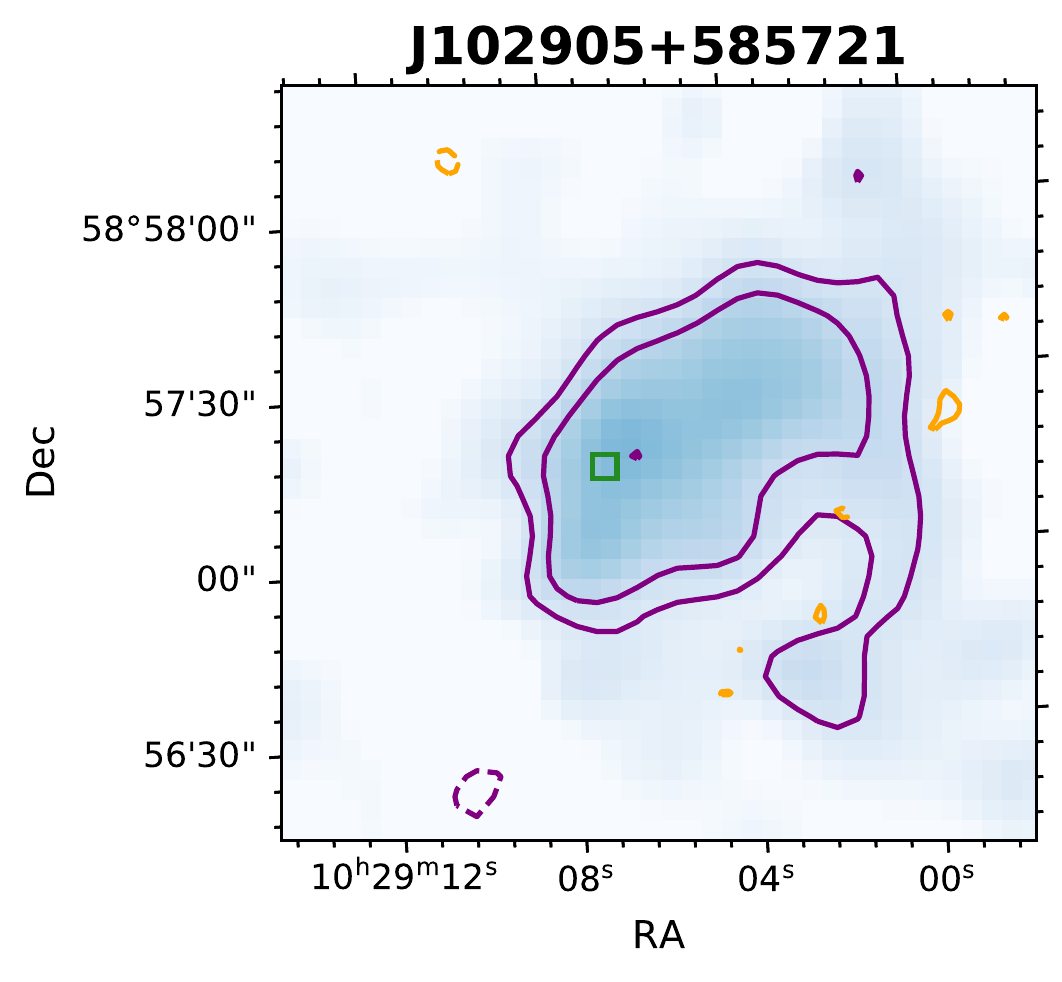}  
        \includegraphics[width=0.195\linewidth] {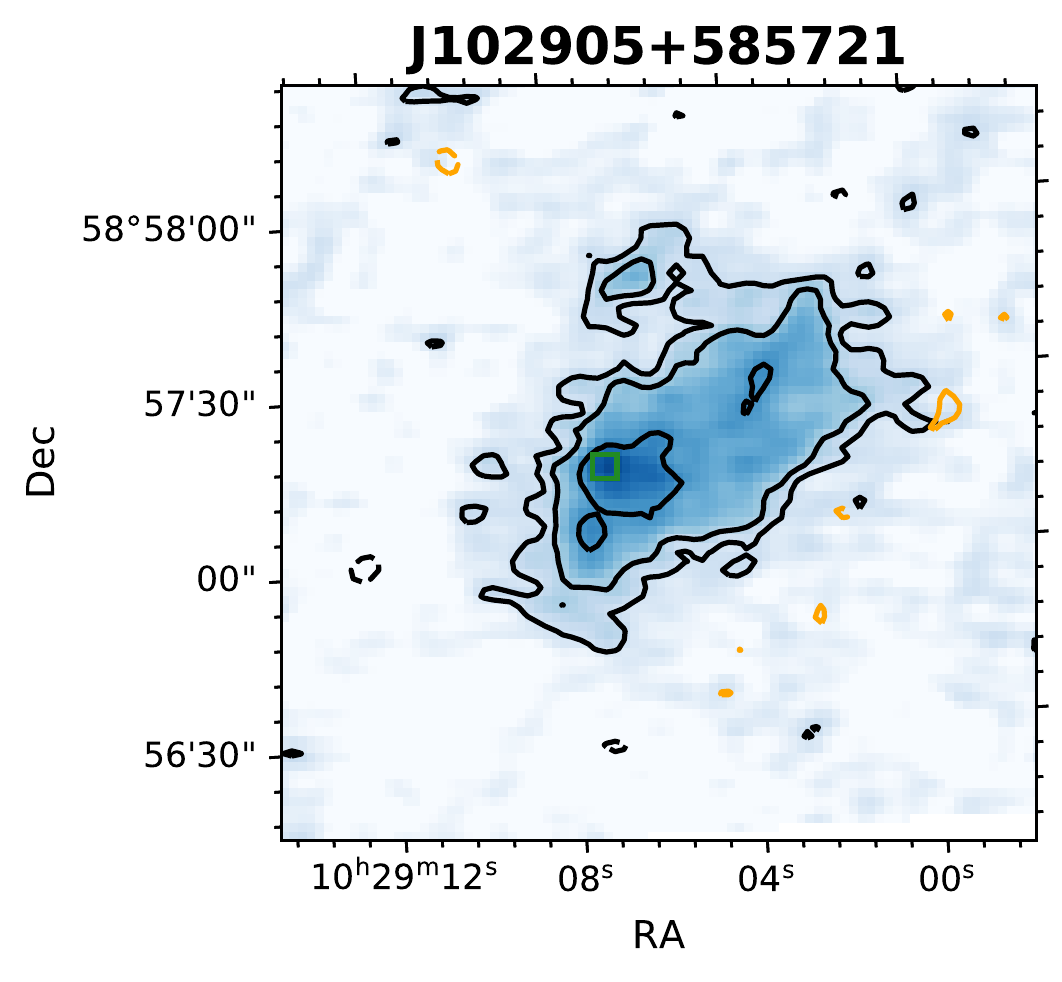} 
        \includegraphics[width=0.195\linewidth] {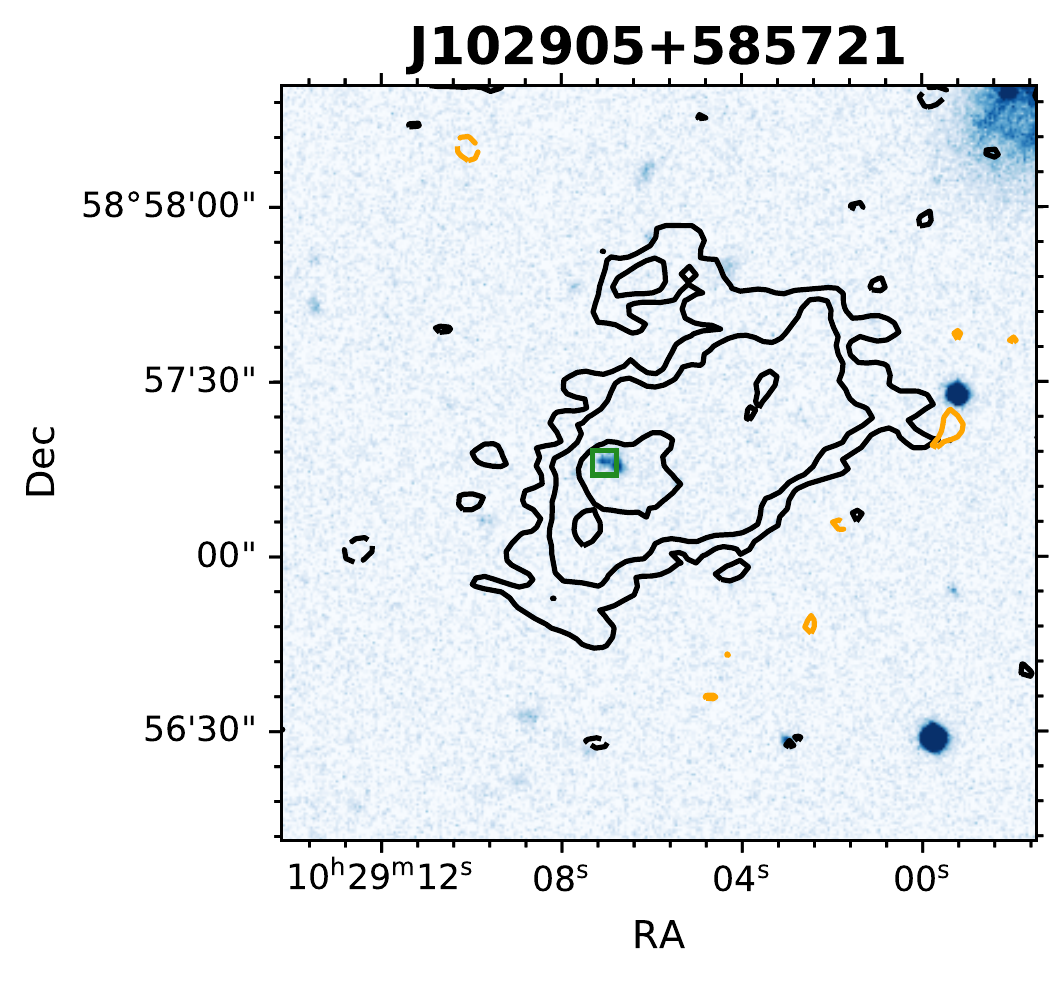}
        \includegraphics[width=0.195\linewidth] {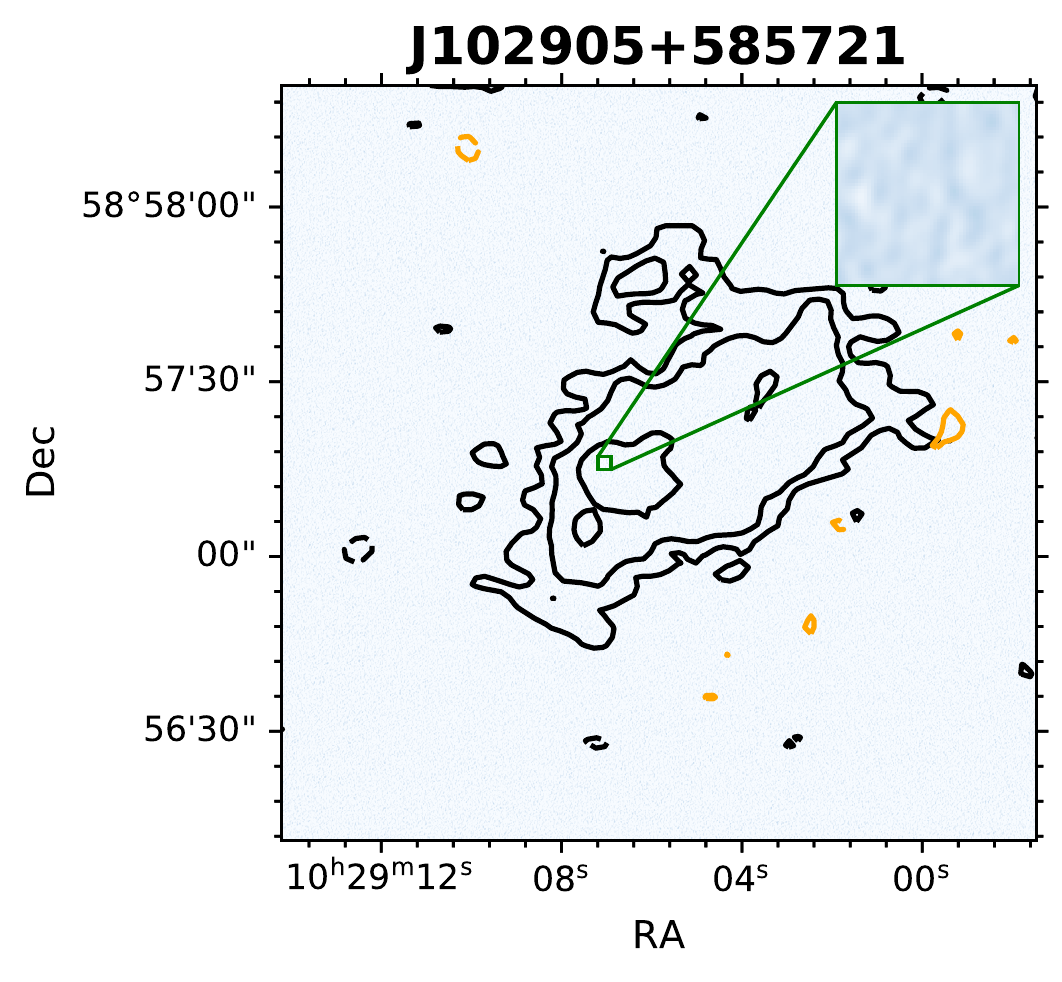}
\endminipage \hfill
\minipage{\textwidth}
        \includegraphics[width=0.195\linewidth] {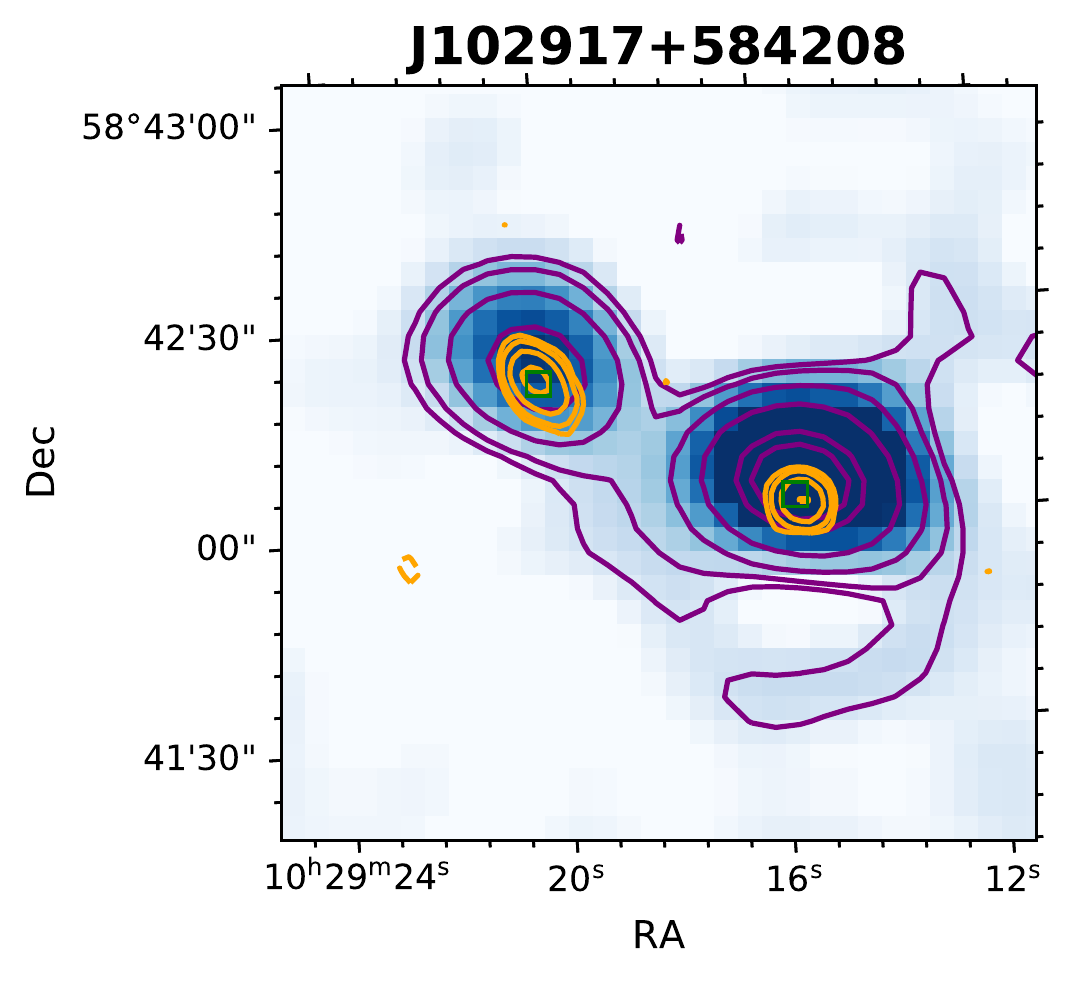}  
        \includegraphics[width=0.195\linewidth] {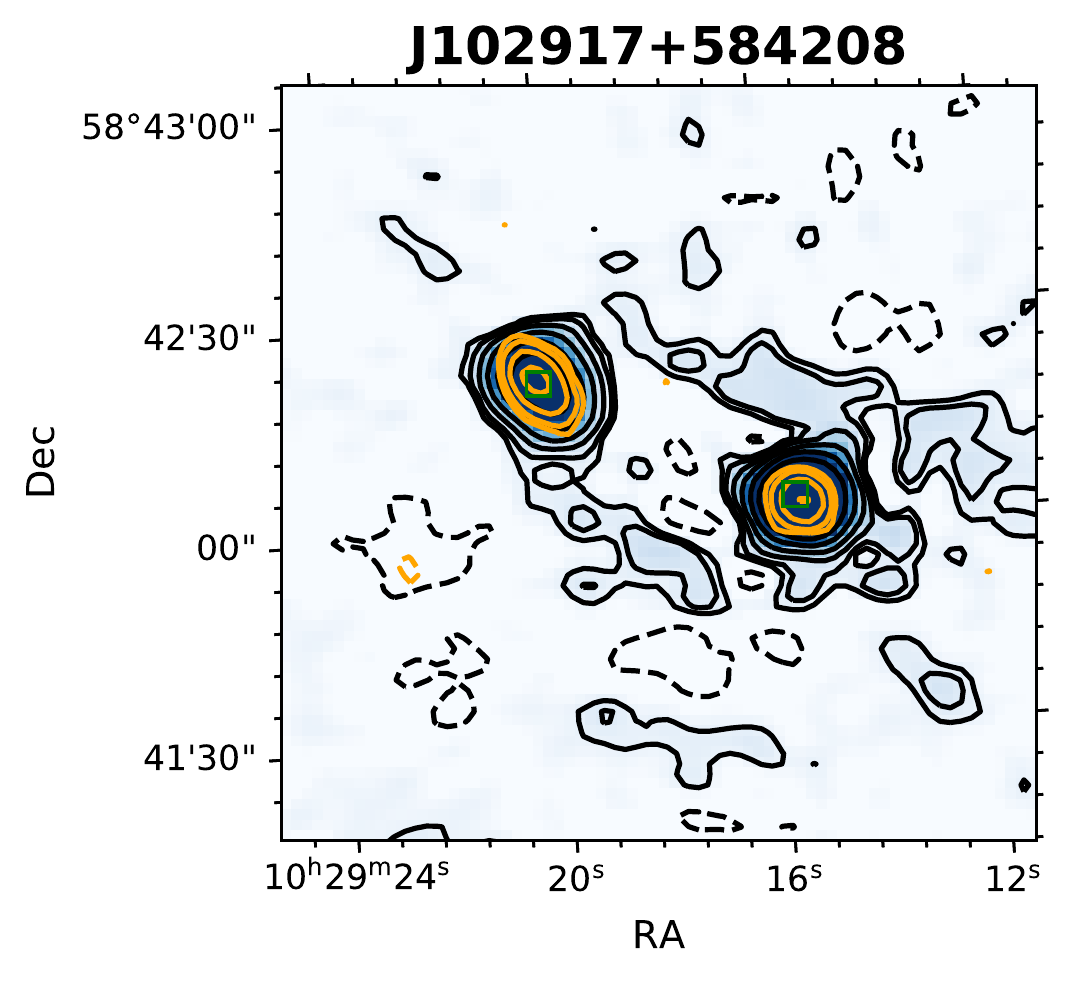} 
        \includegraphics[width=0.195\linewidth] {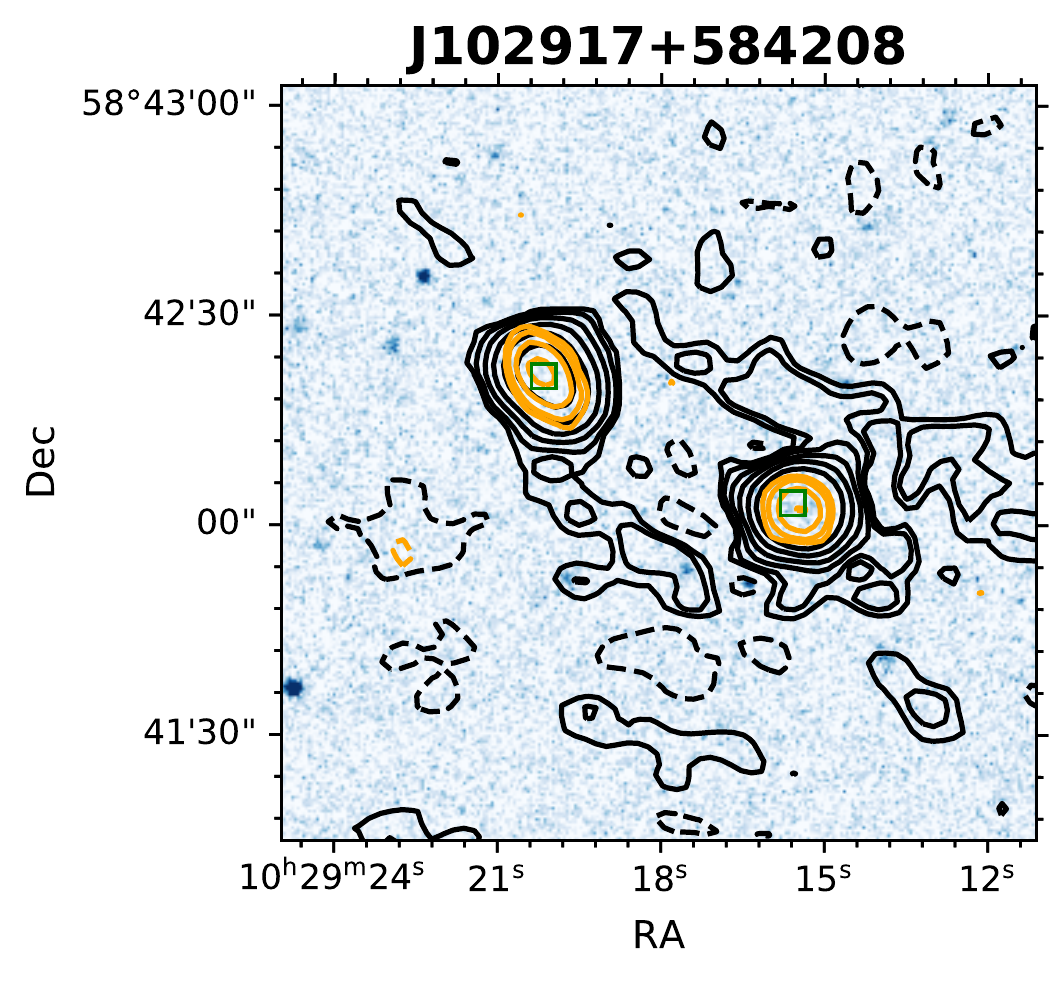}
 	    \includegraphics[width=0.195\linewidth] {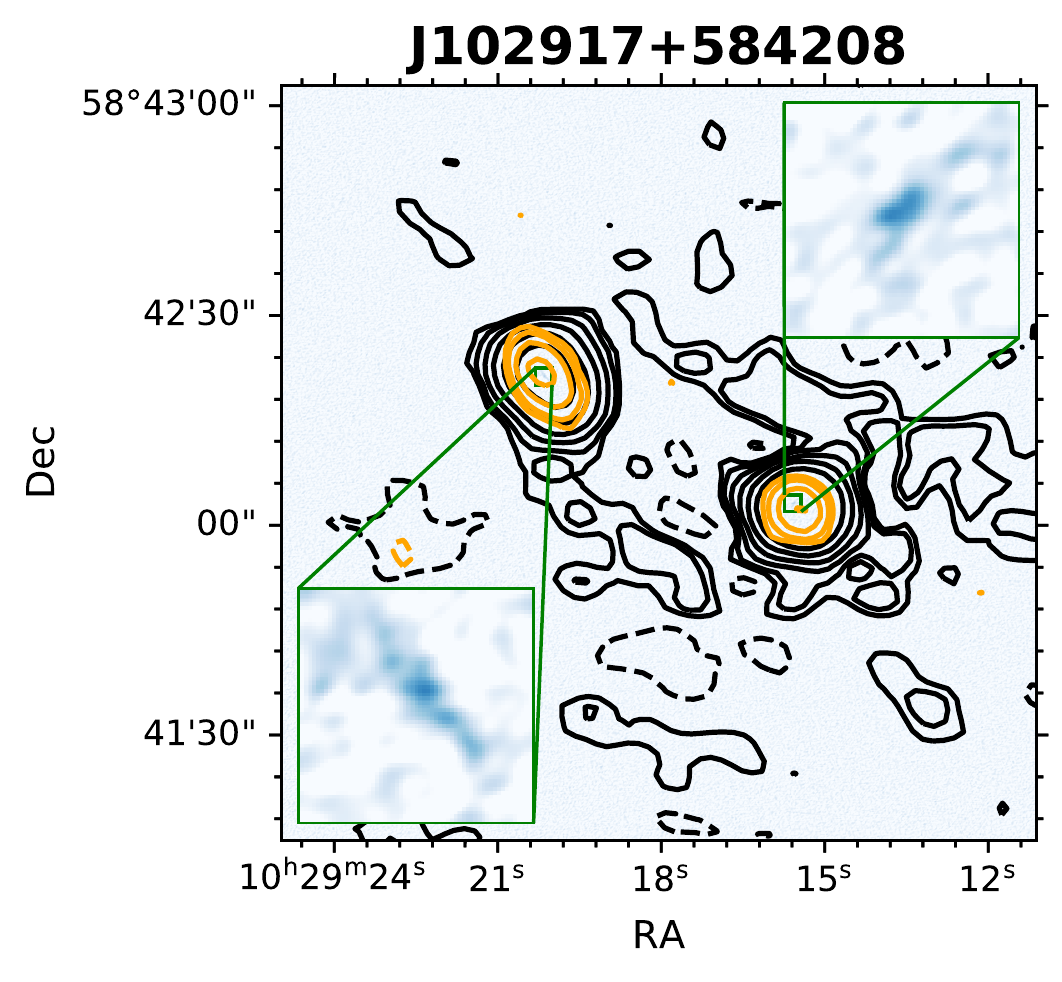}
\endminipage \hfill
\minipage{\textwidth}
        \includegraphics[width=0.195\linewidth] {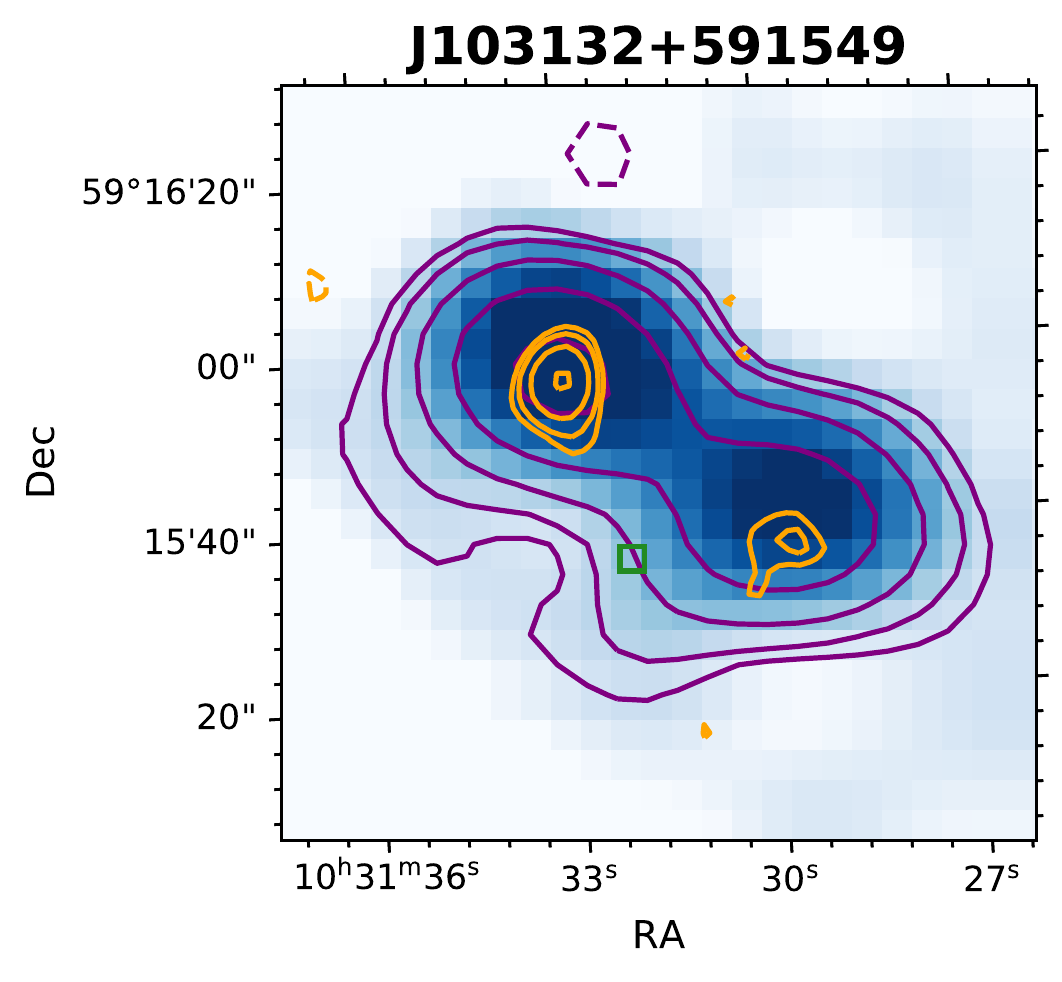}  
        \includegraphics[width=0.195\linewidth] {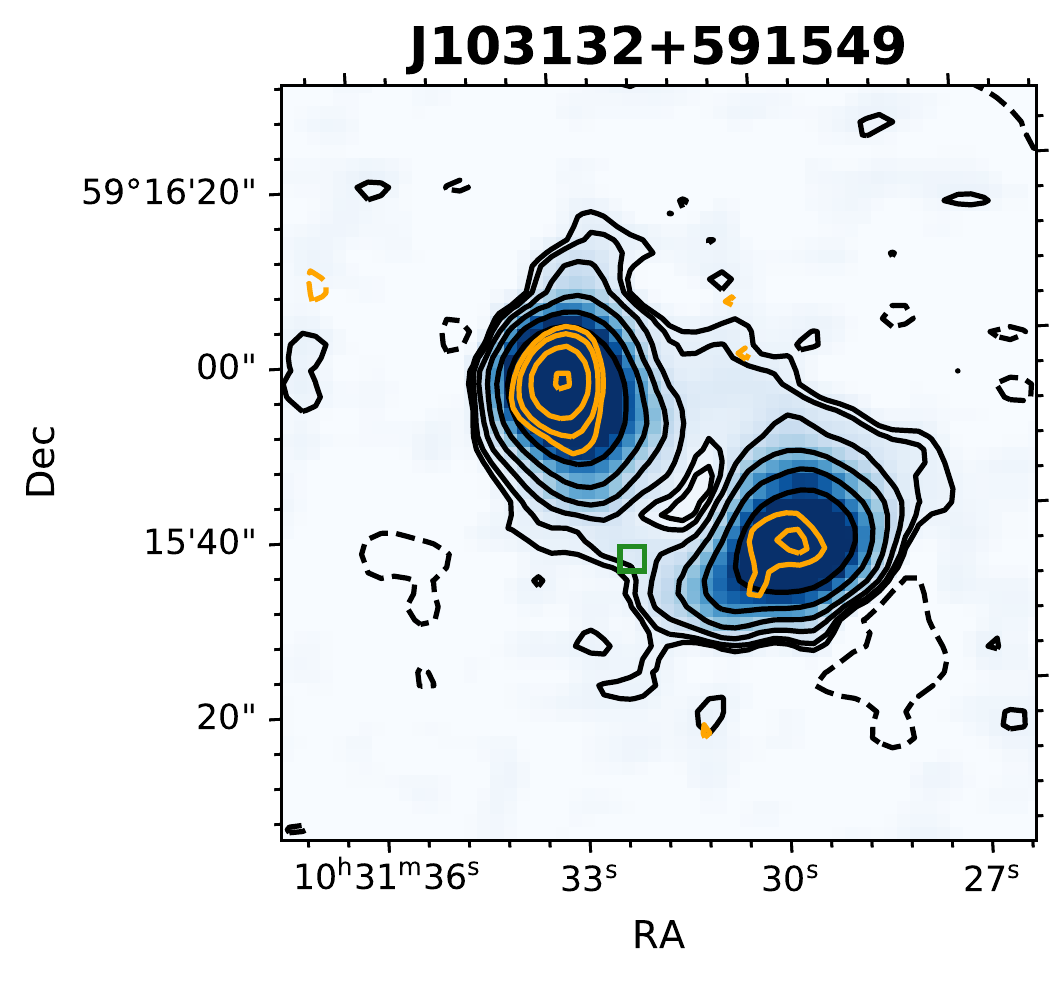} 
        \includegraphics[width=0.195\linewidth] {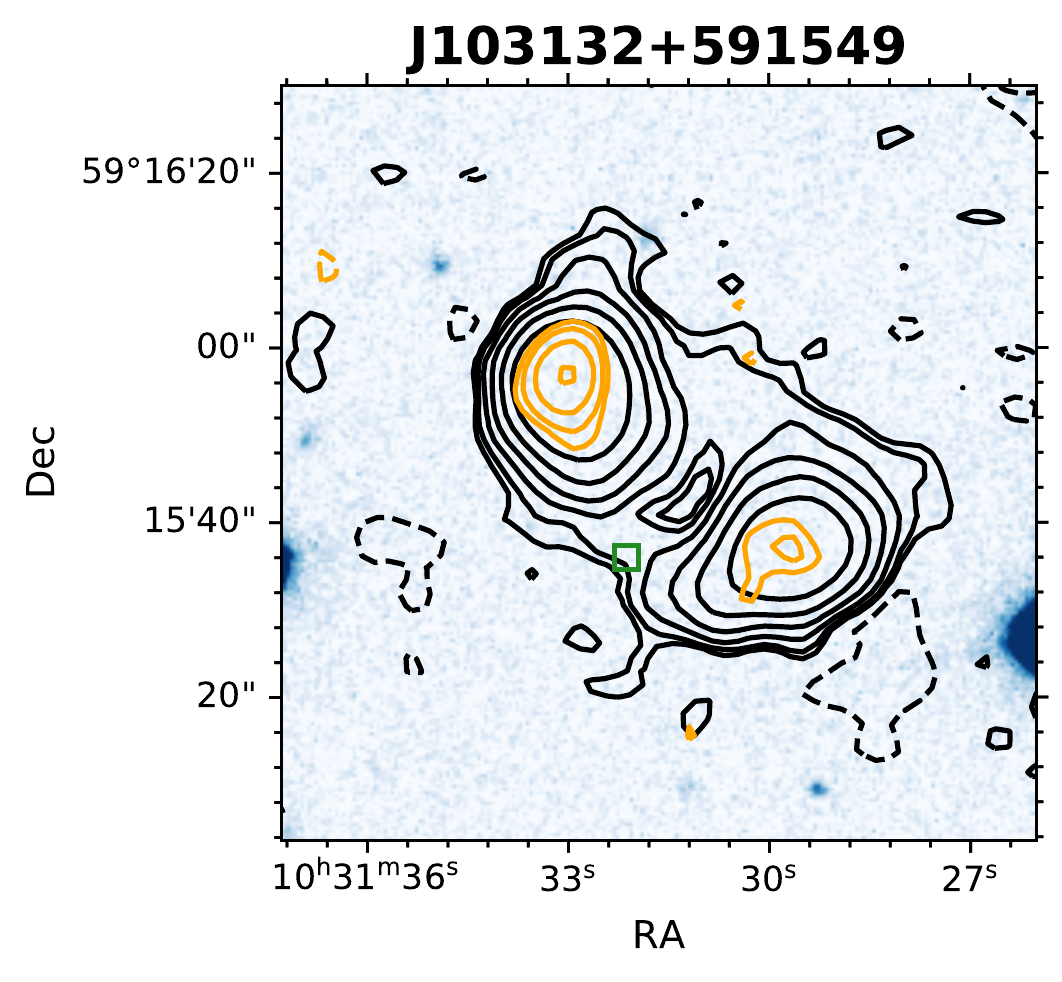}
        \includegraphics[width=0.195\linewidth] {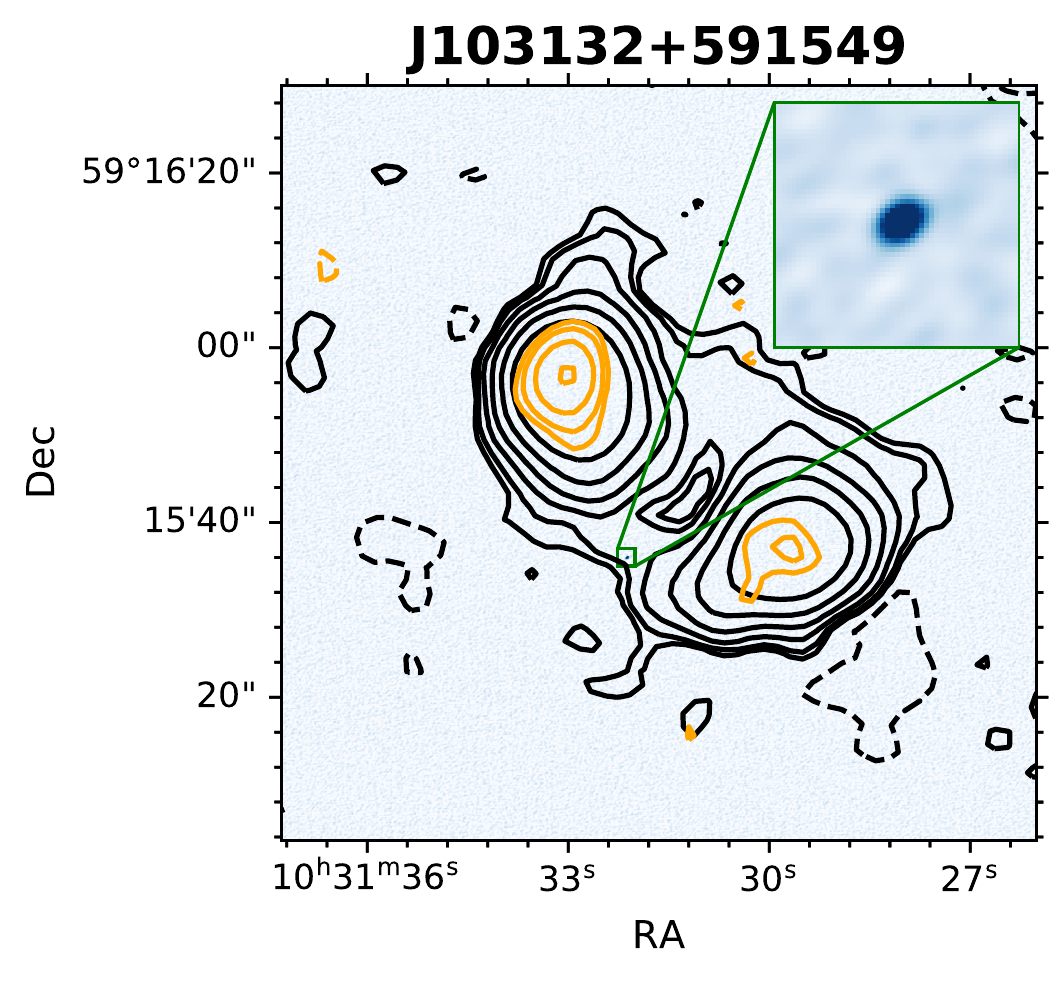}
\endminipage \hfill
\minipage{\textwidth}
        \includegraphics[width=0.195\linewidth] {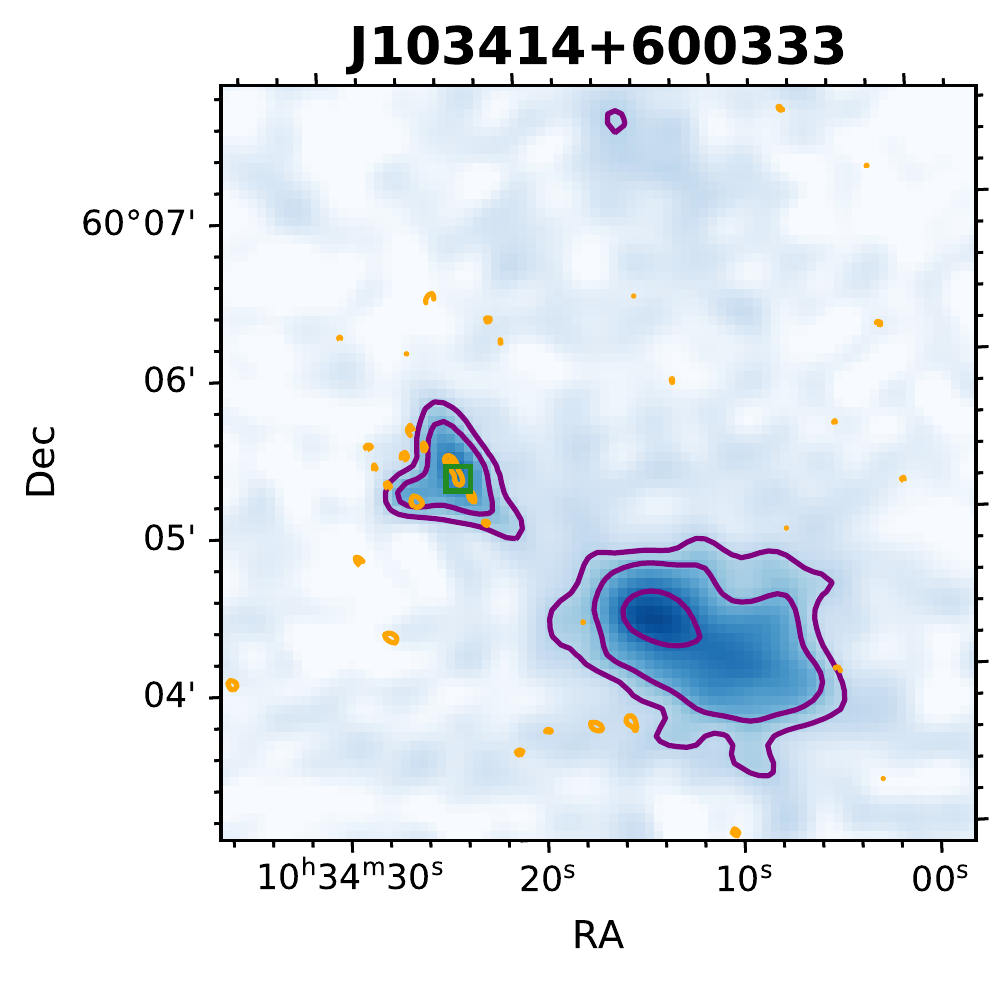}  
        \includegraphics[width=0.195\linewidth] {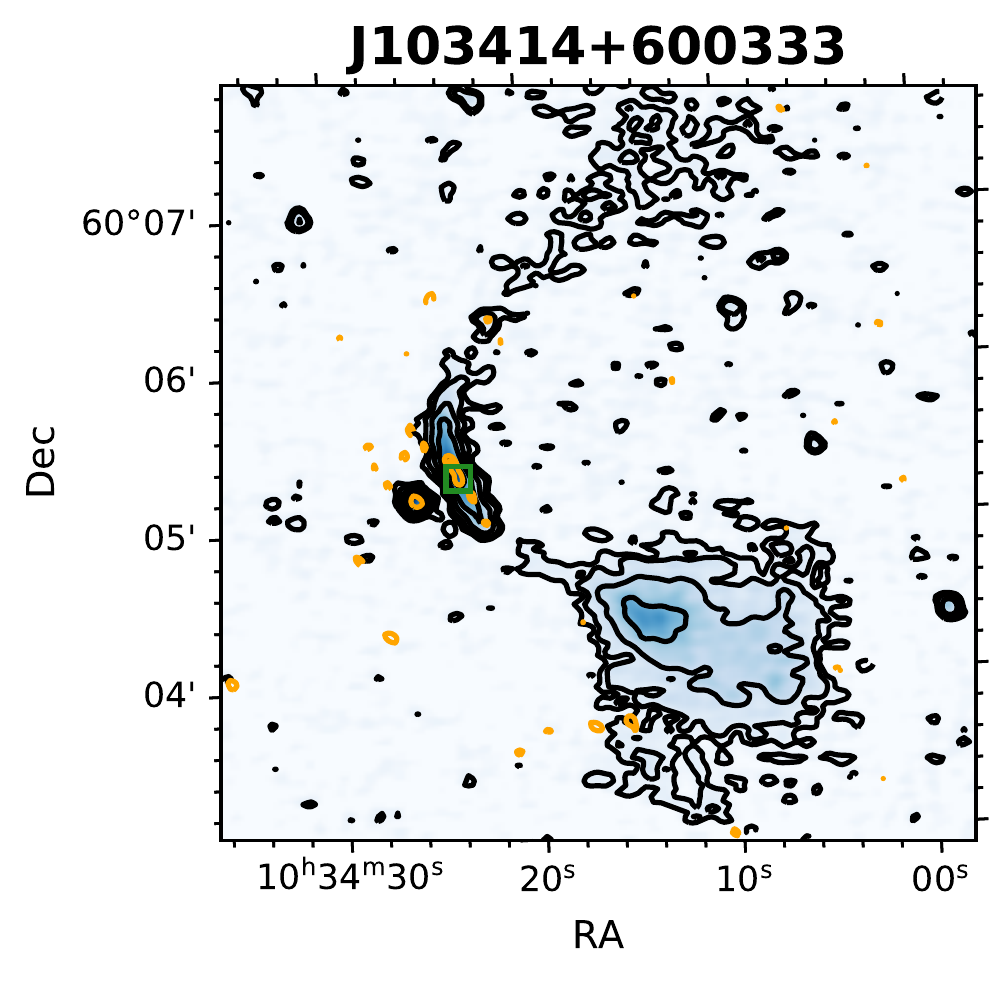}
\endminipage \hfill
\label{fig:remnants1}
\end{figure*}

\begin{figure*} [!h]\ContinuedFloat
\caption{continued}
\minipage{\textwidth}
        \includegraphics[width=0.195\linewidth] {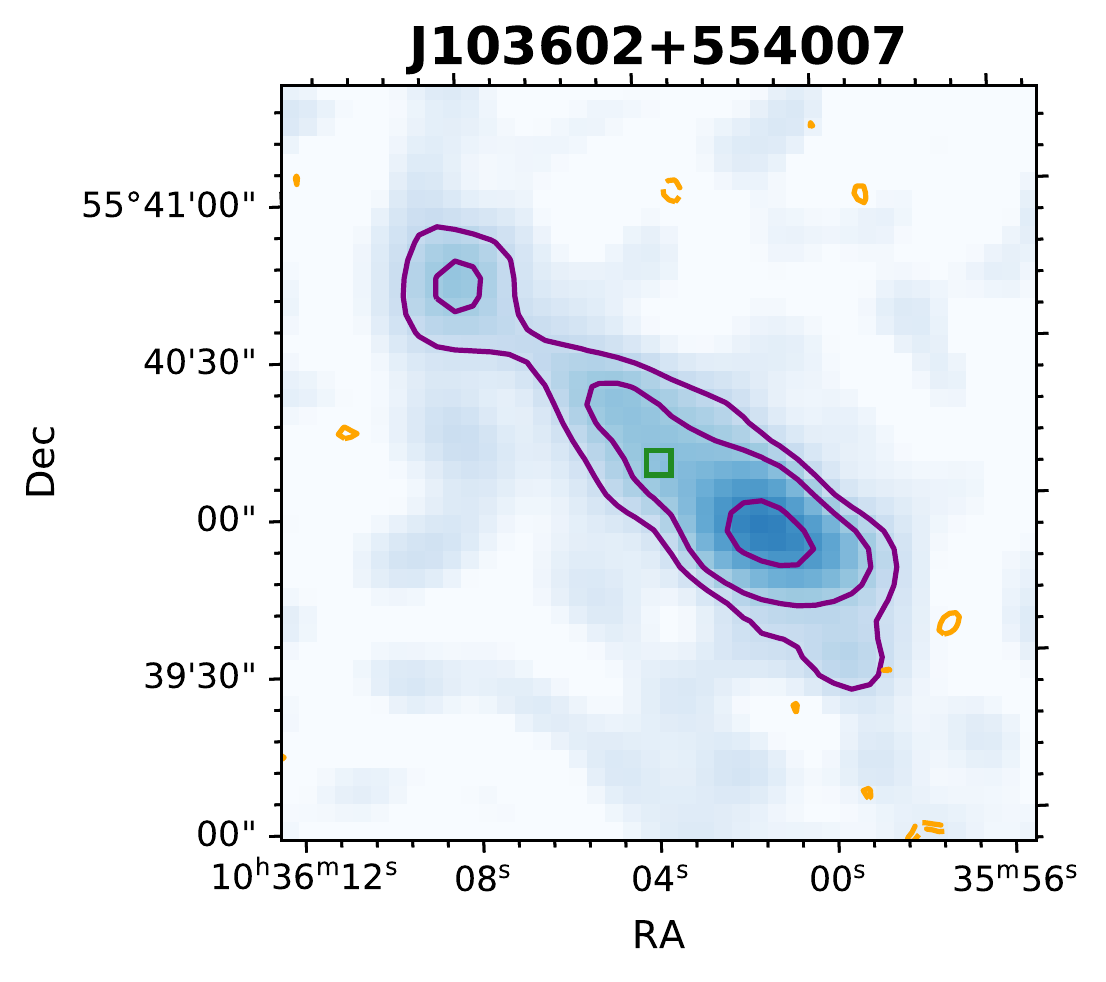}  
        \includegraphics[width=0.195\linewidth] {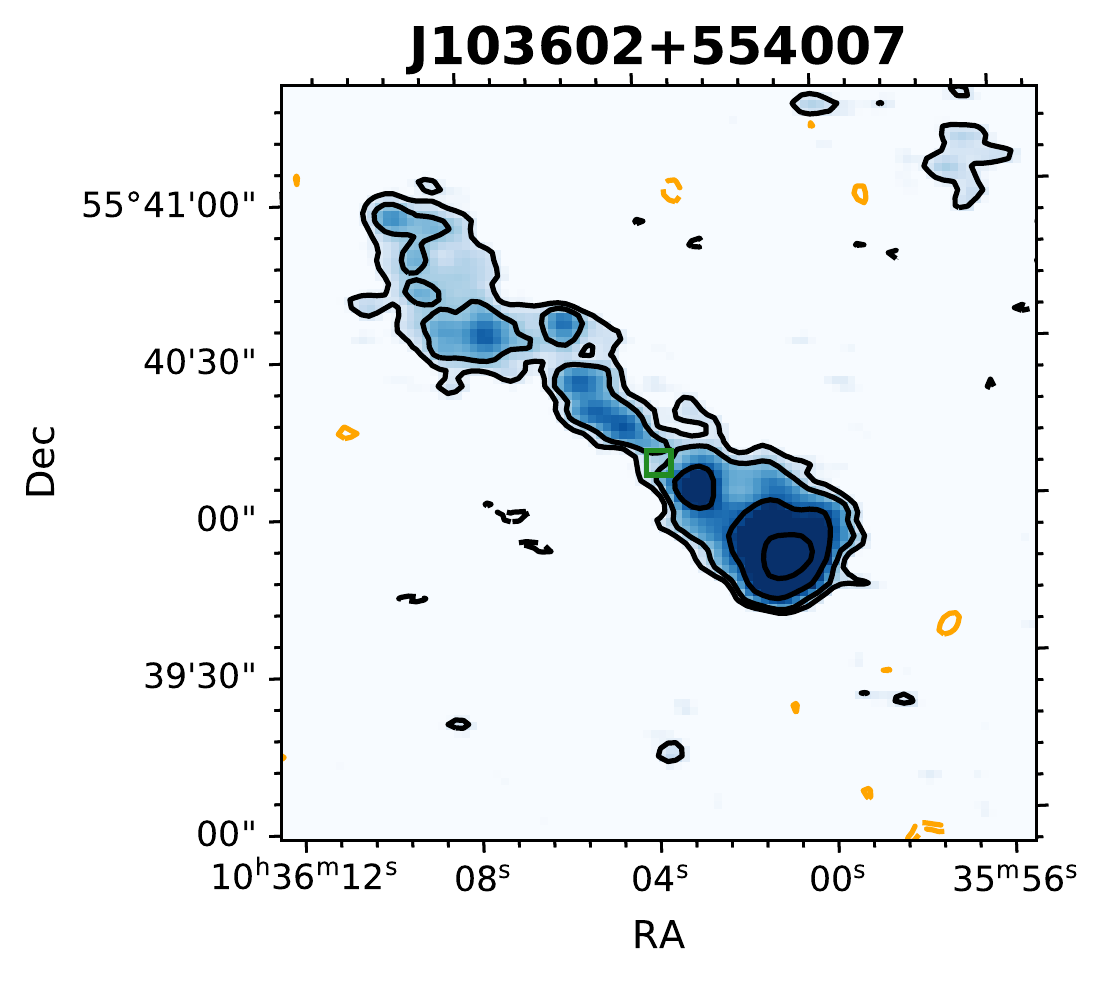} 
        \includegraphics[width=0.195\linewidth] {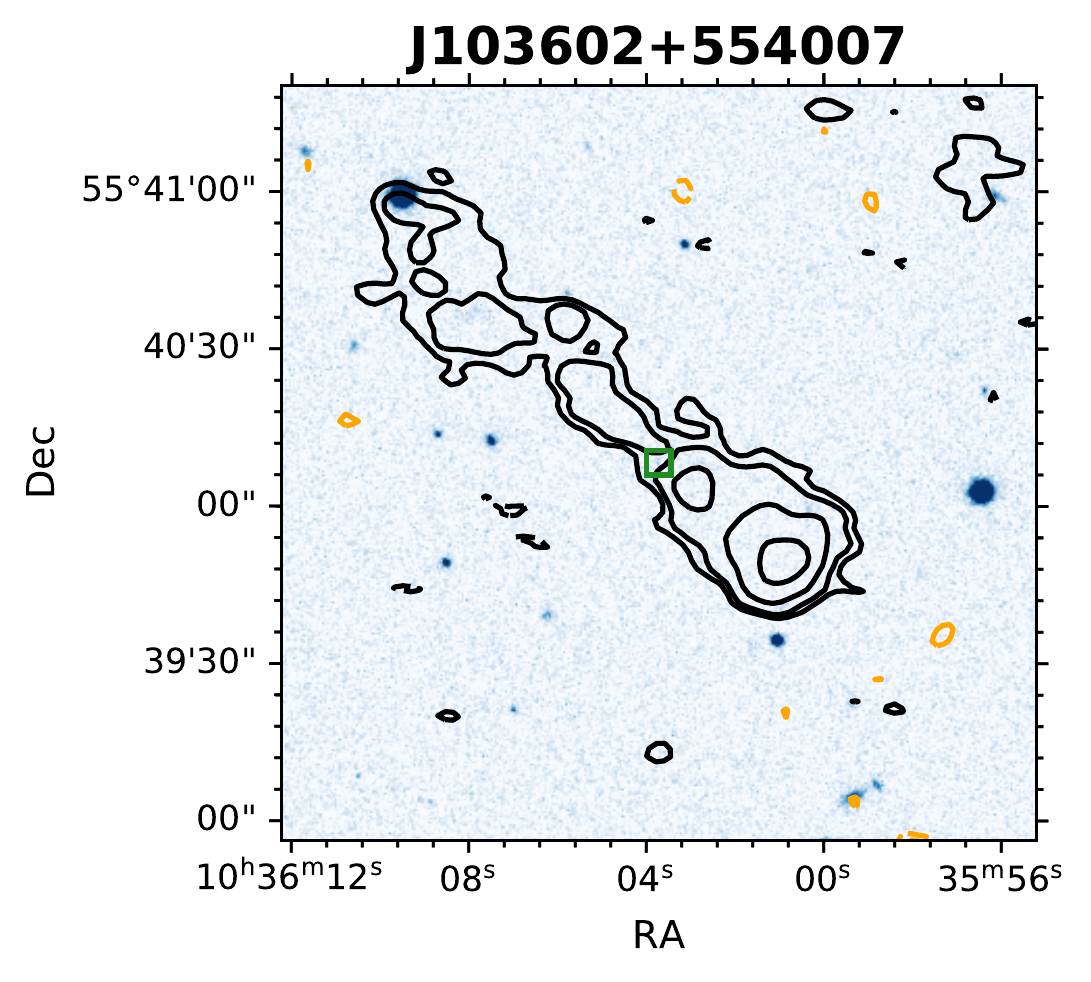}
        \includegraphics[width=0.195\linewidth] {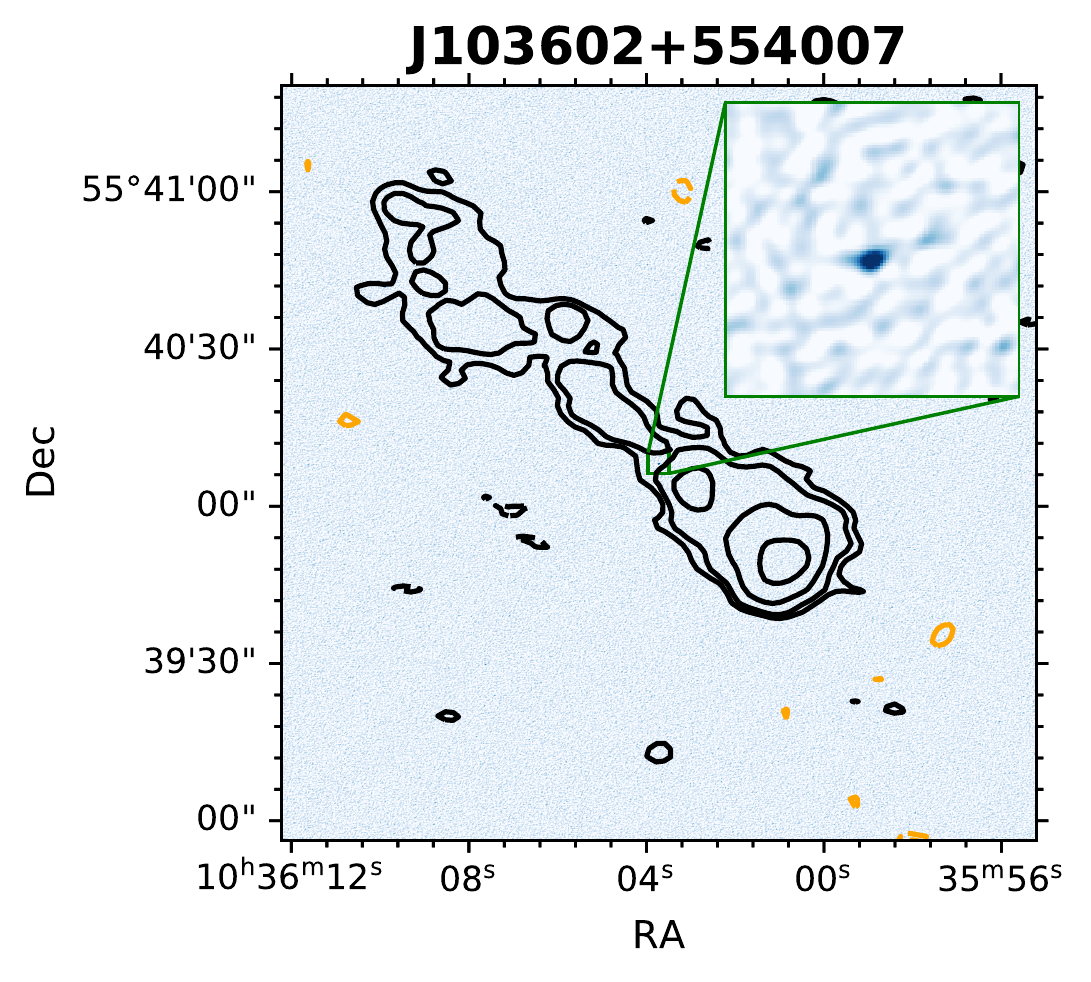}
\endminipage \hfill
\minipage{\textwidth}
        \includegraphics[width=0.195\linewidth] {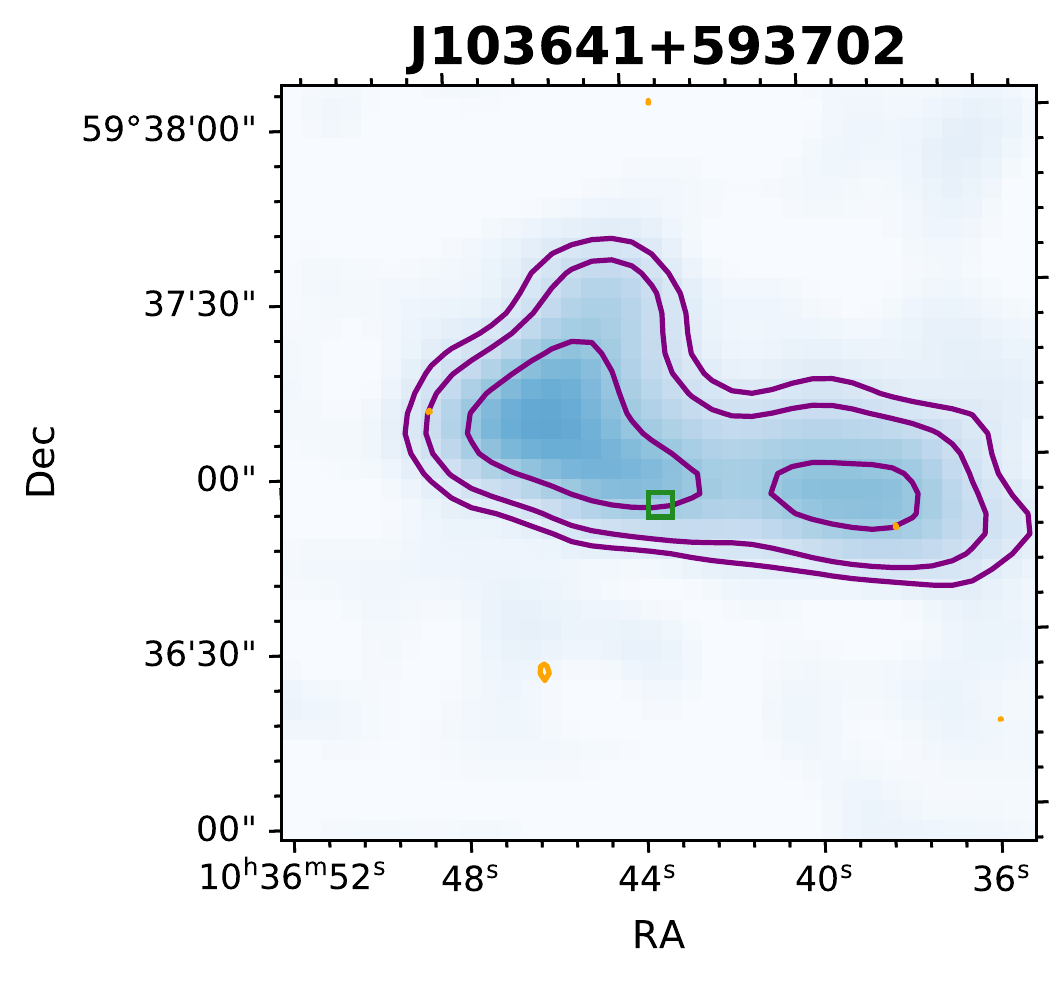} 
       \includegraphics[width=0.195\linewidth]
        {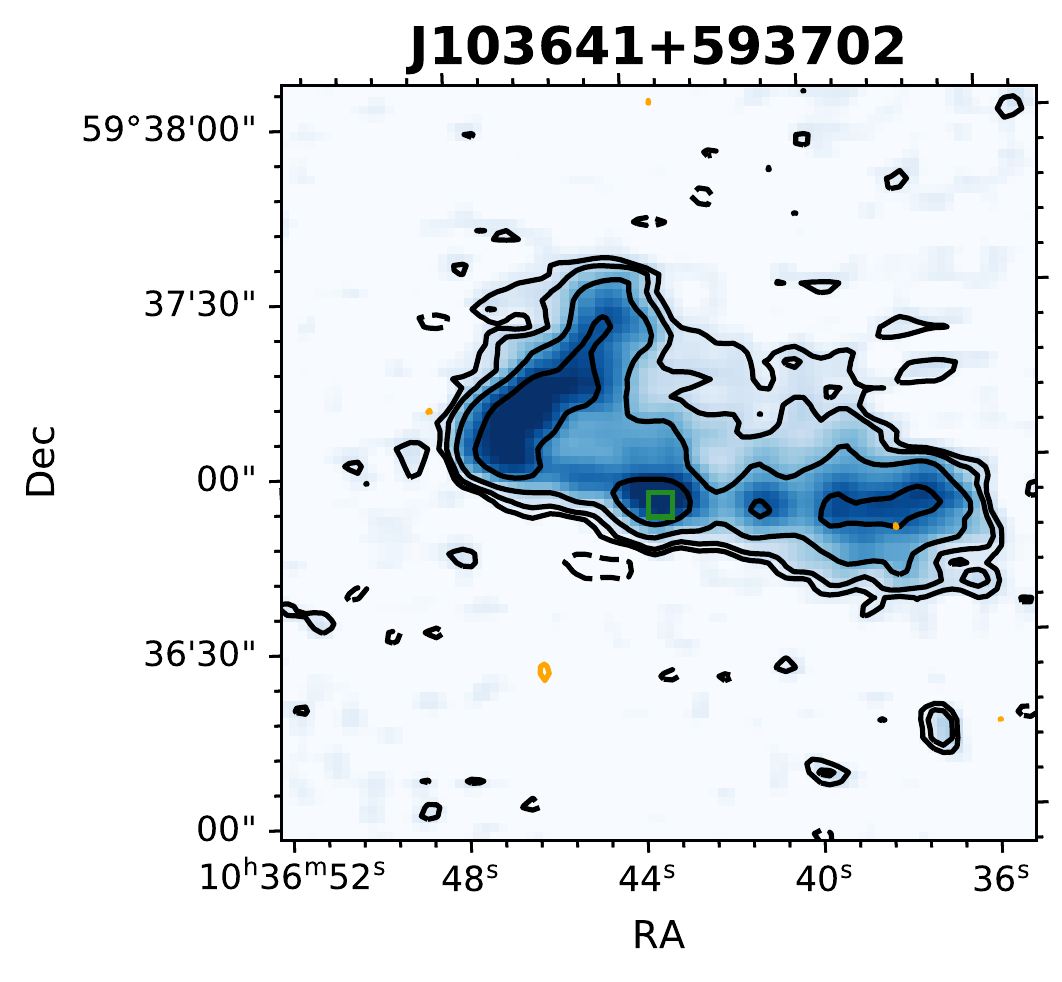} 
        \includegraphics[width=0.195\linewidth] {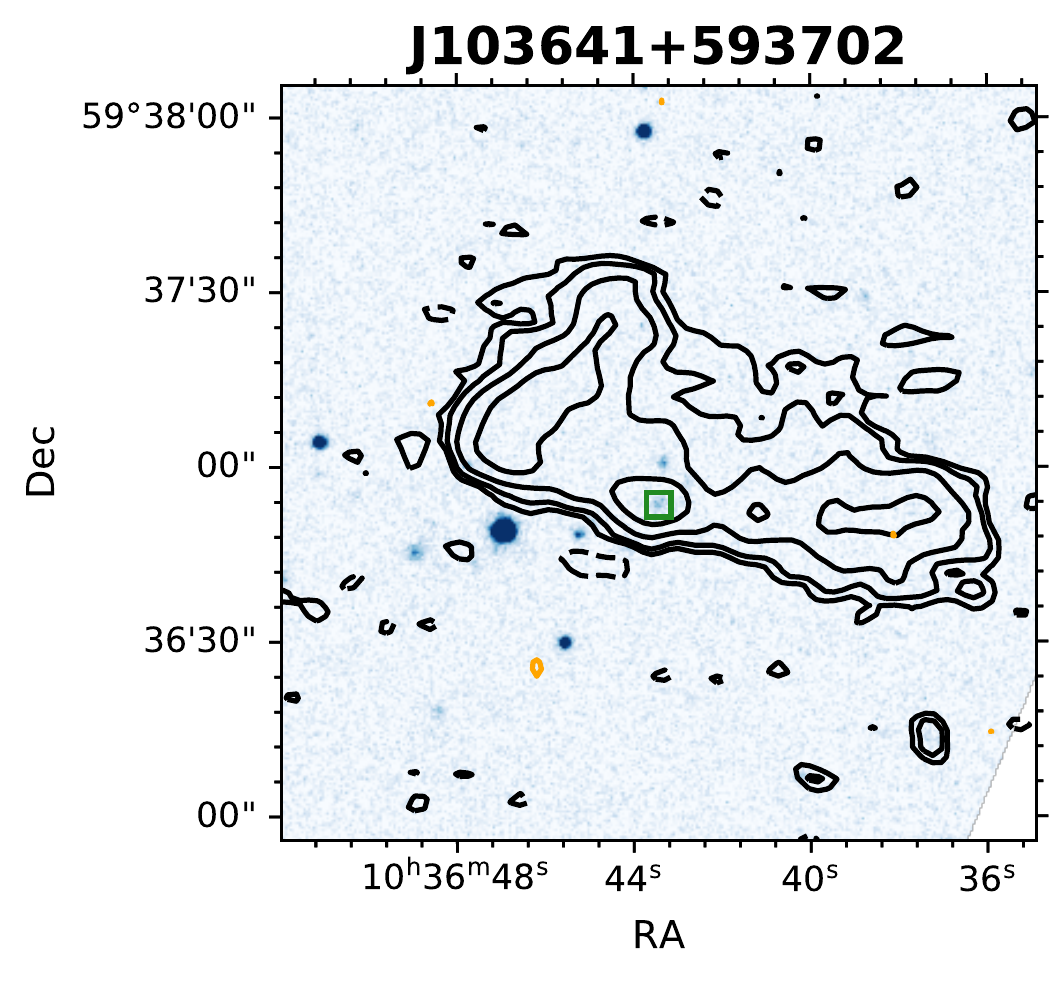}
        \includegraphics[width=0.195\linewidth] {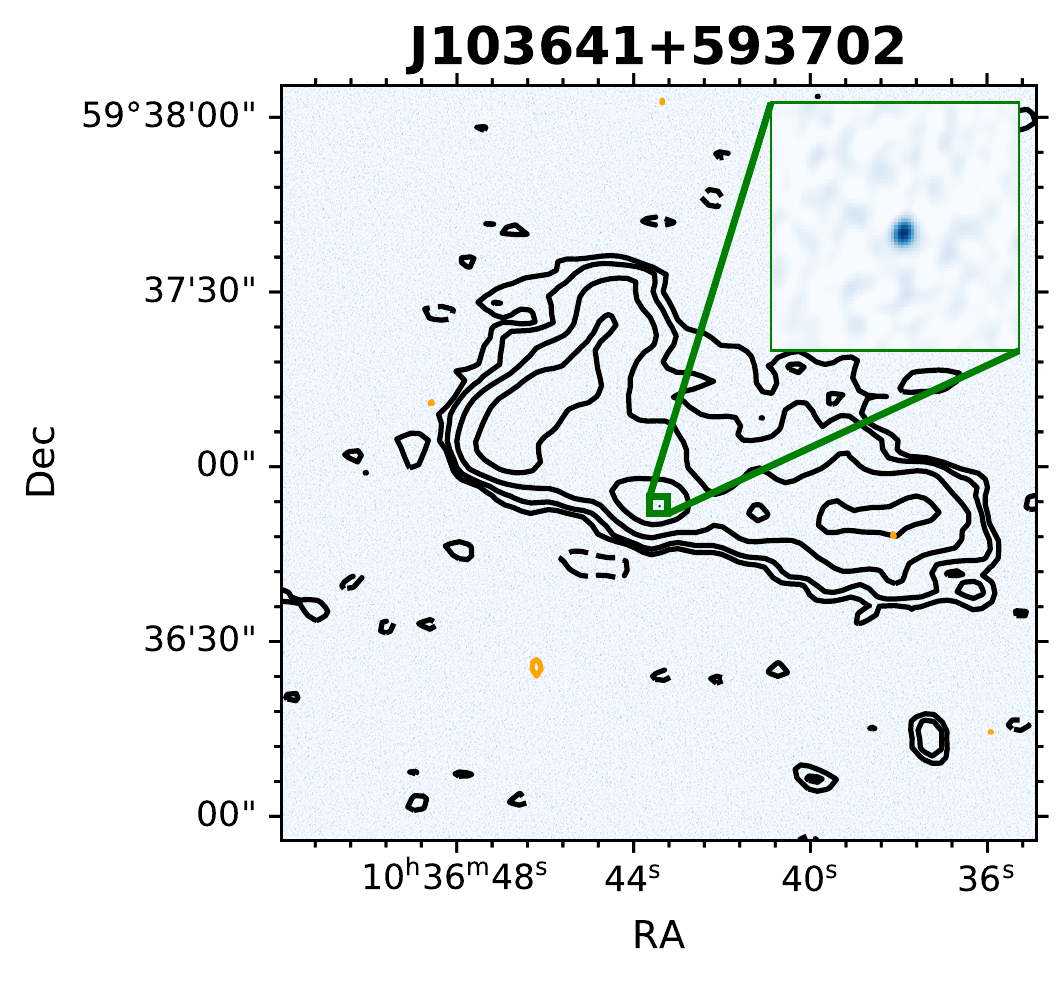}
 	    \includegraphics[width=0.195\linewidth]
 	    {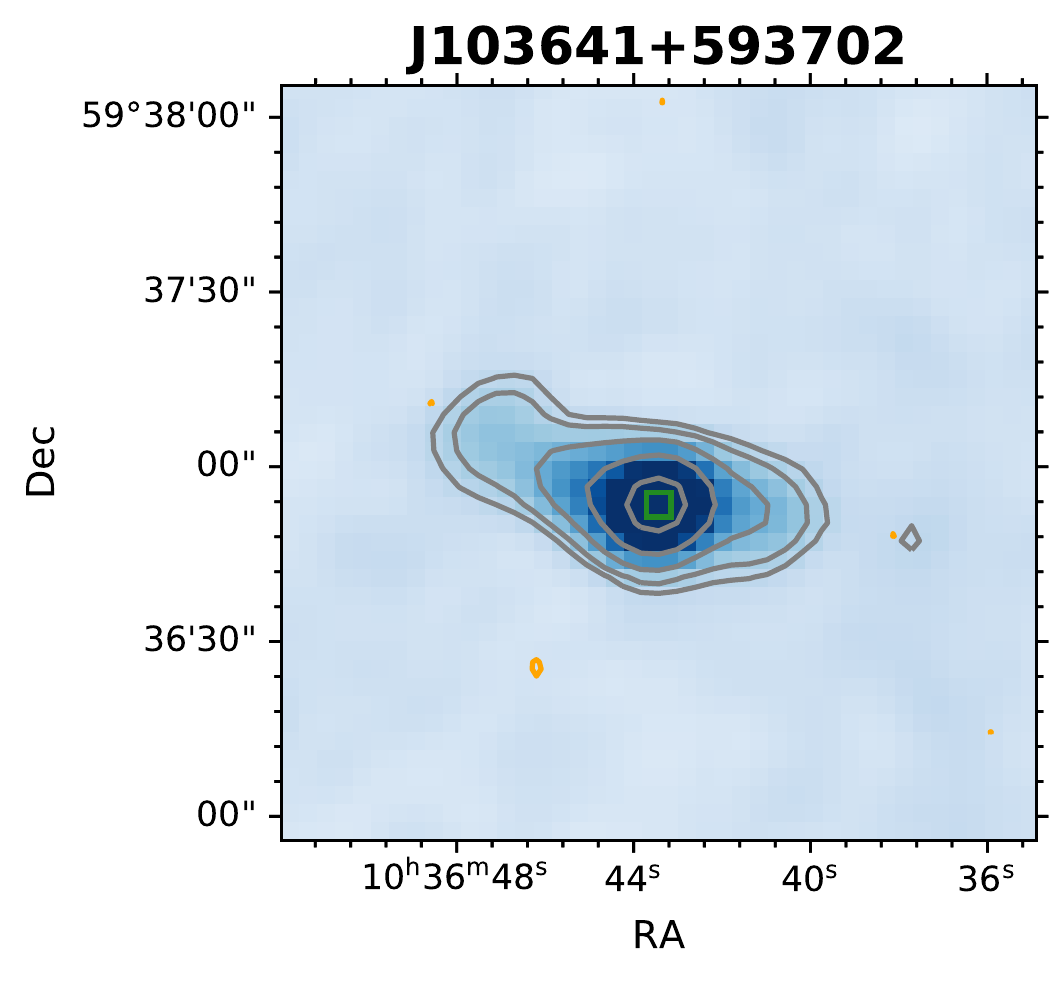}
\endminipage \hfill
\minipage{\textwidth}
        \includegraphics[width=0.195\linewidth] {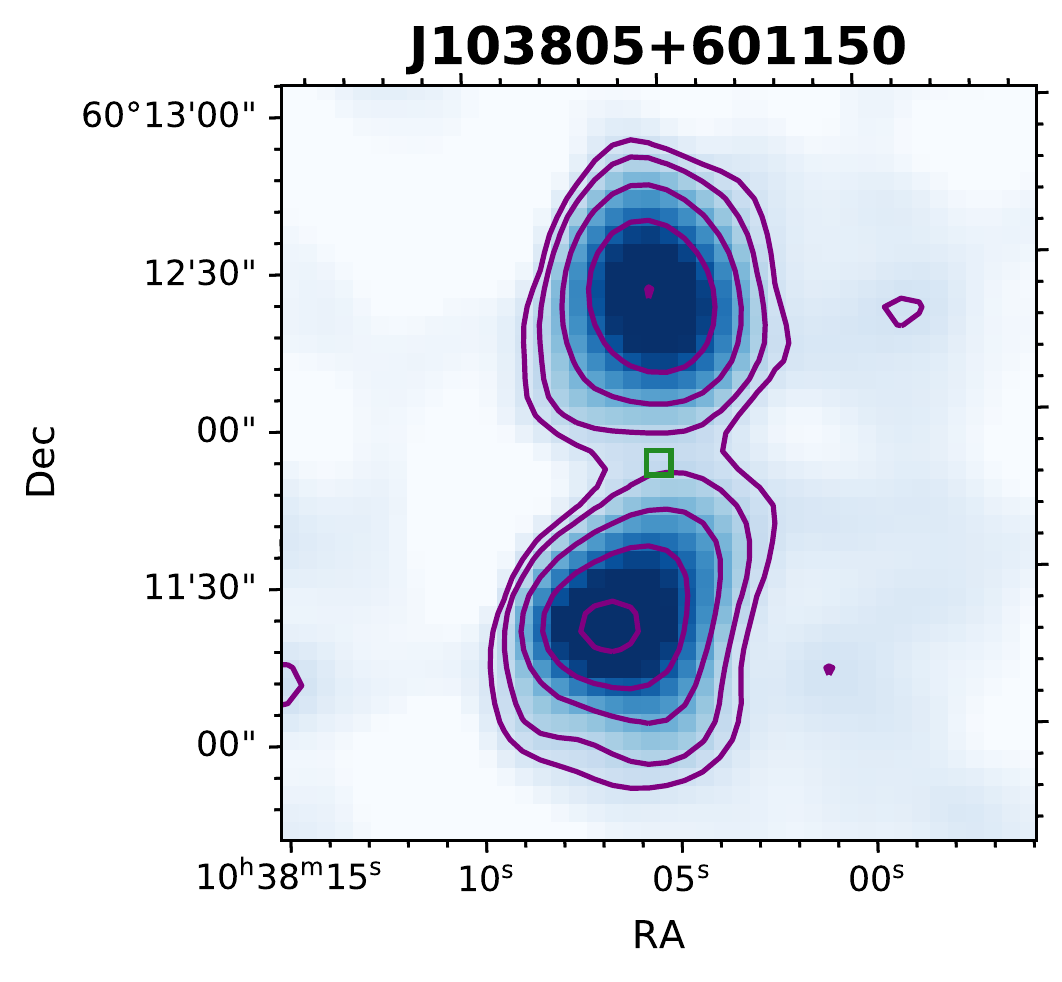}  
        \includegraphics[width=0.195\linewidth] {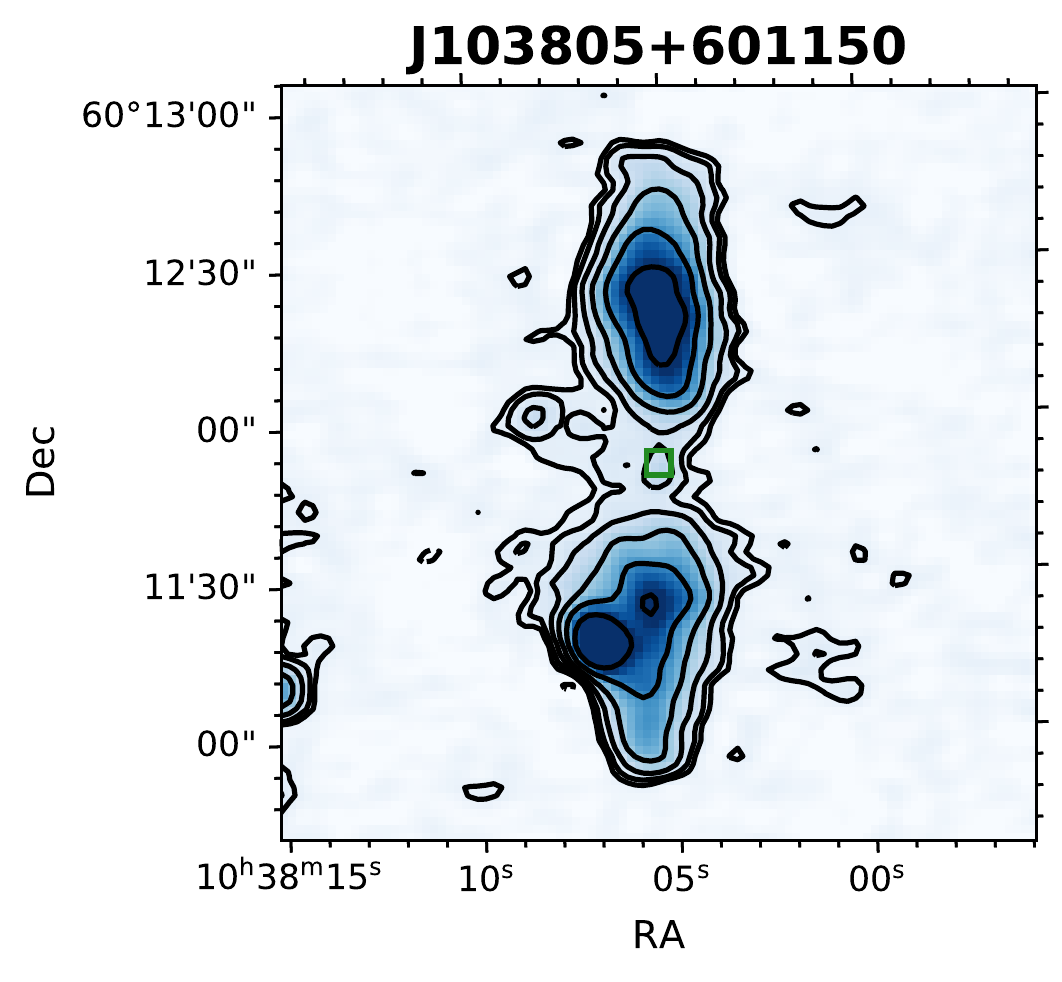} 
        \includegraphics[width=0.195\linewidth] {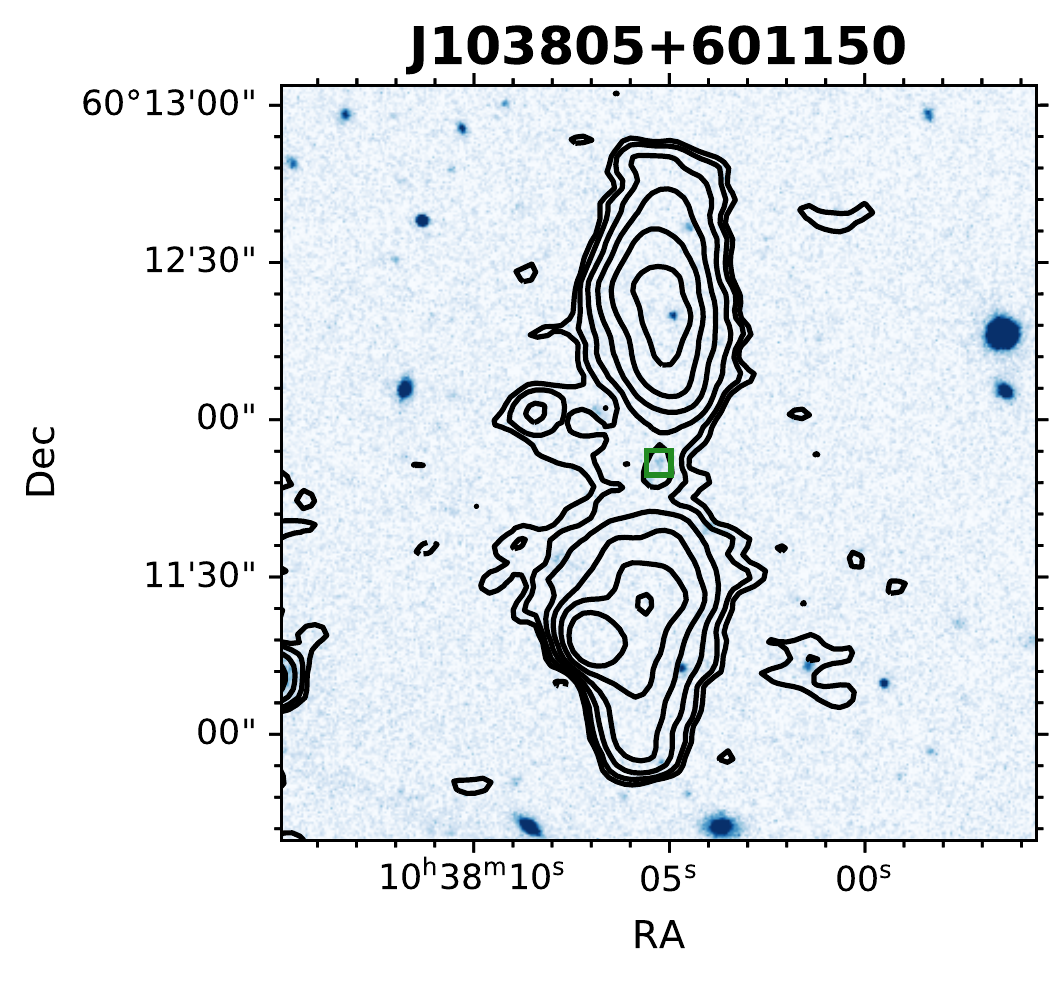}
        \includegraphics[width=0.195\linewidth] {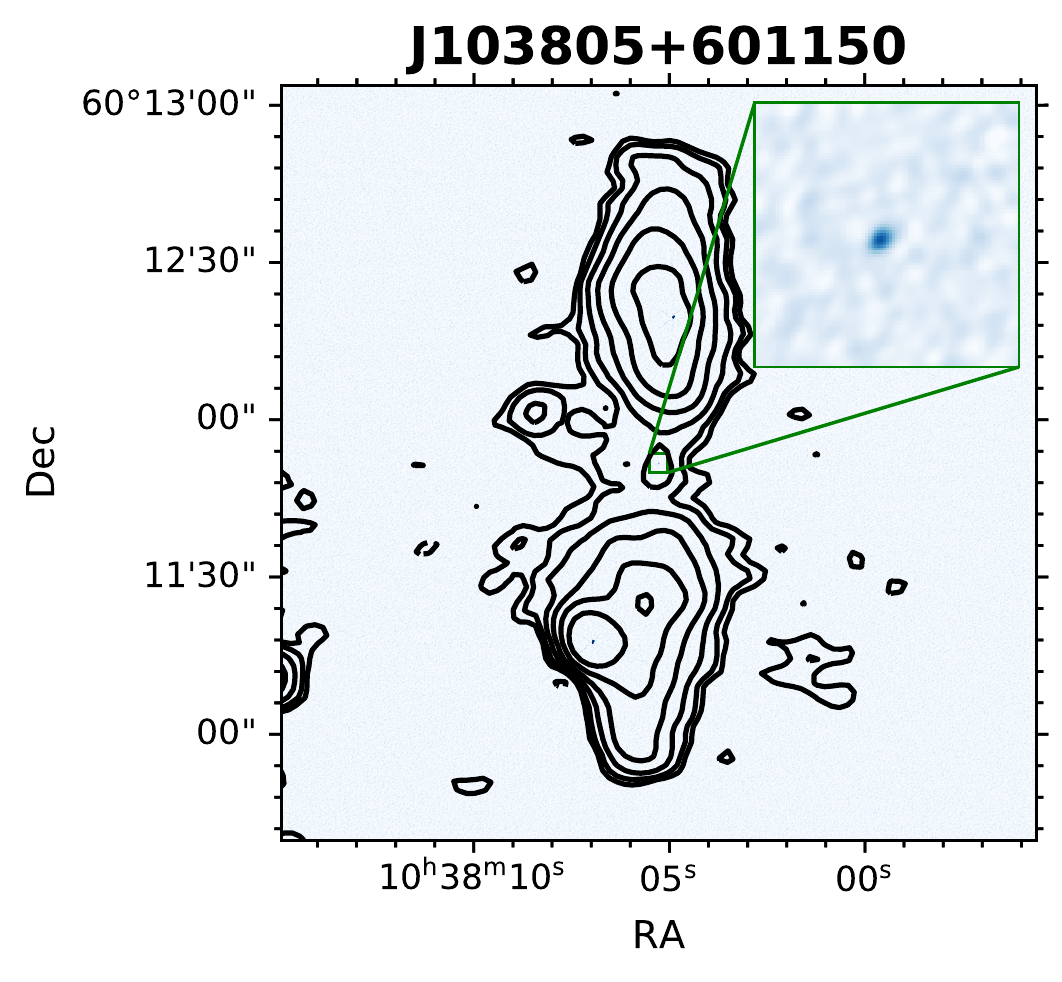}
\endminipage \hfill
\minipage{\textwidth}
        \includegraphics[width=0.195\linewidth] {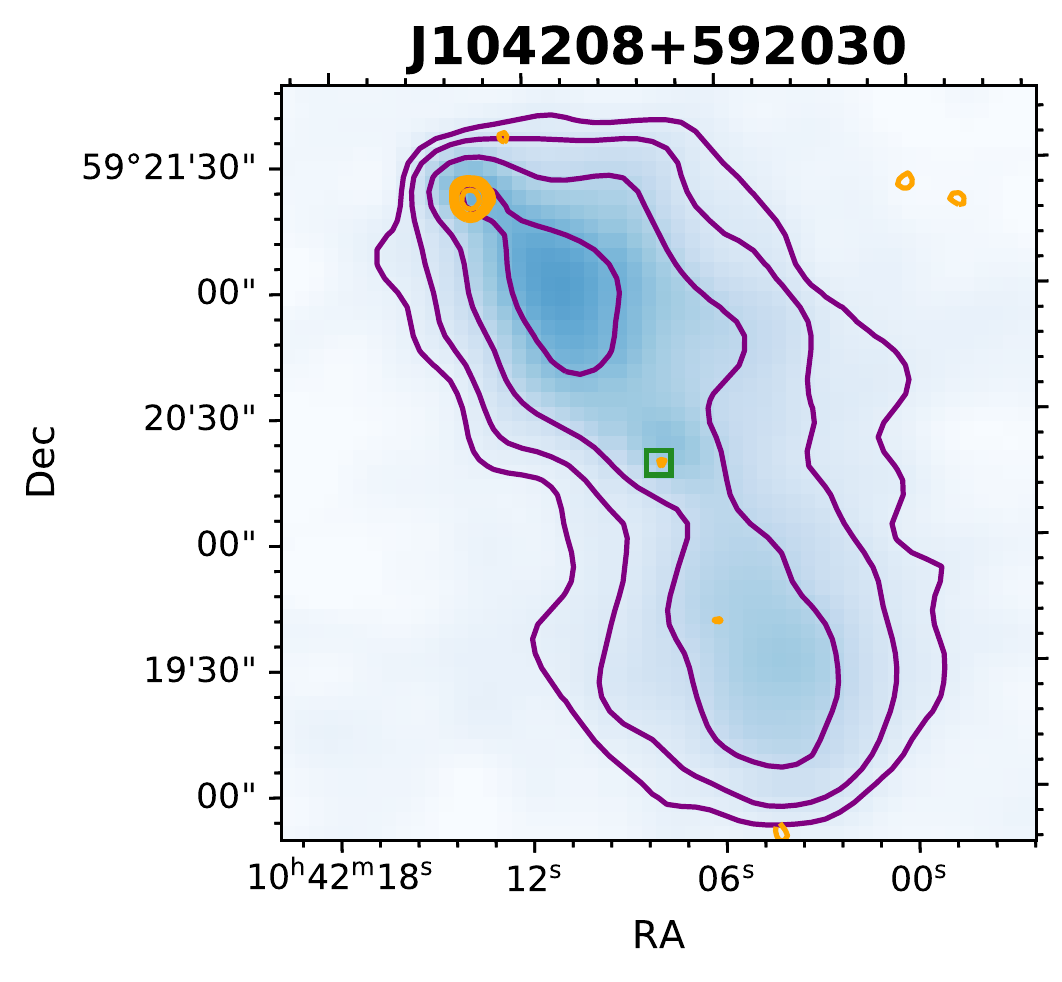}  
        \includegraphics[width=0.195\linewidth] {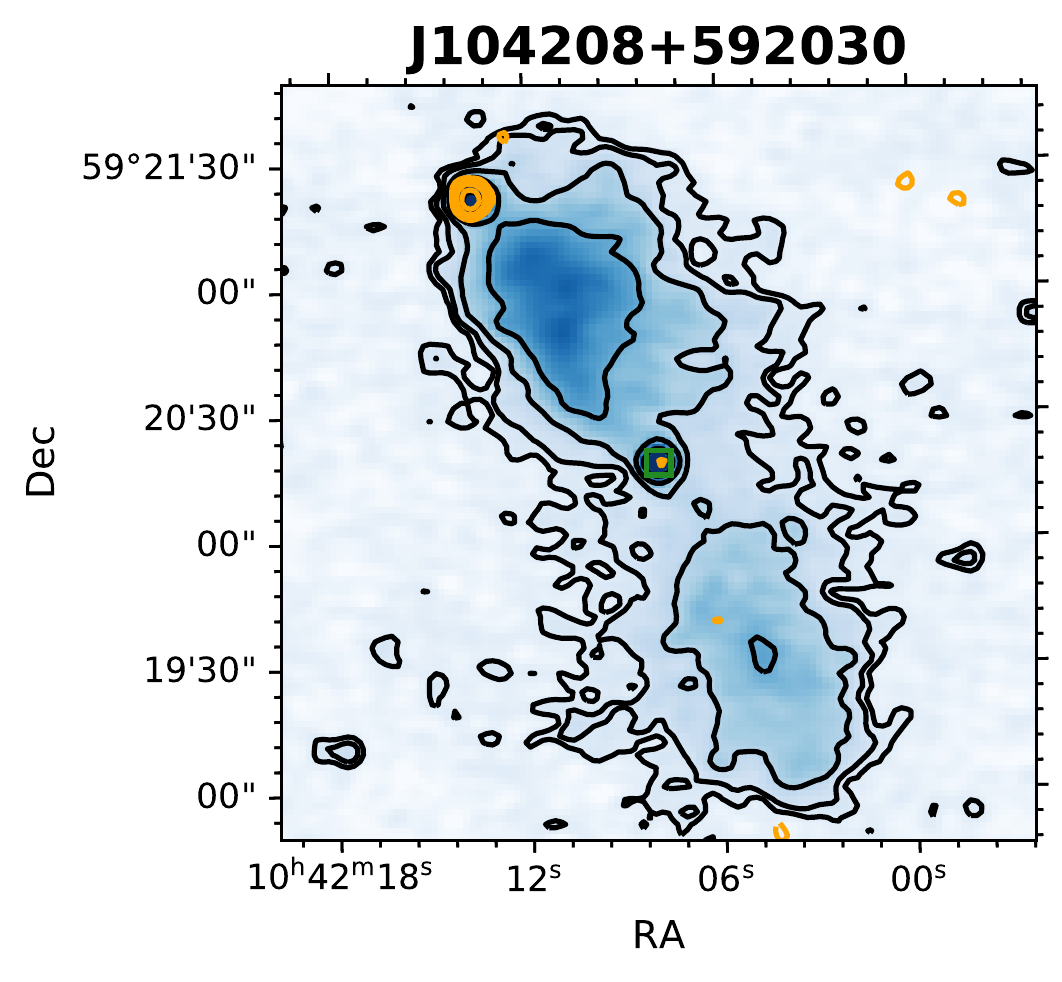} 
        \includegraphics[width=0.195\linewidth] {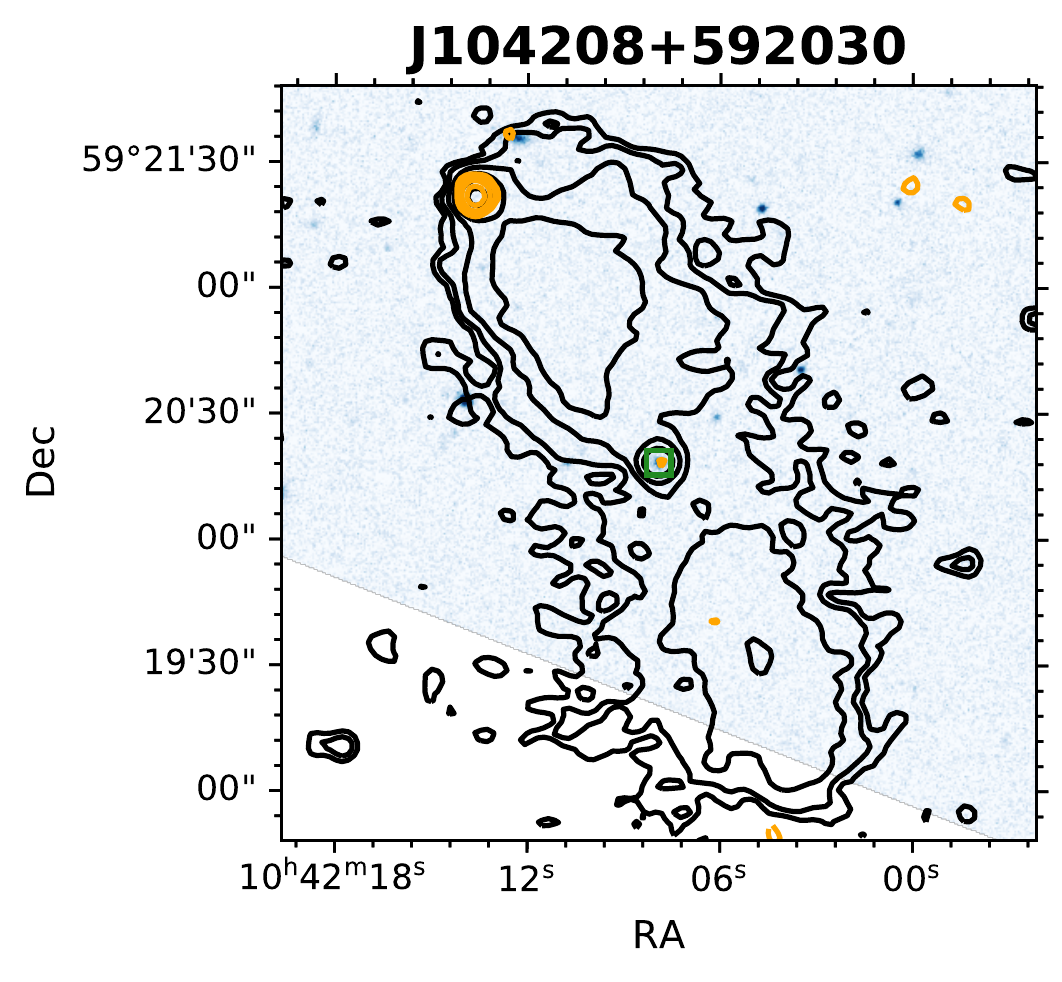}
        \includegraphics[width=0.195\linewidth] {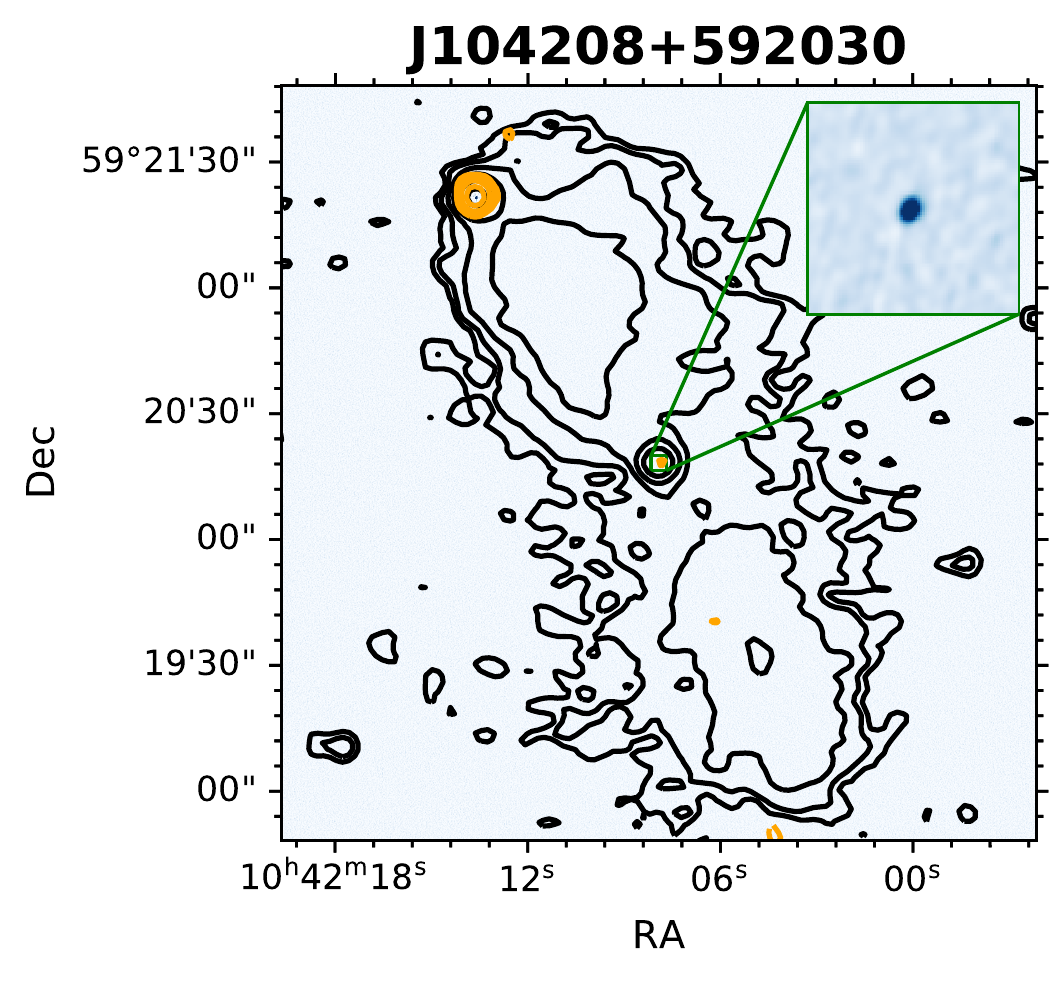}
\endminipage \hfill
\minipage{\textwidth}
        \includegraphics[width=0.195\linewidth] {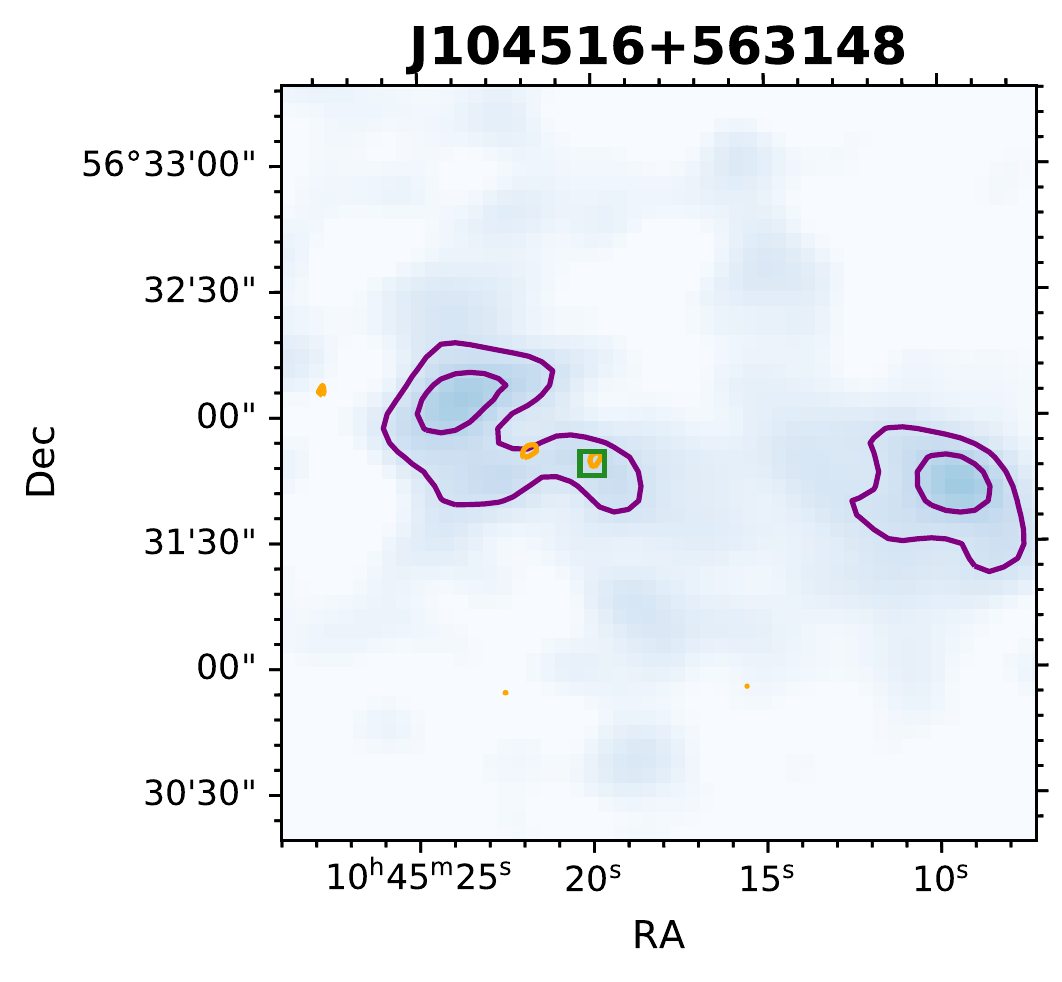}  
        \includegraphics[width=0.195\linewidth] {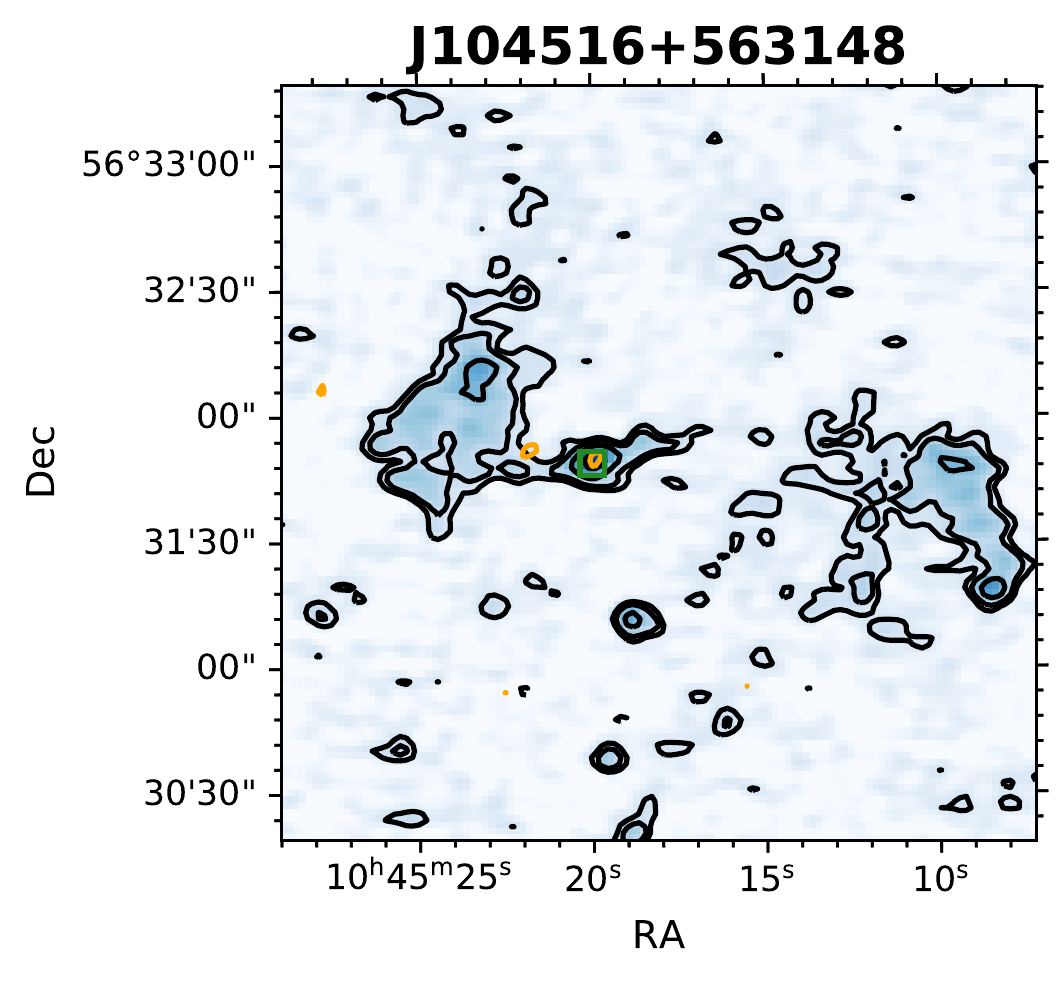} 
        \includegraphics[width=0.195\linewidth] {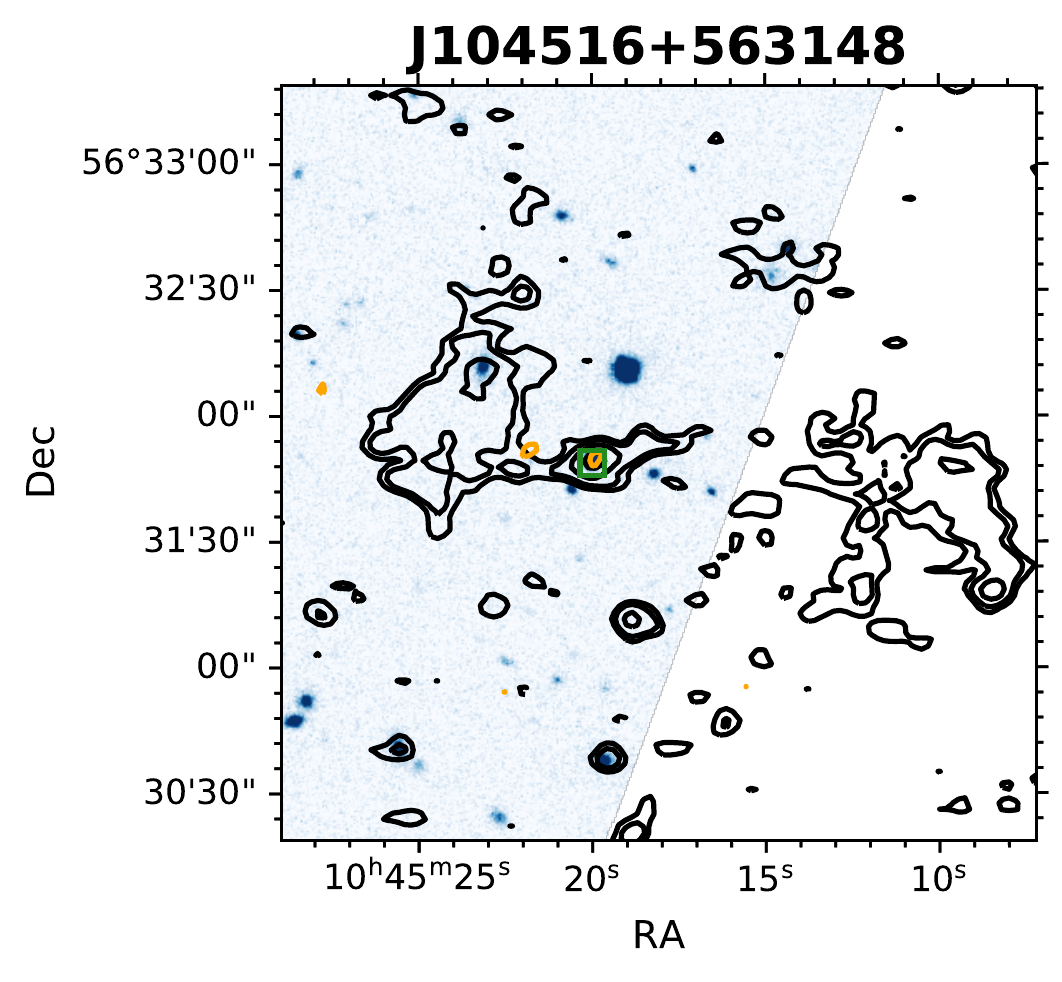}
        \includegraphics[width=0.195\linewidth] {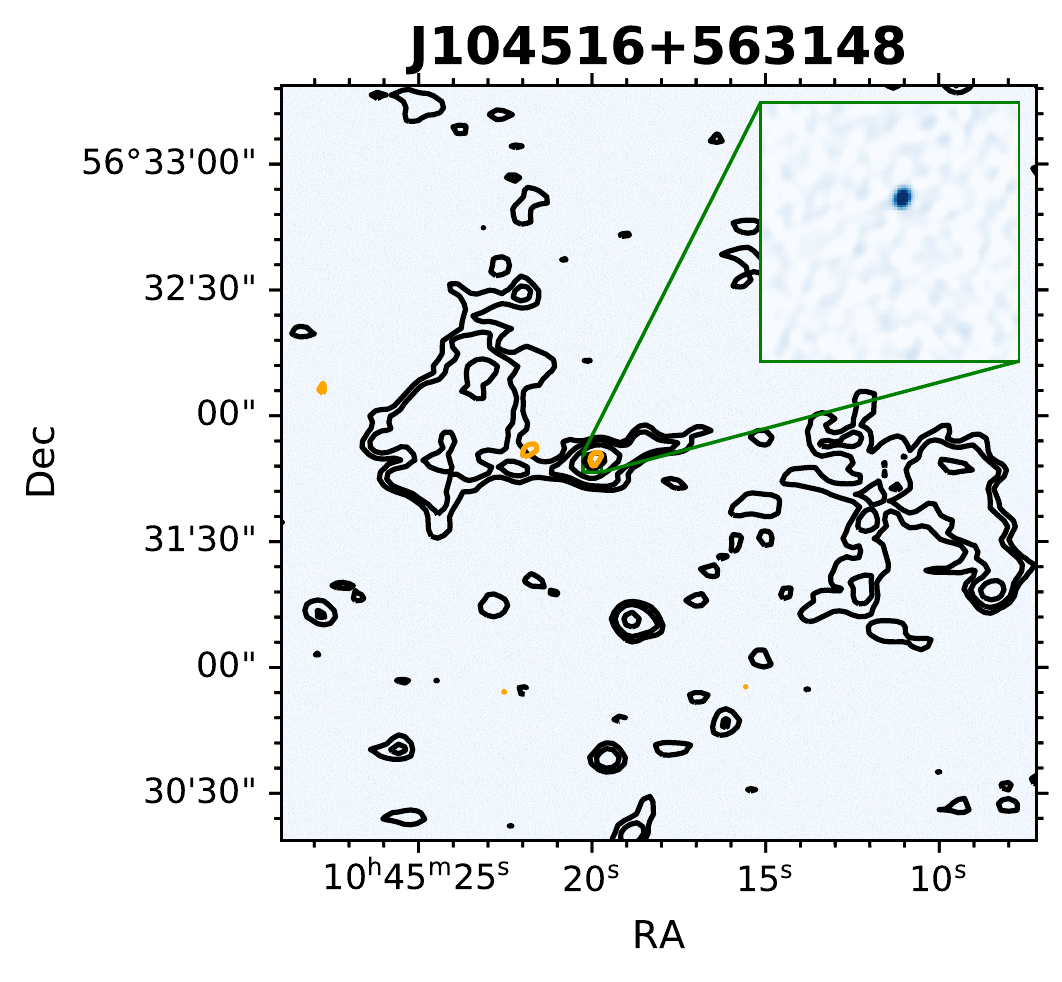}
 	    \includegraphics[width=0.195\linewidth] {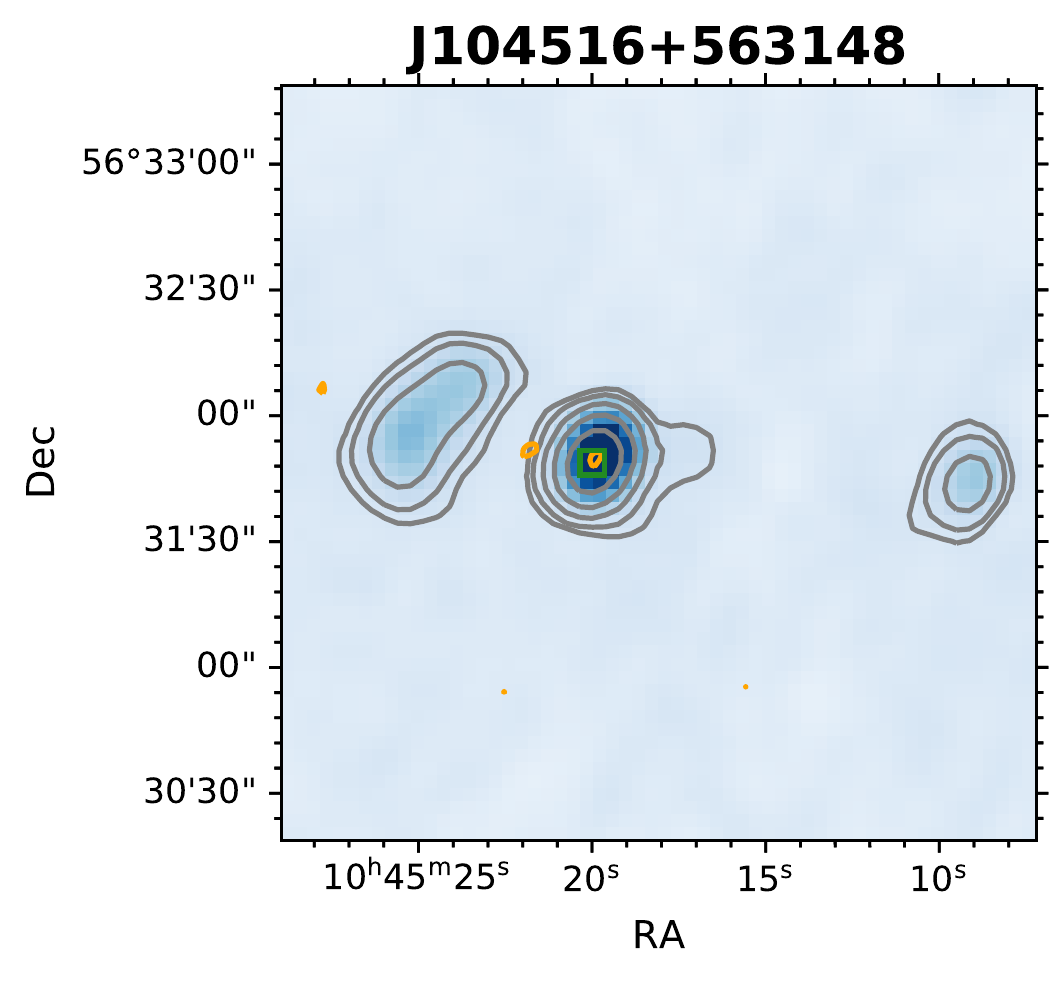}
\endminipage \hfill
\minipage{\textwidth}
        \includegraphics[width=0.195\linewidth] {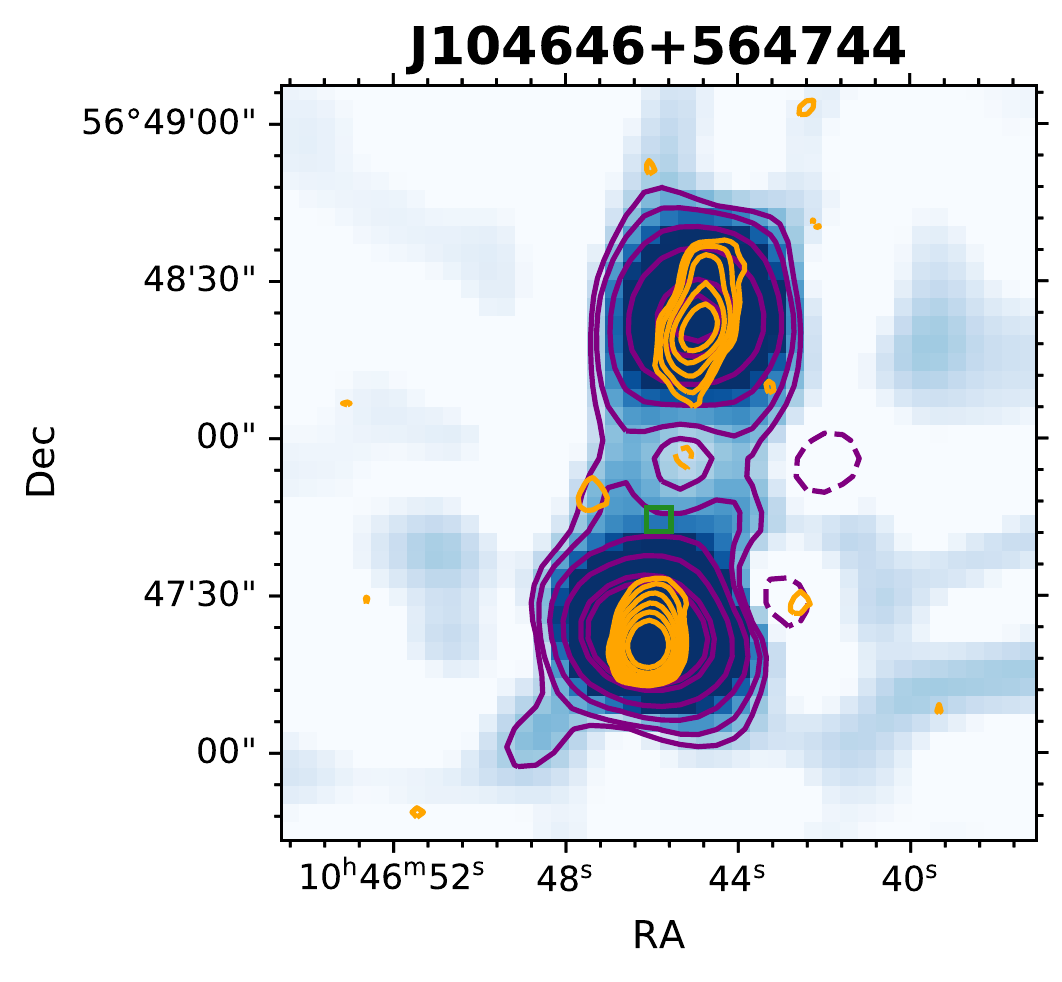}  
        \includegraphics[width=0.195\linewidth] {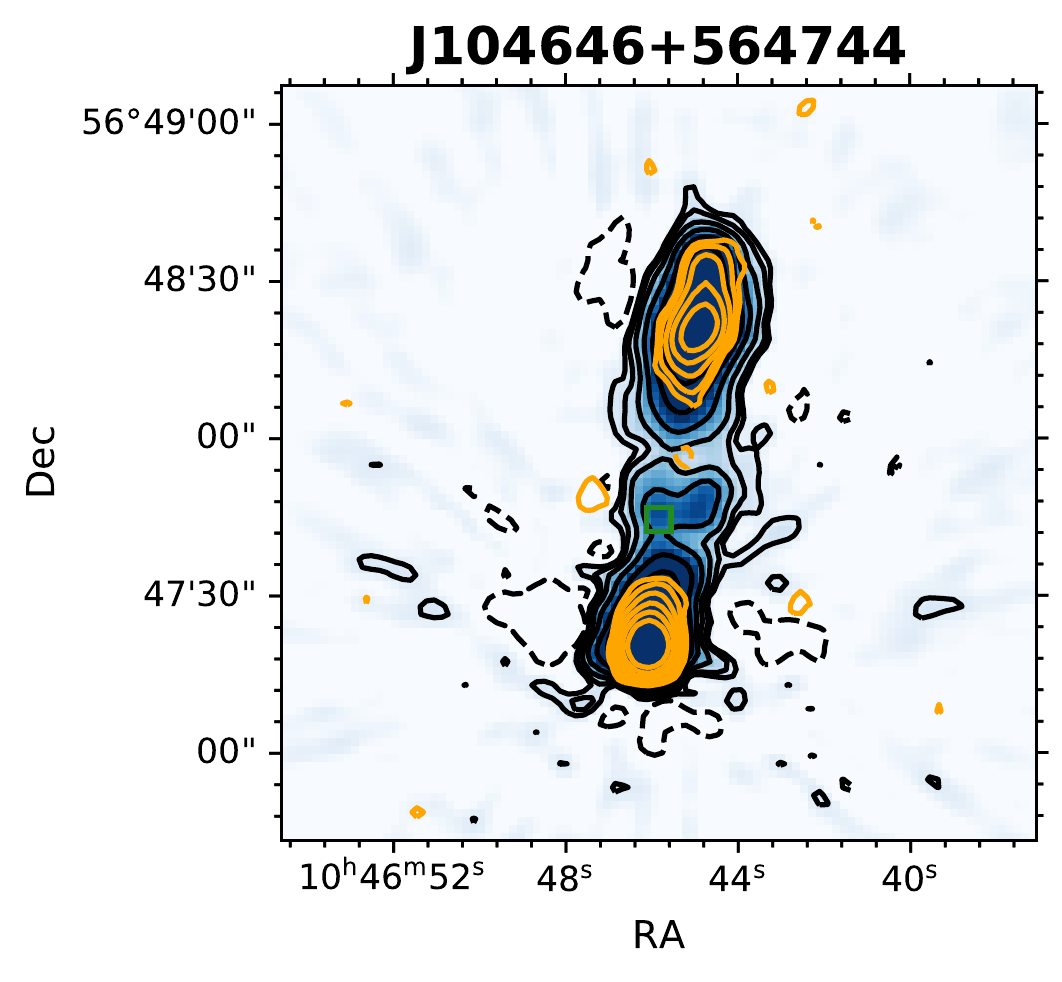} 
        \includegraphics[width=0.195\linewidth] {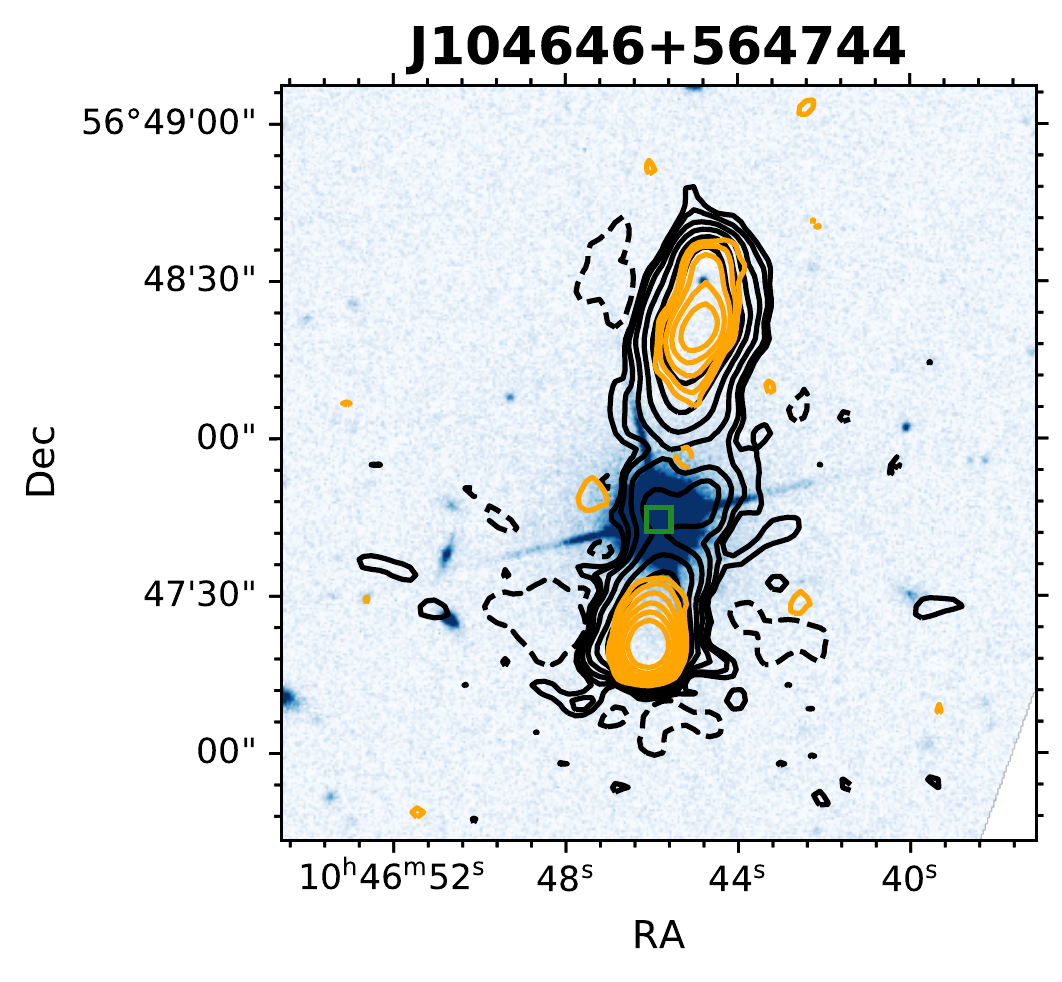}
        \includegraphics[width=0.195\linewidth] {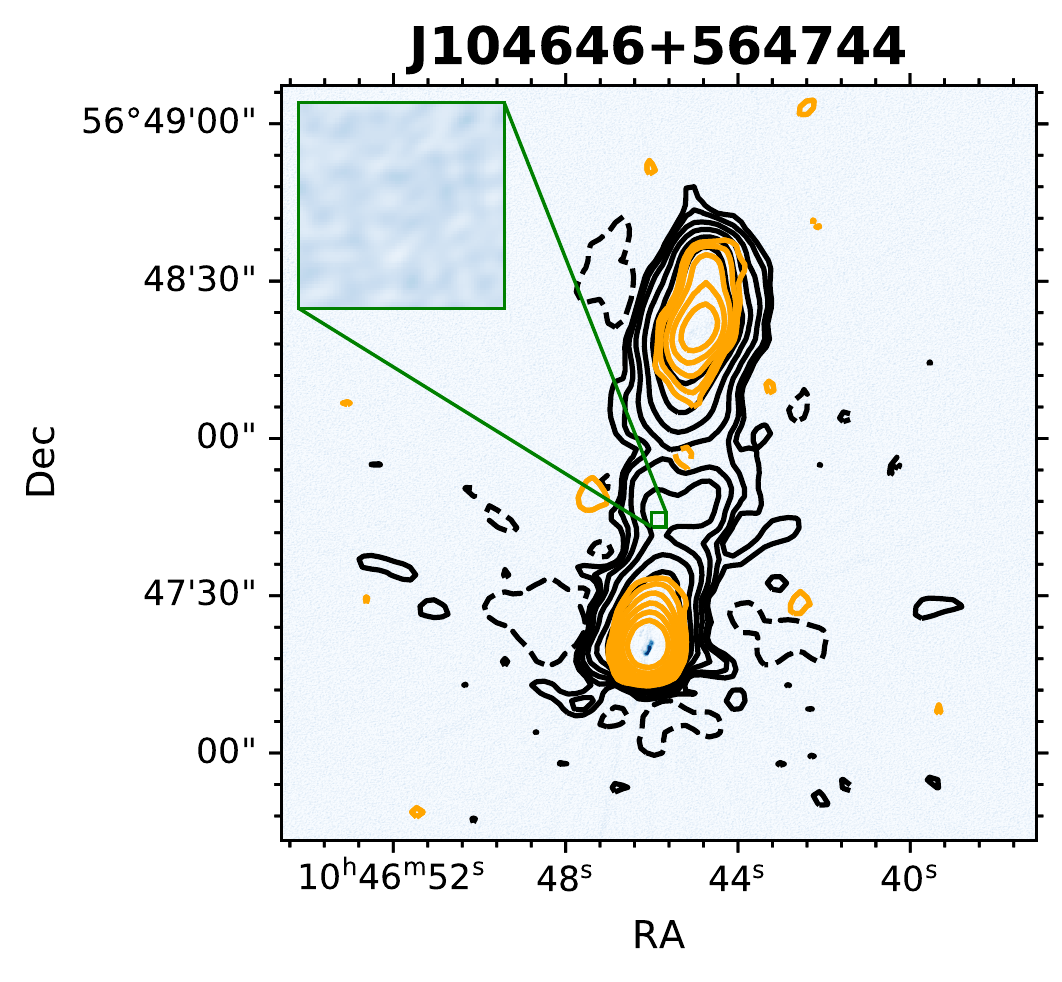}
\endminipage \hfill
\minipage{\textwidth}
        \includegraphics[width=0.195\linewidth] {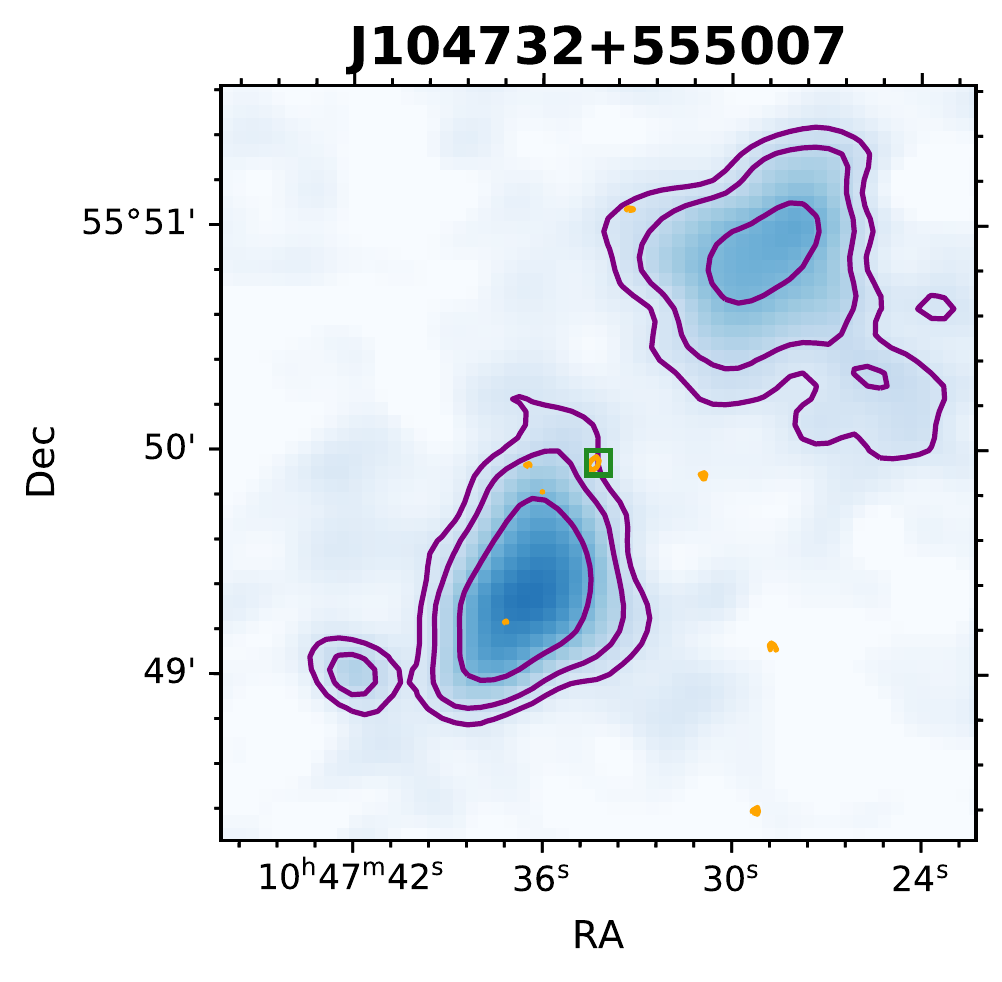}  
        \includegraphics[width=0.195\linewidth] {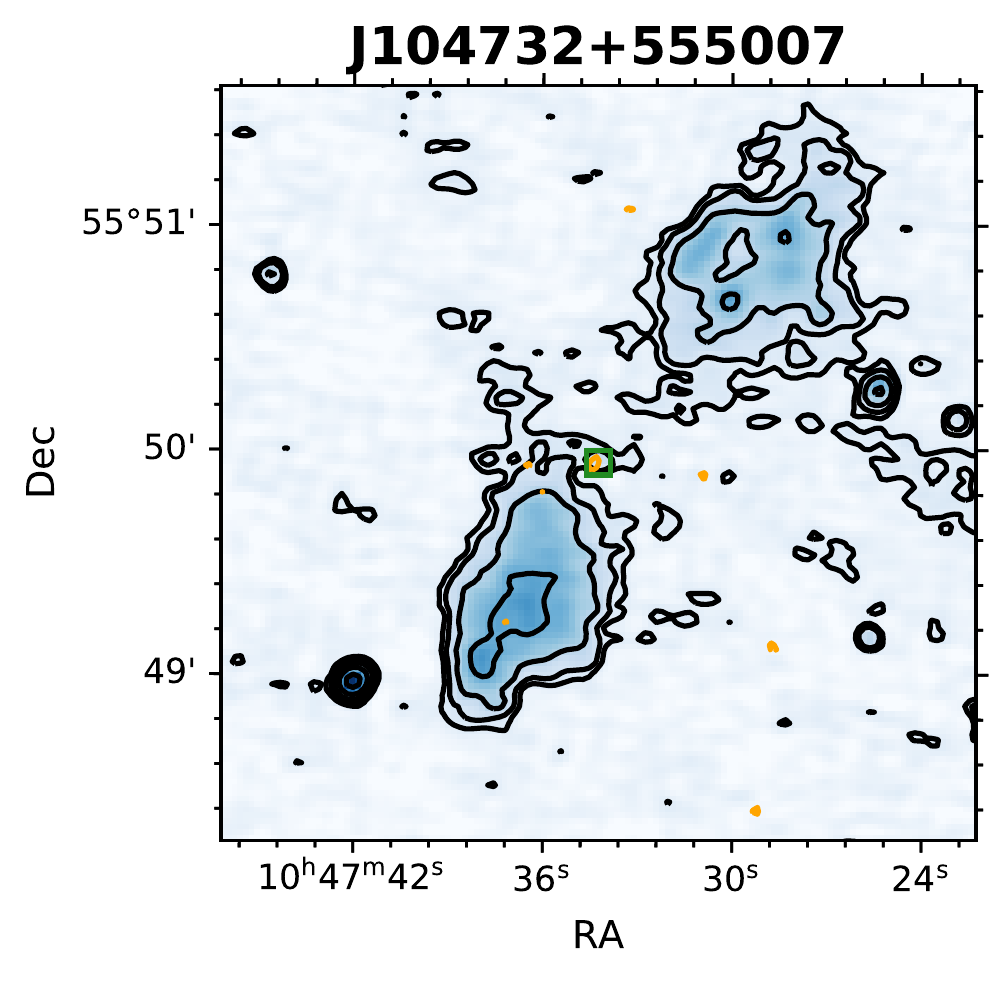} 
        \includegraphics[width=0.195\linewidth] {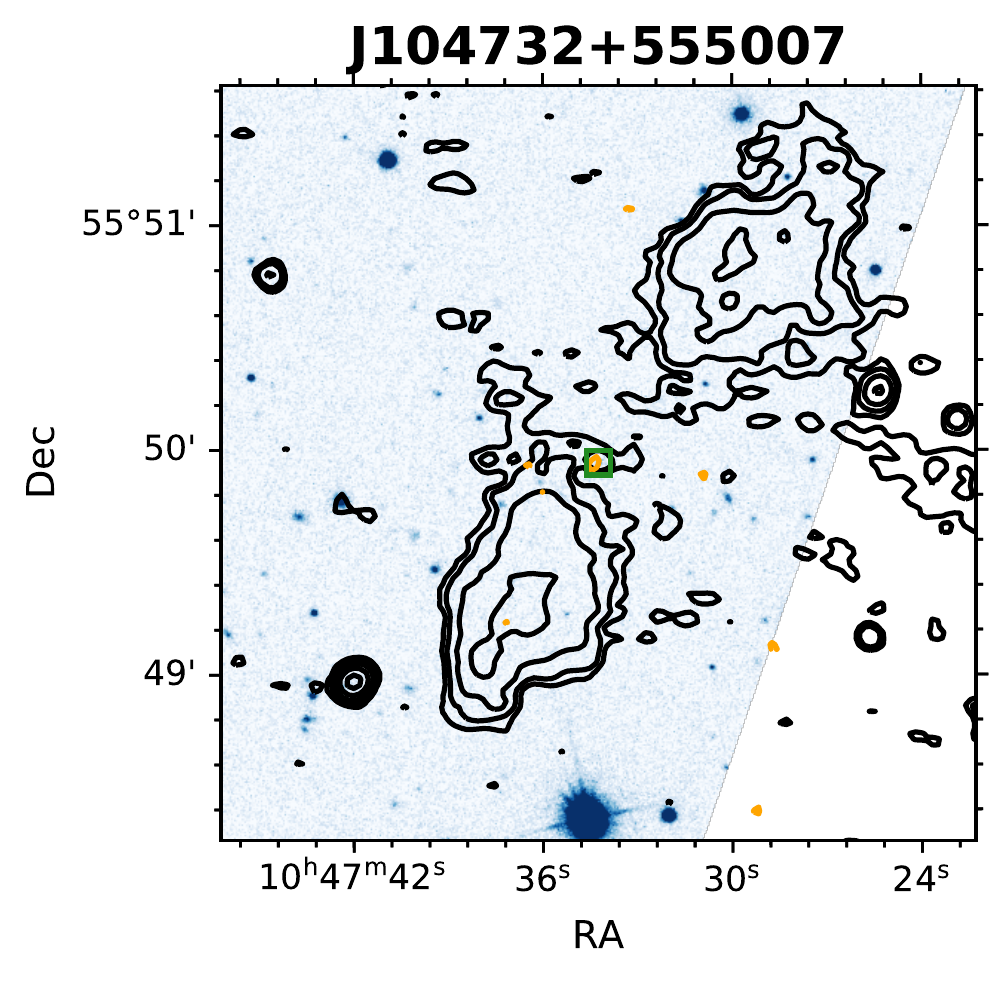}
        \includegraphics[width=0.195\linewidth] {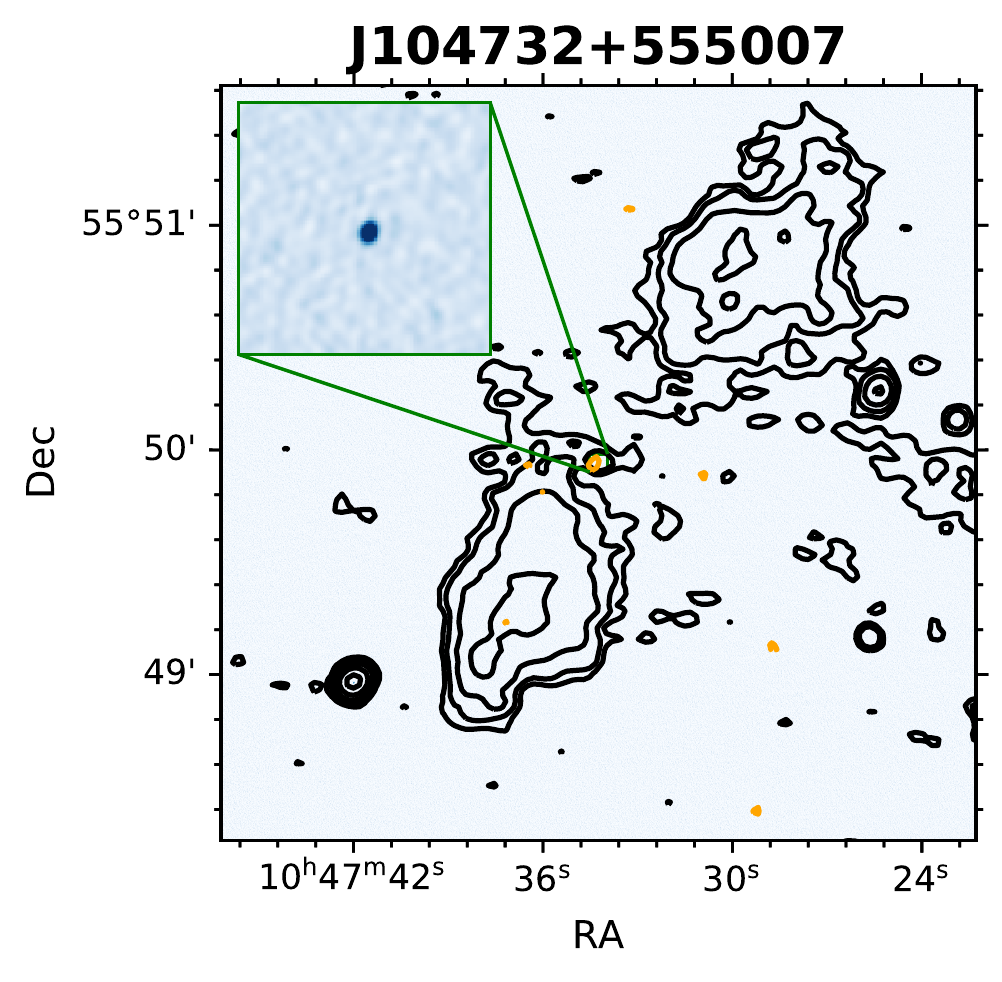}
 	    \includegraphics[width=0.195\linewidth] {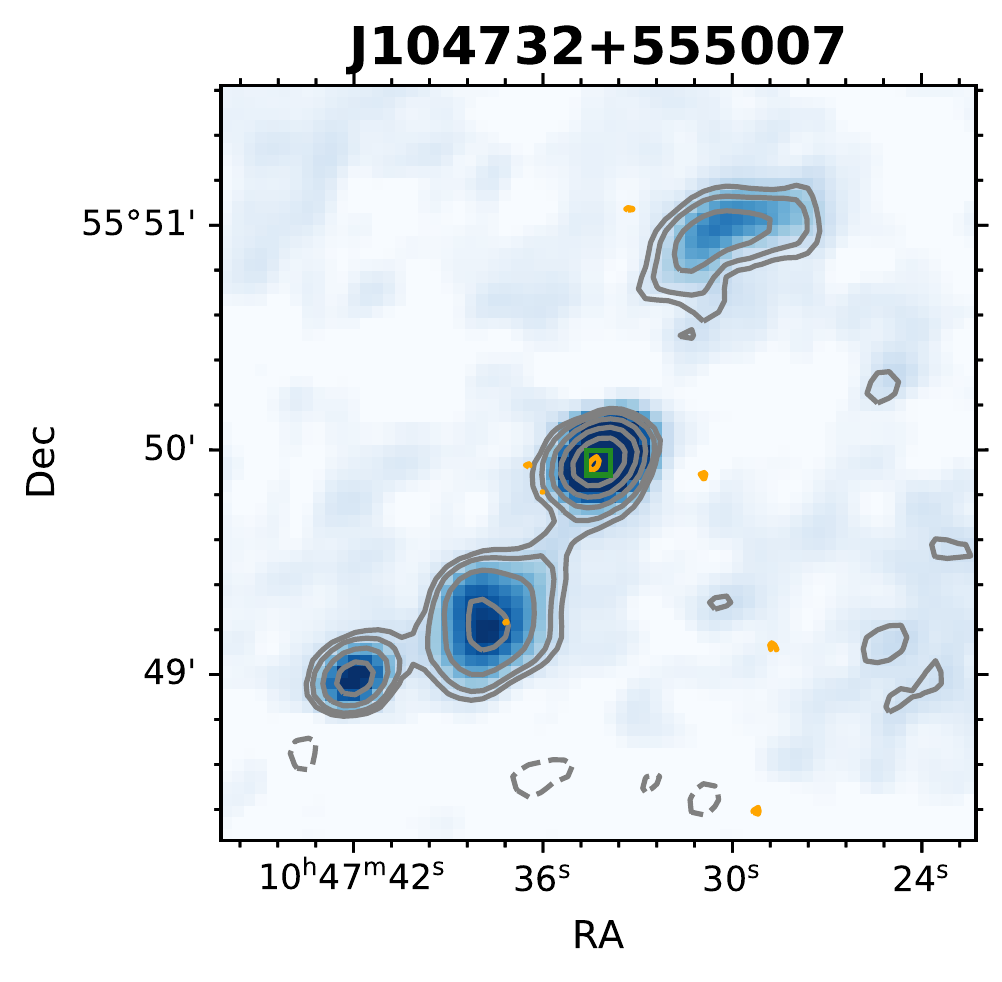}
\endminipage \hfill
\end{figure*}

\begin{figure*} [!h]\ContinuedFloat
\caption{continued}
\minipage{\textwidth}
        \includegraphics[width=0.195\linewidth] {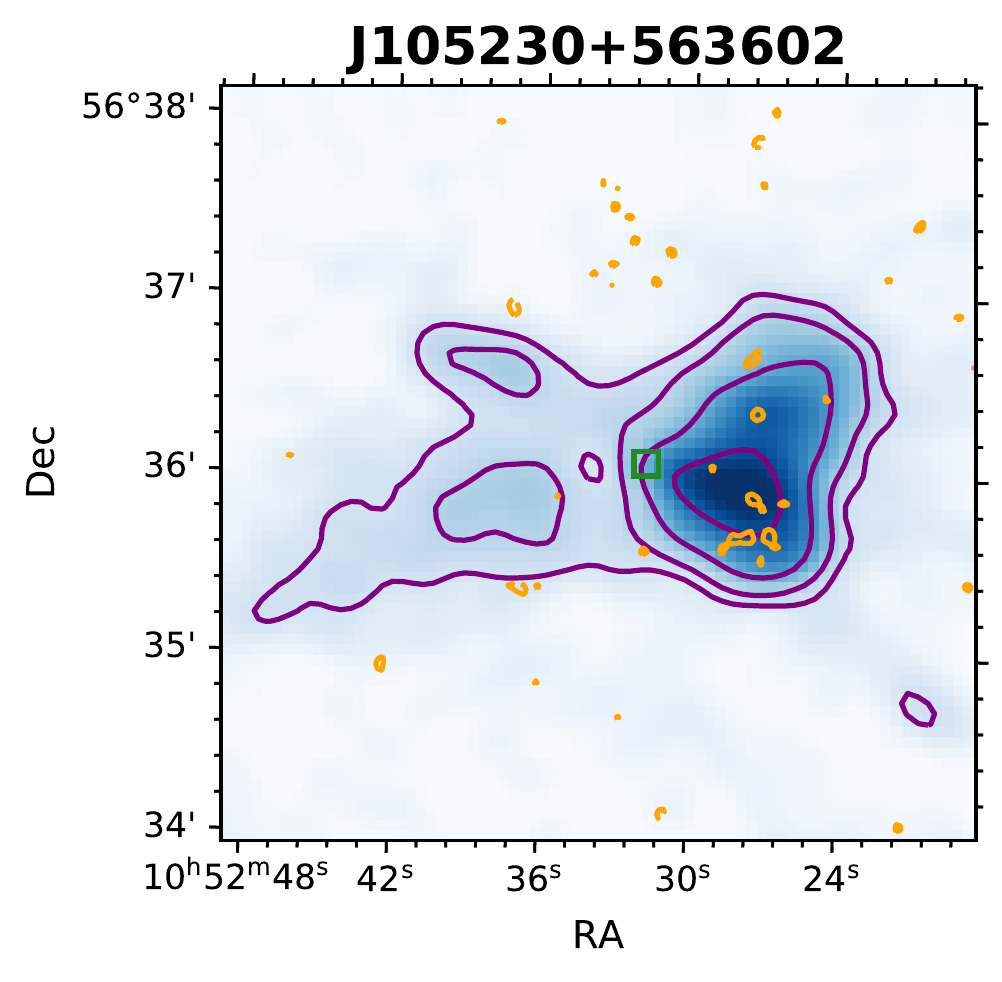}  
        \includegraphics[width=0.195\linewidth] {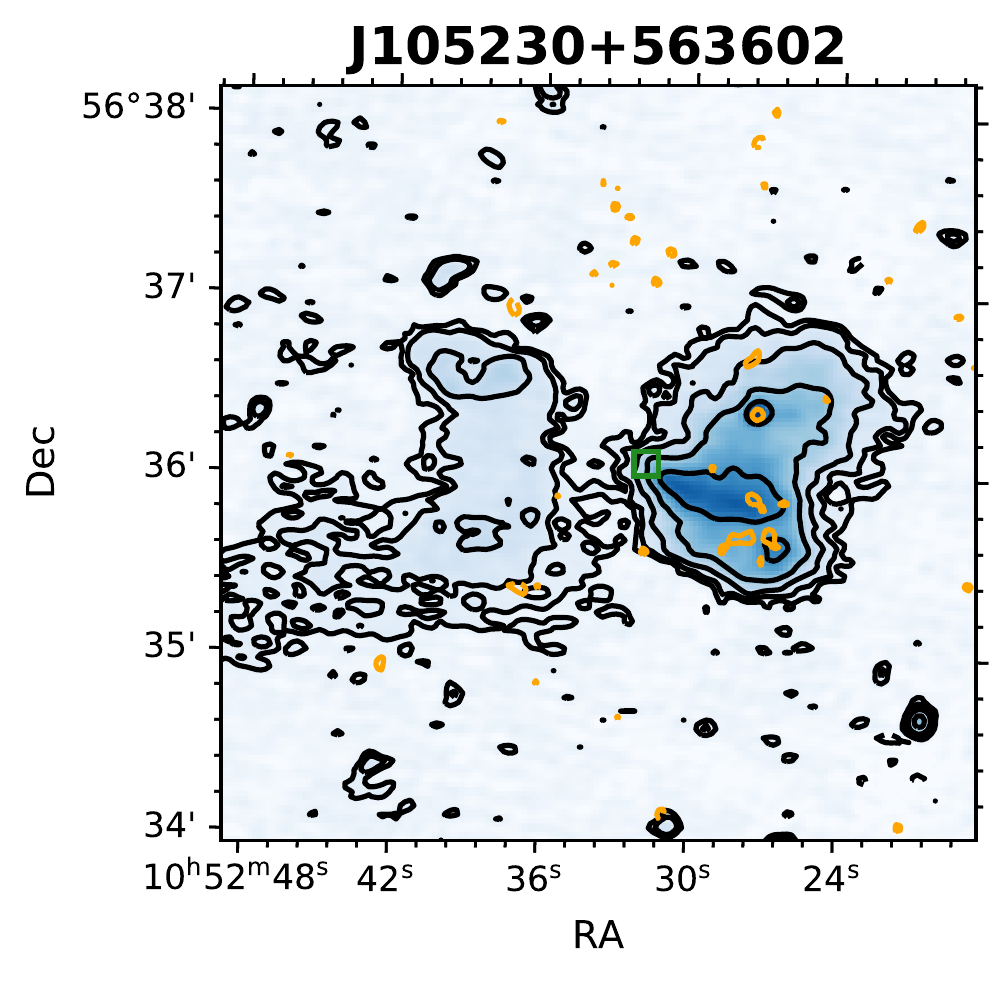} 
        \includegraphics[width=0.195\linewidth] {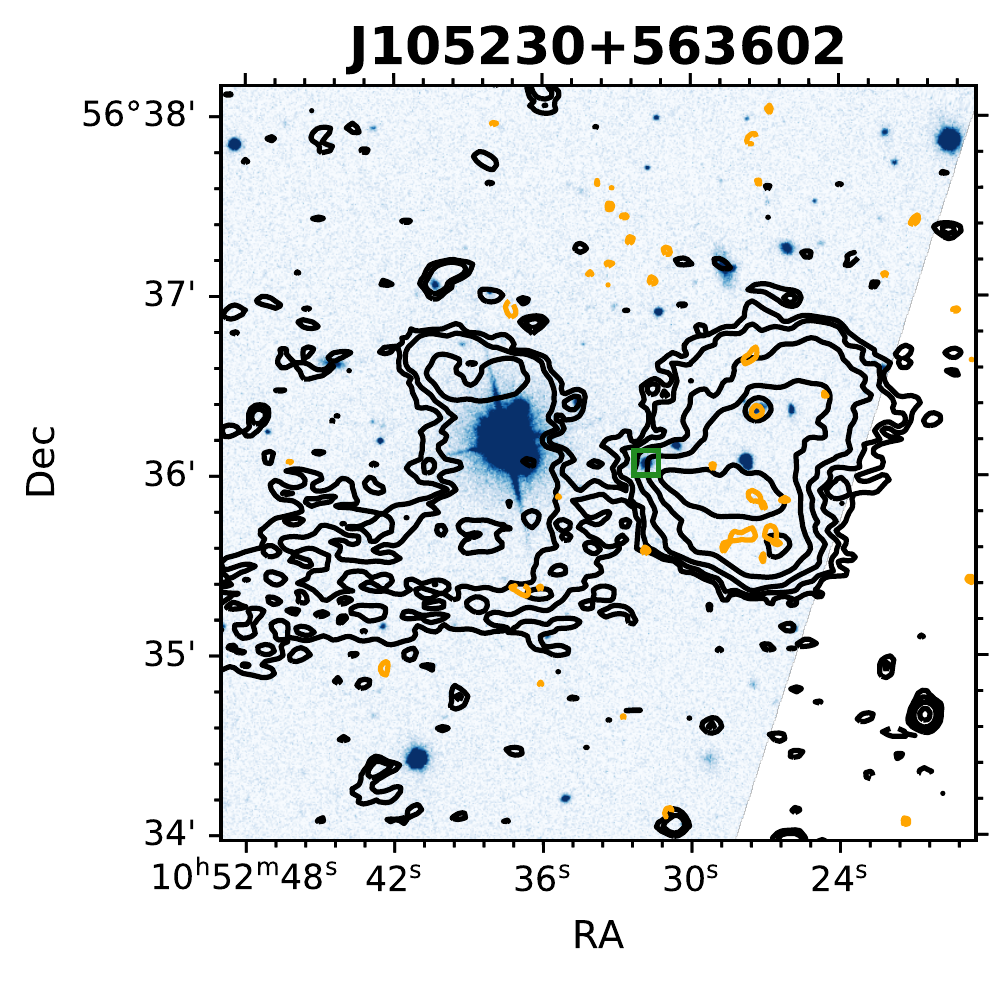}
        \includegraphics[width=0.195\linewidth] {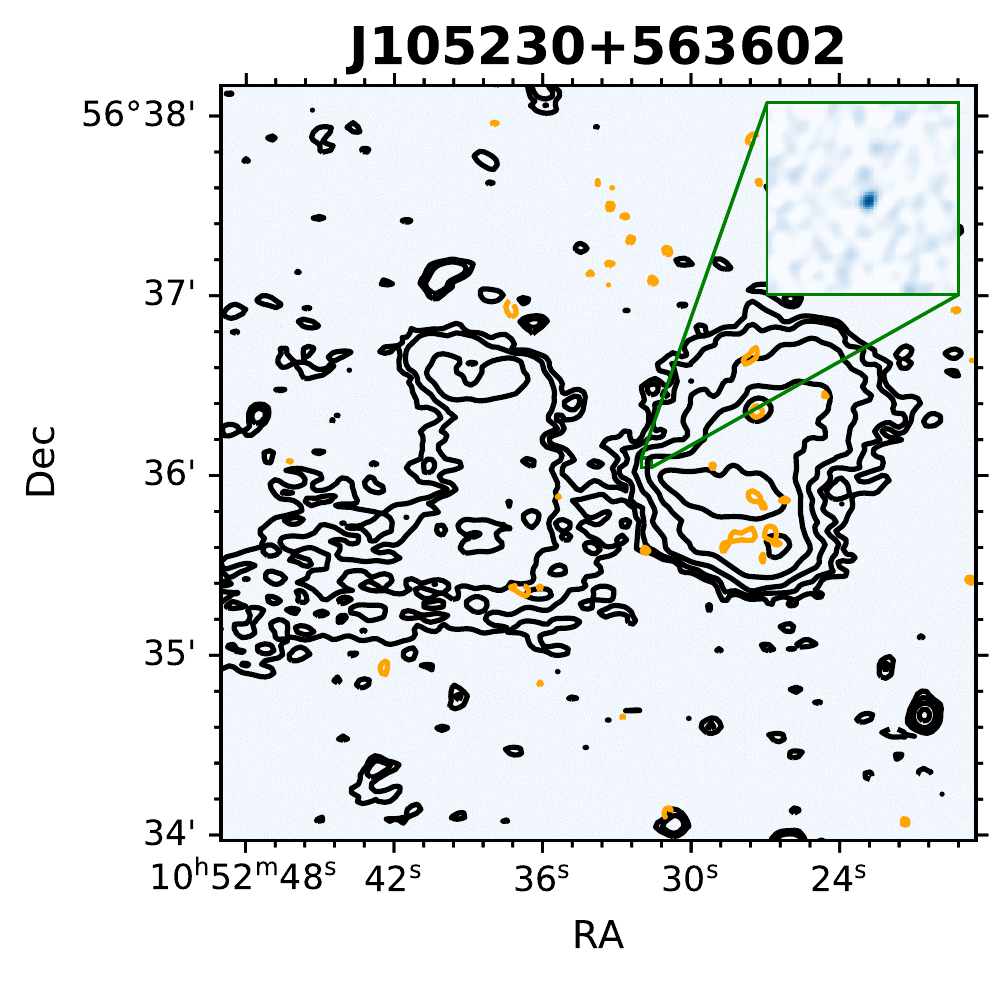}
 	    \includegraphics[width=0.195\linewidth] {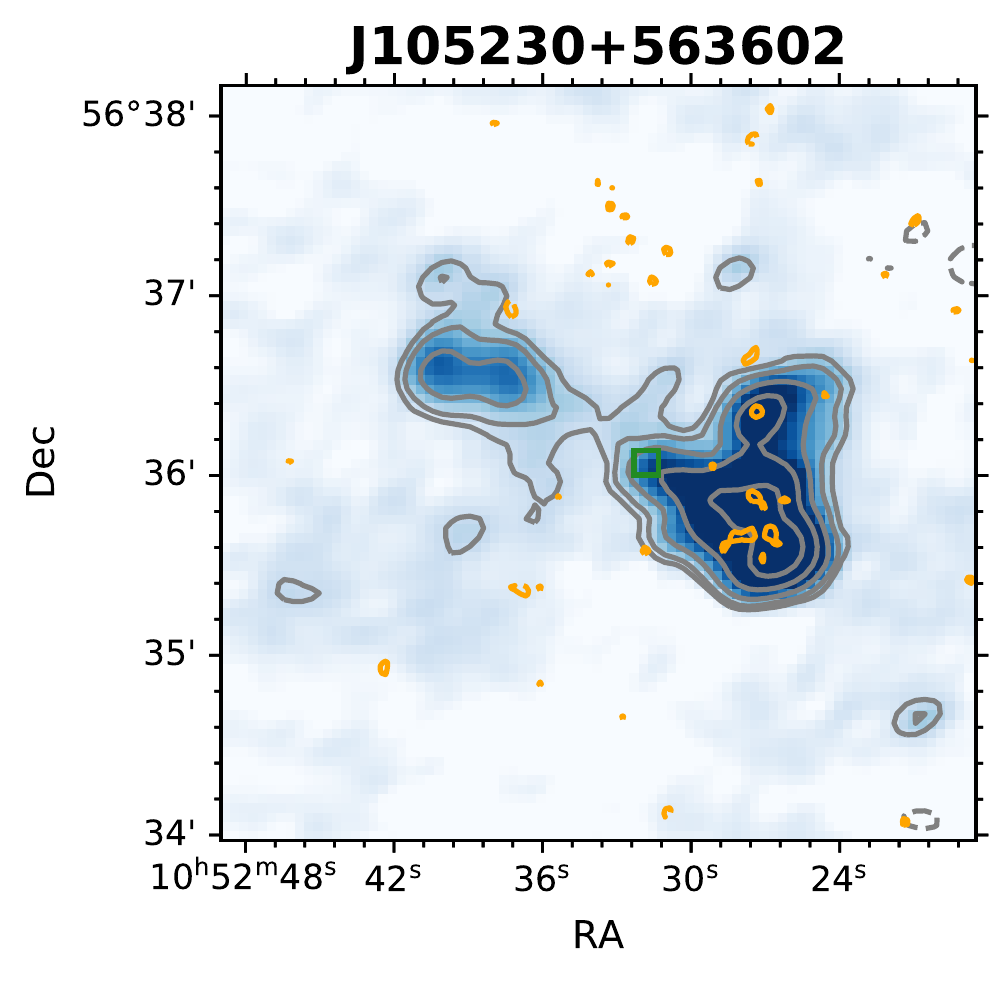}
\endminipage \hfill
\minipage{\textwidth}
        \includegraphics[width=0.195\linewidth] {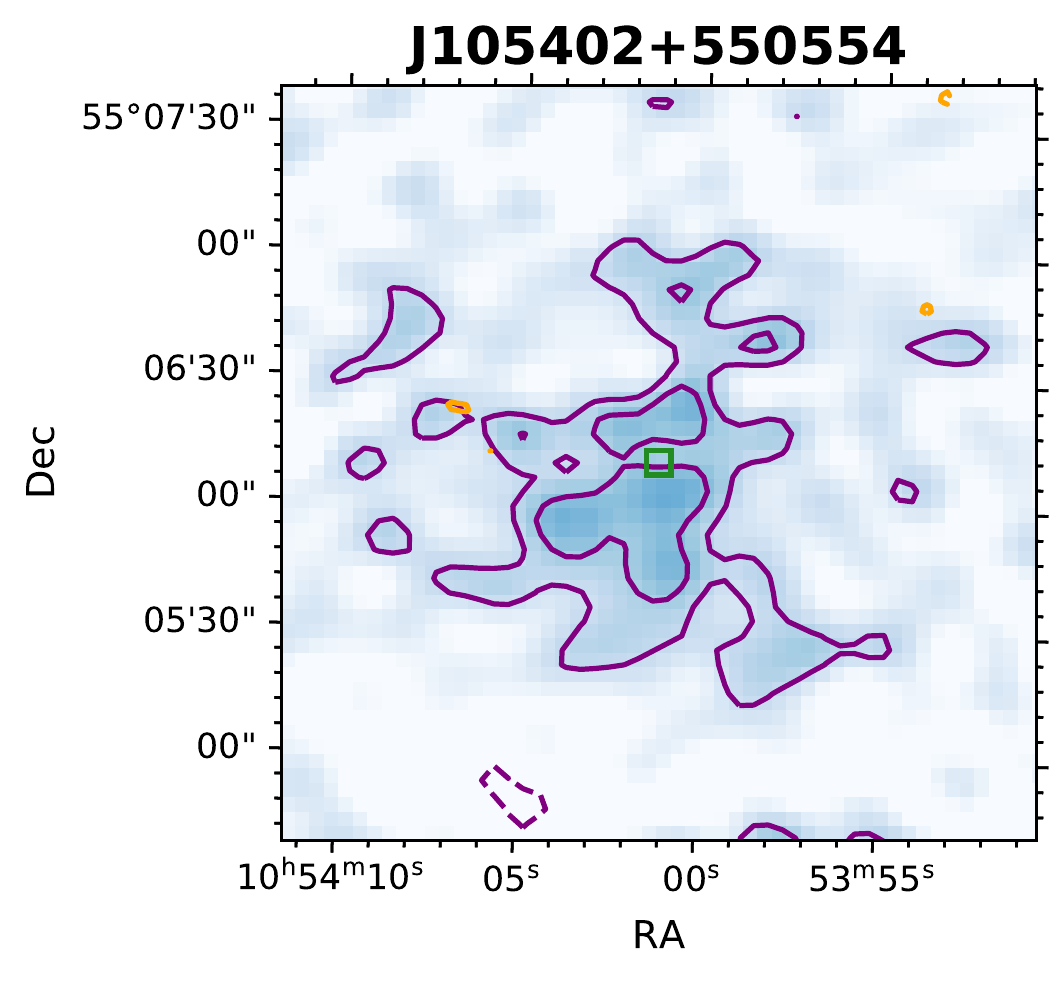}  
        \includegraphics[width=0.195\linewidth] {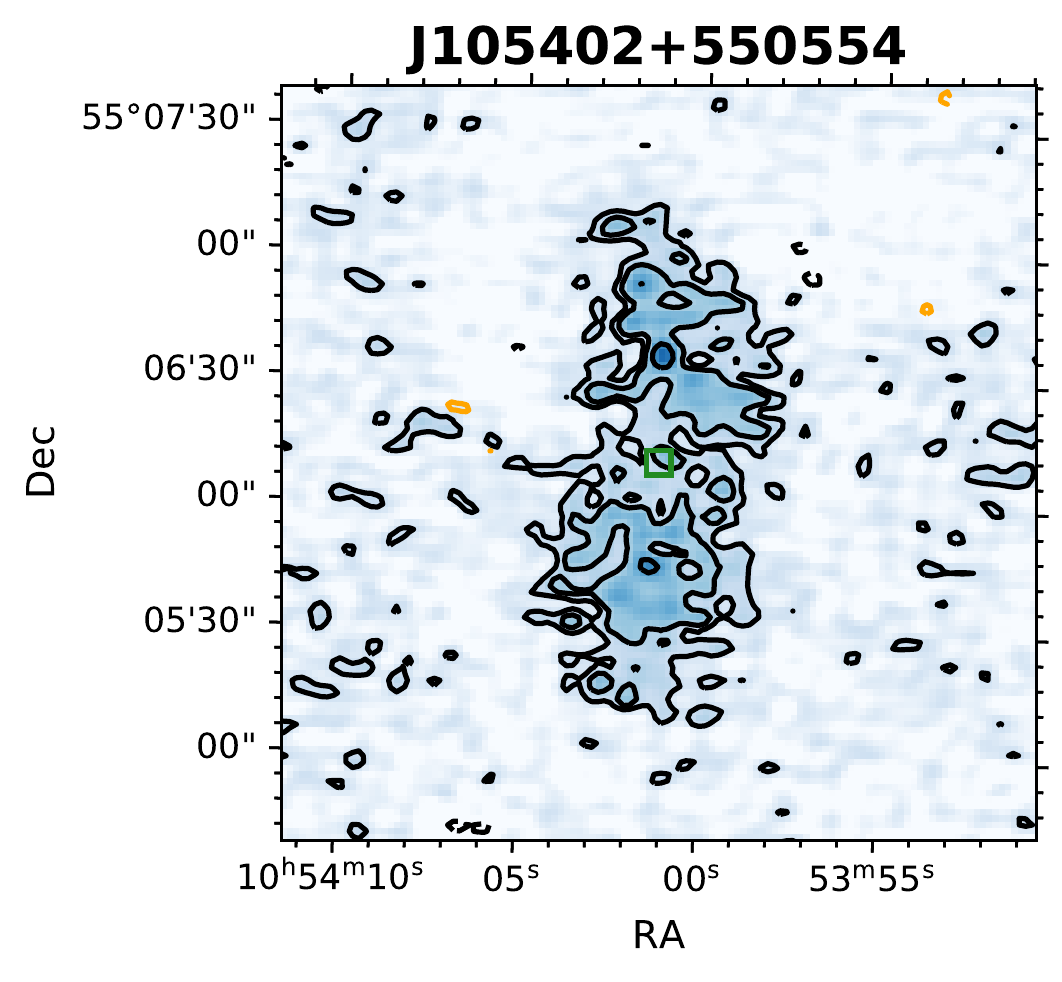} 
        \includegraphics[width=0.195\linewidth] {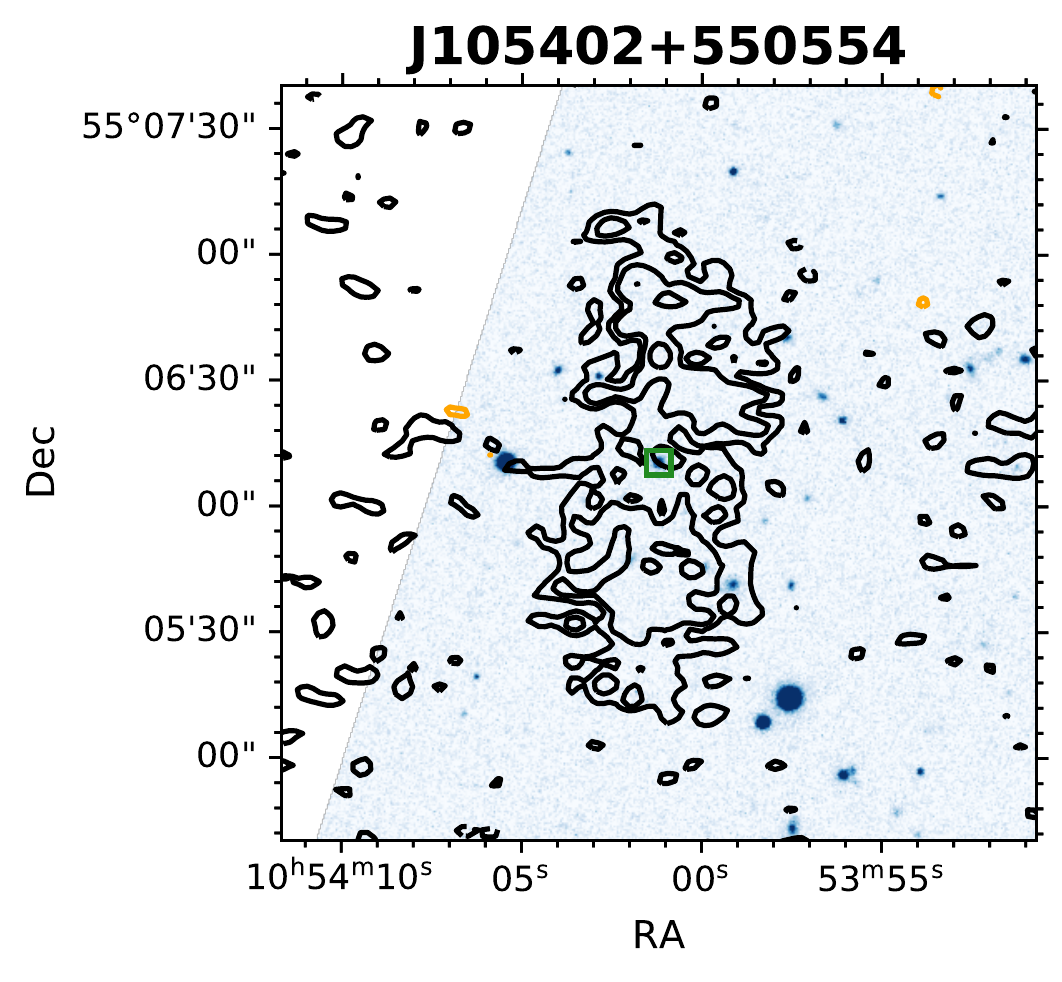}
        \includegraphics[width=0.195\linewidth] {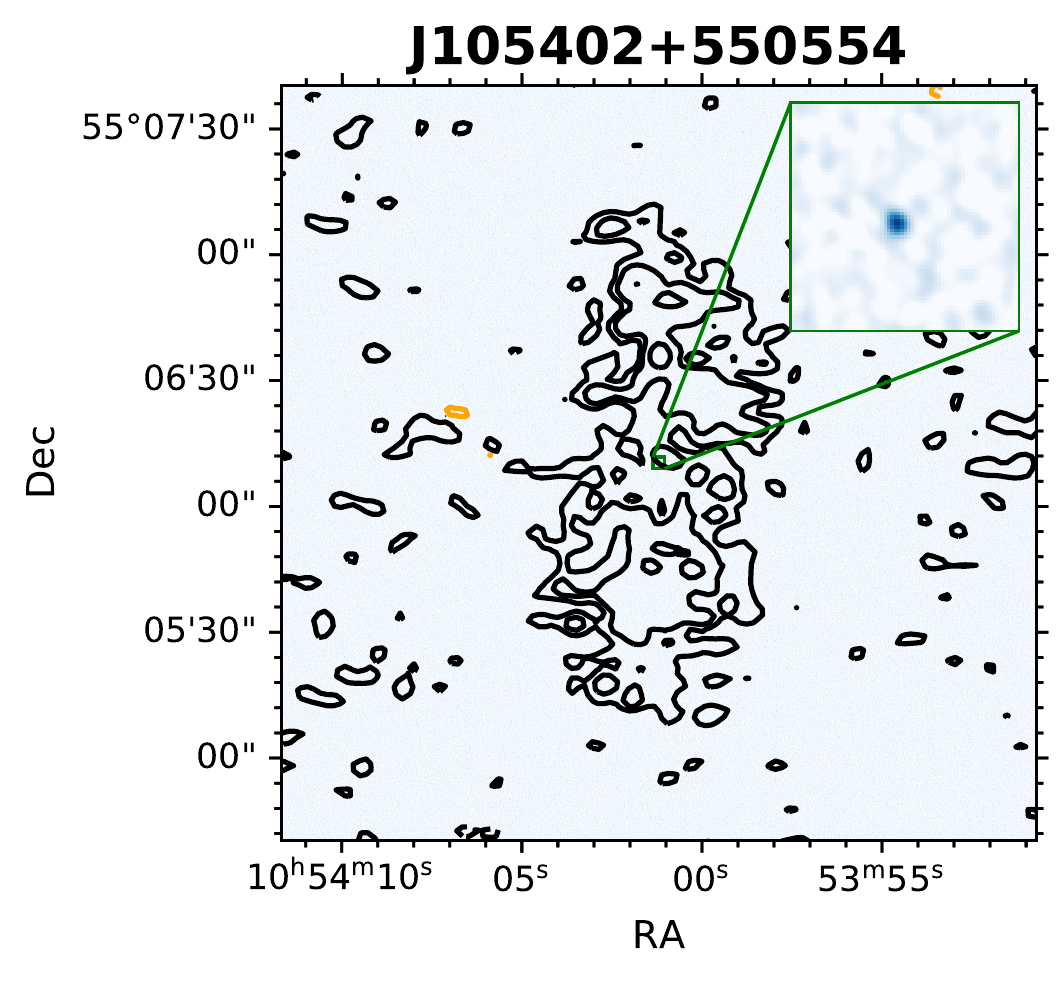}
 	    \includegraphics[width=0.195\linewidth] {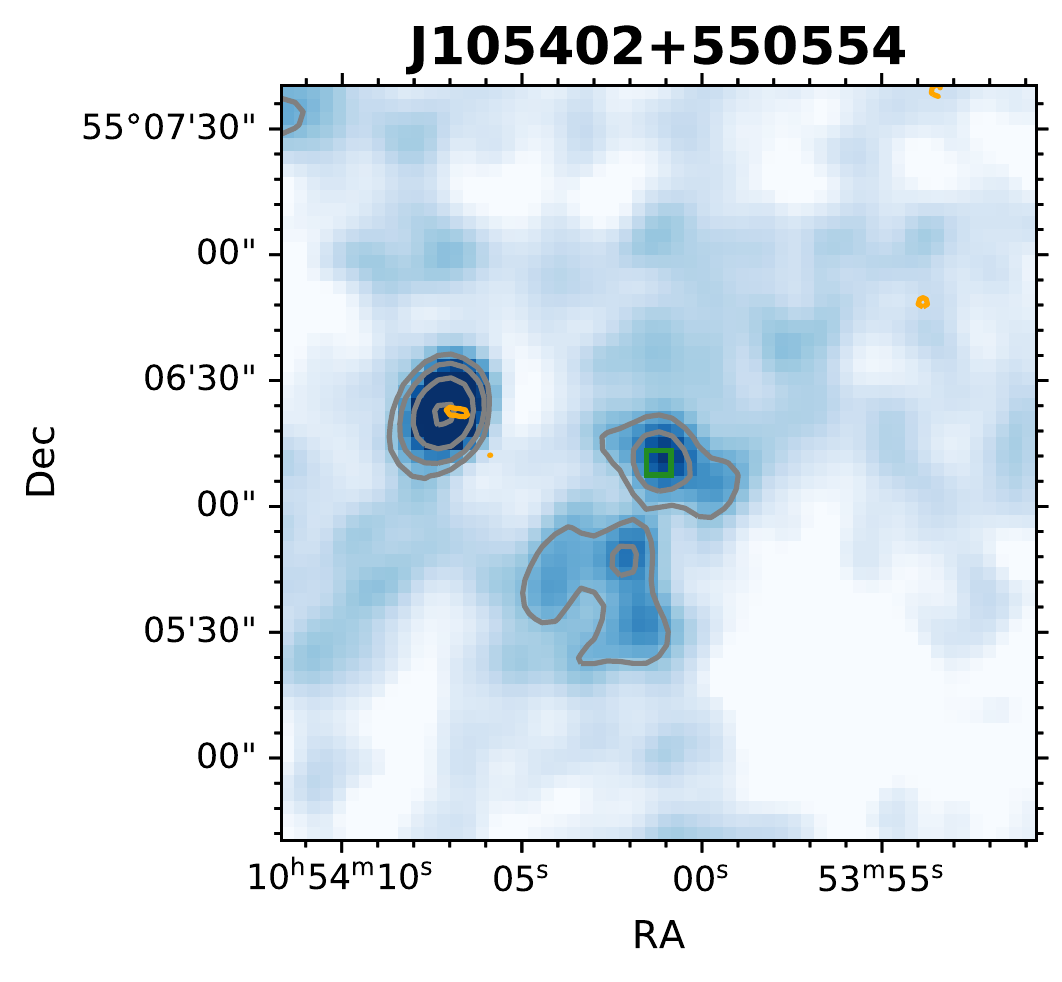}
\endminipage \hfill
\minipage{\textwidth}
        \includegraphics[width=0.195\linewidth] {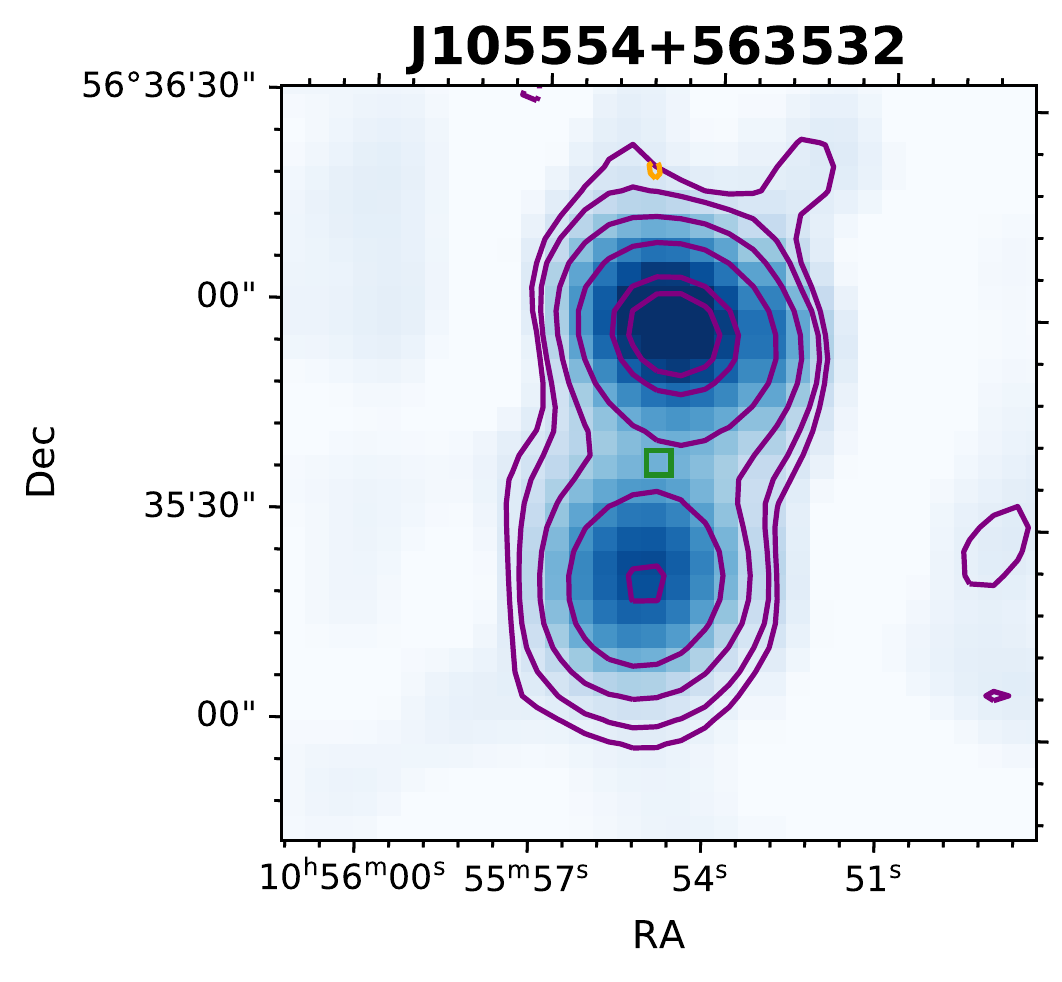}  
        \includegraphics[width=0.195\linewidth] {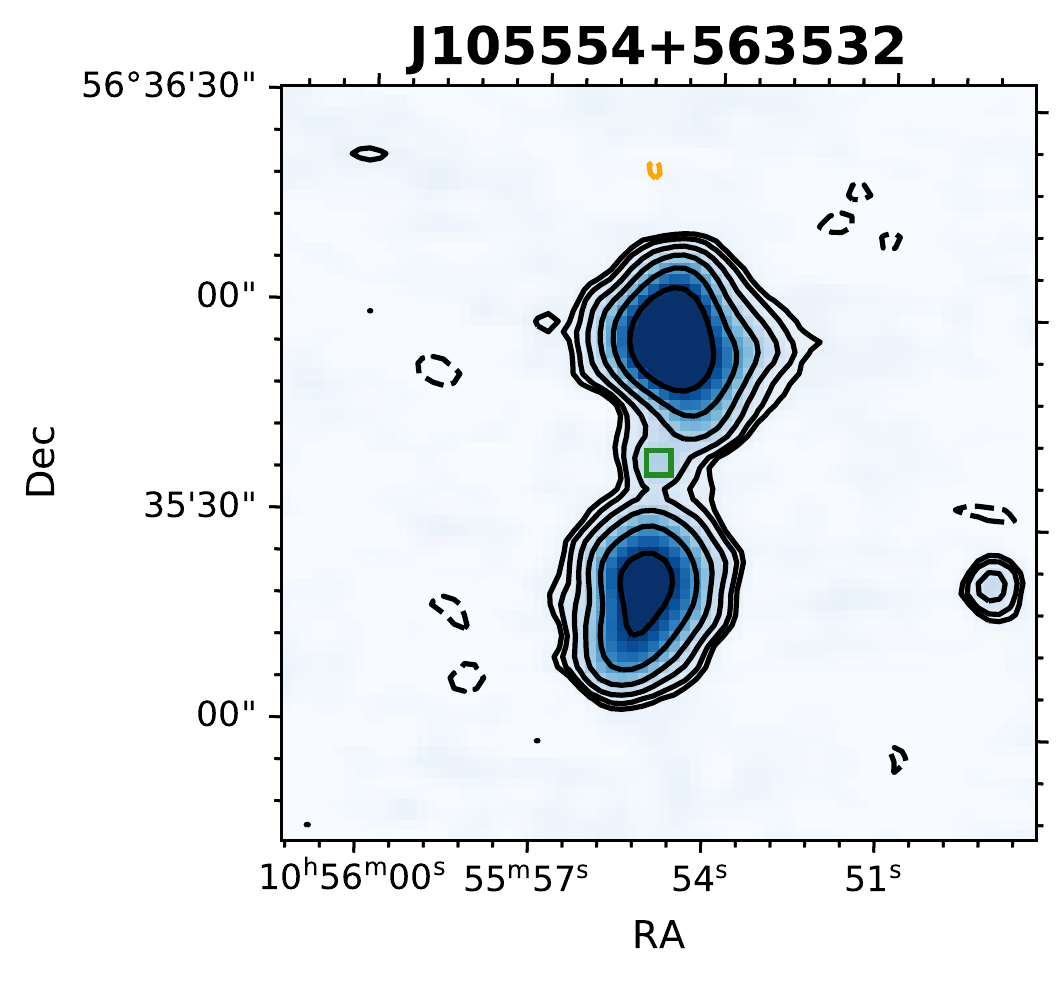} 
        \includegraphics[width=0.195\linewidth] {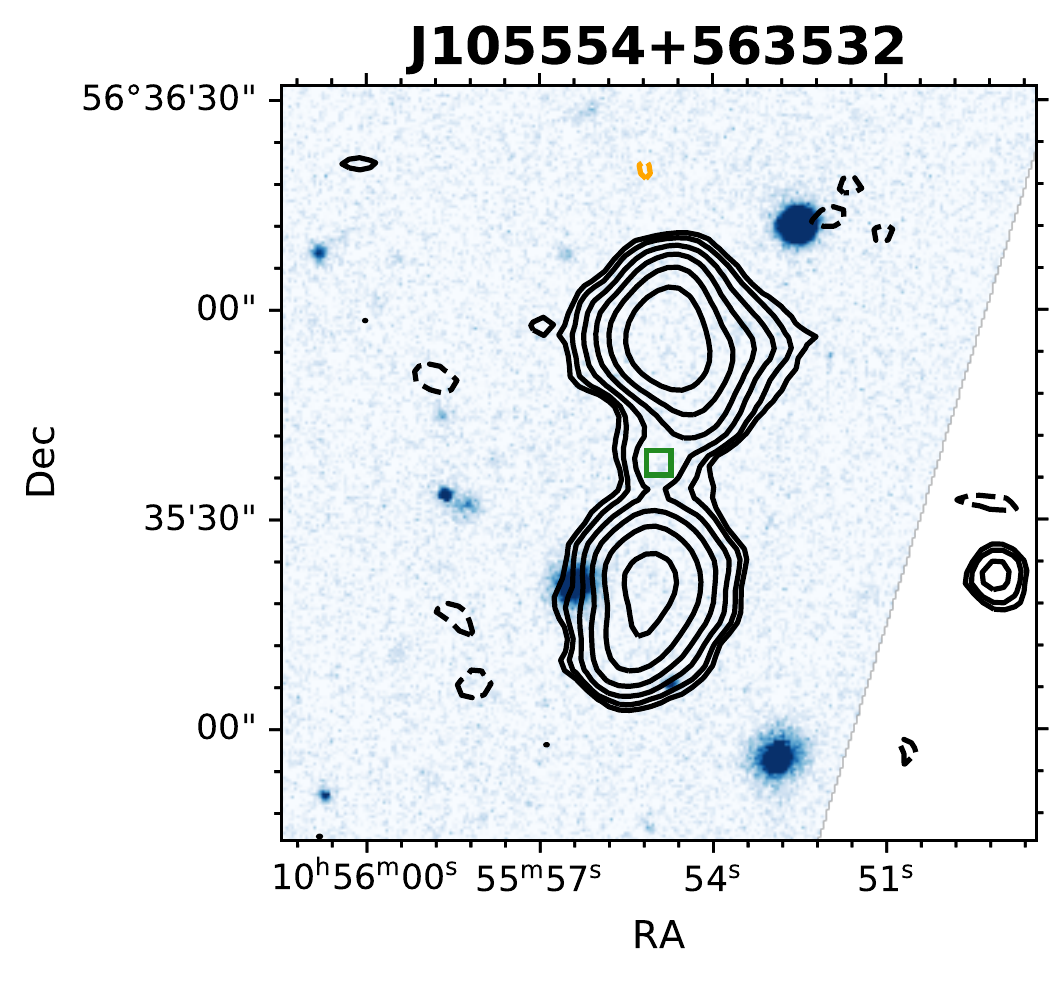}
        \includegraphics[width=0.195\linewidth] {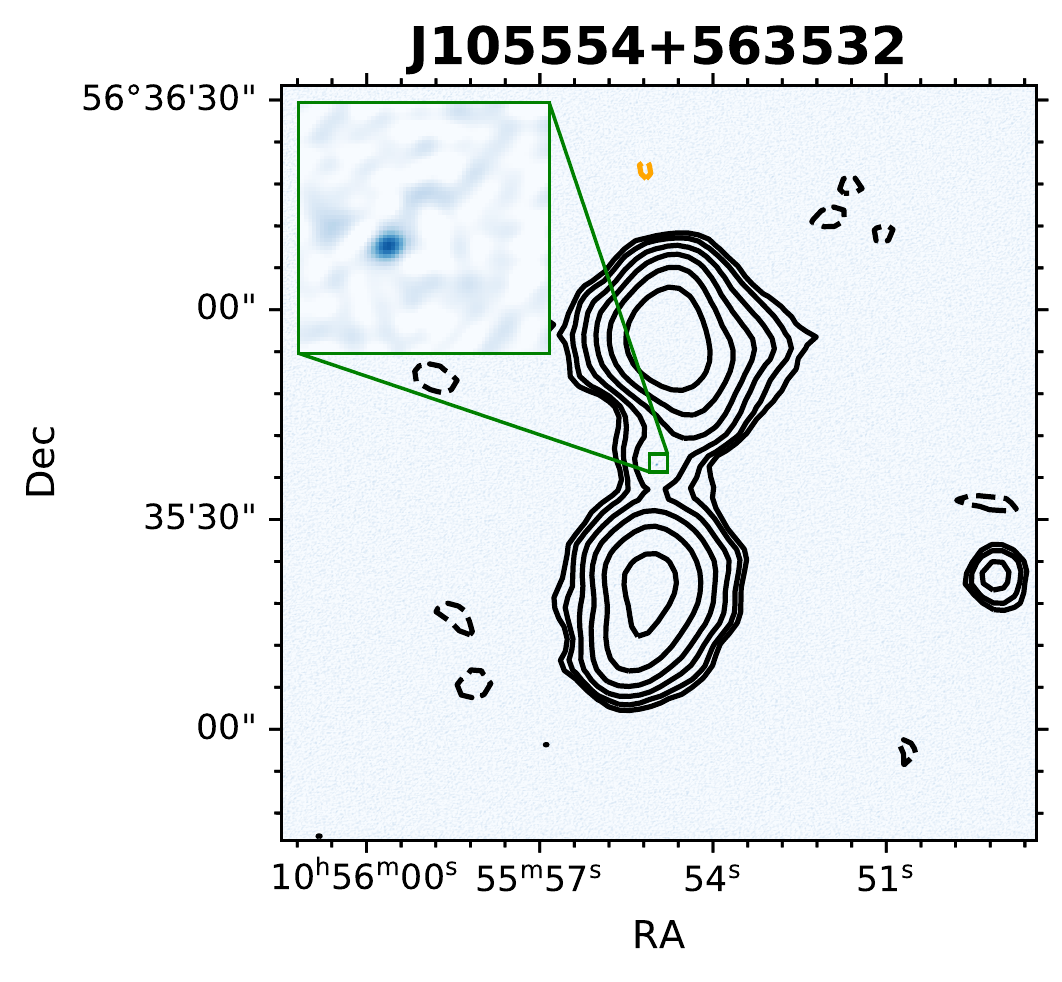}
\endminipage \hfill
\minipage{\textwidth}
        \includegraphics[width=0.195\linewidth] {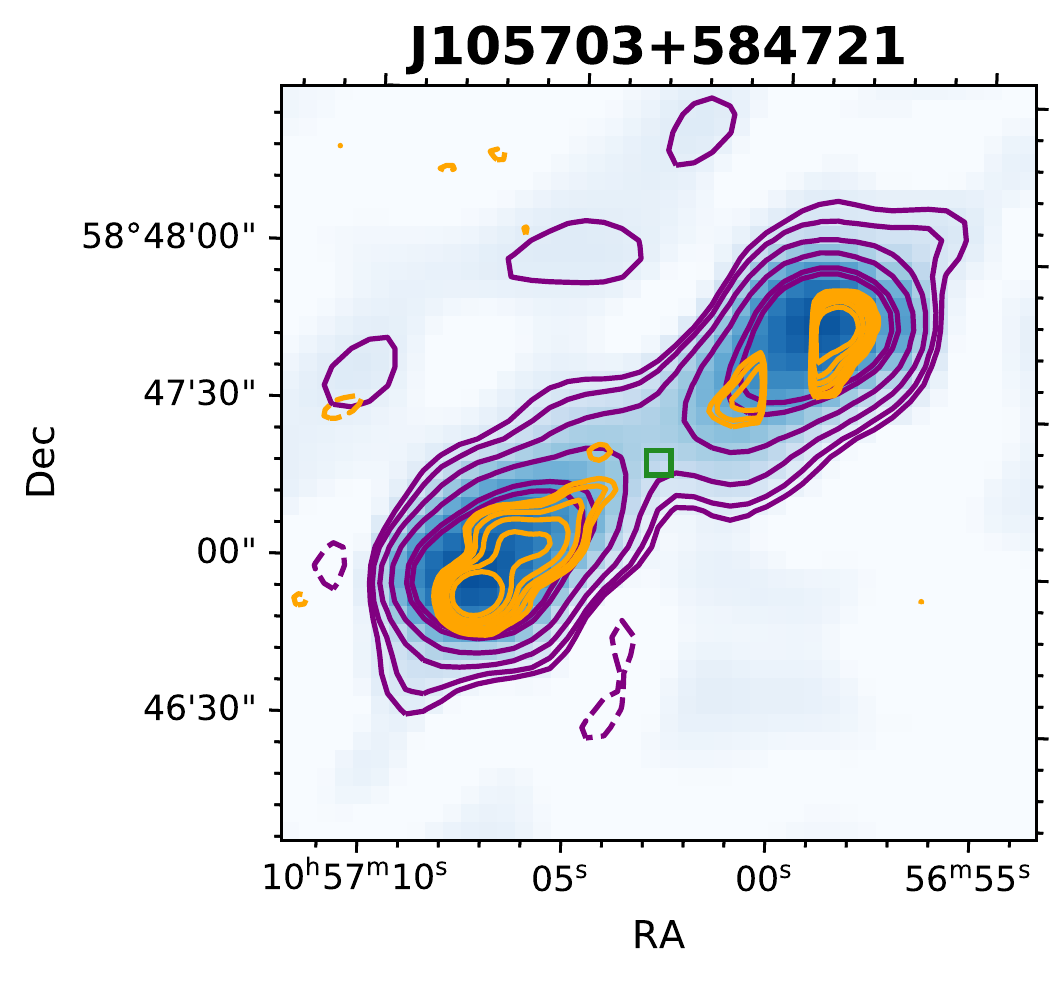}  
        \includegraphics[width=0.195\linewidth] {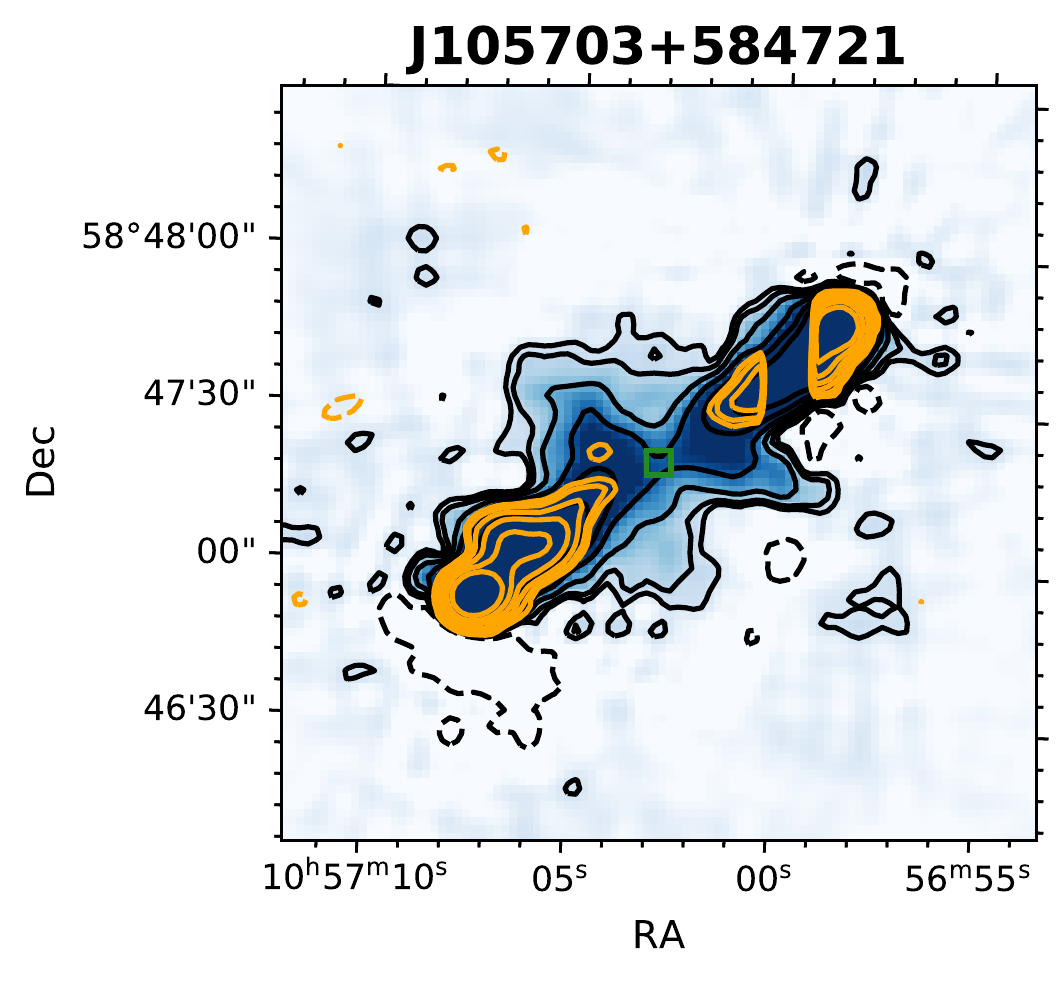} 
        \includegraphics[width=0.195\linewidth] {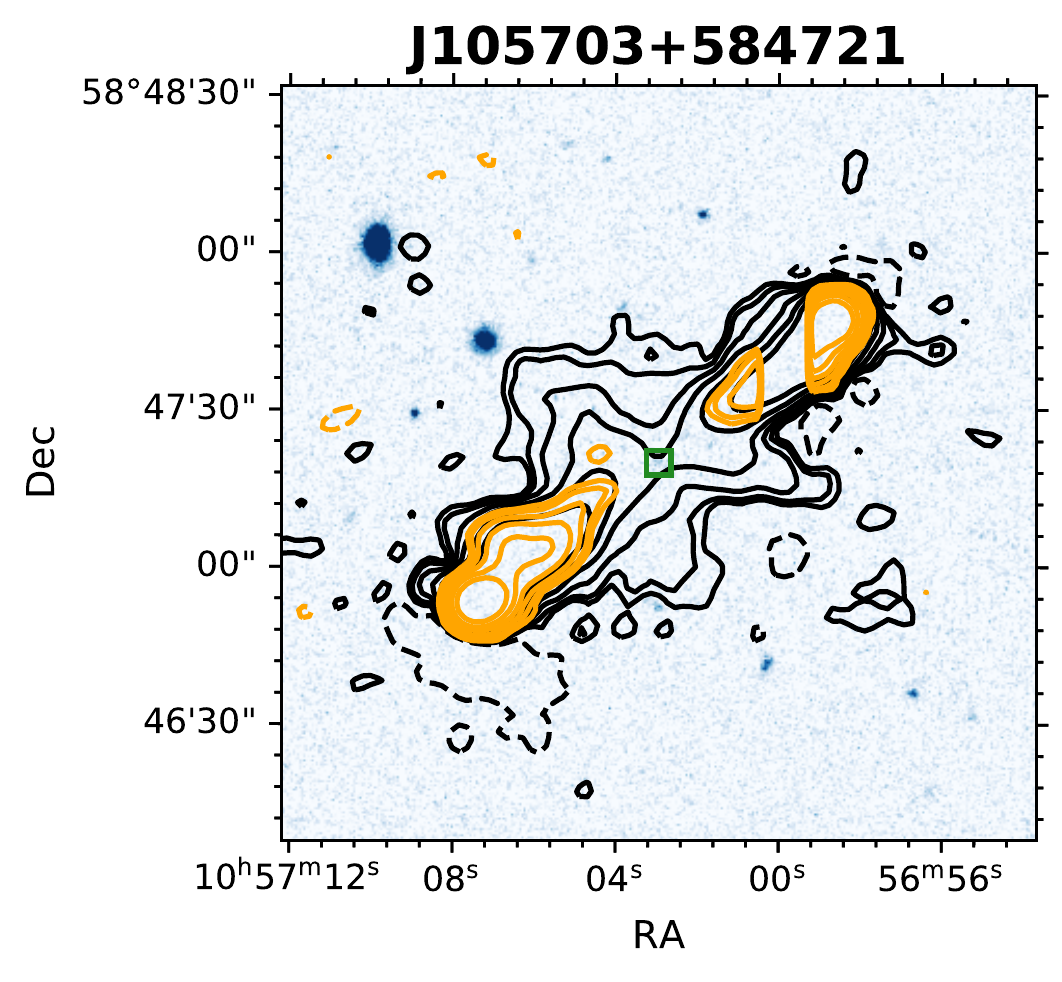}
        \includegraphics[width=0.195\linewidth] {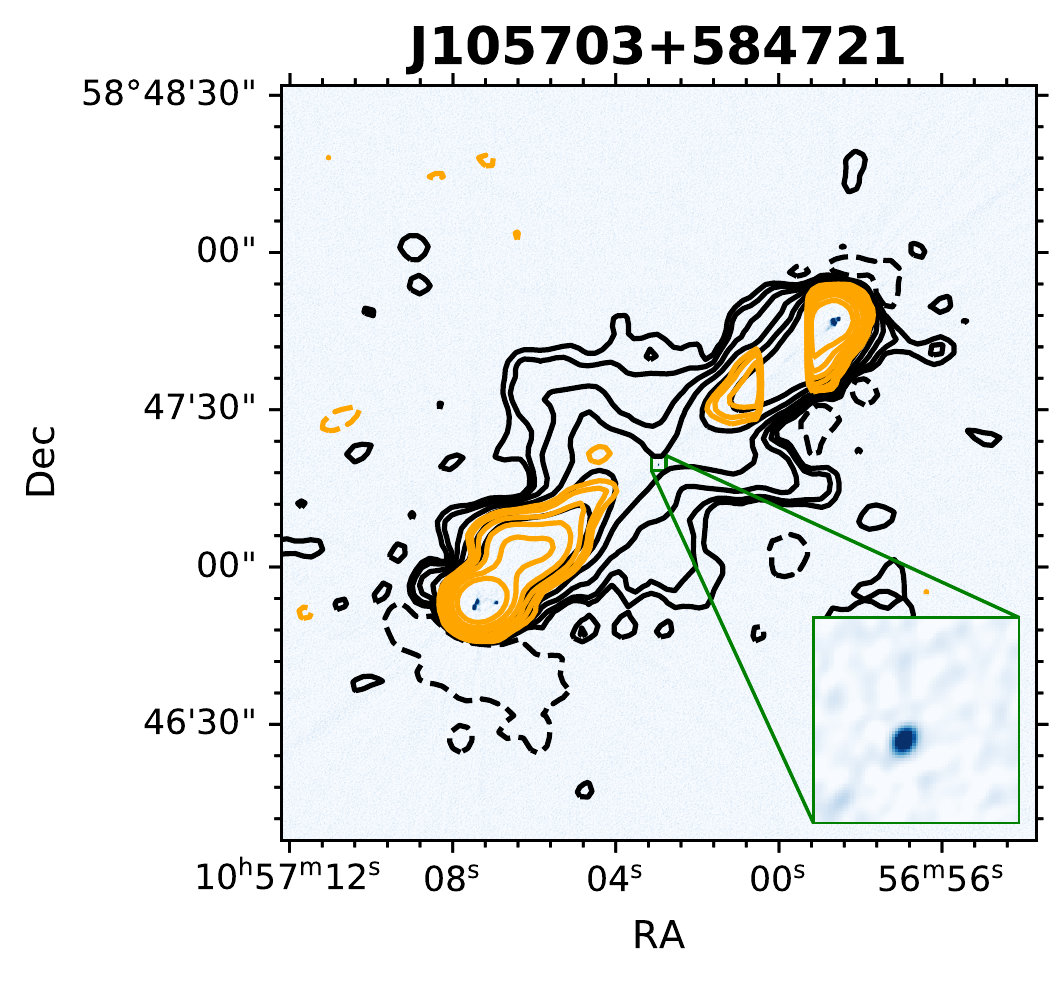}
\endminipage \hfill
\minipage{\textwidth}
        \includegraphics[width=0.195\linewidth] {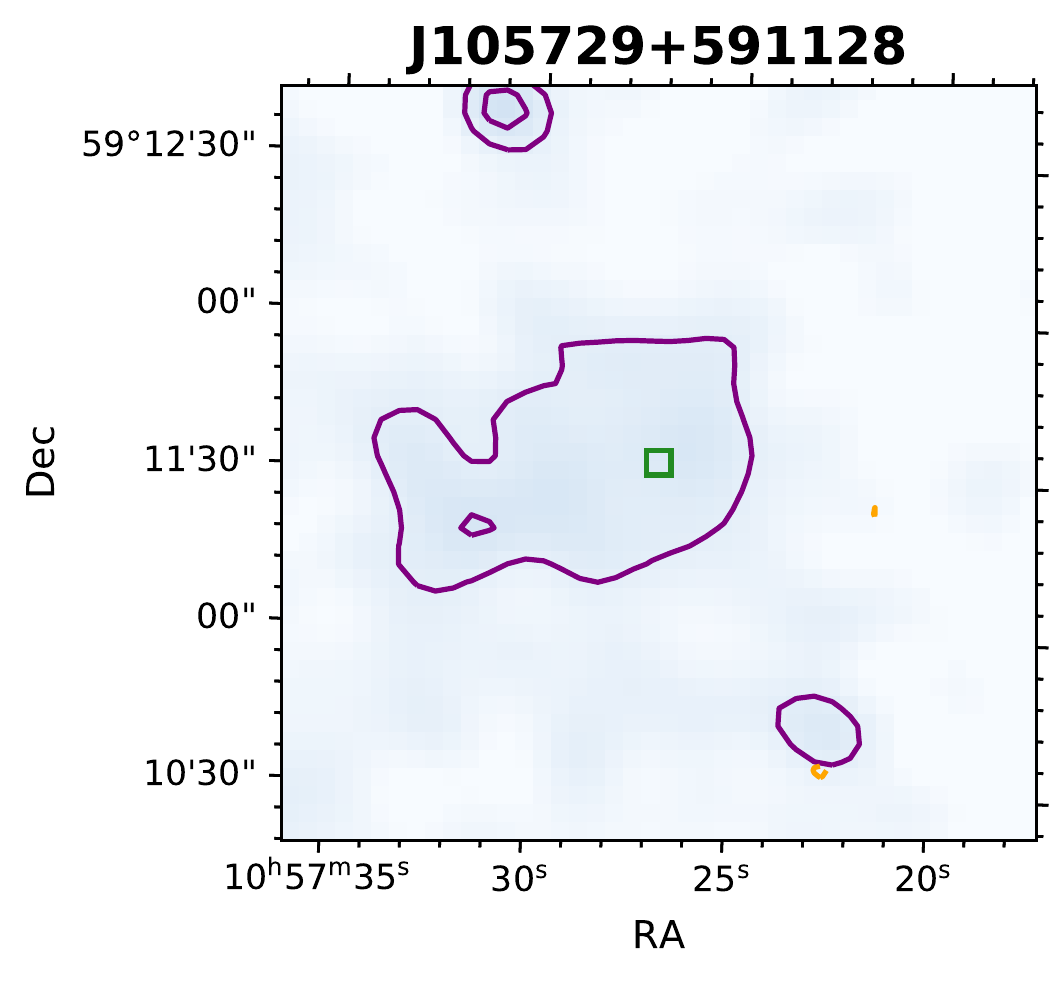}  
        \includegraphics[width=0.195\linewidth] {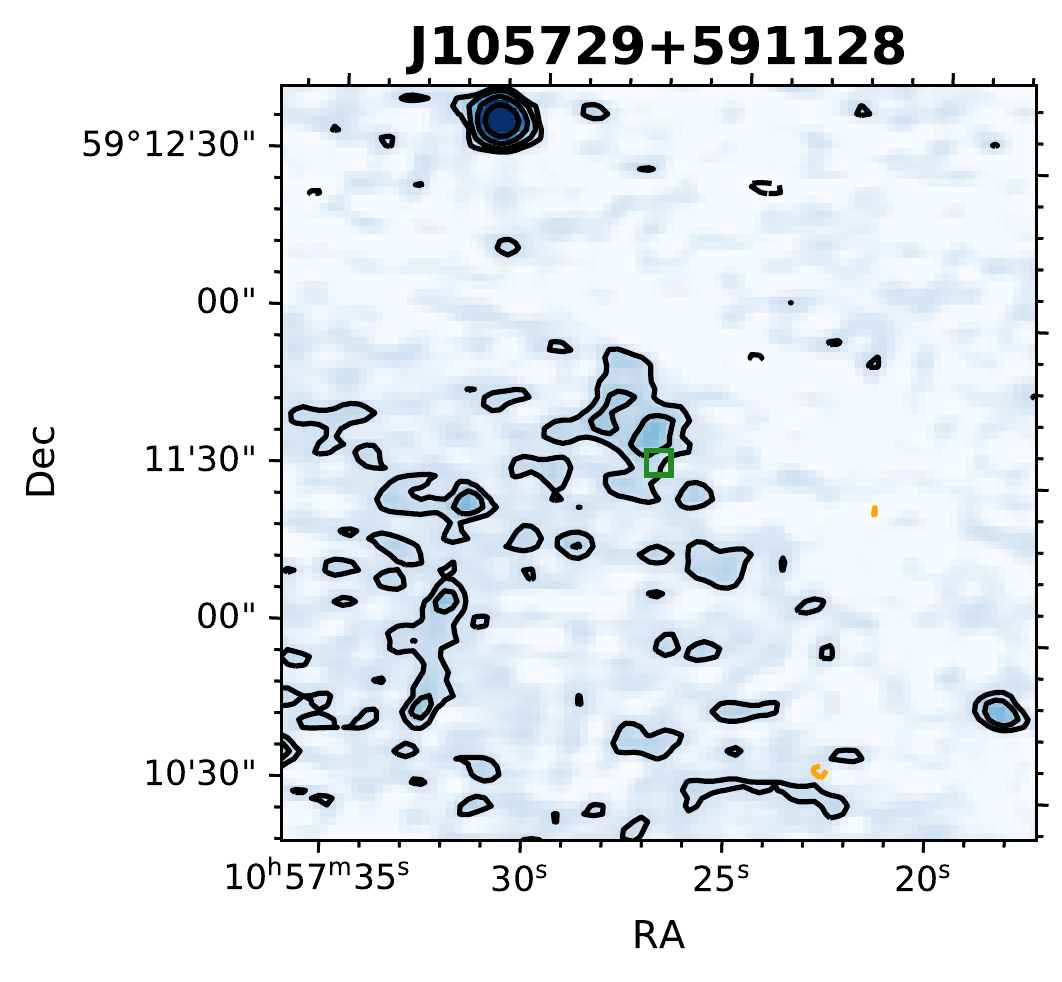} 
        \includegraphics[width=0.195\linewidth] {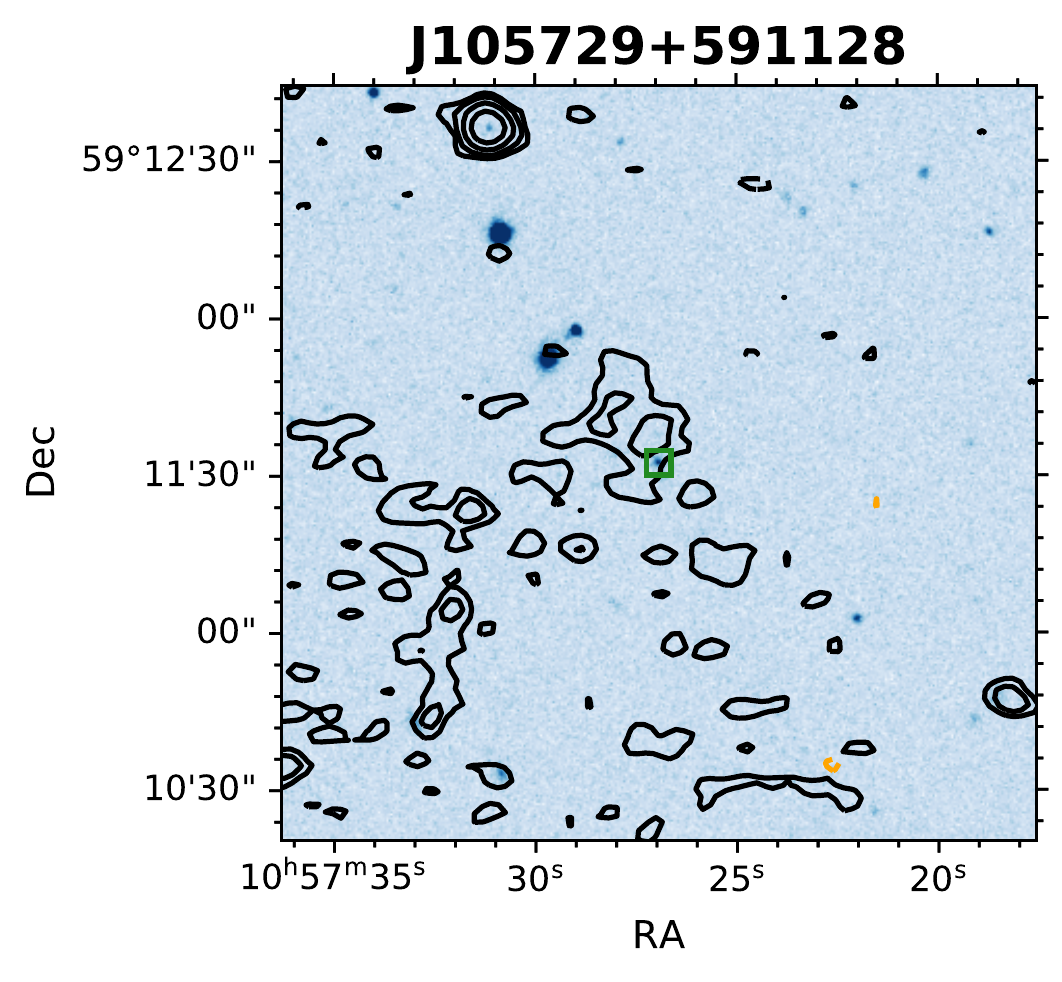}
        \includegraphics[width=0.195\linewidth] {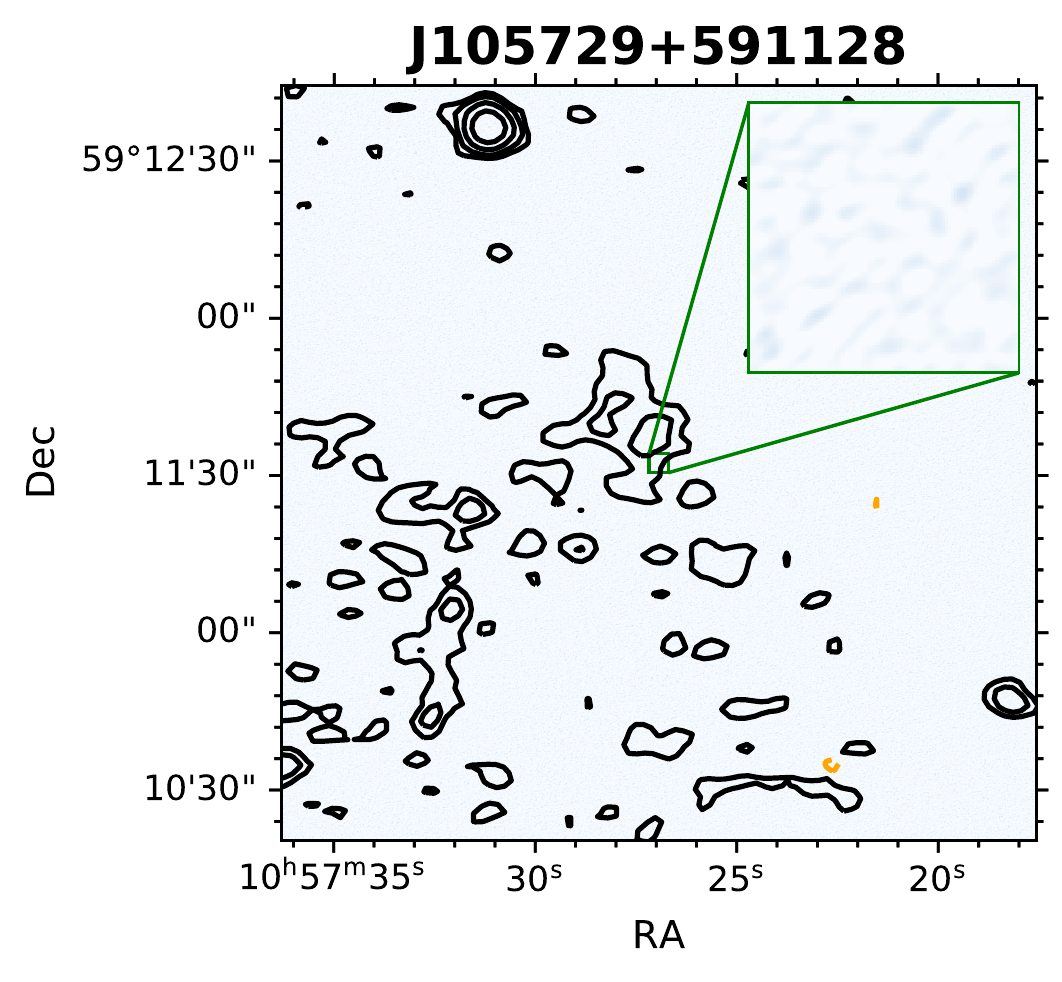}
 	    \includegraphics[width=0.195\linewidth] {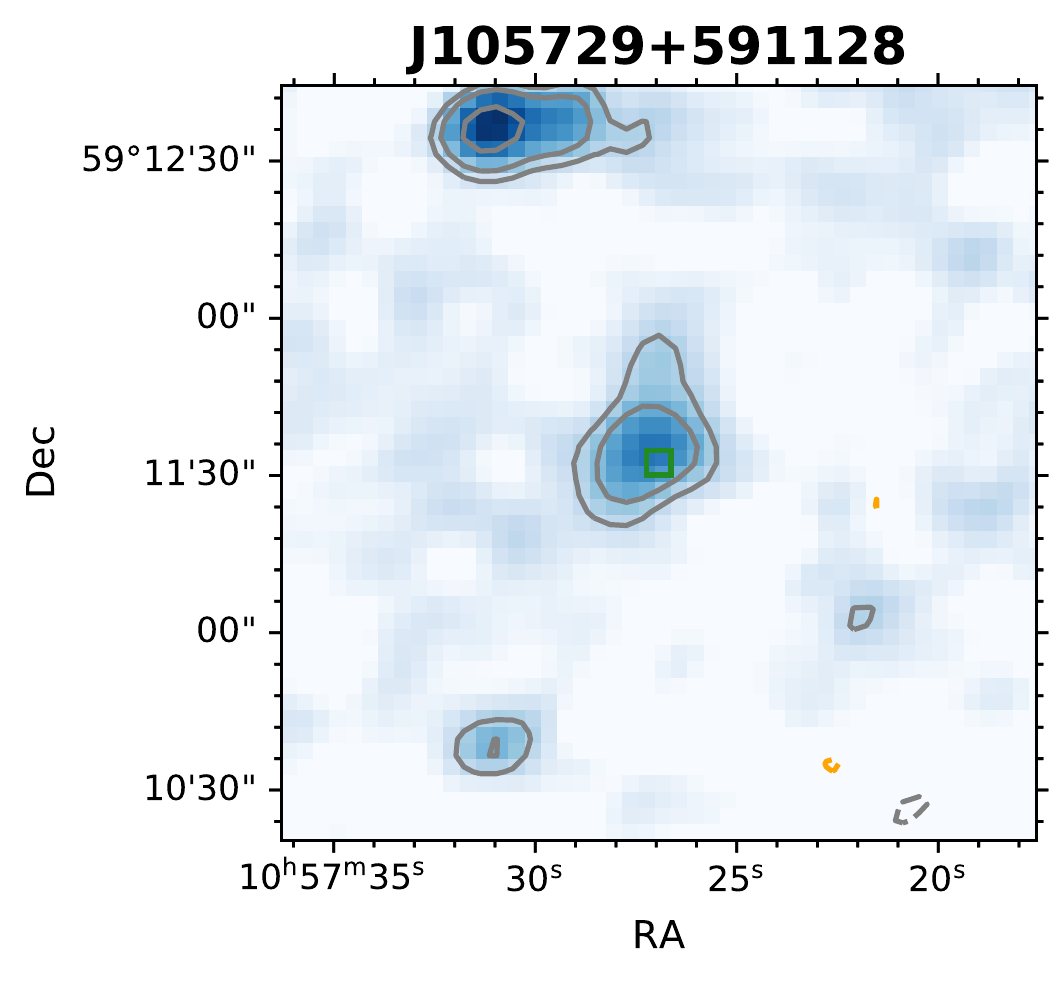}
\endminipage \hfill
\minipage{\textwidth}
        \includegraphics[width=0.195\linewidth] {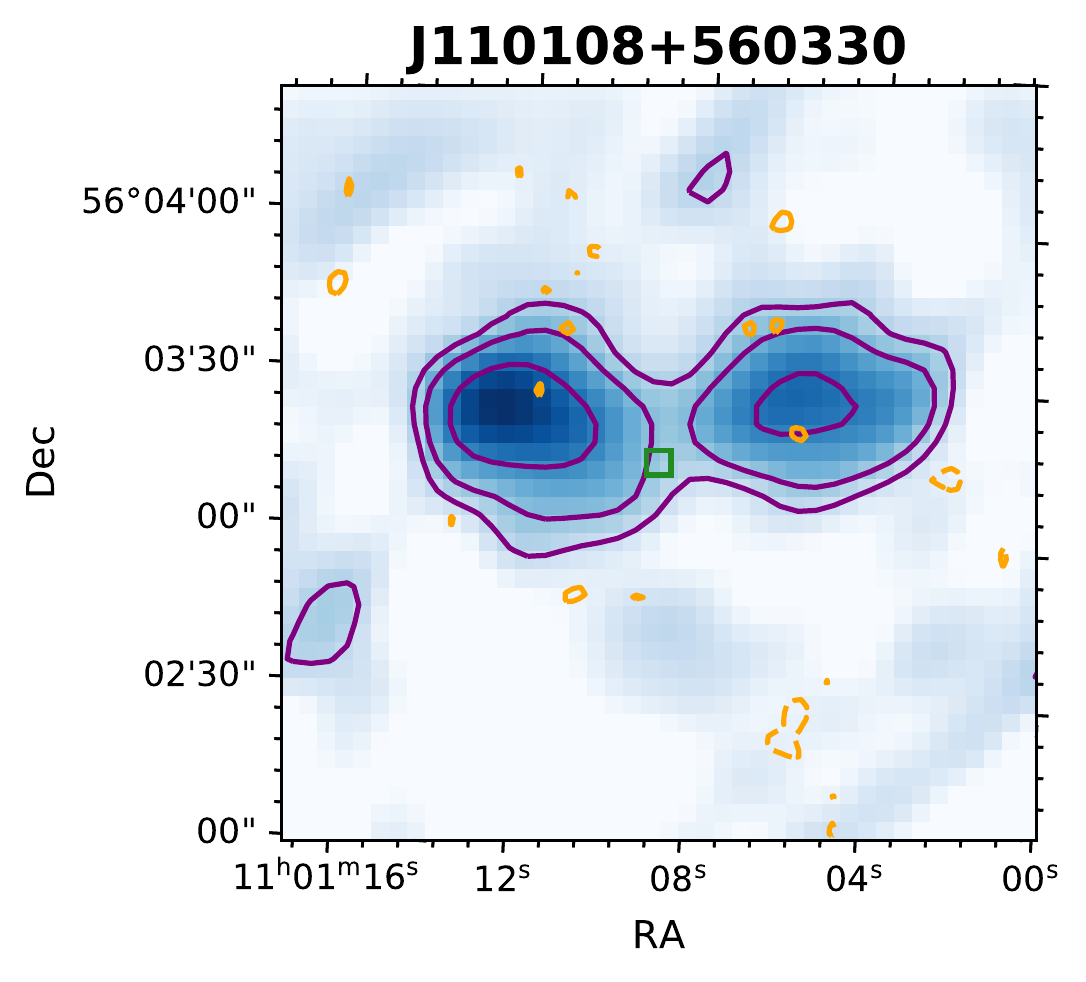}  
        \includegraphics[width=0.195\linewidth] {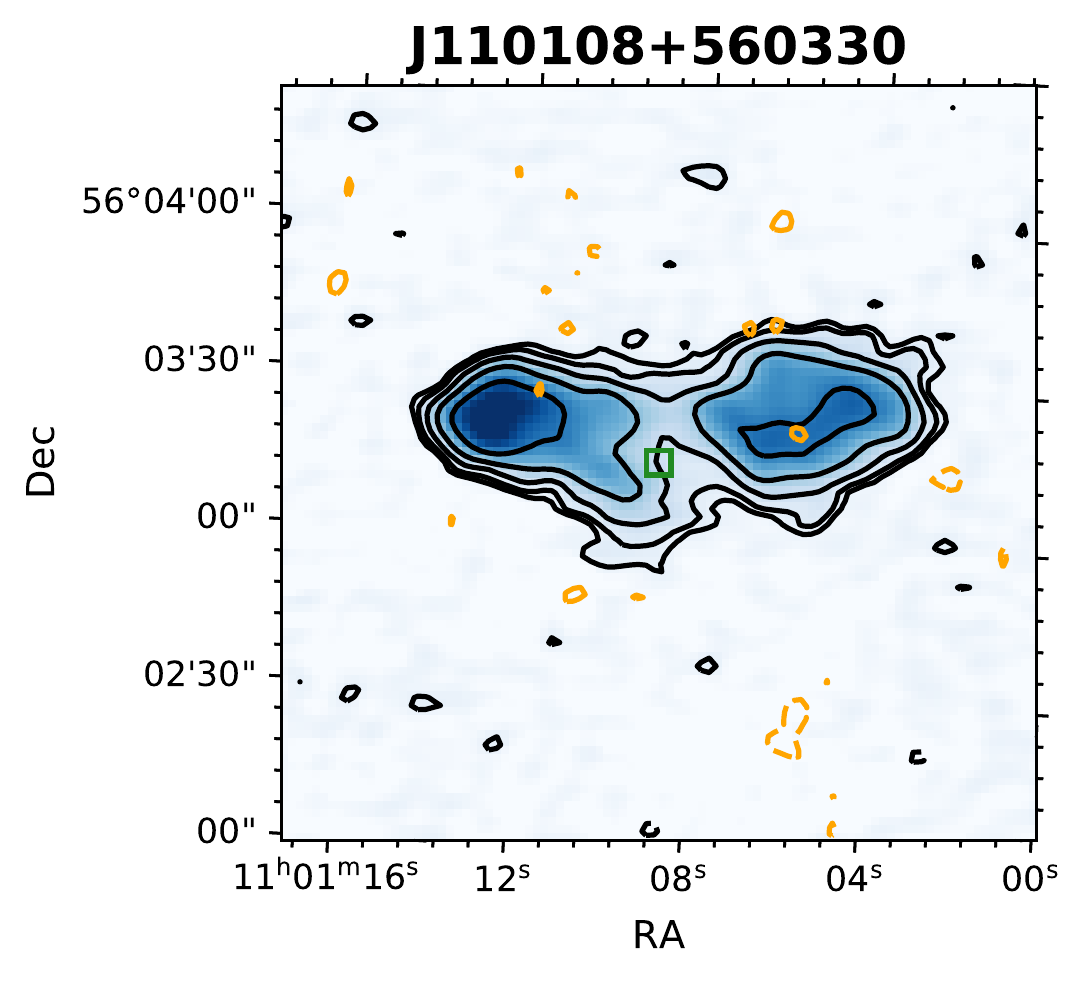} 
        \includegraphics[width=0.195\linewidth] {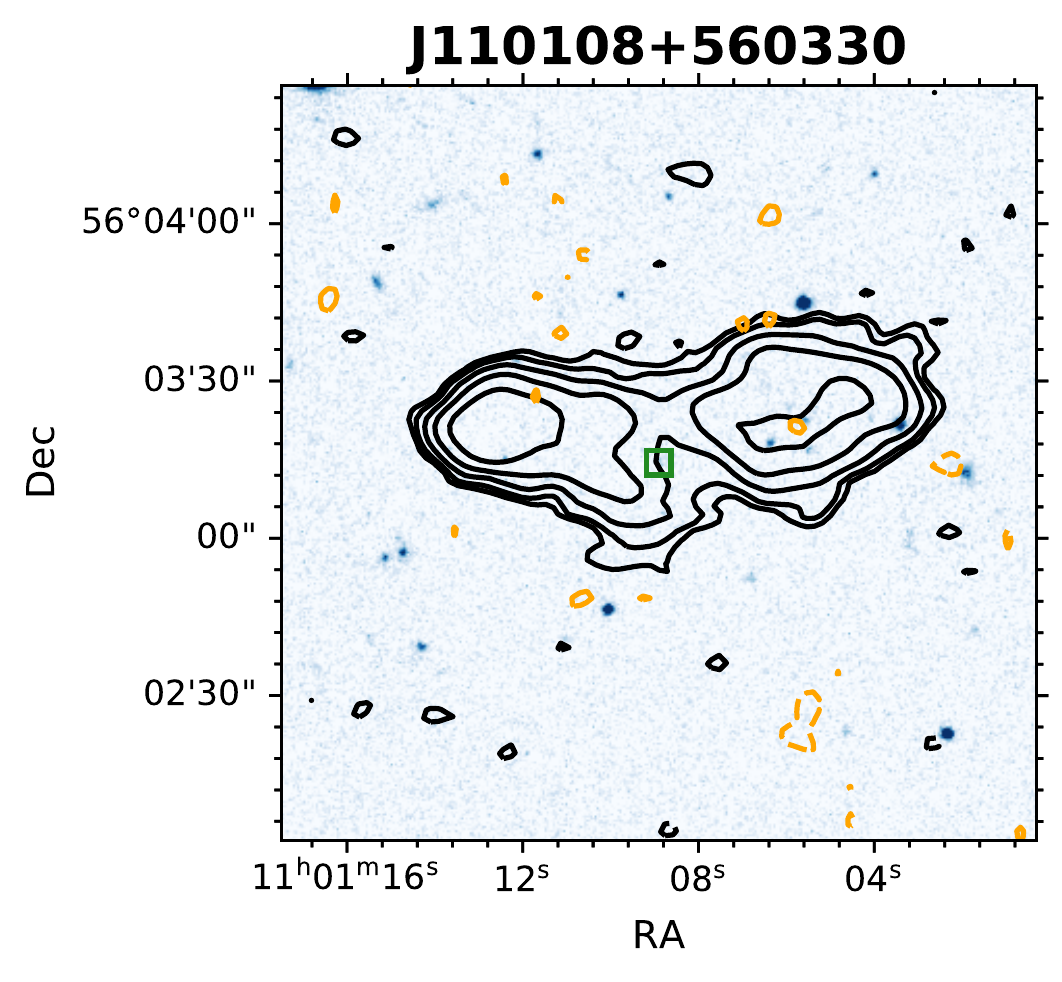}
        \includegraphics[width=0.195\linewidth] {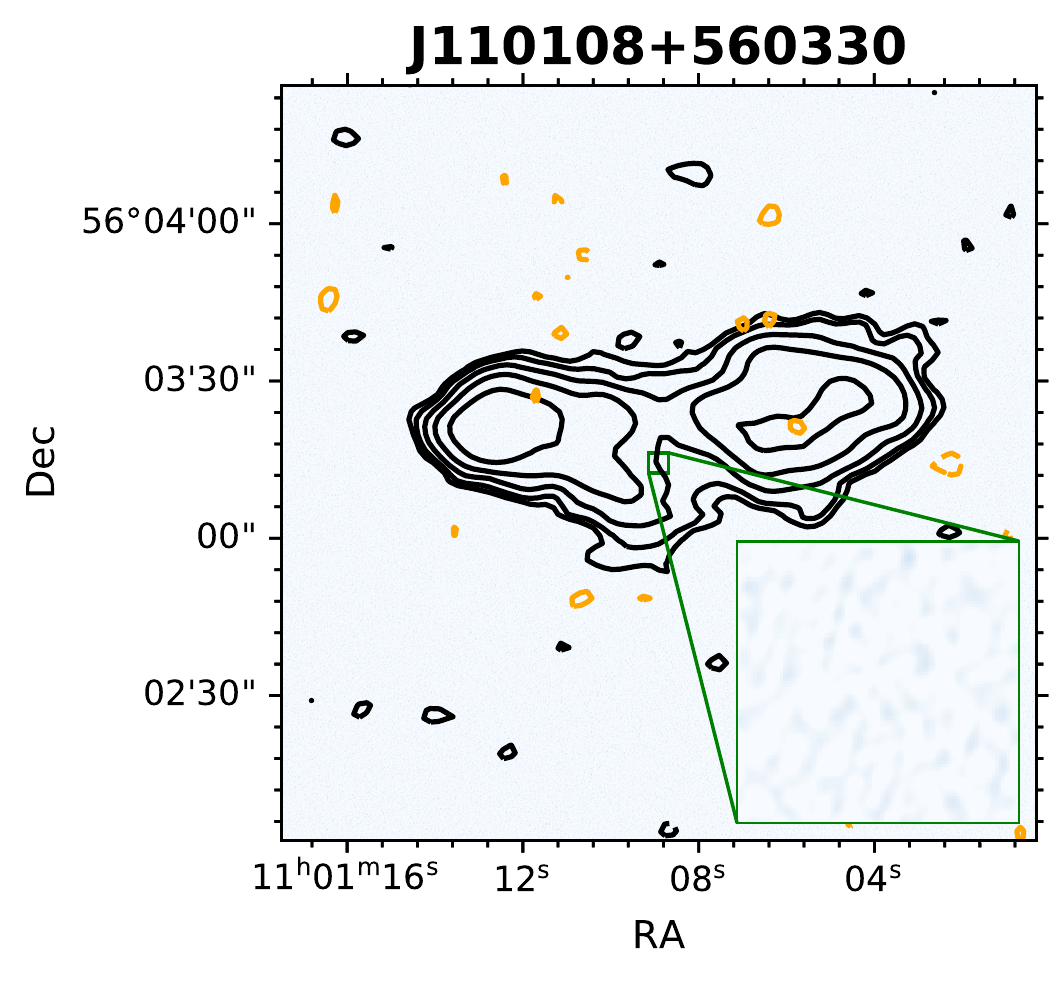}
\endminipage \hfill
\minipage{\textwidth}
        \includegraphics[width=0.195\linewidth] {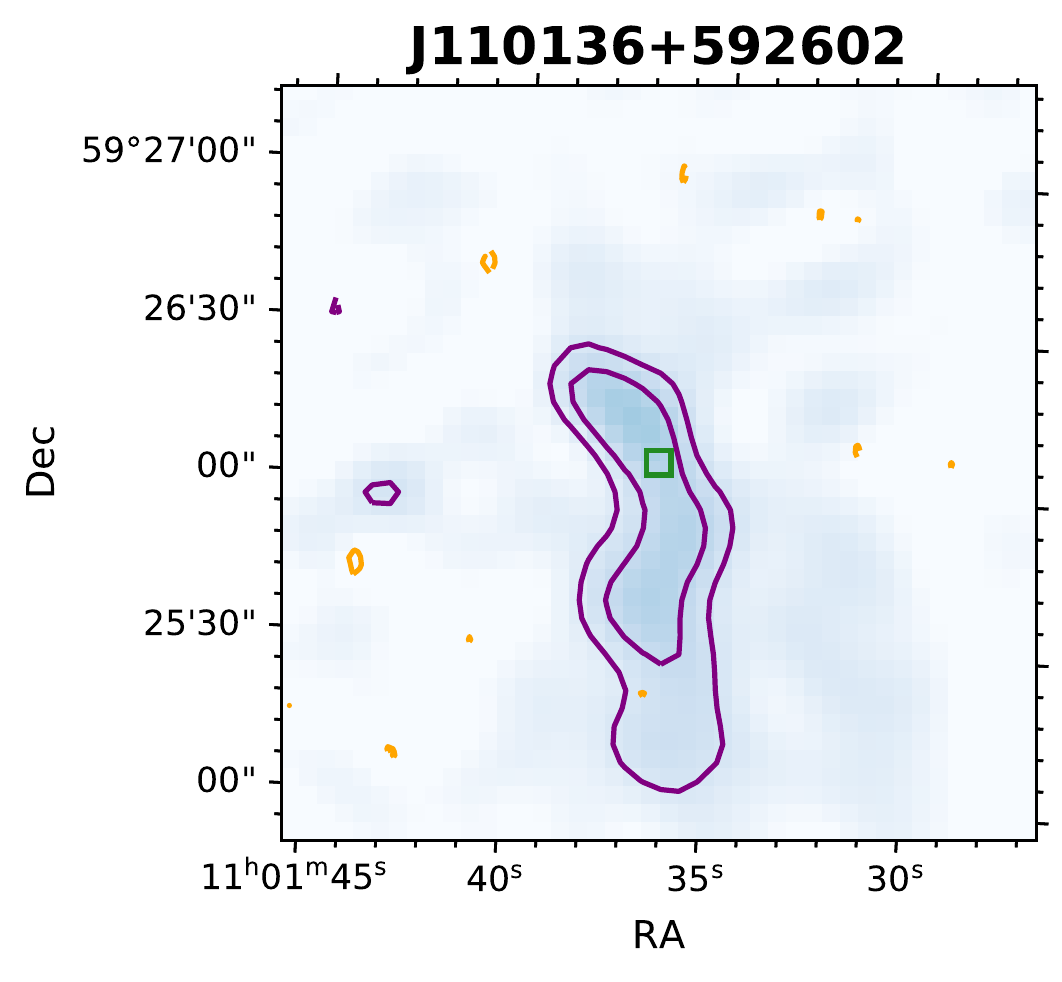}  
        \includegraphics[width=0.195\linewidth] {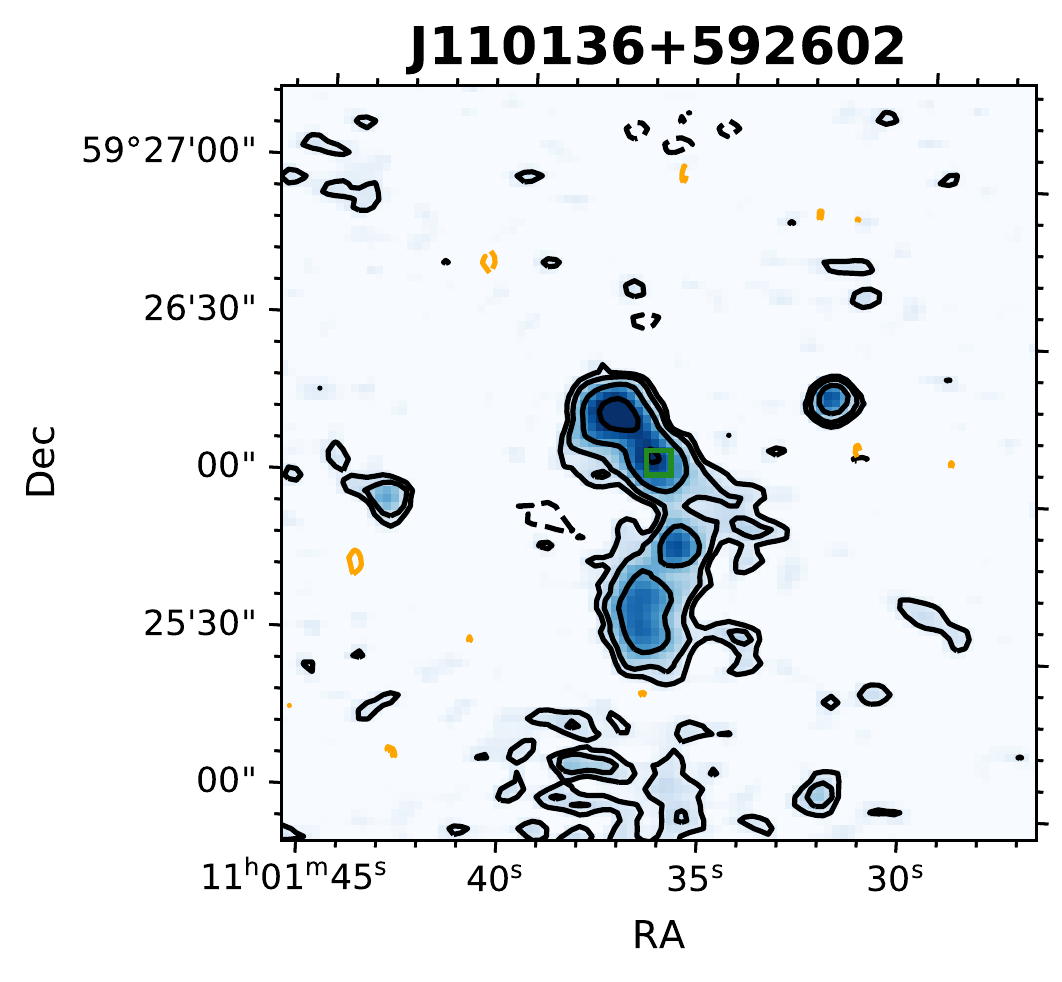} 
        \includegraphics[width=0.195\linewidth] {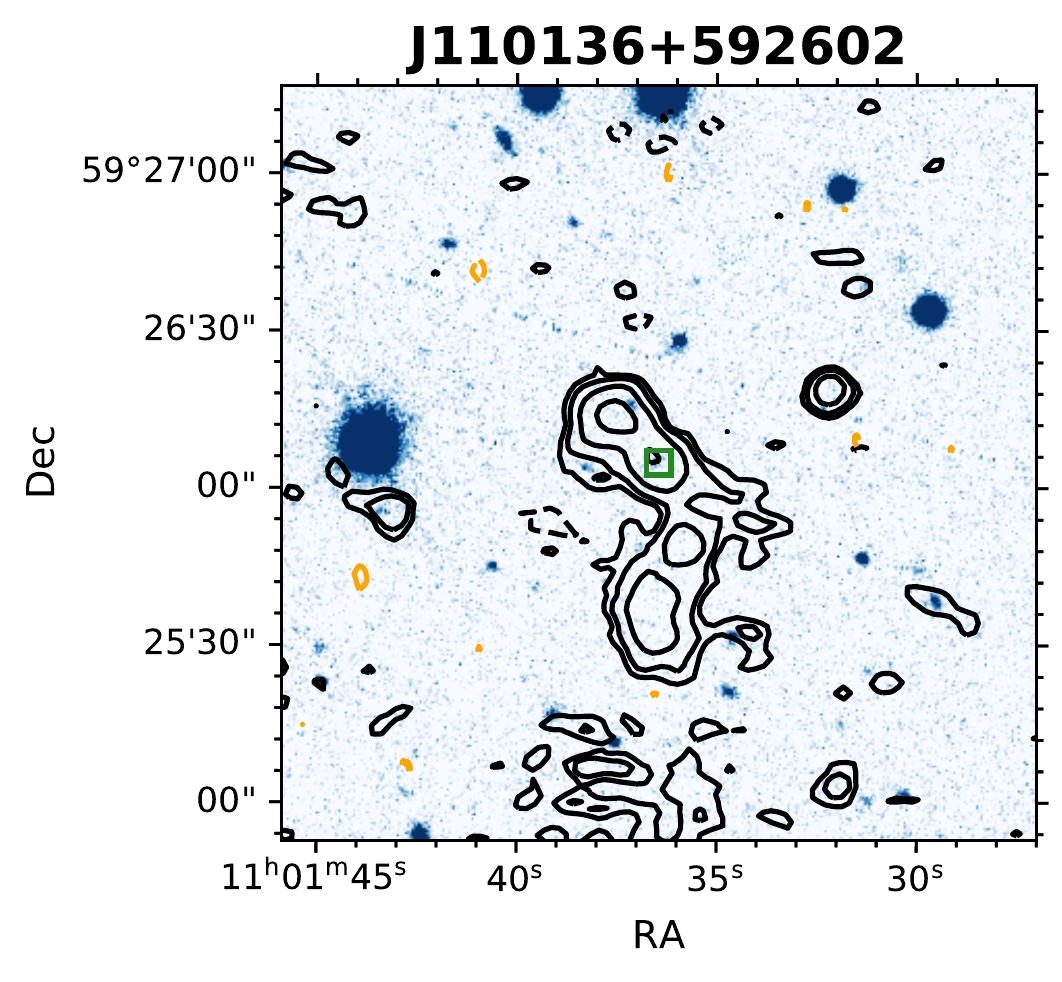}
        \includegraphics[width=0.195\linewidth] {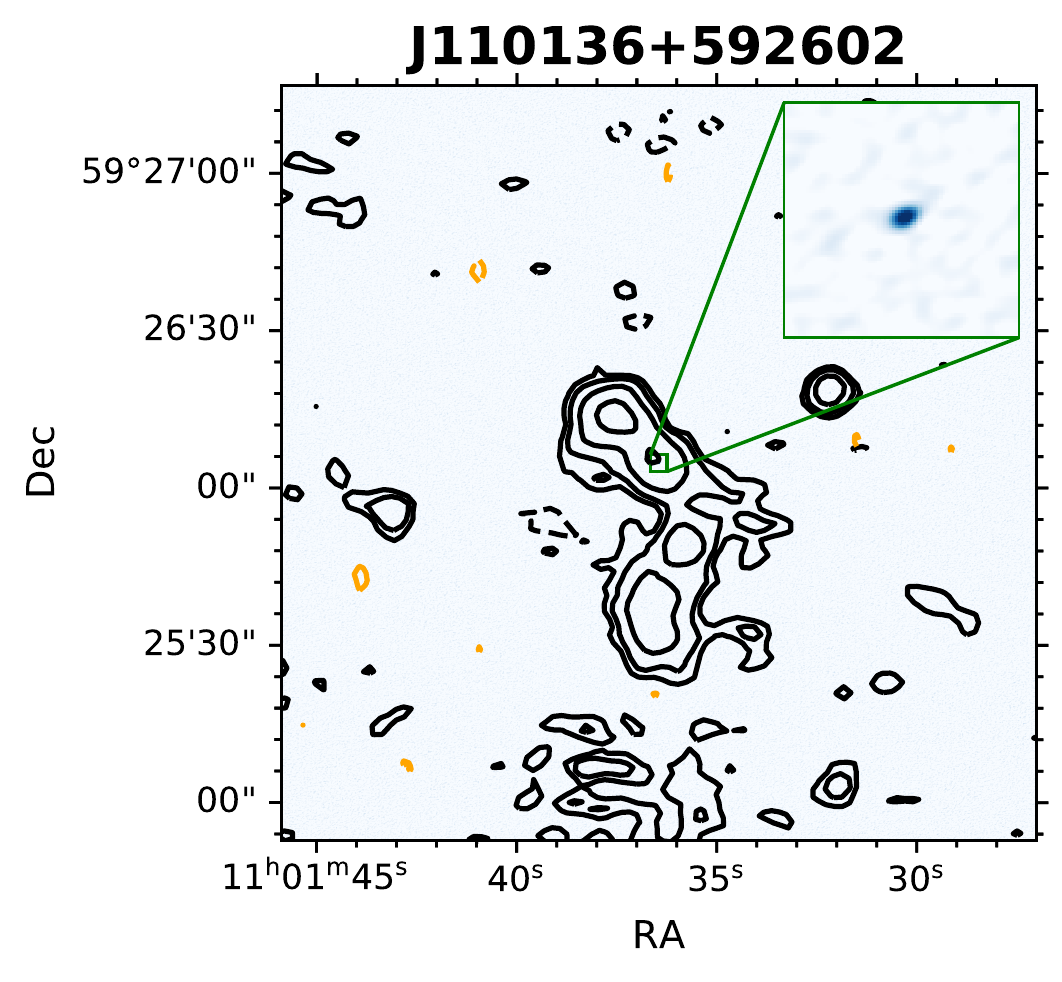}
\endminipage \hfill
\end{figure*}

\begin{figure*} [!h]\ContinuedFloat
\caption{continued}
\minipage{\textwidth}
        \includegraphics[width=0.195\linewidth] {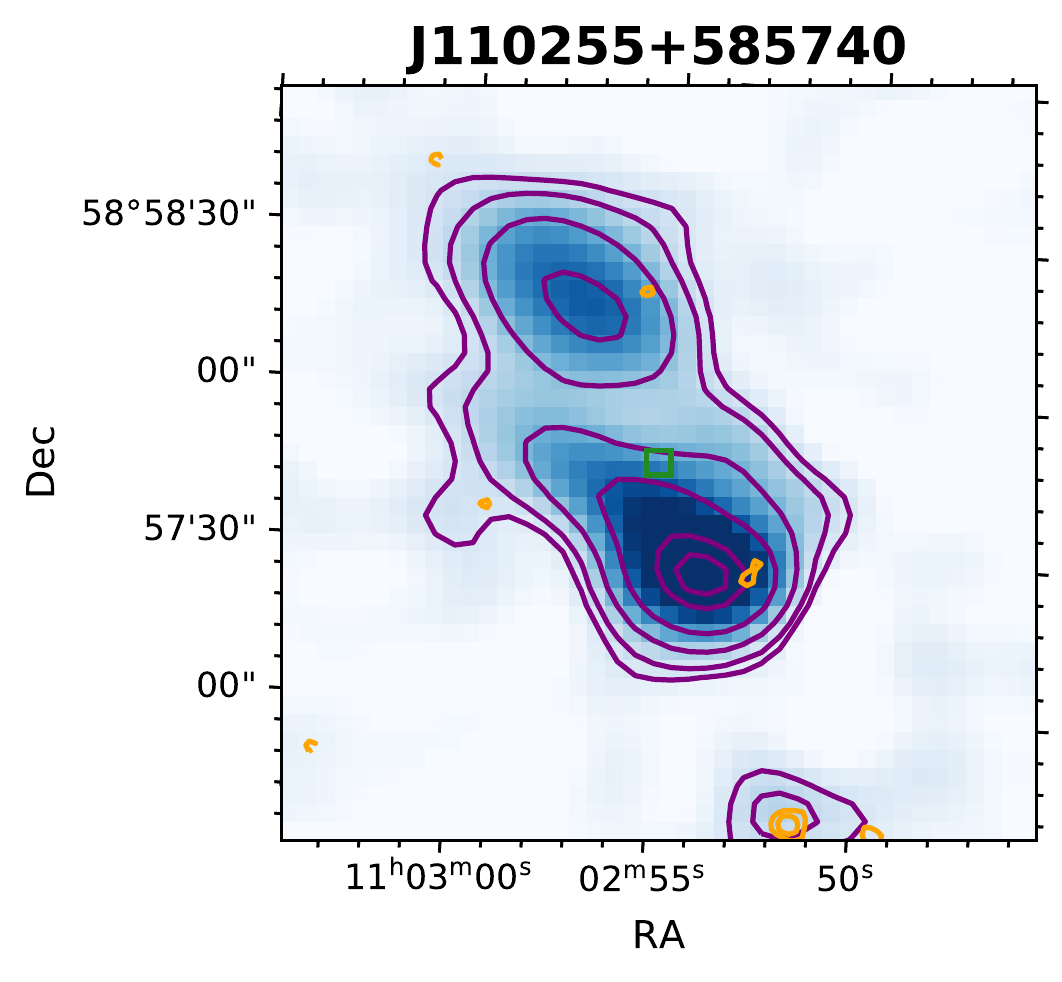}  
        \includegraphics[width=0.195\linewidth] {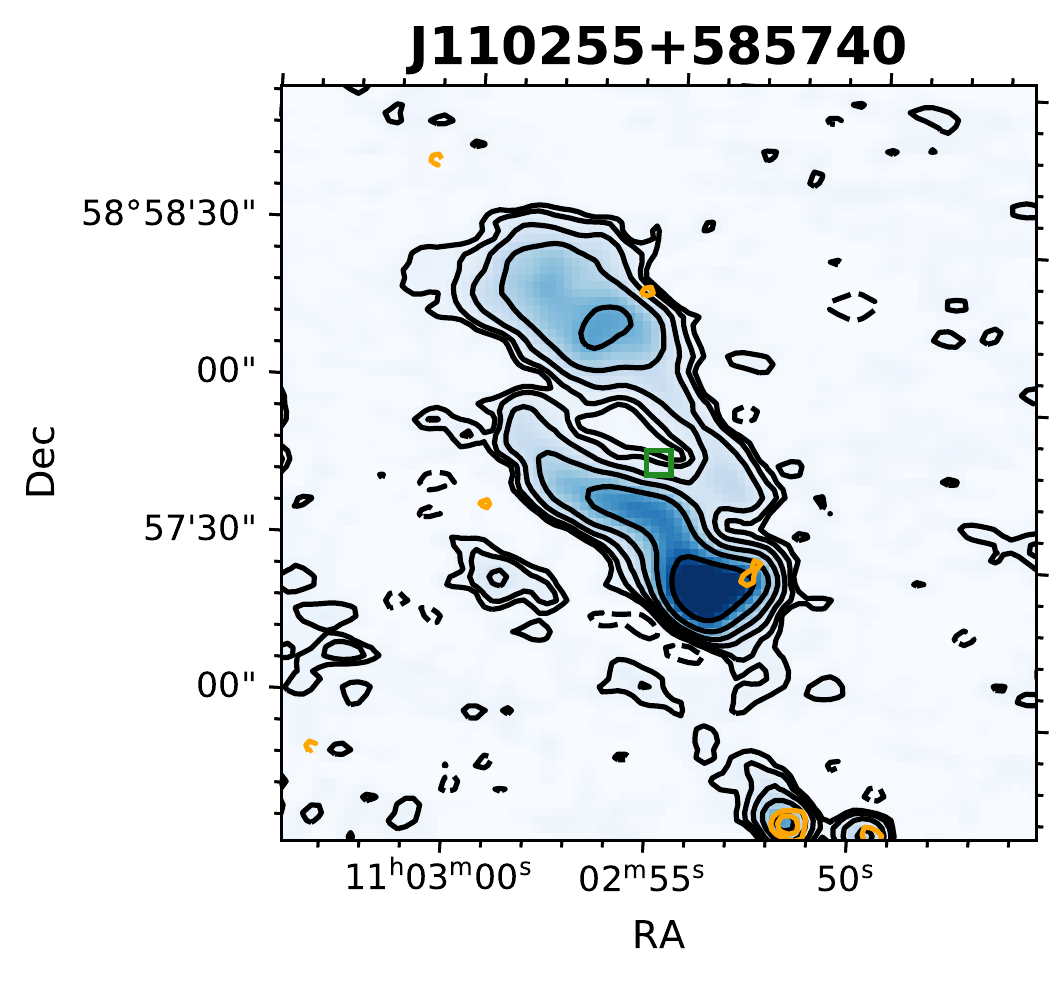} 
        \includegraphics[width=0.195\linewidth] {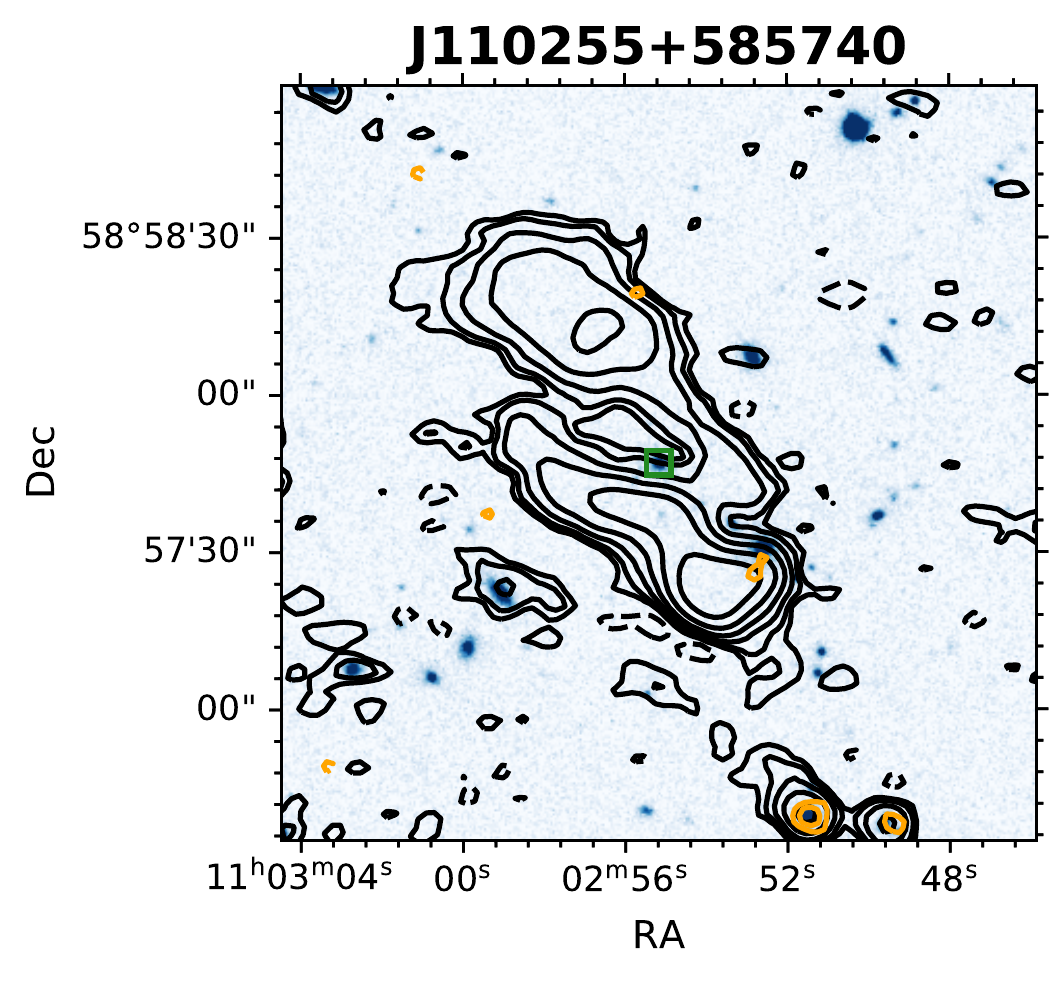}
        \includegraphics[width=0.195\linewidth] {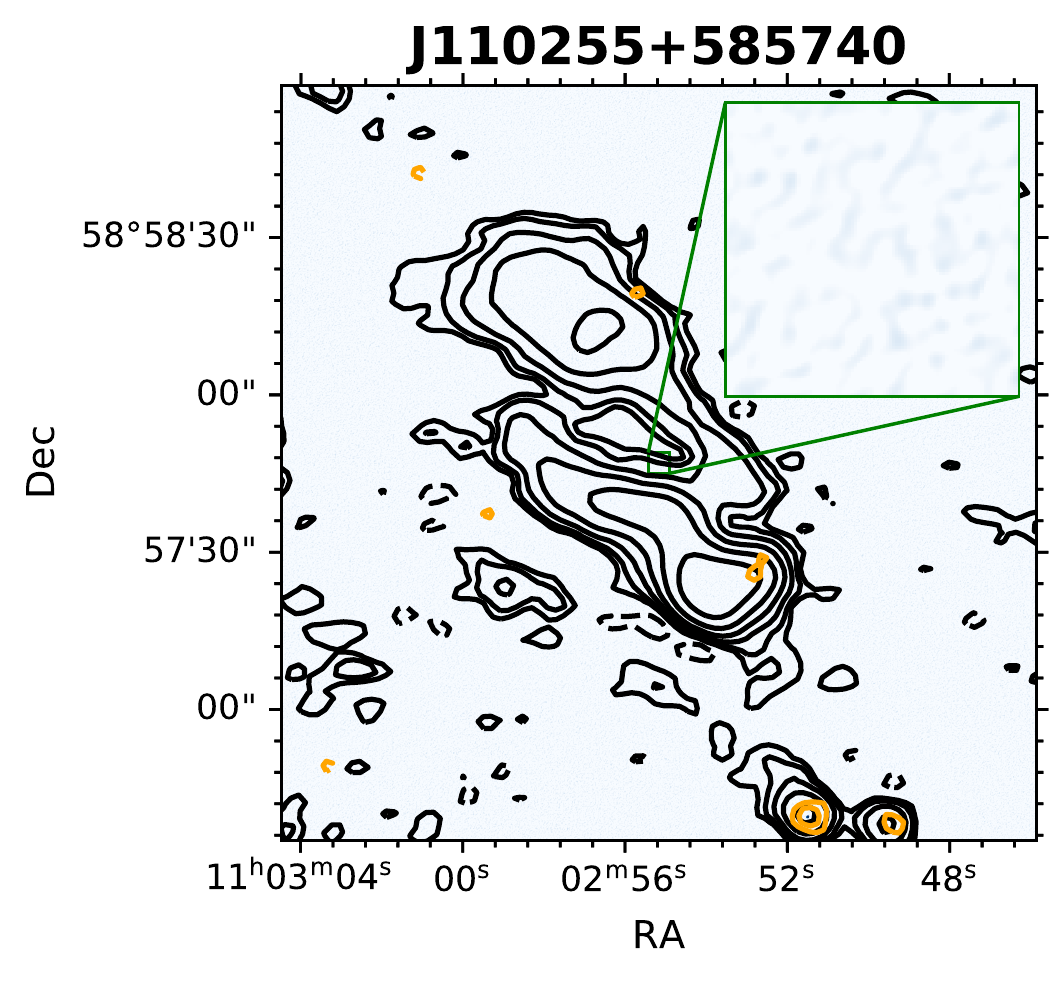}
\endminipage \hfill
\minipage{\textwidth}
        \includegraphics[width=0.195\linewidth] {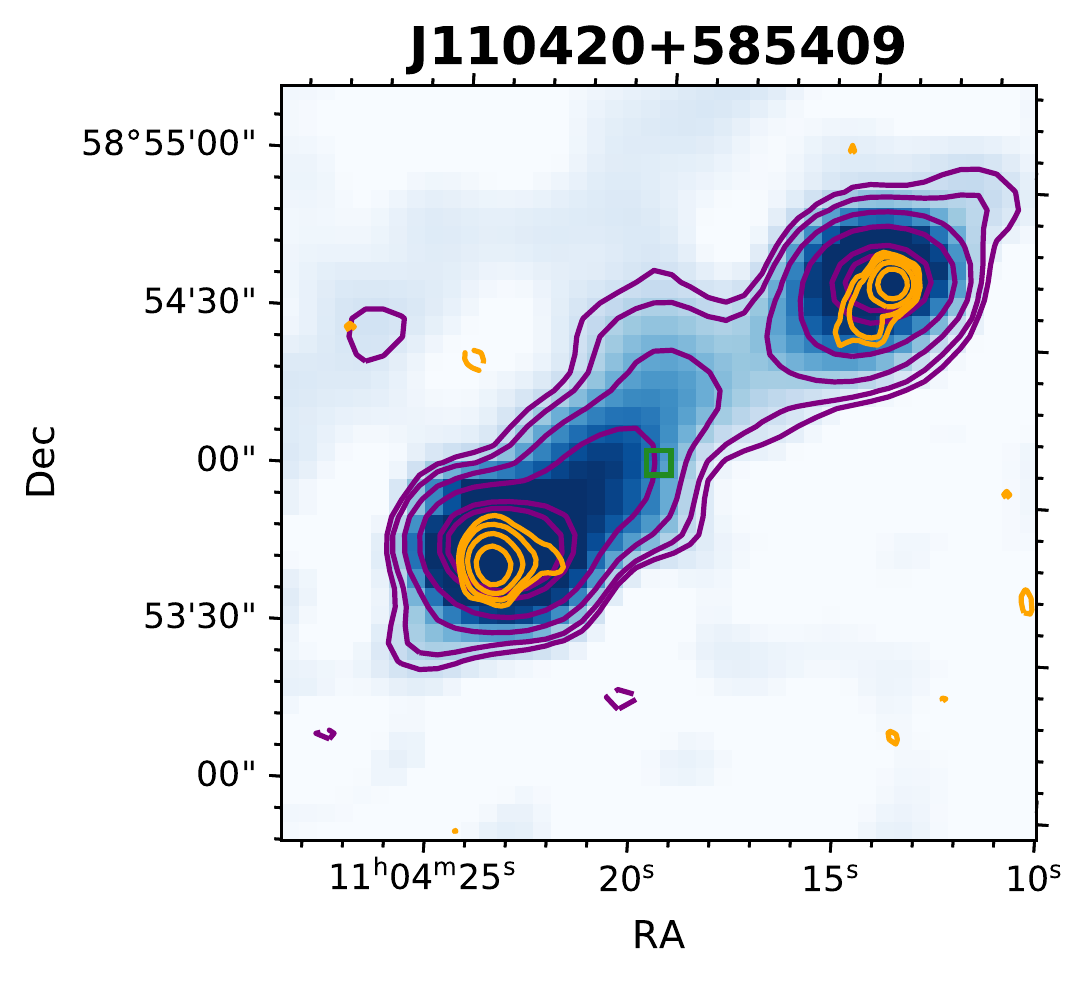}  
        \includegraphics[width=0.195\linewidth] {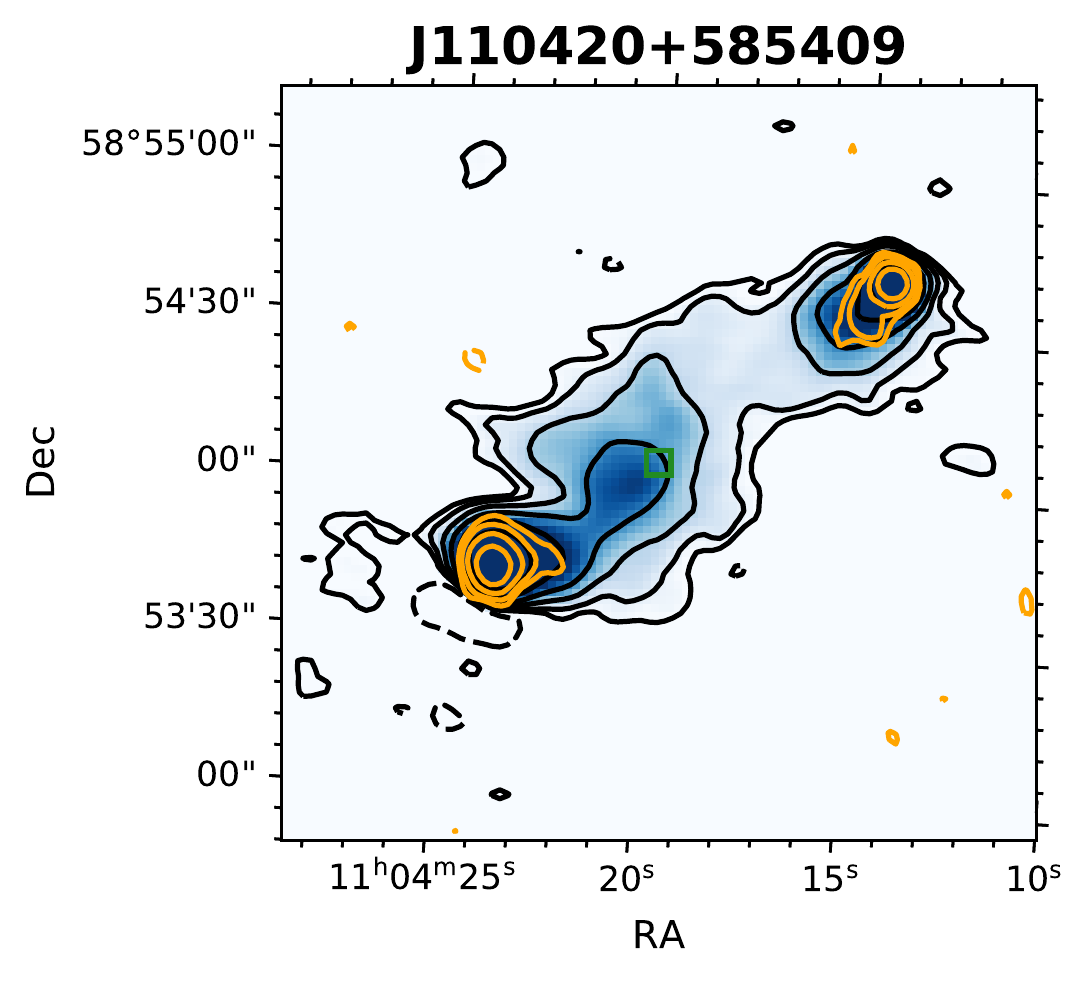} 
        \includegraphics[width=0.195\linewidth] {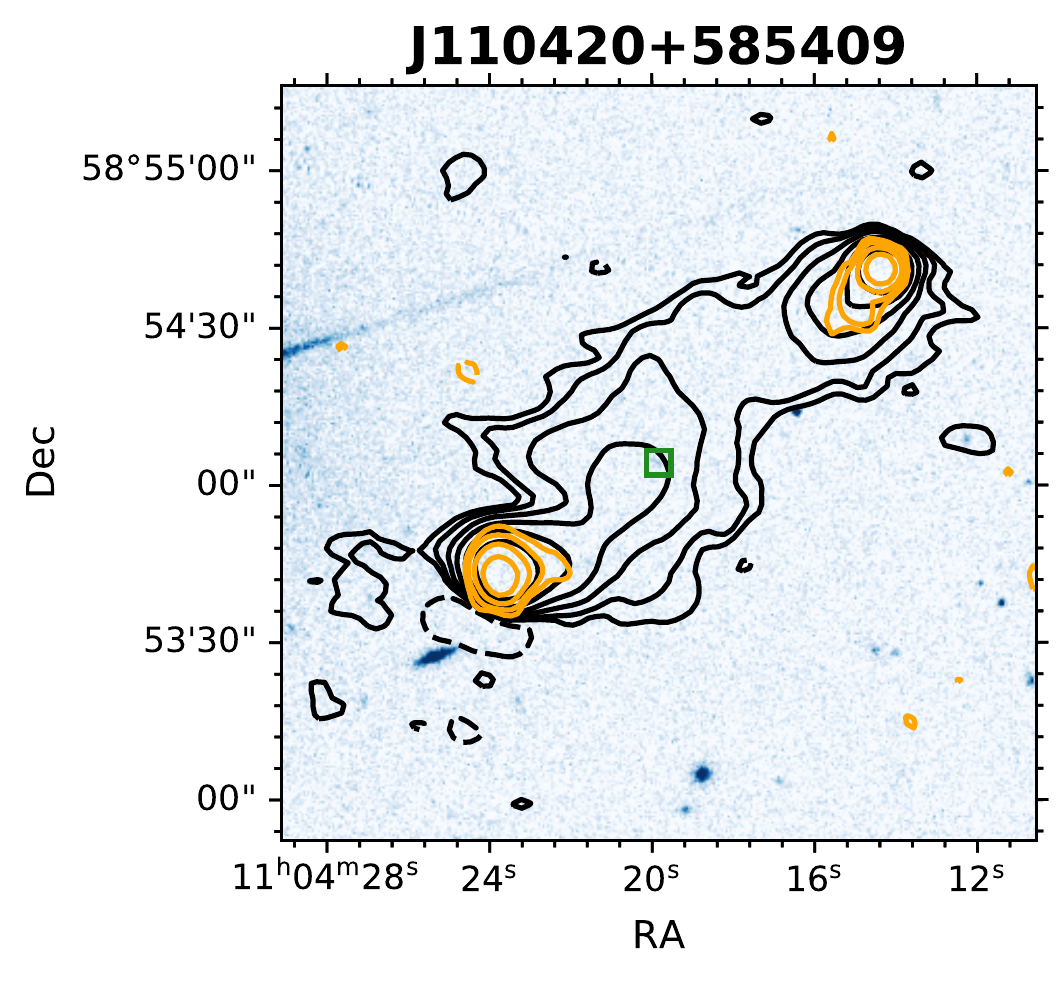}
        \includegraphics[width=0.195\linewidth]  {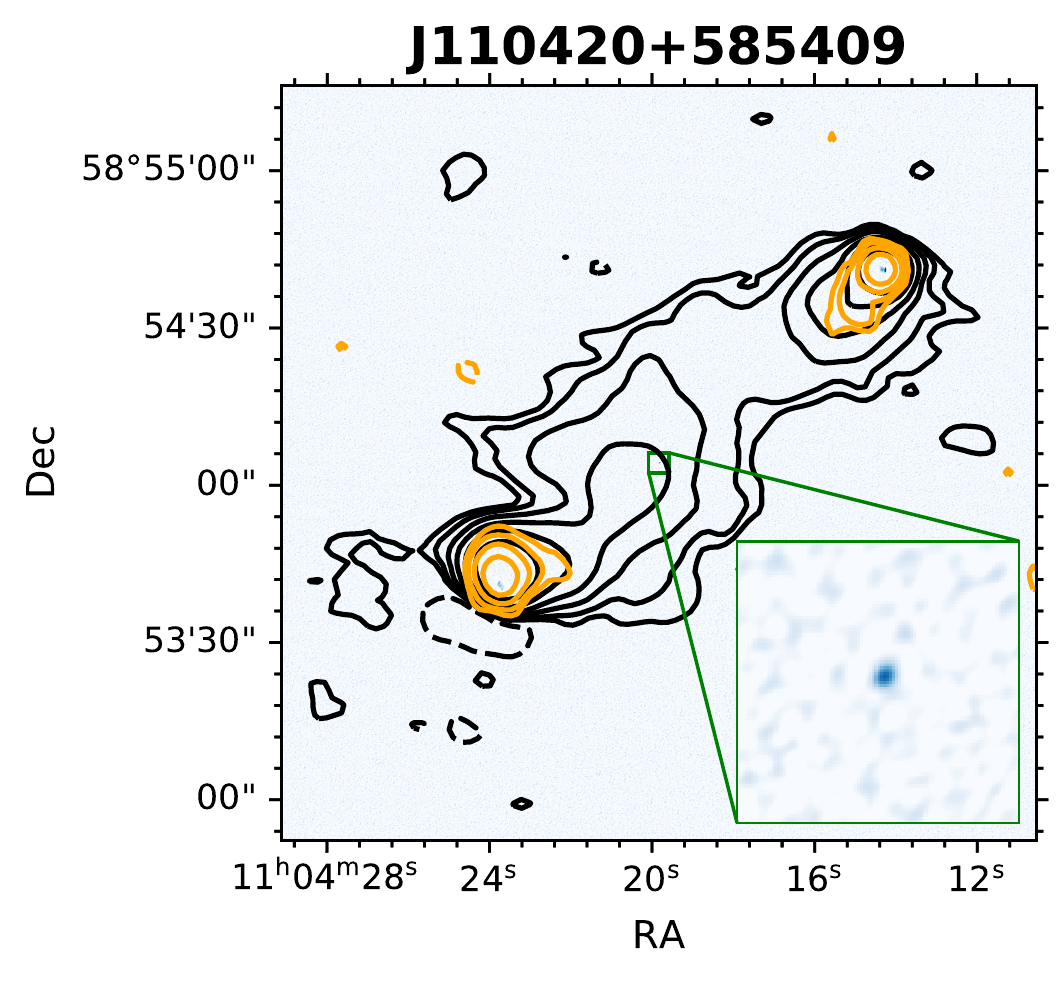}
\endminipage \hfill
\minipage{\textwidth}
        \includegraphics[width=0.195\linewidth] {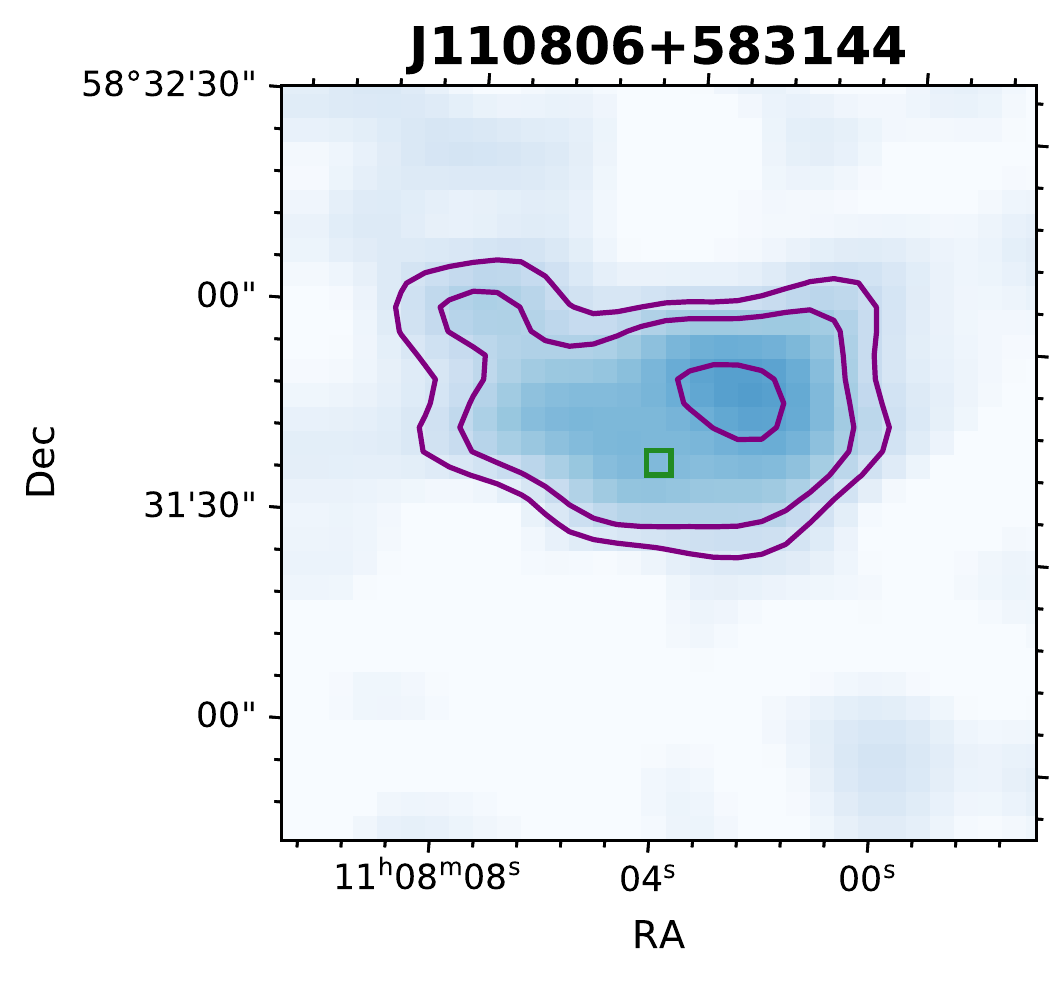}  
        \includegraphics[width=0.195\linewidth] {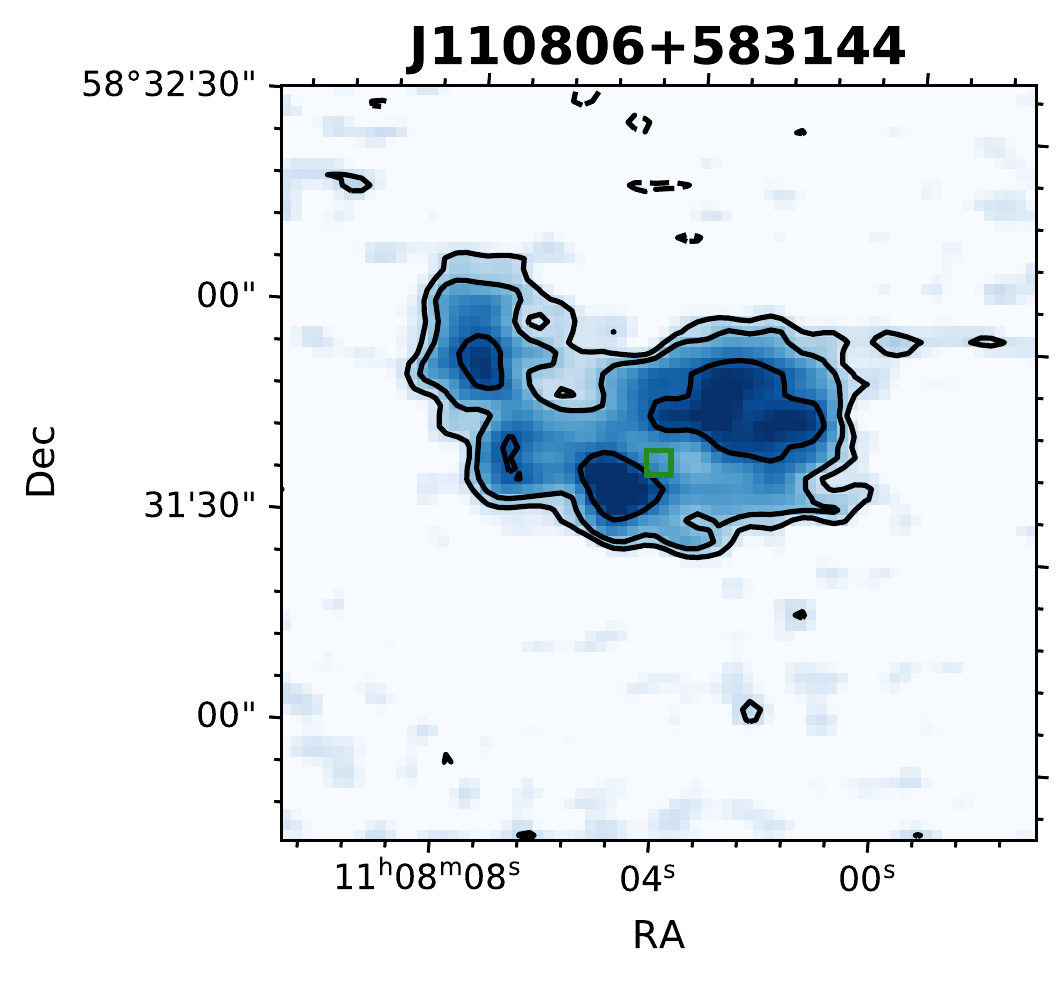} 
        \includegraphics[width=0.195\linewidth] {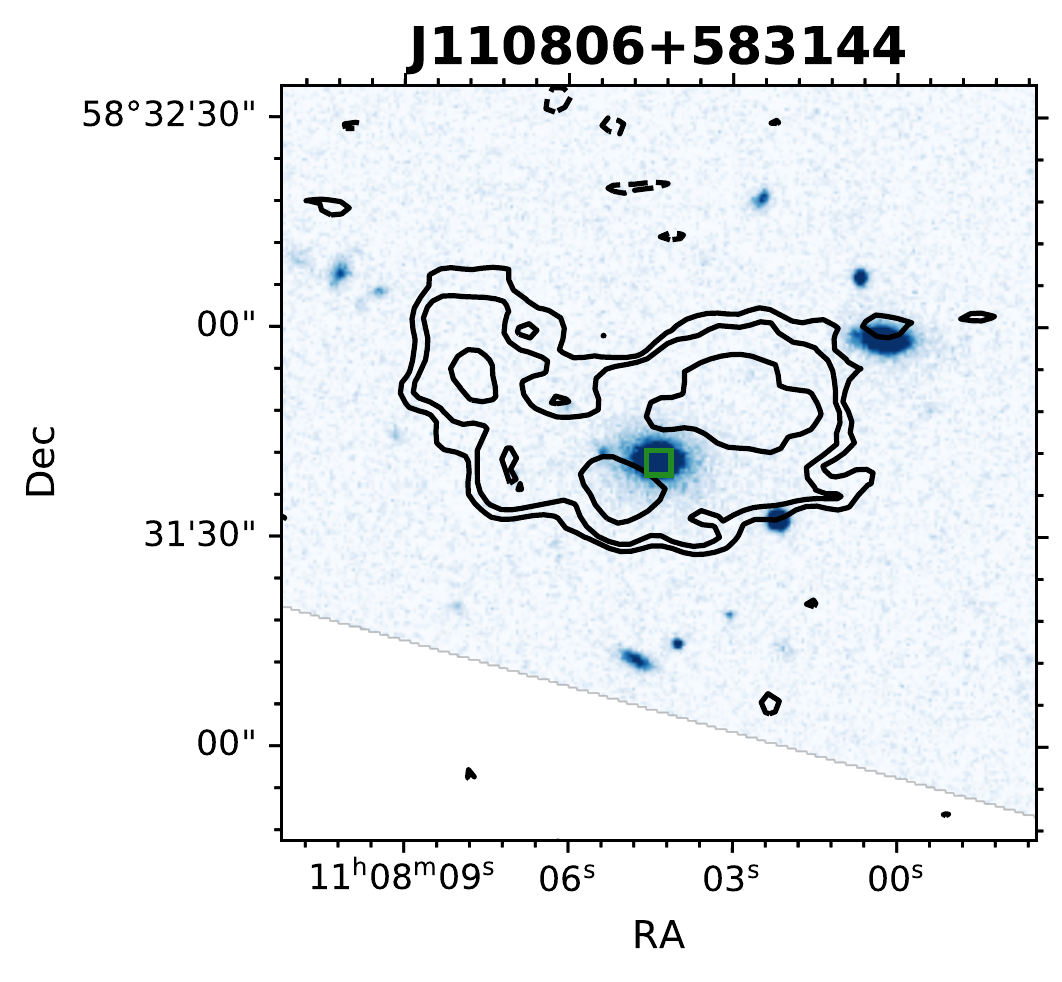}
        \includegraphics[width=0.195\linewidth] {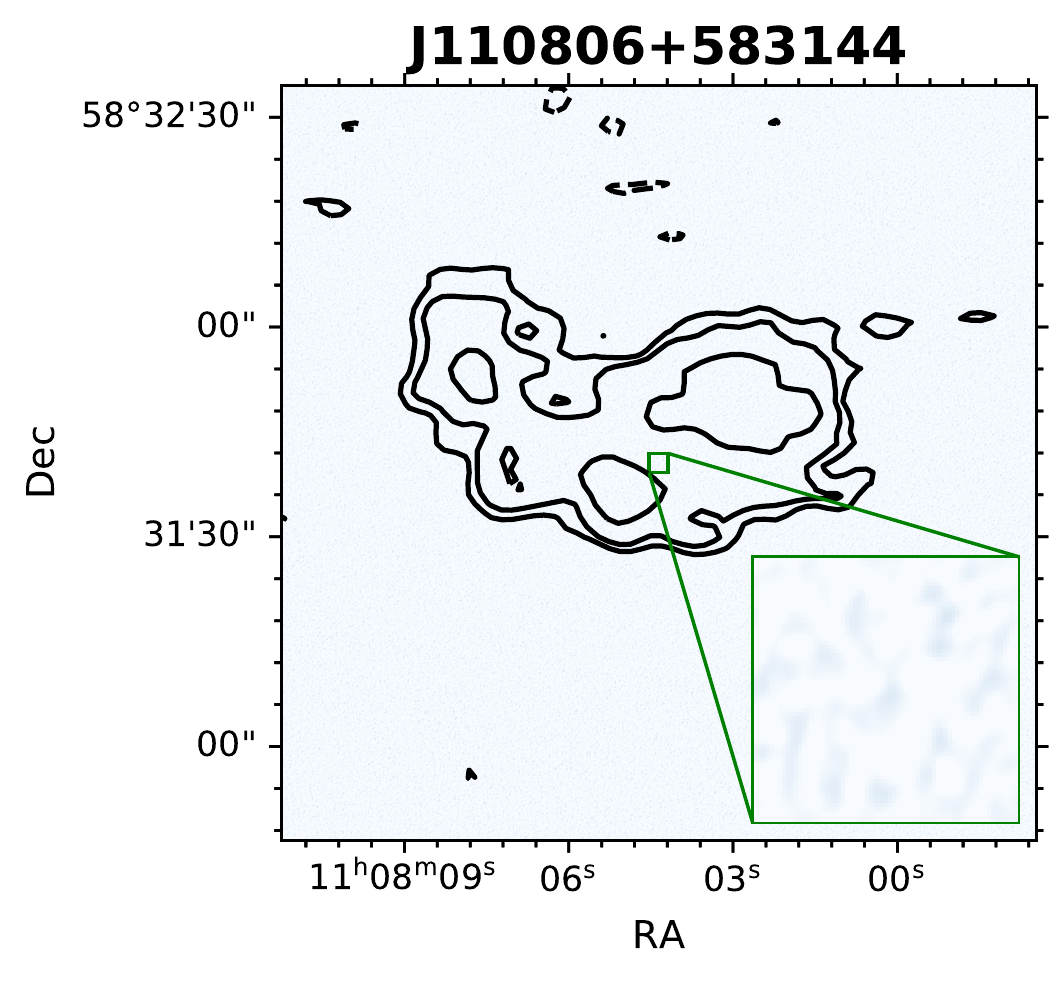}
 	    \includegraphics[width=0.195\linewidth] {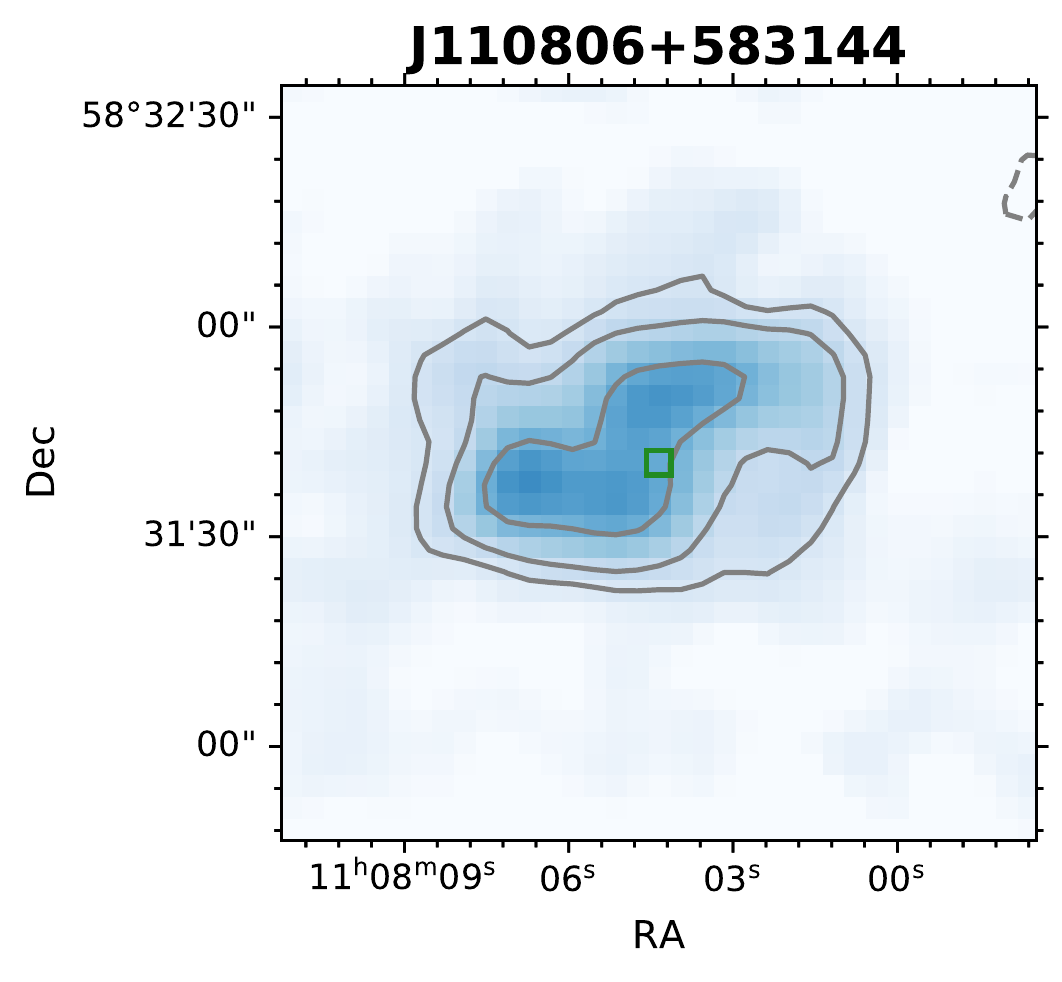}
\endminipage \hfill
\end{figure*}

\section{Radio spectra of the cores and the total emission}
Here we present radio spectra of the cores and the total emission of remnant candidates constructed using measurements from 150 MHz to 6000 MHz (see Sects.~\ref{results_radio/radio_cores} and \ref{results_radio/extended_emission}).
\begin{figure*}
    \includegraphics[width=18cm]{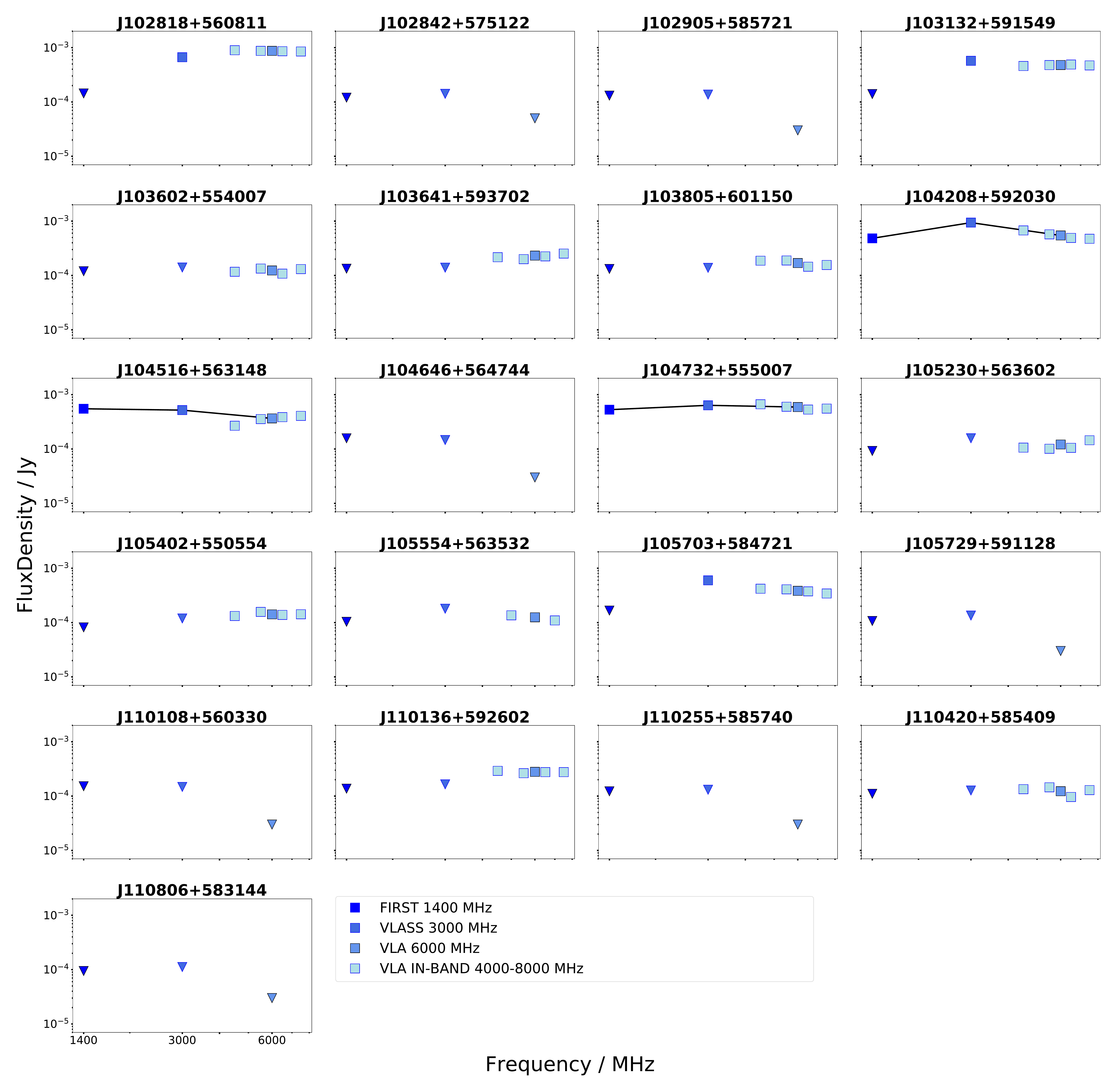}
    \caption{Radio spectra of the cores of the 21 remnant candidates. Flux density measurements from FIRST, VLASS and VLA images centred at 1400, 3000 and 6000 MHz, respectively, are shown with dark blue symbols, while triangles indicate upper limits. In light blue, we show VLA in-band measurements, centred at 4500, 5500, 6500 and 7500 MHz. Black lines connect the flux densities used for the calculation of the spectral index for the sources with detections at all frequencies. Flux density scale errors are smaller than the size of the symbol representing the flux density measurement.}
    \label{fig:spectral_index_core}
\end{figure*}

\begin{figure*} [!h]
        \includegraphics[width=18cm]{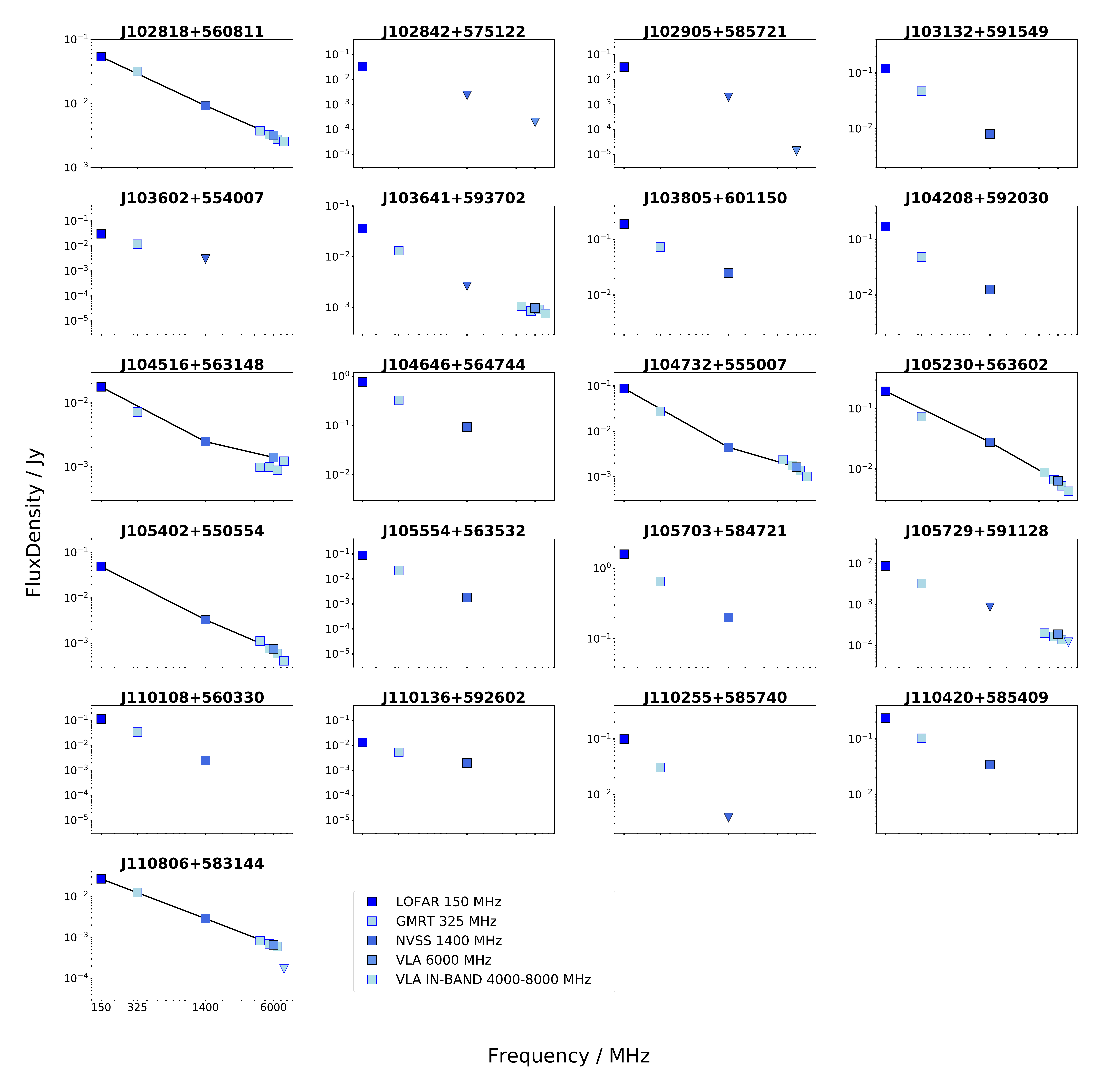} %1
    \caption{Radio spectra of the total emission of the remnant candidates. Flux densities from LOFAR18 at 150, NVSS at 1400 and VLA at 6000 MHz are shown with darker blue symbols, while GMRT at 325 MHz and VLA in-band spectral measurements centred at 4500, 5500, 6500 and 7500 MHz are shown in lighter blue symbols. For the source J1057+5911, we use the WSRT upper limit at 1400 MHz. Upper limits are indicated with triangles. The black line connects the measurements used for the calculation of the spectral index for sources detected at 150, 1400 and 6000 MHz. Flux density scale errors are smaller than the size of the symbol representing the flux density measurement.}
\label{fig:SI_total_plots}
\end{figure*}

\section{Description of individual sources} \label{sec:description_of_individual_sources}
Here we provide individual descriptions of each remnant candidate divided into two sections, according to their final classification.
\subsection{Confirmed remnant radio sources}

Source \textbf{J102842+575122} has a double radio morphology, confirmed also in the LOFAR6 image. It was not detected in both NVSS 1400 MHz and VLA 6000 MHz images and therefore we cannot comment on the shape of its integrated radio spectrum between these two frequencies. However, based on the detection at 150 MHz and the upper limit at 6000 MHz, we conclude that this source has USS in this range. This source does not have a 3$\sigma_{local}$ detection of the core (upper limit being 0.03 mJy; equivalent to the upper limit in the core radio luminosity of 1.08 $\times$ $10^{22}$) and SB of 85 mJy arcmin$\rm ^{-2}$. It was selected as a remnant radio source by MB17 based on its morphology. In this work, we confirm its remnant nature based on no core detection and the USS.\\ 

Source \textbf{J102905+585721} has an amorphous morphology resembling that of blob1 \citep{2016A&A...585A..29Brienza}, indicating that a possible progenitor is an active FRI radio galaxy. The morphology is confirmed in the LOFAR6 image and we can see the structure of the source better. We can see that the spectra at lower frequencies is steeper than 1.2, while we cannot reach a conclusion about the steepness at higher frequencies, given that the source is not detected at both 1400 (NVSS) and 6000 MHz. Based on the detection at 150 MHz and the upper limit at 6000 MHz, we conclude that this source has USS in this range. This source has no detection of the core (<0.03 mJy) and it was selected as a remnant radio source by MB17 based on its morphology. Here, we confirm this source as a remnant based on the lack of core detection, and the USS.\\

Source \textbf{J103132+591549} exhibits a double morphology in both LOFAR18 and LOFAR6 images. In the LOFAR6 image, we can see the orientation of the lobes and position of the core more accurately than in the LOFAR18 image. The VLA observations centred at 6000 MHz detected the core (0.48 mJy) and confirmed that the core is offset from the centre of the emission, as predicted based on the LOFAR6 image. This source was selected by MB17 based on its low CP. Here, we confirm its remnant nature based on the USS.\\

Source \textbf{J104516+563148} was selected as a remnant by MB17 based on its relaxed and amorphous morphology, which is fainter in the LOFAR6 image. In the low-resolution VLA image centred at 6000 MHz, we only detect the central part and edges of the lobes. This source has a detection of the core at 6000 MHz (0.37 mJy). With the availability of the new data, we discovered that this source has high CP, due to its extended emission being extremely faint. However, for the radio source of such low radio luminosity (7.51 $\times$ 10$\rm ^{22}$ W / Hz), the CP is lower than expected for an active radio source. Radio spectrum of this source is not curved and the spectral index is not US. In this study, we confirm source J1045+5631 as a remnant based on its low CP compared to its total radio luminosity. 

We note that in our previous work \citep{2020A&A...638A..34Jurlin} the same source was selected as a restarted radio galaxy, based on its high core prominence as a sign of the newly active radio core. In this study we conclude that the CP is although high, considered to be lower than expected if J1045+5631 was an evolved active radio source. Therefore, the classification of this source remains uncertain and follow-up observations (already on-going) are needed to clarify the nature of this source.\\

A double morphology of a remnant radio galaxy \textbf{J104646+564744} is also confirmed in the LOFAR6 image and this radio galaxy has no detection of the core up to 6000 MHz. It was selected as a remnant by MB17 based on its CP, which is still extremely low. We lack the total flux density information at 6000 MHz and therefore cannot comment on the curvature and steepness of the radio spectra. We confirm this source as a remnant based on the lack of the core detection and low CP.\\

Source \textbf{J104732+555007} shows similar morphology in both LOFAR18 and LOFAR6 images, though a deeper one gives us a better indication on the position of the core. The core position was confirmed at 6000 MHz (0.59 mJy). The total emission detected at 6000 MHz, shows the central region and the edges of the lobes. It was selected by MB17 based on the CP and the morphology. However, CP did not stay as low. Its location in Fig.~\ref{fig:deRuiter_plot} is as well above \citet{1990A&A...227..351DeRuiter} correlation. However, this source shows USS at lower frequencies, and based on it, we confirm it as a remnant radio source.\\

Source \textbf{J105230+563602} has an interesting X-shaped morphology seen already in the LOFAR18 image. It resembles the low SB morphology of the restarted radio source 3C~315 at 150 MHz (see \citet{2020NewAR..8801539Hardcastle_Radio_gal_feedb}). However, source 3C~315 has a much brighter core detection. The availability of the higher resolution, LOFAR6 image pointed us in the direction of the core and the OC in the otherwise dense optical field. The position of the core was confirmed at 6000 MHz (0.12 mJy). This source spectrum is not ultra-steep and has no significant curvature (SPC = 0.14). It was selected by MB17 based on the morphology and the low CP. CP is still low, as is the SB of the extended emission (SB = 70 mJy $\rm arcmin^{-2}$).\\

Source \textbf{J105402+550554} has an amorphous shape where the core position was not clear in the LOFAR18 image. With the availability of the deeper and higher resolution LOFAR6 image, the understanding of the source improved. LOFAR6 image allowed us to locate the core as well as to identify the double morphology of the source. Observations at 6000 MHz confirmed the position of the core (0.14 mJy), but the majority of the extended emission has not been recovered. This source is also detected in the NVSS image at 1400 MHz and it has USS at low frequency ($\alpha_{150}^{1400}$). \\

Source \textbf{J105554+563532} has double radio morphology in both LOFAR18 and LOFAR6 images at 150 MHz, and detection of the core at 6000 MHz (0.12 mJy). It was selected by MB17 and confirmed in this work as a remnant, based on the USS between 150 and 1400 MHz. We lack the information on flux density at higher frequencies.\\

Source \textbf{J105729+591128} was selected by MB17 based on its morphology in the LOFAR18 image. In the LOFAR6 image, we can see some structure of the source, but the majority of the emission seems to be resolved out. This source was not detected in the NVSS image at 1400 MHz but has a detection at 6000 MHz. Therefore we cannot say much about its integrated spectral properties. There is no detection of the core at 6000 MHz (<0.03 mJy). The SB of this object is the lowest one in the sample (SB = 20 mJy $\rm arcmin^{-2}$).\\ 

Source \textbf{J110108+560330} has double morphology, confirmed in the LOFAR6 image, where we can see more structure of the source. It was selected by MB17 and confirmed as a remnant in this work based on the USS of the extended emission up to 1400 MHz. Additionally, this source has no detection of the core (<0.03 mJy) at 6000 MHz, and low CP.\\

Source \textbf{J110255+585740} has double morphology with the inner parts of the lobes pointing in the opposite directions. This is even more prominent in the LOFAR6 image. It resembles the source NGC~326, where the `wings' are discussed to be either due to the backflow coming from the hot spots, or as remnant lobes from the previous epoch of activity \citep{2012ApJ...746..167HodgesKluckReynolds}. This source was confirmed as a remnant radio source due to no detection of the core at 6000 MHz (<0.03 mJy) and based on the USS between 150 and 1400 MHz.

We note that the flux density of this source is contaminated at 1400 MHz. However, we cannot quantify the contamination and therefore advise the reader to take the values derived from 1400 MHz data with caution. \\

Source \textbf{J110806+583144} was selected by MB17 based on its morphology in the 18${^{\prime\prime}}$ image at 150 MHz. The higher resolution 150 MHz image revealed two regions indicating the possible position of the radio core. However, there is no detection of the core in the VLA image centred at 6000 MHz. We also observed total emission of the source at 6000 MHz, but found no significant curvature or USS. The CP of this source is lower than expected for a radio source of that radio luminosity.\\

\subsection{Radio sources rejected as remnant candidates}
Double lobed morphology of a rejected remnant candidate \textbf{J102818+560811} seen in the LOFAR18 image is also confirmed in the LOFAR6 image, though we can see more structure. We also managed to compute the integrated spectrum and we see that the spectrum is not curved and not ultra-steep. This is an indication that this is not old plasma and therefore this source is not an USS remnant. The integrated spectrum is consistent with a power law. This source has a detection of the radio core in the VLA image at 6000 MHz (0.87 mJy). Its SB$\rm _{150~MHz}$ is however not lower than 50 mJy arcmin$\rm ^{-2}$, although it was selected based on its morphology by MB17. The location of this source in Fig.~\ref{fig:deRuiter_plot} is above \citet{1990A&A...227..351DeRuiter} correlation.\\

The double morphology of a rejected remnant candidate \textbf{J102917+584208} was also confirmed in the higher resolution image, though giving the impression that possibly these were in fact two unrelated radio sources. There are elongated detections in each `lobe' at 6000 MHz and the southern lobe has an SDSS OC. There is no VLA core detection at 6000 MHz in the middle of the two `lobes'.\\

Thanks to the availability of a deeper and higher resolution LOFAR6 image at 150 MHz, rejected remnant candidate \textbf{J103414+600333} is found to be an active WAT radio galaxy.\\

Source \textbf{J103602+554007} was selected by MB17 based on its morphology in the LOFAR18 image. The LOFAR6 image gave us a better indication on the position of the core being towards the southern and not the northern part as could be suggested by the LOFAR18 image. Also, it revealed the double-lobe morphology seen in projection and therefore, one lobe is more elongated than the other. The 6000 MHz image confirmed this with a detection of the core (0.12 mJy) closer to the higher flux density regions in the south-west. Its location in Fig.~\ref{fig:deRuiter_plot} is above \citet{1990A&A...227..351DeRuiter} correlation. The SB of the extended structure is $\sim$ 100 mJy arcmin$^{-2}$ but it is not characterised by the USS at low frequencies when using measurements presented in this work. We do not have the high frequency information about this source.\\

Source \textbf{J103641+593702} has a double amorphous morphology with the east part of the extended emission bent. The structure in the LOFAR6 image is more prominent and indicates the position of the core, which was confirmed with the VLA image at 6000 MHz (0.23 mJy). The total emission of this source is detected in the VLA image at 6000 MHz, though only the central part, indicating that the lobes might be steep spectrum. However, we cannot confirm that. It was selected by MB17 based on the morphology. Its location in Fig.~\ref{fig:deRuiter_plot} is above \citet{1990A&A...227..351DeRuiter} correlation.\\

Source \textbf{J103805+601150} has double morphology confirmed in the LOFAR6 image with more details in the structure. There are high flux density knots in the lobes and detection of the core at 6000 MHz (0.17 mJy), indicating that this is an active FRII radio source. This candidate remnant was selected by MB17 based on the CP. CP is still low, as is the radio luminosity of the core.\\

Source \textbf{J104208+592030} has double morphology confirmed in the LOFAR6 image. LOFAR6 image also gave us a clear identification of the position of the core and its host galaxy, that was confirmed with a 3$\sigma_{local}$ detection of the core in both FIRST and the VLA images at 1400 and 6000 MHz, respectively. It was selected by MB17 based on the CP, which is relatively low using FIRST and LOFAR18. However, this source has a rather high CP, calculated using measurements at 1400 MHz (NVSS) and at 6000 MHz, for a source of its radio power considering relation presented by \citet{1990A&A...227..351DeRuiter}.

We note that the flux density of this source is contaminated by a background source (indicated with orange contours in the northern lobe of the source in Fig.~\ref{fig:remnants1}) at all available frequencies and therefore the derived values should be considered with care.\\

Source \textbf{J105703+584721} has double morphology in the LOFAR18 image. The LOFAR6 image revealed some extended emission in the inner region propagating in the direction perpendicular to the lobes. VLA observations at 6000 MHz revealed a detection of the core in the middle of the two lobes (0.38 mJy). This source was selected as a remnant candidate based on its CP, which is still low. However, this source shows regions with higher flux densities in the lobes, indicating hotspots. Therefore, this source is likely an active FRII radio source.\\

Source \textbf{J110136+592602} was selected based on its morphology. However, the deeper 150 MHz image revealed that what was considered to be lobes are two separate sources. As the upper `lobe' resembled a remnant structure and has angular size > 60$^{\prime\prime}$, we followed it up with observations at 6000 MHz and detected a core (0.28 mJy). Its location in Fig.~\ref{fig:deRuiter_plot} is above \citet{1990A&A...227..351DeRuiter} correlation.\\

Source \textbf{J110420+585409} has double morphology with the edges more prominent. This is also the case in the higher resolution LOFAR6 image. It has a detection of the core (0.12 mJy) at 6000 MHz and it was selected by MB17 based on the CP criterion.
CP is still low. Due to the visible hotspots and properties described above, we consider this source an active FRII radio galaxy.\\

\end{document}